**A novel, computationally tractable algorithm flags in big matrices every column associated in any way with others or a dependent variable, with much higher power when columns are linked like mutations in chromosomes.**

Marcos A. Antezana, Independent Researcher; marcos.antezana@gmail.com
Carlos A. Machado, Dept. of Biology, University of Maryland; machado@umd.edu

**ABSTRACT.** When a data matrix DM has many independent variables IVs, it is not computationally tractable to assess the association of every distinct IV subset with the dependent variable DV of the DM, because the number of subsets explodes combinatorially as IVs increase. But model selection and correcting for multiple tests is complex even with few IVs.

DMs in genomics will soon summarize millions of mutation markers and genomes. Searching exhaustively in such DMs for markers that alone or synergistically with others are associated with a trait is therefore computationally tractable only for 1- and 2-marker effects. Also population geneticists study mainly 2-marker combinations.

I present a computationally tractable, fully parallelizable Participation in Association Score (PAS) that in a DM with markers detects one by one every column that is strongly associated in any way with others. PAS does not examine column subsets and its computational cost grows linearly with the number of columns, remaining reasonable even when DMs have millions of columns.

PAS exploits how associations of markers in the rows of a DM cause associations of matches in the rows' pairwise comparisons. For every such comparison with a match at a tested column, PAS computes the matches at other columns by modifying the comparison's total matches (scored once per DM), yielding a distribution of conditional matches that reacts diagnostically to the associations of the tested column. Equally computationally tractable is dvPAS that flags DV-associated IVs by perturbing the matches at the DV.

P values for the scores are readily obtained by permutation and accurately Sidak-corrected for multiple tests, bypassing model selection. A column's Ho and false-positive PAS P values for different orders of association are i.i.d. and readily turned into a single P value.

Simulations show that i) PAS and dvPAS generate uniform-(0,1)-distributed type I error in null DMs and ii) detect randomly encountered binary and trinary models of significant n-column association and n-IV association with a binary DV, respectively, with power in the order of magnitude of exhaustive evaluation's and false positives that are uniform-(0,1)-distributed or straightforwardly tuned to be so. Power to detect 2-way DV-associated 100-marker+ runs disjunct and not is non-parametrically ultimate but that to detect pure n-column associations and pure n-IV DV associations sinks exponentially as n increases.

Important for geneticists, dvPAS power increases about twofold in trinary vs. binary DMs and by orders of magnitude with markers linked like mutations in chromosomes, specially in trinary DMs where furthermore dvPAS fine-maps with highest resolution.

## Introduction

Genomicists are about to determine in millions of people the presence/absence of millions of chromosomal changes (mutations) and the expression level of tens of thousands of genes in thousands of cell types and developmental stages. There are already enormous **data matrices (DMs)** of this kind. Fields like marketing also generate large DMs.

Because of "parsimony" preconceptions and the computational hurdle below, DMs of the "across cell types/stages" and "affecteds vs. controls" type are scanned first for across-group differences in the expression of individual genes and the frequencies of individual mutations, respectively. Significant across-group differences in such **independent variables (IVs)** are used to account for the probability of the various statuses (categories) in the **dependent-variable (DV)** of the DM at hand. This is called scanning for IVs with **marginal effects at the DV**, e.g., for individual mutations or differences in the expression of individual genes that are "**associated**" with disease status.

Sometimes, however, such DMs are also exhaustively scanned for **2-way higher-order effects** at the DV (also called synergistic, epistatic, non-additive, and interaction effects), e.g., for associations of the DV with combinations of expression levels at any possible pair of genes or combinations of the mostly **binary** mutations at any possible pair of genome locations (sites) with mutations. This is also done for diplotypes at pairs of sites, which is more relevant in diploid organisms (the binary mutations 0 and 1 at a site can form the **trinary** diplotypes 00, 01, and 11 or 0, 1, and 2 in a diploid genome). But when a DM has many columns even 2-column scans are at times restricted to columns of *a priori* interest, e.g., mutations in genomic regions that harbor somewhere a mutation with a marginal effect at the DV.

However, exhaustive scans for 2-way associations cannot detect **pure *n*-way associations** like the 3- and 4-column ones in Figure 1, since by definition such associations lack any lower-order signal. Detecting therefore genuine higher-order associations above 2-way requires scans of every 3-column subset, every 4-column subset, etc, which lets the scans become explodingly computationally intractable as DM columns increase. Not last, exhaustive scans of *n*-column subsets cannot fully react to *n*-way associations that extend over $c$ columns when $c>n$, and therefore these scans have lower power to detect such associations ($n \geq 2$; Figure 2).

Ideally, therefore, one would examine the effect of every combination of markers at every distinct subset of columns in a DM, but as intimated this is not computationally tractable when a DM has more than a few dozen columns. Indeed in a DM with $L$ columns the number of *n*-column subsets is equal to the sum of "L choose *n*" over every $n \leq L$, i.e., $\sum_{n \leq L}({}_L C_n)$; the number is the same when a DM has L IVs and a DV and one wants to test the association of every *n*-IV subset with the DV ($n \leq L$; see Fig.1). This number explodes combinatorially with L ($1 \leq n \leq L$), e.g., for L= 10, 100, and 1'000 the number is $\sim 10^3$, $\sim 10^{29}$, and $\sim 10^{300}$, respectively. This combinatorial explosion is deemed insurmountable to the point that most publications mention the existence of higher-order associations only perfunctorily, if at all, and rarely discuss the fact that scanning exhaustively for such associations is not computationally tractable when L is above a few dozen columns.

The aforementioned to-test-first status of marginal effects is not only due to the intimated parsimony preconceptions and the computational tractability of the detection of these effects, but also to the popularity of multivariate statistics and the variety of tools that they offer for describing how IVs contribute "additively" to "explaining" a DV, both of which lets many forget that IVs associated with a DV need not be so only additively. Indeed marginal effects enjoy no any special status in nature. Nothing in molecular genetics, e.g., suggests that specific *n*-way interactions of mutations or gene-expression levels deserve more attention than others.

Furthermore, the results presented here with random models of *n*-IV DV association whose every IV happens to be standardly significantly DV-associated in unconstrained ways, show that on average these IVs' marginal effects contribute to their significance about as much as their higher-order associations do, i.e., that marginal effects are not special combinatorio-frequentistically either.

The above computational hurdle motivated the author **(MAA)** to develop the computationally tractable "Participation in Association Scores" (**PAS**s) presented and studied in this paper. The scores react indirectly to the various associations of a **focal column** with other columns in a DM and require no dedicated evaluation of column subsets. They quantify the non-random co-occurrence of matches in the pairwise comparisons of the rows of a DM that are caused by associations of markers in said rows. Some PASs react at once to all of the associations of a focal-column's markers with markers at other columns. Others react only when focal-column markers are involved in a specific type of pure *n*-way association with others at other columns. The PASs that are optimized for finding DV-associated IVs are called dvPASs and have extraordinary power specially when detecting binary and trinary extended *n*-IV DV associations (Fig.2) as well as trinary *n*-IV DV associations extended and not.

**Richard R.Hudson (RRH)** improved the crude PAS that MAA first developed for an individual column in 2-way association. He also proposed the rationale for the non-parametric contingency-table test in Figure 3 of the cumulated associations of a focal column with the other columns in a many-row few-column DM, after MAA challenged him to propose a test akin to PAS for direct use in the DM rather than in the **pairwise matrix PM** where PASs were calculated from the very beginning.

The above contingency-table test was further developed by MAA for use in finding ("encountering") random **model DMs** whose every column (IV) happens to be statistically significantly associated with a DV by itself and/or in synergy with other IVs. These model DMs are central to most of the assays of PAS and dvPAS power presented below.

The above contingency-table test, furthermore, demonstrates that a single-column test can both replace exhaustive evaluation of column subsets and make model selection unnecessary. Indeed PAS not only overcomes the combinatorial explosion that makes exhaustive-evaluation scans wishful but also recasts searching for associations of columns in a DM as a series of individual-column tests of whether the tested column is significantly associated with others or the DV, again individually and/or in synergy with other columns or IVs.

Only years after discovering PAS, MAA realized that PAS' focal-column approach is analogous to traditional tests in multivariate statistics of whether the DV of a DM is associated with individual IVs and IV subsets. A

PAS scan is indeed akin to carrying out successively at every DM column a standard multivariate test that posits the column as a temporary DV with which the other columns may be associated individually or as synergistic column subsets (Fig.1). Similarly, a dvPAS scan is akin to using standard multivariate techniques to flag every IV that individually and/or in synergy with other IVs is associated with the DV of a DM.

The work presented below shows that the change in testing perspective and output introduced by PAS and dvPAS are powerful and illuminating. As intimated, PAS and dvPAS fully bypass model selection by generating directly a list of every column that is strongly associated in overall and specific ways with others and a list of every IV that individually or in synergy with others is strongly associated with the DV of a DM, respectively, again in overall and specific ways. This is desirable computationally, datamining-wise, and statistically and should facilitate model-selection exercises.

Below the PAS algorithms are introduced and explained and their properties are showcased using simulations that focus on i) the type I error of PAS and dvPAS in null binary and trinary DMs generated under a straightforward $H_o$ and ii) the power and false positives of PAS and dvPAS when detecting columns that are significantly associated according to a variety of models of column association and $n$-IV DV association, respectively, in DMs with increasing numbers of random columns and disparate marker frequencies across model and non-model columns.

The associated columns whose detection by PAS and dvPAS is studied include those in the mentioned randomly encountered binary and trinary model DMs whose columns are all similarly significantly associated with each other *sensu* Fig.3 and whose IVs are all similarly significantly DV-associated, respectively, with $n$ ranging from five to ten columns and four to eight IVs, respectively. Also studied is the detection of binary pure $n$-column associations and pure $n$-IV DV associations up to 7-way, which sheds light on the power of PASs and dvPASs with the randomly encountered models. Much of this work is repeated with pairs of said models co-occurring independently in a DM.

For dvPAS only, the detection of model IVs and the generation of false positives is also studied in DMs in which adjacent IVs are associated into blocks that are independent from other such blocks. The block-defining "background" intra-block associations of IV markers are generated by resampling from a set of binary-marker sequences that summarize the presence/absence of 100 mutations in an actual chromosomal region. These results show a major power boost for dvPAS compared to when non-model columns are random and independent from each other. Most felicitously, power and fine-mapping are additionally boosted when detecting IVs of trinary models in DMs with trinary blocks whose structure is generated by pairing said binary chromosomal sequences.

Finally, simulations are presented about studying a single empirical DM and the detection of 2-way associations that extend over up to 99% of the columns of a DM. The latter results show that dvPAS power is non-parametrically ultimately high when many model IVs are in extended "in-phase" 2-way DV association (Fig.2), a finding that may apply to extended $n$-way associations in general. These two findings launch the Discussion.

**Motivation, history, and overview of PASs, dvPASs, and their datamining properties**.

PAS is the result of a 2004 decision by MAA to try and develop a **focal-column score** that, in a many-column **data matrix DM** of markers, would flag one by one every column whose markers are strongly associated with same-row markers at other columns. It appeared indeed unlikely then as much as now that it would ever become computationally tractable to evaluate exhaustively every one of the say $\sim 10^{300}$ distinct subsets of columns that can be drawn from a 1'000-column DM.

The specific considerations that allowed MAA to zoom on the computationally tractable signal mined by PAS and dvPAS were i) that a non-independence of markers at say two columns in a DM implies that knowing the marker at one of the columns in a row of a DM should allow one to predict to some extent the marker at the other column in that same row, and ii) that a computationally tractable single-column score of association would have to query indirectly the same-row markers at the columns associated with the evaluated focal column, rather than examine those markers directly.

The first successful PAS score was suggested by the case of complete association of binary markers of frequency 0.5 at two columns in a DM, i.e., the case of so-called "perfect" 2-way association or "perfect LD" in which the markers at two columns form, e.g., only the 1-1 and 0-0 same-row combinations shown in Figure 4. MAA realized that in this case the number of matches that can be observed when comparing two marker rows from a DM must be inflated by ~0.5 if a marker match is observed at one of the two associated columns.

Indeed whenever in this perfect-association case a pairwise comparison shows say a 0/0 match at one of the two associated columns, the same pairwise comparison must also show a 0/0 match at the second associated column, etc (Fig.4).

Expressed generally the above meant that in the **pairwise matrix PM** with the pairwise comparisons of DM rows, whenever a PM row shows a match at the first of two associated columns then the probability of observing a match at the second column is higher than in a PM row with a match at a column which in the DM is associated randomly with the other columns. In other words, the number of total matches in the pairwise comparisons with a match at a "focal" column that is 2-way associated with others should be larger than in the pairwise comparisons with matches at a focal column that is less 2-way associated with other columns in the DM.

MAA shared his reasoning with RRH together with a proof of concept that used a low-power initial score. Days later RRH proposed a much more powerful score based on the average number of matches at non-focal columns calculated over all pairwise comparisons with a match at the focal column. But then RRH realized that the average does not change when a column is involved in a "pure" 3-way association (Fig.1a). MAA reacted by showing both that his initial score detected with low power such columns too and that the variance of the non-focal-column matches given a focal-column match detected said columns with much higher power. Through additional work MAA connected pure 3- and 4-way associations to the next two higher moments (skewness and kurtosis) and zoomed so on the general rule that pure $n$-way associations increase the ($n$-1)-th moment of the number of non-focal-column matches when both the focal column is involved in a pure $n$-way association with $c$ others ($c \geq n$) and a match is observed at the focal column. This became the foundation of **Mom$^n$**, the **moment-based PASs** that below are shown to be most powerful when detecting columns that are ($n$-1)-way associated to others in a DM.

Later on, RRH realized that in the case of 2-way "repulsion" of the minor (lower-frequency) binary markers at two columns, i.e., when the minor marker at each of the two columns is paired only with the major marker of the partner column, then there is an inflation of the number of non-focal matches when two minor markers match at one of the two columns or a deflation when two major markers match, two effects that can erase each other. As a solution, MAA proposed scoring the moment of interest separately for every marker-specific match and summing the resulting values. Years later, power studies by MAA showed that when DMs have a dependent variable DV, this approach has both better null behavior and very high or highest power to detect an independent variable IV that is significantly DV-associated in a randomly encountered configuration.

However, it was clear that Mom$^n$s were empirically unsatisfying for detecting associations above 2-way because establishing differences in higher moments requires very large samples. Therefore, MAA developed a column-exclusion score with much higher power but squared computational cost. The score's computational cost prompted MAA to develop the *faux* likelihood- and $\chi^2$-based PAS scores **LKx** and **CHIx** that require a single pass on the data, react to all of the focal column's associations, and have the same ranking by magnitude as do their P values obtained by shuffling the markers at the focal column. These scores assume the (wrong) H$_o$ of independence of PM columns (not DM columns) in order to calculate the probability that matches and mismatches at the focal column pair with tracts of non-focal columns (**"pairwise fragments"**) with a given number $m$ of matches. Under said (wrong) H$_o$, this probability is hypergeometric (formulae below). The probabilities of the possible 1-row 2-part pairings allow one to calculate the expected counts of the pairings in the PM, their $\chi^2$s, and the focal column's total "PAS $\chi^2$", **CHIx**, as well as the PAS "likelihood" (probability), **LKx**, of the H$_o$ given the same-row pairings of pairwise patterns observed in the PM at hand.

Furthermore the high power of the aforementioned column-exclusion score prompted MAA to try permuting only the markers of the DV column of a DM to so flag the IV columns whose Mom$^n$'s react most strongly to the manipulation, which led to the **moment-based dvPAS, dvMom$^n$**. The P values of **dvMom$^2$** obtained through permutation of DV markers (in the DM) are shown below to be very powerful when detecting DV-associated IVs. (The likely equally powerful approach of permuting each IV of interest and scoring how the DV's Mom$^n$ of interest reacts to the permutations is much costlier computationally and was not studied.)

In turn the power of the DV-dependent P values of the dvMom$^n$'s of an IV prompted MAA to develop the "double-focal-column" versions of LKx and CHIx, **dvLKx** and **dvCHIx**, that rely on similarly crude calculations of the probabilities of pairing the pairwise states at two columns of interest (e.g., a DV and an IV) with $m$ matches at non-focal columns (formulae below). Both dvLKx and dvCHIx have very high power to detect IVs that have strong marginal effects at the DV but very low power when the marginal effects of significantly DV-associated IVs are weak.

The last two scores prompted the development of **dvMom$^n$ik**, the double-focal-column moment-based PASs that condition on there being matches at both the DV and an IV of interest. The DV-dependent P values of dvMom$^l$ik are shown below to have equal or highest power when detecting no-marginal-effects models. (Simulations showed that type I error as well as power and false positives suffer when also mismatches are considered by Mom$^n$ and dvMom$^n$ so this possibility was not further explored.)

Summarizing, for any column of interest in a DM (Figure 5a) the PAS method generates several scores that quantify indirectly in the matrix PM (Fig.5b) how strongly the markers of a focal column are associated with those at other columns in the DM. The scores quantify aspects of the distribution of the number of total matches across the pairwise comparisons in the PM.

This distribution is perturbed diagnostically by the presence in a DM of the different types of association in which the markers of the focal column can participate (2-way, pure 3-way, etc.; Fig.1) but some PASs and dvPAS are meant to react at once to all of the perturbations that both are present and involve the focal column, and they quantify these with a single value.

When the goal is detecting DV-associated IVs, the distribution of the number of non-focal-column matches of a focal IV can be additionally perturbed diagnostically and very powerfully by shuffling vertically the markers of the DV (in the DM).

The arguably ultimate all-associations PAS and dvPAS is the probability of the H$_o$ or H$_1$ that is postulated to have generated the DM at hand whose PM shows the observed counts of the intimated 2- or 3-part pairwise-comparisons that are PAS and dvPAS relevant, respectively, i.e., the probability of observing the counts of the various groups of pairwise comparisons that can be distinguished by their having specific pairwise states at the focal-column site(s) and a number $m$ of matches at other sites (see Fig.5c for the PAS case). Below it will be shown that calculating this probability will most likely be very arduous combinatorially but does not require exhaustive evaluation of every applicable distinct subset of DM columns since the calculation is not column-position-explicit, very much like the various PASs and dvPASs (and akin to the intended result when one multiplies by a multinomial coefficient in some cases?).

While considering the PASs and dvPASs presented in detail below, the reader should also keep critically in mind a few additional expected properties of all PASs and dvPASs that are confirmed by the simulations below, namely i) that because of their single-column nature these scores should in principle make model selection unnecessary because each column significantly associated with others is directly individually flagged; ii) that for the same reason, one should be able to correct straightforwardly the scores' P values for multiple tests when P values are estimated for several columns of a DM (e.g., with the Sidak method); iii) that the scores' power should be unaffected by multiple tests that are *de facto* frequentistically irrelevant, e.g., it should be higher when DM columns form independent blocks of strongly intra-block associated columns than when DM columns are independent, unlike the power of exhaustive evaluation that sinks when the number of distinct column subsets to be tested explodes combinatorially as the number of columns of any kind increases.

**Labels used for different PAS conditionings.** As intimated, labels are used below i) to distinguish the PASs that are conditional on a generic match at the focal column from the PASs that sum every separate score which is conditional on a match for a specific marker $i$ at the focal column, over all $i$'s, and ii) to distinguish the latter PASs from those that add up the separate scores only after turning each of them into a Z value (using a mean and a variance estimated by permuting a suitable DM column; below). When labelling the various Mom$^n$'s, e.g., the suffix "M" in **Mom$^n$M** indicates conditioning on a generic focal-column match; the "i" in **Mom$^n$i** indicates a sum of the scores given a match for the i-th marker over all $i$'s; and the "i$_Z$" in **Mom$^n$i$_Z$** indicates a sum of the Z values of said marker-specific scores. The "MM" in **dvMom$^n$MM** and **dvLKx-MM** indicates conditioning on two generic matches, one at the DV column and one at the second-focus IV column, whereas the "ik" in **dvMom$^n$ik** indicates conditioning on both a match for the $i$-th marker at the DV column and a match for the $k$-th marker at the IV column, over every possible $i,k$ pair (with $i,j$ and $k,l$ mismatches being ignored). For its part, the label "ijkl" in **dvLKx-ijkl** indicates evaluation over every possible combination of marker pairs at the DV and IV columns. And "ik$_Z$" indicates a sum of the "ik" scores after Z-valueing each of them. Finally, the "**dv**" prefix in dvCHIx-ijkl and dvMom$^n$ik is redundant given that the "ij" and the "i" in the two affixes already indicate that the markers at the DV are considered explicitly, but the prefix is added nonetheless to stress that IV associations with the DV are at stake and that the DV is permuted (etc; below).

# PAS algorithms in depth.

This section presents in detail the PASs and dvPASs that are most powerful and thus most relevant for empirical data mining. The all-associations scores LKx-M, LKx-ij, dvLKx-MM, and dvLKx-ijkl take center stage, which may seem unjustified when one thinks that dvLKx-ijkl and dvCHIx-ijkl are shown below to be specially powerful only when detecting IVs with strong marginal effects at the DV. However, the PM patterns whose counts appear in the formulae of these PASs (below) are relevant to every PAS and dvPAS presented here, including the most powerful ones for models without marginal effects.

Furthermore, the focus on these LKx's and their merits and demerits prepares the reader for the following section on the actual combinatorics and probability distributions that govern the PAS- and dvPAS-relevant PM patterns when markers across DM columns are independently randomly vertically ordered, a section that is meant to help other workers develop LKx versions that are correct combinatorio-probabilistically.

**Patterns in a PM that determine the various PASs and dvPAS.** Assume that a data matrix DM has R rows and $L$ columns and that each row is a sequence of $L$ markers (e.g., binary 0 and 1 markers; Fig.5a) that are said to be at sites numbered from 1 to $L$ (the L positions in each sequence). The rows need not be different from each other and are numbered from 1 to R.

As intimated, PAS exploits the fact that associations of same-row markers in a DM create associations of matches and mismatches in the rows of the pairwise matrix PM that lists all the pairwise comparisons of the marker sequences in the DM (Fig.5b). Indeed all PASs quantify in one way or another how such same-row associations of matches and mismatches perturb the distribution across pairwise comparisons of the total number of matches of each pairwise comparison.

All of the PASs and dvPASs studied here probe these perturbations by quantifying in various ways how in individual pairwise comparisons the matches and mismatches at one or two focal-column sites co-occur with "pairwise fragments" of non-focal sites that show $m$ matches, where $m$ ranges from 0 to L-1 or L-2 matches. Importantly, i) scoring the $m$ of a pairwise fragment does not require re-scoring the matches of pairwise comparisons every time a new focal column is evaluated, since the $m$'s are calculated by taking directly the total matches of the pairwise comparison at stake when there is a mismatch at the focal-column spot or decrementing the total by 1 when instead there is a match at said spot (Fig.5c); and ii) said numbers of total matches are scored for each pairwise comparison only once at the beginning and then they are used as described in (i) every time a new focal column is interrogated.

More formally, to quantify such perturbations for a focal column S (to-be-<u>S</u>huffled) of interest, one groups the W pairwise comparisons in the PM according to i) the pairwise state $S_i$ found at their focal-column site (e.g., by using $S_1$ and $S_2$ to signify a generic match and mismatch, resp.) and ii) the type of non-focal-column pairwise fragment $M_m$ that is paired to said $S_i$, where the subscript "$m$" specifies the number $m$ of non-focal-column matches of the $M_m$ at hand. This gives the set $\{S_i..M_m\}$ of the counts of the $S_i..M_m$ pairings in the PM (Fig.5c). When scoring the counts in $\{S_i..M_m\}$, one also scores the set $\{M_m\}$ of the observed $M_m$'s and the set $\{S_i\}$ of the observed $S_i$'s.

Once the sets $\{S_i..M_m\}$, $\{S_i\}$, and $\{M_m\}$ are computed, one can calculate any applicable **Mom$^n$**, i.e., the average, variance, skewness, kurtosis, etc. of $m$, over the pairwise comparisons that have a given $S_i$ at their focal-column site (e.g., that show there a match for any marker). As intimated, Mom$^{n-1}$M (generic match) has highly targeted power to detect pure $n$-way association of $n$ or more columns at which binary markers have frequency 0.5 ($n>1$).

However and as intimated, when the frequencies of same-column markers are not identical (e.g., unlike 1/3 for three markers), the Mom$^n$ given a match for a given marker $i$ can vary strongly across $i$'s, reverse direction, etc., making it useful at times to score and Z-normalize separately every Mom$^n$ that is conditional on a match of type $S_i$, before summing over the $S_i$'s to get the final **Mom$^n i_Z$** of the focal column at stake. Depolarizing the Z values before adding them up was tried but it lowered the power to detect randomly encountered models of column association, so this is not done. The average and standard deviation used to calculate the Z values are estimated by permuting the focal column in the case of Mom$^n{}_Z$ and the DV in the case of dvMom$^n{}_Z$ and dvMom$^n ik_Z$. The power studies further below show that among the PASs studied below Mom$^1 i_Z$ has highest power to detect the columns of randomly encountered models of significant $n$-column association.

Figure 6i shows the perturbation of the distributions of Mom$^n$M for $n$=1 and 2 when two and three columns in pure 2- and 3-way association, respectively, are present in a DM with binary markers of frequency 0.5. The 2- and 3-way patterns of substracted probability mass in panels (b,d) are distinctly specific and suggest that the power to detect a column's *n*-way associations may increase vs. that of PAS moments if one could contrast, say with likelihood ratios, the distribution of the conditional matches under the null model of independent random columns to the distribution under models where some columns are in pure *n*-way association and the other columns are random and independent. (Note that both kinds of models ignore the order and positions of the columns along the DM since the number of matches in pairwise comparisons is insensitive to where matches occur, e.g., when *n* columns are pure *n*-way associated and L-*n* are independent.)

In panel (d) in particular it is easy to see that the pattern of substracted probability of the 3-column pure 3-way association is the result of summing the matches from comparing the markers at random columns and those from comparing the markers at the two non-focal columns in pure 3-way association with the focal column (the latter matches can only be zero or two; Fig.1).

Indeed, Figure 6ii presents for the visually oriented reader the distribution of row matches in PMs from many-row *n*-column DMs whose columns are either pure *n*-way associated with each other or perfectly non-associated in the sense that every *n*-marker sequence is equally present (i.e., in these DMs there are no added random columns). The reader is invited to try and find more powerful ways than $Mom^{n-1}$ and $dvMom^n$ to detect the shown perturbations of the patterns of row matches, specially those caused by higher-order associations. MAA's intuition is that $Mom^1$ and $dvMom^2$ will be hard to beat when detecting 2-column associations and 2-IV associations with a DV that involve many columns and IVs, respectively.

When dealing with higher-order associations, however, contrasting distributions by means of likelihood ratios may turn out to be substantially more powerful while still being computationally tractable since, as just adumbrated, using such ratios under the PAS and dvPAS frameworks does not require specifying the positions of the associated columns postulated by an $H_1$, i.e., requires no exhaustive evaluation of subsets of model columns. The likelihood section under "Theoretical Results..." further below discusses the basics of the calculation of the likelihood of the null model of independent random columns given an observed distribution of matches across pairwise comparisons; it also illustrates why such likelihoods are not explicit about column positions.

Figure 7 shows, again for binary-marker frequency 0.5 throughout, the reaction of $Mom^1$ to in-phase 2-way associations that extend for up to seven columns (the pattern in Fig.2 sans DV). Unfortunately when *n*>2 pure *n*-way associations that extend over more than *n* columns are diagrammatically unclear to MAA and therefore the detection of such associations was not studied.

The sets $\{S_i\}$, $\{M_m\}$, and $\{S_i..M_m\}$ allow one to calculate also the "all-associations" **KS**, **CHIx**, and **LKx**. These PASs react at once to all types of column association that involve a focal column. As intimated, however, they have much lower power when detecting columns involved in higher-order *n*-way associations (*n*>2) than say $Mom^{n-1}$'s do.

The **KS**s are straightforward Kolmogorov-Smirnov PASs that make no assumptions about how the $S_i$'s pair with the $M_m$'s (unlike the CHIx's and LKx's). These scores are sums over every given $S_i$ of the Kolmogorov-Smirnov departure by the c.d.f. of an observed set $\{S_i..M_m\}$ of interest from the null c.d.f. of the expected $\{S_i..M_m\}$ given independence of the $S_i$'s from the $M_m$'s. The latter null c.d.f.s are robustly estimated for each evaluated column by permuting the markers at the S column (in the DM), but an analytical derivation of these c.d.f.s is desirable.

As intimated above, **CHIx** and **LKx** are 100%-correlated all-signals PASs that assume (wrongly) both that the vertical order of $S_i$'s and $M_m$'s in the PM (i.e., not that of the markers in the DM) is random and that the vertical order of the $S_i$'s is independent from the order of the $M_m$'s. **CHIx**, e.g., is a $\chi^2$ that quantifies the departure of the observed set $\{S_i..M_m\}$ from the "expected" set that under this wrong assumption of independence can be calculated by first multiplying the frequencies (in the PM) of the $S_i$'s by those of the $M_m$'s and then multiplying each such 2-term product by W.

The above scores –and the highly related **dvKS**, **dvCHIx**, and **dvLKx** for detection of IVs associated with a DV– are presented in detail immediately below, followed by a combinatorio-probabilistic study of the non-random vertical order of matches and mismatches in the columns of PMs derived from null DMs in which markers are independently randomly vertically ordered across columns, a study that is backed by both numeric and DM-to-PM simulations. (But see also the simulation results on type I error, power, and false positives further below).

Before presenting the main statistics in detail, however, one can note that the legitimacy and data-mining effectiveness of describing pairwise comparisons and single-column pairwise states using the sets $\{S_i..M_m\}$, $\{S_i\}$, and $\{M_m\}$ depends neither on matches and mismatches being independent across PM columns nor on the

$S_i$'s being independent from the $M_m$'s. KS and dvKS, e.g., assume neither but will be shown below to generate uniform-(0,1)-distributed type I error and false positives and be powerful at times. Indeed albeit matches and mismatches in the columns of the null PM are arguably non-independent and clearly non-randomly vertically ordered, one can still take advantage of how same-row marker associations in the DM cause the observed $\{S_i..M_m\}$ set to depart from the true null $\{S_i..M_m\}$ set, if say one probes the departure non-parametrically by permuting the vertical order of the markers in appropriate columns in the DM, which queries the true null $\{S_i..M_m\}$ set.

**CHIx and LKx assume that across PM columns matches are independently randomly vertically ordered**. The just mentioned probability LKx of observing the counts of the distinct $S_i..M_m$'s in the set $\{S_i..M_m\}$ under the "permutational" $H_o$ of random pairing of $S_i$'s with $M_m$'s in the rows of the PM, is called **LKx$_{SM}$** here. This is a hypergeometric probability where the number of "favorable cases" is equal to the multinomial coefficient for the observed counts in $\{S_i..M_m\}$, e.g., 15!/2!3! in Fig.5d (1!'s are ignored) while the number of "possible cases" is equal to the multinomial coefficient for the set $\{S_i\}$, i.e., 15!/(3!3!9!) for the pairwise states 0/0, 1/1, and 0/1(1/0) that are present 3, 3, and 9 times, respectively, in Fig.5c, multiplied by the multinomial coefficient for the set $\{M_m\}$ of the counts of the observed $M_m$'s, which for the sums of 1, 2, 3, 4, and 5 non-focal-column matches that are present 3, 3, 2, 3, and 3 times, respectively, in the right margin of Fig.5c, is equal to 15!/(3!3!2!3!3!. This gives Formula (1).

$$LKx_{SM} = \frac{\frac{W!}{\prod_{i,m}^{S,M}(S_i..M_m)!}}{\frac{W!}{\prod_i^S S_i! \prod_m^M M_m!}} = \frac{\prod_i^S S_i! \prod_m^M M_m!}{W! \prod_{i,m}^{S,M} S_i..M_m!}$$

(1)

The likelihood **LKm$_{SM}$** of the "multinomial $H_o$" of independent sampling of the $S_i$'s and $M_m$'s with replacement (i.e., two independent but jointly carried out bootstrapping exercises) is simply the multiplication of (1) by the frequencies $P_i$'s of the various $S_i$'s and $M_m$'s (calculated in the PM), raised to the observed counts of the $S_i$'s and $M_m$'s. It is shown in Formula (2).

$$LKm_{SM} = \frac{\prod_s^S S_s! \prod_m^M M_m!}{W! \prod_{i,m}^{S,M} S_i..M_m!} \prod_i^S P_i^{S_i} \prod_m^M P_m^{M_m}$$

(2)

The maximum-likelihood PM-level $H_1$ model for LKx has probability 1.0 and is hence trivial but that for LKm in Formula (3) is not and it can be used to calculate likelihood ratios (e.g., for $H_1/H_o$) that are uninteresting when only column S is assumed to be independent in the DM but may become useful when marker order is independent (permuted or bootstrapped) across all columns of a DM (as assumed when calculating the "overall $\chi^2$" in Fig.3 in a similar context).

$$maxLKm_{SM} = \frac{W!}{\prod_{i,m}^{S,M} S_i..M_m!} \prod_{i,m}^{S,M} P_{i,m}^{S_i..M_m}$$

(3)

However, Figure 8 shows that CHIx-M and CHIx-ij, which at S distinguish generic matches/mismatches and fully specified 2-marker pairwise states, respectively, are not $\chi^2$-distributed neither i) under the DM-level $H_o$ of marker independence across DM columns nor ii) under the other DM-level $H_o$ of the markers at the focal column being randomly paired to the marker-sequence fragments which, when pairwise-compared, yield the pairwise fragments with the observed $m$'s at non-focal sites. The figure also shows that, obviously, when one permutes directly in the PM the pairwise states at the focal column the two CHIx's become $\chi^2$-distributed with straightforward d.f. Given that the $H_o$ in (ii) is the one of interest here and that it is at the DM level, it would be highly desirable to derive the theoretical distribution of the CHIx's under this $H_o$ so one could simply look up the P value of a column's CHIx from a table and avoid having to permute the column in the DM, etc.

Moreover and felicitously, Figure 9 shows that ranking columns directly by the magnitude of their CHIx-ij is, under the DM-level $H_o$, almost equivalent to ranking them by the smallness of their CHIx-ij P values estimated by permuting the markers at the focal column (in the DM), even when marker frequencies differ across focal columns. Therefore one can safely compare CHIx-ij values across columns. However, this is not so for CHIx-M values that are strongly affected by marker-frequency differences across focal columns (Fig.9d,f). To use CHIx-M directly one could group the columns by marker frequency and pick an equal number of columns with largest CHIx-M's from each group. However, these picks would not necessarily be similarly extreme frequentistically (e.g., in terms of P values).

Be the above as it may, the differences in power presented below between CHIx-ij and CHIx-M when detecting columns associated in violation of the DM-level $H_o$ should have the last word in deciding which score should be preferred, provided of course that the accompanying false positives at random columns are uniform-(0,1)-distributed.

Finally and as we just saw, the observed sets $\{S_i..M_m\}$, $\{S_j\}$, and $\{M_m\}$ allow one to both calculate the LKx's in (1) and (2) and compute likelihood ratios of the most likely S-M pairing scheme over those of the corresponding $H_o$ models. Therefore also LKx-ij and its ratios can be used to flag columns associated in a DM. Simulations of DMs where across columns the markers have independently random vertical order show that i) LKx-ij and the ratios of LKx-ij's have almost identical distributions across columns with different marker frequencies and that ii) their ranking by magnitude coincides with that of their P values obtained by permuting the S column (in the DM; neither is shown). And as it was the case for CHIx-M above, also LKx-M and its ratios react strongly to changes in marker frequency across focal columns and therefore LKx-M values, like CHIx-M ones above, cannot be directly compared across focal columns that have different marker frequencies either (not shown). The two trends also apply to DMs with sequences of trinary "diploid" markers (not shown).

**PASs for associations involving the focal column, a second focal column, and any other columns.**

As mentioned further above, the power of $Mom^{n-1}$ to detect columns in pure $n$-way association increases markedly if one permutes in the DM (or removes from the DM) every column one at a time and scores at every column the largest change in $Mom^n$ caused by any such individual manipulation of another column. When exactly $n$ columns are in pure $n$-way association in a DM, it is obvious that whenever one of them is manipulated, the $Mom^{n-1}$ values of its $n$-1 "companion" columns must decrease.

Figure 10 shows that by removing individual columns and recalculating each time the $Mom^3$s of the other columns, one can flag four columns that are in pure 4-way association in a 4'000-row 1'000-column DM, because every one of the four columns ends up showing a very large "largest change in focal column's $Mom^3$ caused by the exclusion of another column".

This "maximal exclusion effect" score is called **meePAS** and is referred to in several occasions below, but it is not further studied with simulations. Indeed, meePAS requires recalculating the PASs of all the non-manipulated columns every time a column is excluded and therefore meePAS' computational cost is about the square of that of a single-pass PAS analysis. This makes meePAS impractical computationally (if still tractably so) when a DM has say a million columns. However, meePAS is not further studied here beyond contrasting the results in Fig.10 to those below when using $Mom^n$ and $dvMom^n$ to detect pure 4-way associations and pure 3-IV DV associations, respectively.

As intimated and very felicitously, however, meePAS's added computational cost is negligible when in a DM there is a **dependent variable column DV** that can be excluded, allowing so one to flag IV columns whose PASs react strongly to the manipulation and declare them to be strongly associated with the DV individually and/or in synergy with other IVs. This highly computationally tractable application of meePAS leads to a family of DV-focussed PASs that are called **dvPASs** and below are studied in great detail with simulations. However, all of these dvPAS permute rather than exclude the DV.

There is indeed a major additional advantage in permuting rather than excluding the DV: The power of excluding columns individually is likely to be low, e.g., when both there are $c$ columns connected in-phase by extended pure $n$-way association and $c>>n$ (as in Fig.2 for $n=2$ and $c=6$).

With increasing *c*, the exclusion of any one of the *c* columns reduces the cumulated *n*-way PAS of the remaining *c*-1 associated columns by increasingly smaller 1/*c* amounts of a portion of said PAS (while the contributions by other columns mostly remain frozen). Fig.7 shows such 1/*c* increments when one lengthens an in-phase 2-way association by up to five additional columns.

Specifically for dvPAS, excluding columns individually is likely not to be powerful when both *c*>> and the IVs to be detected are in extended 2-way DV association (like in Fig.2), since in this case it is easy for an excluded random IV to match the 1/*c* decrements in these IV's dvMom$^2$ that are caused by excluding any of the associated IVs individually.

When instead one permutes the binary markers in the DV and scores how this changes say the dvMom$^2$ of an individual IV, it is unlikely that at the same time one of the two DV markers become by chance markedly associated with the affecteds DM rows that contain the in-phase runs of IV markers at stake (while the other DV marker becomes similarly associated with the rows lacking such runs), independently of whether the contribution by the DV to the dvMom$^2$ of the tested IV is large or small.

The power simulations further below will show indeed that dvMom$^2$ P values estimated by permuting the markers at the DV have very high power when detecting IVs whose markers participate in in-phase extended 2-way runs that are DV-associated. Moreover, the power of such dvMom$^n$ P values is shown below to be comparable to meeMom$^n$'s even when only *n* IVs are in pure *n*-way association with the DV.

**Several dvPASs can flag IVs that alone and/or in synergy with others are associated with a DV.** As mentioned above, shuffling a column S is specially powerful when the goal is to find other columns that alone or in synergy with further columns are associated with S in the DM, e.g., when S is the DV column of a DM with several IV columns (Fig.1, Fig.3). Specifically for PAS, when an evaluated column E belongs to a pure *n*-way association that includes said S column, permuting S in the DM should affect strongly the Mom$^{n-1}$ of E. As intimated, these scores are called **dvMom$^n$** and each is a simple Mom$^n$ that sets E as the focal PAS column and S (the DV) as the permuted column. By permuting S in the DM one randomizes the effect of S on the set {M$_m$} and estimates so a distribution of Mom$^n$ that i) is dependent on the vertical order of the markers at S (in the DM) and that ii) allows one to estimate a P value for any Mom$^n$ of interest, turn the latter into a Z value, both, etc. Note again that a case of pure (*n*-1)-IV association with a DV is also a case of pure *n*-way association involving a DV and *n*-1 IVs (Fig.1).

Below it will be shown that permuting the DV is very powerful when detecting any IV involved in a pure higher-order *n*-IV association with the DV, even if the IV is DV-associated in additional ways, e.g., the IV may both have a marginal effect at the DV and participate in a pure 3-IV association with the DV. As intimated, the same power should be achievable by permuting the IV of interest and surveying how the Mom$^n$'s of the DV react to the manipulation, but this is much more costly computationally and was not studied here. Two additional moment-based dvPASs along these lines are **dvMom$^n$MM** and **dvMom$^n$ik** that are conditional on the two focal columns S and E showing two generic matches M and M or two fully specified matches for two not necessarily equal markers *i* and *k*, respectively. The corresponding scores for KS-M and KSi are called **dvKS-M** and **dvKSi**.

In the case of LKx and LKxm, the effect of permuting the markers at column S on the score of column E can in principle be assessed purely formulaically for the little additional computational effort that is needed to calculate the double-conditional versions of these two scores, **LKx$_{SME}$** and **LKxm$_{SME}$**, for columns S and E simultaneously, i.e., for the DV column and the IV column of interest (below). Everywhere in the paper, however, except in the formulae immediately below, these two scores are called **dvLKx-ijkl** and **dvLKxm-ijkl** in order to stress that IV associations with the DV are at stake, that the DV is permuted, and that every distinct pairwise combination of markers that is possible at columns E and S is individually considered.

Calculating dvLKx$_{SME}$ and dvLKxm$_{SME}$ (and dvMom$^n$ik for that sake) requires scoring in each PM row the pairwise states at the S and E sites and the matches at non-focal sites. Insignificant additional computational cost is needed in order to ascertain the pairwise states at E when scoring the set {M$_m$..E$_k$} of the observed "hybrid" M$_m$..E$_k$ pairwise fragments that show both *m* matches at non-focal spots and the *k*-th pairwise state at the column-E spot.

If one assumes that in the PM the **S$_i$**'s pair randomly with the observed hybrid M$_m$..E$_k$ pairwise fragments in the individual pairwise comparisons, then the hypergeometric dvLKx$_{SME}$ is given by Formula (4) and the multinomial dvLKxm$_{SME}$ is given by Formula (5). Obviously, in the hybrid M$_m$..E$_k$ pairwise fragments the M$_m$'s do not consider eventual matches at columns E and S. The multinomial LKm$_{SME}$ for the maximum-likelihood PM-level H$_1$ is in Formula (6).

$$LKx_{SME} = \frac{\prod_i^S S_i! \prod_{i,m}^{S,M} M_{m..}E_k!}{W! \prod_{i,m,k}^{C,M,E} S_{i..}M_{m..}E_k!} \tag{4}$$

$$LKm_{SME} = \frac{\prod_s^S S_s! \prod_{i,m}^{S,M} M_{m..}E_k!}{W! \prod_{i,m,k}^{S,M,E} S_i M_{m..}E_k!} \prod_i^S P_i^{S_s} \prod_{m,k}^{M,E} P_{m,k}^{M_{m..}E_k} \tag{5}$$

$$maxLKm_{SME} = \frac{W!}{\prod_{i,m,k}^{S,M,E} S_{i..}M_{m..}E_k!} \prod_{i,m,k}^{S,M,E} P_{i,m,k}^{S_{i..}M_{m..}E_k} \tag{6}$$

Formulae (4) and (5) are explicit about the pairwise states at S and E and may therefore deliver more power than just using say $LKx_{SM}$ with column E set as the only focal column and letting eventual matches at column S influence merely the $M_m$'s, but the latter "reduced explicitation" option is not studied here.

The formulae above are not explicit about how double mismatches (i,j)_(k,l) at S_E are "phased" in individual pairwise comparisons, i.e., about how the markers at the two focal sites are arranged in the two DM rows being compared. In a binary DM, e.g., the two mismatches could be phased (0,1)_(0,1) or (0,1)_(1,0), i.e., the S_E markers in the two compared DM rows could be 0_0 and 1_1, or 0_1 and 1_0, respectively, etc.

It might appear desirable *a priori* to be as explicit as possible about same-row markers, given that the purpose of PAS is to take maximal advantage of any tractably computationally assessable non-randomness created in the PM by associations in the DM among same-row markers. However, cursory results with binary DMs showed that considering the phase of paired markers at S and E does not increase power, so this was not further pursued.

On the contrary and as intimated, $dvMom^n i$, $dvMom^n ik$, and their Z-valued versions, the moment-based PAS scores that below are shown to be most powerful when detecting columns in randomly encountered models of significant *n*-IV DV association without marginal effects, ignore every PM row that shows a focal-column mismatch. However, several additional single- and double-conditional $dvMom^n$ versions, Z-valued and not, were tried that condition on mixtures of generic and marker-explicit matches and mismatches. but were found to have lower power.

Going back to $dvLKx_{SME}$ and $dvCHIx_{SME}$, simulations show that $dvCHIx_{SME}$ and the ratios of $dvLKx_{SME}$'s, like the ratios of $CHIx_{SM}$ and the $LKx_{SM}$ above, are not $\chi^2$-distributed (not shown) so that deriving their theoretical distribution is desirable to avoid having to estimate their P values by permuting the DV. On the plus side, Figure 11 shows that the P values of $dvCHIx_{SME}$ (and thus $dvLKx_{SME}$'s) that are estimated for different columns by permuting S in the DM, have a smallness ranking that matches closely the ranking by magnitude of the $dvCHIx_{SME}$ values of the same columns. This is so in binary and trinary DMs and applies regardless of marker-frequency differences across IV columns. Finally, Figure 11 shows that the two rankings coincide not only when conditioning on fully specified marker pairs at S and E but also when conditioning on generic matches and mismatches there, unlike what Fig.9 shows for $CHIx_{SM}$.

Therefore when one performs a DV-focussed PAS analysis, one can choose say 5% of the IVs with most extreme dvCHIx of either type and be certain that they are those with lowest dvCHIx P values, regardless of marker frequencies across columns, were said P values to be estimated by permuting the DV.

However and as was noted for CHIx and provided that type I error and false positives are generated in correct amounts, differences in power should have the last word in deciding whether conditioning the various dvPASs on generic matches is to be preferred over conditioning them on each marker-explicit pairwise match, Z-valueing the latter scores, summing over marker pairs, etc.

**DM and PM patterns and the power of $Mom^1 i_Z$, $dvMom^1 i_Z$, and $dvMom^1 ik_Z$.**

The next section presents further simulation assays of PAS and dvPAS type I error, followed by extensive simulation assays of power and false positives. The focus is on the performance of $Mom^n i_Z$, LKx-M, $dvLKx$-ijkl, $dvMom^n i_Z$, and $dvMom^n ik_Z$, i.e., the PASs and dvPASs that detect with highest power the randomly encountered models of column association and $n$-IV association with a DV, respectively, that are at the center of the power assays in this paper.

Beyond studying the reaction of $Mom^{n-1}M$ to pure $n$-column associations and the reaction of $dvMom^n i$ and $dvMom^n i_Z$ to pure n-IV DV associations when binary markers have frequency 0.5, very few attempts are made below at connecting same-row marker associations in the DM with associations in the rows of the PM of focal-site matches with pairwise fragments with given numbers of matches.

Therefore, to help the reader engage more critically the simulation results further below, some additional results and intuitions are included already here about why $Mom^1 i_Z$, $dvMom^1 i$, $dvMom^2 i_Z$, and $dvMom^1 ik_Z$ have a power edge with said models and which connections between marker associations in the DM and PAS-detectable patterns in the PM may be responsible. This is done using null DMs with binary-marker frequency 0.5 across their independent random columns because results are then easier to connect with the patterns of markers and matches in the DM and the PM, respectively. The focus is on $Mom^{n-1}M$ and $dvMom^n M$ that are most powerful when binary marker frequency at the focal column if 0.5.

A first indication to mention is that in such null DMs the $Mom^{n-1}M$ of a focal column correlates perfectly or very strongly with the sum of the "pure $n$-column $\chi^2$s" that can be calculated (below) starting with the standard $\chi^2$ of any possible $n$-column contingency table (e.g., the 5-column "Table $\chi^2$" in Fig.3) that involve both the focal column and $n$-1 other columns from the DM at hand. Three such extreme correlations are shown in Figure 12 and discussed below (the affix "M" in $Mom^n M$ is omitted).

Figure 12 shows that in a 100-column such DM a focal column's $Mom^1$, $Mom^2$, and $Mom^3$ are almost perfectly positively linearly correlated ($R^2 = \sim 1.0$) with the sums of all possible 2-, 3-, and 4-column pure $\chi^2$s that involve said focal column, respectively. To calculate say the pure 3-way $\chi^2$ of a specific subset of three columns that includes the focal column, the three "internal" 2-column $\chi^2$s are substracted from the plain 3-column $\chi^2$. Similarly, the pure 4-way $\chi^2$ of a subset of four columns that includes the evaluated column is equal to the plain 4-column $\chi^2$, minus the four internal pure 3-way $\chi^2$s, minus the six internal 2-way $\chi^2$s. This is done for every distinct $n$-column subset that can be drawn from the DM at hand and includes the focal column, and the sum of the resulting pure $n$-way $\chi^2$s becomes the score of the focal column at stake. Note, however, that the correlations sink rapidly when same-column markers differ in frequency and when a DM has fewer than 100 columns (neither is shown). Note also that the sums of pure 3-way $\chi^2$s calculated using the expectations for the 3-marker combinations derived by Geiringer (1944) give a lower 95% $R^2$ with $Mom^2$ (not shown).

These ~100% correlations mean that in binary DMs with more than 100 columns with binary markers of frequency 0.5, the columns that by chance have the largest sums of pure $n$-column $\chi^2$s are also those with the largest values of $Mom^{n-1}$, so that when flagging columns with the two methods one should generate the same type I error and possibly the same false positives, as well as perhaps have similar power. A second property that these sums share with PAS is the fact that in the data used in Figure 12 the $R^2$s are ~0.0% for every pairing of any two sums of same-focal-column pure $m$- and $n$-way $\chi^2$s and same-focal-column $Mom^m$s and $Mom^n$s ($m \neq n$). Therefore the P values of these scores, say estimated using permutations, are independent and likely even i.i.d., which is what $Mom^n$'s and $dvMom^n$'s are argued to be further below. A third behavior shared with PAS is that these sums are independent of all column associations that do not involve the focal column of the sum at stake, so that "background" associations cannot affect them. These properties make the sums interesting in their own right, but are also major properties of PAS and dvPAS as will be shown below.

Unfortunately for a better understanding of $dvMom^n ik$, i.e., of the dvPAS that say as $dvMom^1 ik_Z$ has highest power in the binary case to detect IVs in DV association without marginal effects when IVs do not form blocks (see power section further below), no correlations could be found between $dvMom^n ik$ or $dvMom^n ik_Z$ and any presumably applicable sum of pure $n$-column $\chi^2$s involving the DV and two IVs. Contrary to intuition, e.g., the sum of every pure 3-column $\chi^2$ is not noticeably correlated with any $dvMom^n ik$. This is consistent with the fact that the power of $dvMom^1 ik$ P values when detecting any of $c$ IVs that are in-phase 2-way DV-associated (Fig.2) increases faster with growing $c$ than does the detection of the same IVs by the sum of pure 3-column $\chi^2$s (not shown).

**Predicting PAS moments and sample variances when DM columns are independent from each other.**

Above it was stated that the single- and double-conditional first- and second-moment PAS and dvPASs are among the most powerful PASs and dvPASs. A few connections were made between same-row marker associations in a DM and match numbers in the rows of its PM, to shed some light on why PAS and dvPAS can detect PM columns at which, in the DM, same-row markers are associated in various ways.

However, before presenting both additional simulation results about type I error and extensive simulation results about the power and false positives of the most powerful PASs and dvPASs, some analytic results are presented here to help the reader develop further insight into the combinatorics and probability distributions governing the **mean m$^1$** and the **variance m$^2$** of the **number of matches *m*** in the rows of a PM as well as those governing the "**sample variances**" of m$^1$ and m$^2$ (the variances across PMs derived from DMs) under the H$_o$ that an *r*-row *n*-column null DM has independent columns in which markers are randomly vertically ordered. See M&M for the generation of the null DMs used in this section.

Towards the end of this section, a computationally faster, possibly new way to generate the PM column of a DM column through stereotypical direct copying of the corresponding DM column successively apically decremented by one marker (with markers at times "boolean-toggled"; below) is presented. The copying proves that a PM column's vertical order of matches and mismatches *de facto* repeats in a way the order of the markers in the generating DM column, straight-jacketing thereby the vertical order of matches and mismatches in said PM column and making many vertical orders impossible in the columns of a given PM. The copying also allows one to index systematically the matches created in the PM when pairwise-comparing the markers of any DM column of interest, i.e., it makes the matches addressable for algebraic operations. Finally, to exemplify the algebraic results that said indexing should ultimately deliver, exact expressions are presented for the likelihood of the mentioned H$_o$ given that one observes one of the possible vectors of *m*'s (the *m*'s of the PM, sorted from smallest to largest) for a couple of DMs with very few columns and rows.

**Theoretical results about PAS moments when DM columns are independent.**

This section intends to help the reader develop an intuition about the combinatorics and probability distributions that are relevant to the passage from the marker sequences in a DM to the sets of pairwise comparisons with distinct *m*'s in the PM of said DM. The predictions are confirmed with simulations but may fail in situations that are not simulated.

Several expressions are presented for calculating, under the aforementioned H$_o$ that markers across DM columns are independently vertically random, the joint expected occurrence of said sets of pairwise comparisons. Knowing this occurrence allows one to calculate the expected values of m$^1$ and m$^2$ but not their sample variances (across null PMs).

The first case considered is that of a DM with many more rows than columns. DM-to-PM simulations show, however, that these expectations for m$^1$ and m$^2$ also apply when DMs have as few as three and 100 rows, respectively. Numeric simulations are presented therefore that calculate m$^1$ and m$^2$ and their sample variances when a DM has very few rows and many columns. These numeric results are shown to match perfectly the m$^1$s and m$^2$s and their sample variances obtained from DM-to-PM simulations.

Some probabilistic considerations are then made for the PMs from DMs with similar numbers of rows and columns. For this case and the previous ones, results are presented from an any-DM numeric simulation that uses fast copying of at times boolean-toggled marker tracts of DM columns to generate the corresponding PM columns. These results are compared to results from numeric and DM-to-PM simulations.

The entirety of the work in this section is for a PM derived from a DM with L independent random columns. This type of DM is PAS-relevant: In a DM with L+1 columns where say the "+1" column is the focal PAS column, the sub-DMs formed by the rows sharing a given marker at their "+1" column site (rows whose pairwise comparisons must show a match at that site) should be qualitatively and quantitatively no different from L-column DMs with the same marker frequencies and number of rows as have the subDMs of interest. By extension, when the "+1" focal column has S markers of frequency 1/S, PASs that are conditional on say any match at the focal column are equivalent to combining (e.g., summing) the *m*'s in S independent L-column sub-DMs with 1/S as many rows as the original DM has.

R.R.Hudson and P.McCullagh (both University of Chicago) derived in 2005 the expectations of the first two PAS moments (and their sample variances?), but MAA does not understand their notation, procedures, and results. Theoretical statisticians and combinatorists interested in these results are strongly advised to contact these two scholars.

**Results for a DM with many more rows than columns.**

For this type of DM three main naive applied-mathematical intuitions are confirmed for some DM types but falsified in important ways for others. The intuitions are:

A) that the probability of a pairwise comparison of two L-marker sequences with *m* total matches is approximated by the binomial probability P(*m*) of getting *m* successes over L trials with success probability in each trial equal to the frequency of the matches in the corresponding PM column, i.e., this binomial assumes that the vertical order of matches and mismatches is both random and independent across PM columns.

B) that the probability of *c* pairwise comparisons having *m* matches in a PM with W rows may be binomial with W trials and success probability $p={}_LC_c P(m)^m Q(m)^{L-m}$ and $q=1-p$, with P(m) coming from (A), so that the expected $c= W {}_wC_c p^c q^{W-c}$, etc.

C) that the set $\{c_m\}$ of the counts in a PM of the pairwise comparisons with the various *m*'s may behave like the set of the counts from a multinomial event with W total trials and L+1 distinct outcomes (the number of distinct possible *m*'s), each outcome with probability equal to the expected frequency in the PM of the applicable *m*.

Below it is shown that the binomial prediction in (B), that as stated is partly based on (A), of the count of each *m*-defined group of pairwise comparisons is always accurate, whereas the binomial variance of this count is equal to that across (simulated) PMs only when at each column the distinct markers have equal frequency. Then to address intuition (C), the average variance-covariance matrix of the counts of the *m*-defined groups of pairwise comparisons calculated over $10^5$ PMs obtained from as many simulated DMs is compared to the average variance-covariance matrix of the corresponding counts obtained from as many numerically matched multinomial simulations. These contrasts show that the multinomial covariances (always negative) match those in matched PMs only when same-column markers have equal frequency (again), and that otherwise some of the PM covariances are markedly larger than the multinomial ones, including many that become positive. Therefore for DMs with "even" same-column marker frequencies, one can use this multinomial approximation to calculate any $m^n$ and its sample variance.

**1. Case in which every column has two markers of frequency 0.5.** If the L columns of a DM have binary markers of frequency *p* 0.5, every one of the $2^L$ distinct L-marker sequences ("sequence types") has identical probability $p^L$. In a DM with many more rows than columns, the count of every distinct sequence type should therefore tend to be the same, so that the frequencies of pairwise comparisons that show a given number of total matches should tend to match those observed when pairwise-comparing the sequences of a DM that contains every distinct sequence exactly once. If in this minimal "full-representation" DM one compares the sequence consisting of only 0's to one of the others, it is easy to see that the others can differ by having up to L markers of type 1 at its L positions. More specifically, in the DM there are L sequences that differ from the all-0s sequence by exactly one marker (of type 1), "L choose 2" sequences that differ by exactly two 1s, etc. Generally stated, there are ${}_LC_j$ sequences that differ by exactly *j* markers from the all-0s sequence ($0 \le j \le L$), which expressed for *m* matches is equal to ${}_LC_{L-m}$ and is combinatorially identical to ${}_LC_m$. If one divides opportunistically the actual counts of the pairwise comparisons with various *m*'s in the PM derived from the minimal full-representation DM by the corresponding ${}_LC_m$ terms, one easily zooms on Formula (7), where: $2^L(2^L-1)/2$ is the number of pairwise comparisons of $2^L$ sequences, $\frac{1}{2} 2^L$ is the (constant) number of distinct sequences that generate pairwise comparisons with a given *m*, sequences are not re-used, the approximation is valid when L>>10, *m* is smaller than L, and prob[L] is 0.0.

$$prob[m] = \frac{\frac{1}{2} 2^L \binom{L}{L-m}}{2^L (2^L - 1)/2} = \frac{\binom{L}{m}}{2^L - 1} \approx \frac{\binom{L}{m}}{2^L}$$

(7)

If one knows the counts of the pairwise comparisons with *m* matches in the PM derived from a DM in which every distinct L-marker sequence is present exactly twice, one can again divide the counts opportunistically by the "L choose m" terms in (7) and modify (7) to match the counts in a DM in which every sequence type is present *n* times. This gives Formula (8) where $n 2^L(n 2^L-1)/2$ is the number of pairwise comparisons of the $n 2^L$ sequences and $\frac{1}{2} n 2^L$ is the number of pairwise comparisons with a given *m*, again without re-using sequences and with both approximations being valid when both L>>10 and *n*>>.

$$prob[m] = \frac{\frac{1}{2} n^2 2^L \binom{L}{L-m}}{n 2^L (n 2^L - 1)/2} = \frac{n \binom{L}{m}}{n 2^L - 1} \approx \frac{\binom{L}{m}}{2^L}$$

(8)

$$prob[L] = \frac{\frac{1}{2}n(n-1)2^L \binom{L}{0}}{n2^L(n2^L-1)/2} = \frac{n-1}{n2^L-1} \approx \frac{n-1}{n2^L} \approx \frac{1}{2^L}$$

(8.1)

**2. Case in which every DM column has S markers each of frequency 1/S.** The formulae for a DM with many more rows than columns where at every column there are S markers of identical frequency 1/S, is only slightly more complex since also in this case every one of the $S^L$ distinct sequences has equal probability. In this case too there are L sequence types that differ by exactly one marker of type 1 from the all-0s sequence, etc. But one must consider additionally the L sequences that differ from the all-0s sequence by one marker of type 2, etc, so that there are (S-1)L different sequences that differ from the all-0s sequence by exactly one marker of any type (since there are S-1 types of non-0 markers). Similarly, there are $(S-1)^2$ (L choose 2) sequence types with two markers of non-0 type that differ at two spots from the all-0s sequence, $(S-1)^3$ (L choose 3) with three differences, etc. Generally expressed, there are $(S-1)^j {}_LC_j$ sequence types with $j$ differences vs. the all-0s sequence, which for $m$ matches is $(S-1)^{L-m} {}_LC_m$. After adding this complexity to Formula (8) plus adjustments one readily obtains Formula (9), where $m<L$, $n \geq 1$, and approximations are valid if L>>10 and $n$>>.

$$prob[m] = \frac{\frac{1}{2}n^2 S^L \binom{L}{L-m}(S-1)^{L-m}}{nS^L(nS^L-1)/2} = \frac{n\binom{L}{m}(S-1)^{L-m}}{nS^L-1} \approx \frac{\binom{L}{m}(S-1)^{L-m}}{S^L}$$

(9)

$$prob[L] = \frac{\frac{1}{2}n(n-1)S^{L-1}\binom{L}{0}(S-1)^0}{nS^L(nS^L-1)/2} = \frac{(n-1)S^{-1}}{nS^L-1} \approx \frac{n-1}{S^{L+1}-S} \approx \frac{1}{S^{L+1}}$$

(9.1)

At this point the mathematically inclined reader has certainly realized that when the binary marker probability $p=q=0.5$ then the ${}_LC_m/2^L$ in (7) is equal to the binomial probability of getting $m$ successes (matches) over L trials with success probability P per trial equal to $p^2+q^2$ and Q= 1-P= $2pq$. This binomial probability is indeed equal to ${}_LC_m P^m Q^{(L-m)}$ that for $p=0.5$ becomes ${}_LC_m (0.25+0.25)^m (2 \cdot 0.5 \cdot 0.5)^{(L-m)} = {}_LC_m/2^L$, i.e., Formula (7). Similarly, if for the trinary case $p=q=r=1/3$ and $p+q+r=1.0$ then $p^2=q^2=r^2=1/9$ and $2pq=2pr=2qr=2/9$ so P(match)= 1/3 and P(mismatch)= 2/3 and then p(m)= ${}_LC_m (1/3)^m (2/9)^{L-m}= {}_LC_m 2^{L-m}/3^L$, which is Formula (9) for S=3. Both results are confirmed with simulations further below.

**3. Case in which every column has two markers of frequencies $p$ and $q$ not necessarily equal.** Also in this case one can use a binomial to obtain the probability P(m) of a pairwise comparison showing $m$ matches. There are now two different probabilities of a match, i.e., $p^2 \neq q^2$ and one is truly dealing with a trinomial trial. Setting $p^2+q^2$ as the probability of a generic match at a site and using it as success probability in a binomial trial delivers accurate estimates of P(m) but simulations show that when $p$ is other than 0.5 this approach fails to deliver accurate results for other important quantities.

Expression (10) below allows one to calculate P(m) for any $p+q=1.0$. It follows the strategy in (7) of comparing exhaustively every possible pair of marker sequences. The expression can be readily evaluated numerically with a computer when L<100. The adopted evaluation logic is to calculate combinatorially the matches shown by, and on that basis the occurrence probability of, every possible pairwise comparison involving a first sequence that has a given number of 1s and a second sequence with the same number of 1s or higher. After evaluating this subset of pairwise comparisons, the next evaluated subset is the one in which the first sequence has exactly one more 1 and the second sequence has the incremented number of 1s or more, etc, until one reaches the 1-element subset of the all-1s sequence that is compared to itself.

The combinatorics of the pairwise differences are akin to those encountered above when comparing the all-0s sequence to sequences that differ from it by one difference, two differences, etc. Additional book-keeping is necessary because one needs to i) visit orderly the subsets and each of the two classes of sequences that are pairwise-compared at any given stage of the enumeration; ii) keep track of the number of differences in each visited pairwise comparison; and iii) keep track of whether the differences are caused by 0s or 1s so one can calculate the occurrence probability of the two sequences of the pairwise comparison of interest and the probability of the two sequences being randomly encountered as a pairwise comparison.

In every subset of pairwise comparisons considered in Expression (10), the sequences in the first sequence class have L-$r$ spots with 0s and $r$ spots with 1s ($0 \leq r \leq L$). The "L choose $r$" term is therefore the count of all possible first-class sequences with $L-r$ 0s and $r$ 1s in any order across columns. The "$r$ choose $k$" term is the count of the distinct second-class sequences in which one finds 0s at spots that in first-class sequences show a 1 (spots that are of $r$-type there), i.e., "$r$ choose $k$" is the count of all possible second-class sequences that differ from first-class sequences because they have $k$ 0s at the spots where first-class sequences have their $r$ 1s. Therefore in second-class sequences there are only $r-k$ 1s at spots in which first-class sequences have 1s. The "L-$r$ choose $j$" term is the count of all the pairwise comparisons in which second-class sequences differ from first-class ones by having $j$ 1s at the L-$r$ spots that have 0s in first-class sequences. Since the definition of the subset requires that the 1s in second-class sequences be at least $r$, $j$ must be at least as big as $k$. All of this makes the total number of matches $m$ equal to L-$k$-$j$.

Therefore the two classes of compared sequences have a total of 0s equal to L-$r$ +L-$r$-$j$+$k$ and a total of 1s equal to $r$ +$r$+$j$-$k$, i.e., 2(L-$r$)-$j$+$k$ 0s and 2$r$+$j$-$k$ 1s. These pairwise comparisons have probability $p^{L-r}q^r p^{L-r-j+k} q^{r+j-k} = p^{2(L-r)-j+k} q^{2r+j-k}$. This probability needs to be multiplied by A=2.0 only if $j \neq k$, since when $j=k$ the two compared sequences are generated twice by the 2$^{nd}$ and the 3$^{rd}$ "X choose y" terms (from the left) that account for the permutation of column succession order in the pairwise comparison at stake. When L=4, e.g., the comparison 0001 vs.1000 is a permutation of the order of the columns in the comparison of 1000 to 0001 ($r$=1, $j$=1, $k$=0). Therefore, Formula (10) visits each such comparison individually through said "X choose y" operators and these comparisons' probabilities need to multiplied by A=1.0. Expression (10) too is shown to be accurate by the simulations further below.

$$prob[m] = \sum_{r=0}^{L} \binom{L}{r} \sum_{k=0}^{r} \binom{r}{k} \sum_{j=k}^{L-r} \binom{L-r}{j} A p^{2(L-r)-j+k} q^{2r+j-k} \mid L-j-k = m \qquad (10)$$

$m$= matches
L= DM columns
$p$= frequency of markers of type 0
$q$= frequency of markers of type 1
$r$= number of 1 markers in the compared pair's first sequence
L-$r$= number of 0 markers that can be markers of type 1 in the second sequence
$k$= number of 0 markers in the second sequence's "$r$ spots"
$j$= number of 0 markers at the L-$r$ spots that can become markers of type 1 in second sequence
$j+k$= number of differences between the two sequences
L -$j$-$k$= number of matches between the two compared sequences
A= 2.0 if $j \neq k$, and 1.0 otherwise.

4. **Case in which every column has three markers of frequencies $p$, $q$, and $r$ not necessarily equal.** For this case too one can follow the strategy in (7) and expand (10) to get an expression that can be evaluated numerically, but this was not done here. If one sets P(*match*)= $p^2$ +$q^2$ +$r^2$ and P(mismatch)=2$pq$ +2$pr$ +2$rs$ then the naive binomial P(m)= $_L C_m$ P(*match*)$^m$·P(mismatch)$^{L-m}$ works fine when estimating m$^1$ and m$^2$ (L>100), but fails to deliver accurate results for other important quantities when any of $p$, $q$, or $r$ is different from 1/3. See simulations below.

**Simulated and combinatorially calculated pairwise matches vs. expectations from naive binomials and multinomials.**

This section presents simulation results that confirm the accuracy of the formulae and expressions introduced above for calculating the expected frequency of a pairwise comparison with $m$ total matches. The simulated and predicted frequencies as well as the sample variance (across PMs) of the count of each $m$-defined group of pairwise comparisons are compared to those calculated using the naive intuitions in (A) and (B).

Next one estimates m$^1$ and m$^2$ using the empirical and theoretical estimates of the distinct m's and compares the result to both the average empirical estimates of m$^1$ and m$^2$ over simulated PMs and those obtained using the naive assumption in intuition (C) of multinomially joint-distributed counts of the pairwise comparisons with different *m*'s.

Then the departure of the distribution (joint occurrence) of the counts of the *m*-defined groups in a PM from the distribution of counts generated by a multinomial trial that uses as multinomial-category probabilities the expected probabilities of the pairwise comparisons with each possible *m* is studied by comparing the covariance of the counts of any two different *m*'s within a PM, in the two cases.

Panel (a) in Table 1 shows, for a 5-column DM in which every one of the 32 possible 5-marker sequences is present exactly once (*n*=1 in Form.8), that the scored counts of the six possible *m*'s (0 to 5 matches) coincide with the counts predicted by Formula (7) and (8).

Panel (b) shows the counts and the relative frequencies of the eight possible *m*'s in the PM from a 7-column binary DM in which every one of the 128 distinct 7-marker sequences is present exactly three times. The observed frequencies are compared to both those predicted by Formula (8) and the expectations from the aforementioned naive binomial. The match of combinatorio-exact and observed counts (and frequencies) is perfect and both coincide closely with the binomial predictions. Panel (b) also shows that, non-surprisingly, the m$^1$s and m$^2$s of the observed and predicted *m*'s match perfectly (bottom right).

Panel (c) is for a 6-column DM with four different markers per column where every one of the 4'096 6-marker sequences is present twice. Also here the three predictions match very well.

Table 2 shows the frequencies of the possible *m*'s in the PM from a 10-row 40-column DM with minor-marker frequency 0.2. The counts of the various *m*'s (the $c_m$'s in intuition (C) above) in $10^5$ PMs from as many simulated DMs are used to calculate the mean simulated count (and frequency) of each *m* and the variance of this count across PMs. The three sets of predicted frequencies match again very well, specially those from numerical evaluation of Expression (10) and the naive binomial. But for most *m*'s the rightmost binomial variance of the per-PM count of each *m*-defined group is bimodally between 10 to 60% bigger than that across simulated PMs.

Therefore the naive binomial in intuition (B) above delivers an overestimate of the variance across PMs of the count of most *m*'s (despite delivering the expected count very accurately). This discrepancy is consistent with that observed when comparing the corresponding covariances below, where additional light is shed on the reasons.

Table 2 (bottom left) shows that the three sets of m$^1$s and m$^2$s calculated using the predicted frequencies of the various *m*'s match perfectly, albeit the three m$^2$s differ clearly from the observed and numeric m$^2$s (the latter numeric-simulation estimates are introduced further below).

However, in DMs with binary-marker frequency 0.5, the across-PMs variance of each *m* count becomes equal to the corresponding binomial variance. This is also so when same-column markers have identical frequency in a DM and the binomial sample variance is multiplied by a constant (neither is shown). For a 10-row 40-column DM (i.e., a 45-pairwise-comparison PM) with trinary markers of frequency 1/3, e.g., the across-PMs variance is equal to $\sim 110.45 \cdot P(m) \, Q(m)$.

Figures 13 and 14 present the announced contrasts between the empirical covariances of any two *m*'s in a PM and the corresponding covariances within a set of m's generated by the multinomial in intuition (C). At issue is whether the covariances in a real PM behave like those in a set of *m*'s generated by a multinomial trial in which the W pairwise comparisons of the PM are assigned different *m*'s according to probabilities equal to the average frequencies of the various *m*'s over $10^5$ PMs from as many simulated DMs of the type at stake.

Figure 13 shows, on the left, the covariances between pairs of *m*'s in the PM from a 10-row 40-column DM whose every column has two markers, averaged over $10^5$ PMs from as many independently simulated DMs. On the right the corresponding multinomial covariances are substracted from the empirical covariances on the left. To calculate the two covariances for each PM the simulated mean is used rather than the mean estimated from each PM at stake.

The top plots show that with binary marker frequency 0.5, the two variance-covariance matrices behave very similarly, i.e., that both sets of covariances are similarly negative so that subtracting one from the other yields differences that are randomly positive or negative. Therefore with marker frequency 0.5, the joint behavior of the counts of the *m*-defined groups is very close and possibly identical to the joint behavior of as many counts generated by a multinomial event with category probabilities equal to the expected frequencies of the various *m*'s (below).

The bottom plot on the left shows that with minor-marker frequency 0.2, the empirical covariances between large and small *m*'s (many and few total row matches) are positive while those between small *m*'s tend to be negative (whereas in the frequency-0.5 case all of the covariances are negative and the only strong ones are those between intermediate *m*'s). The similarity between the bottom left and right plots shows that the empirical covariances dominate the substractions, i.e., are bigger than the multinomial covariances, negatively or positively. Therefore in binary DMs the empirical covariances are approximated by the multinomial covariances only when markers have frequency 0.5.

Figure 14 shows results for trinary DMs that, from top to bottom, have trinary-marker frequencies 0.25 0.5 0.25, 0.04 0.32 0.64, and 1/3 1/3 1/3, respectively. Only the covariances within PMs from DMs with "even" 1/3 marker frequencies behave multinomially (as do above those from DMs with binary markers of frequency 0.5). Otherwise the empirical covariances are larger and dominate the substractions (as observed with binary-marker frequency 0.2). The covariances between large and small *m*'s are positive and those between intermediate *m*'s are negative, with both being larger in absolute terms than the corresponding multinomial covariances. Therefore in PMs from trinary DMs with markers present in the genomics-relevant H&W frequencies $p^2$ $2pq$ $q^2$ the multinomial covariances do not approximate the empirical covariances of the *m*'s and hence the naive intuition in (C) cannot be used in this important case in genetics.

Obviously, for PMs from DMs with "even" marker frequencies, the above multinomial should allow one to estimate numerically in a straightforward manner any $m^n$ and its sample variance. The estimates of $m^1$($m^2$) and their two samples variances, e.g., across $10^5$ PMs from as many independent simulations of the 40-row 100-column DM with 0.5 binary marker frequencies are **20.0000(10.00101)**;0.0020(0.03939) vs. **20.0000(10.01814)**;0.0020(0.03950) from as many iterations of said multinomial, whereas with trinary frequency 1/3 the DM-to-PM results are **13.3334(8.8880)**;0.0018(0.03098) vs. **13.3332(8.8878)**;0.0018(0.03069) from the multinomial, i.e., the two sets of estimates are nearly identical in both cases.

We will see below that very subtle perturbations of PAS moments underlie their detection power, so that theoreticians are warned not to place much hope in the two binomials and the multinomial in (A), (B), and (C), respectively (albeit the binomials' perfect fit when marker frequencies are "even" in every column of a DM may be of interest to combinatorists).

**Simulations of the mean and variance of *m* and their sample variances.**

This section presents simulation results about $m^1$ and $m^2$ in PMs from DMs that have very few rows and columns, much fewer rows than columns, and about equal numbers of rows and columns. The results allow one to check the accuracy of the $m^1$s and $m^2$s calculated using the expectations presented above for the frequencies of every possible *m* in the infinite-row DM. Several attempts are made at exploring the probability distributions that govern the *m*'s of individual PMs and connecting them to the probabilities of marker rows and marker columns in the DM.

These attempts serve as introduction to two probabilistically justified numeric simulations that calculate the across-PMs sample variances of $m^1$ and $m^2$ for various types of finite-row DMs. The obtained numeric estimates are found to be accurate when compared to across-PMs sample variances estimated using DM-to-PM simulations; they thus shed light on the actual combinatorics and probability distributions that govern $m^1$ and $m^2$ as well as their sample variances.

An important additional result is also presented, namely the observation that any vertical marker pattern in a column of a DM with R rows can be near-instantly transformed into the vertical pattern of matches and mismatches of the corresponding column in the PM, by performing only R-1 pairwise comparisons rather than the usual ½ R(R-1). This transformation should allow one to order the 1-column marker patterns in the pattern-combinatorially comprehensive DM in a way that lets the patterns of contiguous columns in the PM form blocks in which systematically and straightforwardly, if tediously, matches can be indexed by their position and be assigned the probability of the DM column that generates the PM column at stake. This can be done nearly as easily for PMs from DMs with trinary markers, etc. The indexing should help in deriving formulae for calculating the across-PMs sample variances of $m^1$ and $m^2$ for the PM of a DM with many columns (see below) and even the likelihood of the $H_o$ given the observed set of *m*'s in a PM at stake. Indeed, this section ends with the presentation of the exact algebraic formulae of such likelihoods (obtained *brute force*) for PMs from selected DMs with very few rows and columns.

Figure 15 show most of the numeric and simulation results of this section but the reader is advised to engage these results as the corresponding generating methods are successively introduced below. The left plots in Figure 15 show that the empirical $m^1$ averaged over $10^5$ individual PMs from as many simulated DMs coincides with the values calculated on the basis of the expectations presented above for the set of *m*'s in the PM from an infinite-row DM. However, the plots on the right show that the average empirical $m^2$ over the same $10^5$ individual PMs coincides with the infinite-row expectation of $m^2$ only when DMs have 100 rows or more and is otherwise overestimated by the infinite-row expectation. For the case of a 3-row 4-column DM with binary-marker frequency 0.2, e.g., the average $m^2$ is ~74% of the $m^2$ estimated using the *m*'s in the PM of the infinite-row DM (Fig.15a,b), suggesting that studying the 3-row many-column DM may shed light on the combinatorio-probabilistic fundaments of $m^1$, $m^2$, and their sample variances across PMs.

Both sample variances of $m^1$ and $m^2$ in Figure 15 are shown using double-log plots (right vertical axes) and seem to decrease about linearly with the number of DM rows, but in non-log plots they sink non-linearly and very fast with the number of rows. Also here it appears promising to study combinatorio-probabilistically the sample variances of $m^1$ and $m^2$ across PMs from 3-row DMs.

As intimated, Figure 15 includes results from two different probabilistically justified numeric simulations that estimate the sample variances of $m^1$ and $m^2$ in a PM, namely across PMs derived from few-row many-column DMs and across PMs derived from many-row many-column DMs, respectively. Both sets of numeric values match the empirical variances of $m^1$ and $m^2$ across simulated PMs. These numeric simulations are described further below. Note just in case, that above say 100 DM rows the multinomial estimates of $m^1$ and $m^2$ are very accurate but the estimated sample variances are very different from the empirical and numeric ones in Figure 15, albeit as noted the two multinomial sample variances for binary and trinary cases with $p= 0.5$ and $p= q= r= 1/3$, respectively, match the empirical estimates very closely.

**Numeric estimation of $m^1$ and $m^2$ and their sample variances when L is much larger than $S^R$.** In this section several accurate numeric estimates of the $m^1$ and $m^2$ of PMs from few-row many-column DMs are successively presented and their combinatorio-probabilistic assumptions are discussed and shown to be insufficient in the general sense when used isolatedly (below).

When an L-column binary DM has only two rows and L>> (Figure 16, left), a DM column can show any of four different 1-column 2-marker patterns, i.e., the two same-column markers at DM rows 1 2 are, vertically, 00, 01, 10, and 11. The L-column PM from this DM has only one row with pattern M n n M from left to right, where M and n stand for a match and a mismatch, respectively. When DM columns are independent and binary marker frequencies are *p* and *q* at every column ($q=1-p$), the expected counts of the four 1-column patterns in the DM and PM are L times $p^2$, $pq$, $qp$, and $q^2$, respectively, i.e., $Lp^2$, $Lpq$, $Lqp$, and $Lq^2$ (with sum equal L·1.0=L). The four counts are indeed multinomially joint-distributed with L trials and four multinomial probabilities $p^2$, $pq$, $qp$, and $q^2$. Since DM columns are independent, the four types of 1-column 2-marker DM patterns and 1-column 1-marker PM patterns appear randomly along the DM and the PM, respectively. And since matches in the only row of the PM are found only at columns spots where the markers of the two compared sequences are 0/0 and 1/1, one can sum the probabilities $p^2$ and $q^2$ of these two 1-column DM patterns to get the probability P of getting a match at one of the L columns. Column independence in the DM means that the probability of getting *m* matches over the L PM columns is binomial with probability of success P and L binomial trials, giving estimates for $m^1$ and a $m^2$ equal to LP and LPQ, respectively (Q= 1-P).

When an L-column binary DM has three rows and L>> (Fig.16a, right), there are eight different 1-column 3-marker binary patterns in the DM, i.e., $S^R$ for S=2 markers and R=3 rows. The distinct 1-column patterns formed by markers in DM rows 1 2 3 are, vertically: 000, 001, 010, 011, 100. 101, 110, and 111, with probabilities $p^3$, $p^2q$, $p^2q$, $pq^2$, $p^2q$, $pq^2$, $pq^2$, and $q^3$, respectively. The count expectations are obtained by multiplying each probability by L. The eight 1-column 3-marker PM patterns MMM, Mnn, nMn, nnM, nnM, nMn, Mnn, and MMM for the three pairwise comparisons (1,2) (1,3) (2,3) belong to only four distinct types, namely MMM, Mnn, nMn, nnM. Like above they too appear randomly along the L columns of the PM. The counts of the eight 1-column patterns in the DM are also here multinomially joint-distributed with L trials and eight multinomial probabilities equal to the probabilities of the eight 1-column patterns in the DM. In the PM, however, the multinomial trials are still L but the multinomial categories are only four.

In the 3-row PM, $m^1$ and $m^2$ involve three pairwise comparisons, to each of which only four of the said eight 1-column patterns contribute matches. For the (1,2) row of the PM, e.g., matches are contributed by the 1-column PM patterns 1, 2, 7, and 8. Indeed the *m* of every one of the three rows has matches contributions from patterns 1 and 8 as well as two "internal" patterns of probability $p^2q$ and $pq^2$, respectively. Therefore the three PM rows have all the same expected *m* and, unlike the DM, the PM is bilaterally symmetrical when suitably ordered (albeit not so probabilistically when $p \neq q$).

Figure 15 shows that both in the binary and trinary case already with L>100 the binomial variance of the *m* of a single pairwise comparison from a 3-row many-column DM coincides with the average $m^2$ over $10^5$ PMs from simulated DMs. For L<100, however, $m^2$ is over-estimated by the binomial variance. In the case of the (1,2) pairwise comparison, e.g., the binomial P for the 3-row 40-column binary DM (Fig.15b) with 0.2 marker frequency, is the sum of the probabilities of the aforementioned 1-column patterns 1, 2, 7, and 8, i.e., 0.008 +0.032 +0.128 +0.512= 0.68, *Q is* 0.32, and L is 40, which if one uses LP and LPQ as mean(variance) gives

27.2(8.704). The $m^1(m^2)$ values averaged over $10^5$ simulated PMs are instead 27.2(6.373), i.e., the binomial mean $m$ matches the average simulated $m^1$ but the binomial variance of $m$ is 26% bigger than the average simulated $m^2$. This is also the case for a trinary 3-row 40-column DM with the H&W trinary frequencies for $p$ 0.2 ($p^2$ $2pq$ $q^2$ equal to $r$ $s$ $t$ equal to 0.02 0.64 0.32), a DM in which the probability P of any of the three pairwise comparisons is the sum of the probabilities of nine 1-column patterns ($3^3$), i.e., $r^3$ $+s^3$ $+t^3$ $+p^2q$ $+p^2r$ $+pq^2$ $+q^2r$ $+pr^2$ $+qr^2$= 0.000064 +0.262144 +0.032768 +0.001024 +0.000512 +0.016384 +0.131072 +0.004096 +0.065536= 0.5136. With L equal 40 and this $P$, the binomial LP(LPQ) are 20.544(9.926), where the binomial variance of $m$ again does not match that in the 20.534(8.878) obtained from simulated PMs but it matches quite well the 9.975 in the 20.545(9,975) from the PMs of 40-column DMs with 100 rows or more. Therefore the $m^2$ of a PM from a few-row DM cannot be estimated using the above binomial expectations for the mean and variance of the $m$ of an individual row of the PM derived from a few-row many-column DM, except when at each column all marker frequencies are equal. Above we saw that several other estimates based on naive binomials are accurate when same-column marker frequencies are equal. And the general point was made in Figures 13 and 14 that with such frequencies the covariances within a PM between any two $m$'s can match those between the counts of the $m$'s generated by a multinomial experiment with ½ R(R-1) trials and L+1 different outcomes (the $m$'s from 0 to L matches) each of probability $_LC_m P^m Q^{L-m}$. So it is not surprising that here too such a multinomial generates sets of $m$'s that yield accurate estimates of $m^1$ and $m^2$, e.g., for 3x3, 10x10, and 10x100 binary DMs with marker frequency 0.5, but when binary frequency is 0.2 it does so for $m^1$ but not $m^2$ (not shown).

A final (trivial?) possibility considered here for estimating $m^1$, $m^2$, and the sample variances of $m^1$ and $m^2$ for a few-row few-column PM is to simulate PMs by sampling multinomially L times any of the $S^R$ possible 1-column DM patterns, evaluate the $m$'s of the rows of each simulated PM over the L resulting 1-column PM patterns, and calculate $m^1$ and $m^2$ for each set of $m$'s obtained from each PM. Over multiple such $m^1,m^2$ pairs one can calculate the sample variances of $m^1$ and $m^2$. Figure 15 (big circles) shows that the average $m^1$ and $m^2$ and their sample variances over $10^5$ such multinomial simulations match perfectly the values estimated using $10^5$ PMs from as many simulated 6- and 40-column binary and trinary DMs with three to ten rows (above 40 columns $S^R$ becomes too large for the computer).

**Estimating $m^1$ and $m^2$ and their sample variances when L is much smaller than $S^R$.** The paragraphs above addressed both the few-row many-column DM and the infinite-row few-column DM but the approximations that were tried were inaccurate for PMs from few-row few-column DMs. However, only when L>>$S^R$ can every distinct 1-column pattern be expected to appear multiple times in the DM. (The case with R>>L is approximated by the infinite-row few-column DM).

In fields like genomics, however, large DMs may have R>$10^4$ and L ~$10^7$, making relevant the "intermediate" case in which only very few out of very many possible 1-column patterns ever show up together in a DM and none is likely to show up more than once. When DM columns are independent, the 1-column patterns that show up in such a DM can be modelled as a very sparsely populated multinomial event, i.e., one in which the number of distinct outcomes is much larger than the number L of multinomial trials and the expected count of each type of outcome is much smaller than 1.0. Also in this case the $S^R$ 1-column patterns are sampled independently from each other and show up at most once in a DM (like independent Poisson trials with innumerous different outcomes of very low probability that succeed each other until L outcomes accumulate). A numeric simulation that performs direct multinomial sampling from amongst the $S^R$ possible 1-column DM patterns (like the one yielding the big circles in Fig.15) is not viable in this case because $S^R$ is mostly too large for a computer to handle in reasonable time. However, a 2-step strategy for numeric evaluation is viable that simulates the random sampling of L 1-column patterns for a DM by using two rounds of multinomial sampling. In the binary case, e.g., the first round generates the DM's count of every possible L+1 1-column pattern with $i$ minor markers in any vertical order (each of probability $_LC_i$ $p^i q^{R-i}$; $0 \leq i \leq R$). The second round of multinomial trials samples random 1-column vertical patterns with $i$ and R-$i$ minor and major markers, respectively, for every $i$-marker 1-column type sampled in the first round with non-zero count, by choosing among $_RC_i$ multinomial categories of identical (uniform) probability $1/_RC_i$.

The $m^1$s and $m^2$s and the two sample variances of $m^1$ and $m^2$ obtained from this 2-step numeric simulation of DMs with binary-marker frequency 0.2, match closely those from simulated PMs in the case of square DMs with 3, 10, 100, and 1'000 rows and columns. The numeric $m^1(m^2)$ and the two sample variances are **2.0408**(0.47929);0.33339(0.23687), **6.7990**(1.9680);0.25251(0.36144), **67.991**(21.5428);0.23380(2.7670), and **680.00**(217.34);0.22828(28.121), respectively ($10^5$ iterations; only $10^4$ PMs in the 1'000-row 1'000-column case), while the values over $10^5$ simulated PMs are **2.0381**(0.48033);0.33156(0.23734), **6.8031**(1.9648);0.25446(0.35801), **67.989**(21.5480);0.22237(2.7149), and **680.01**(218.56);0.2290(27.320),

respectively (1'000-case: $10^4$ PMs). For the 100-row 100-column case with H&W trinary frequency 0.2 ($10^5$ iterations), the numeric values are **51.352**(24.880);0.128914(0.98031) and the values over $10^4$ simulated PMs are **51.364**(24.863);0.17465(0.99802).

While implementing the 2-step numeric simulation, a possibly new algorithm was conceived by MAA that in principle should speed up in a major way the generation of the PM column of a DM column. Panel c.2 of Figure 16 shows how this algorithm works for a binary DM. Given a data column with say five binary markers in the vertical order 1 0 0 1 0, the first vertical "tract" of pairwise comparisons between the first marker and the others is equal to 0 0 1 0, if in the PM a 1 and a 0 symbolize a match and a mismatch, respectively. One notes both that the first marker is a 1 and that one could have used directly the markers 2, 3, 4, and 5 in that DM column as the pairwise comparisons 1vs.2, 1vs.3, 1vs.4, and 1vs.5. The second tract of the PM column are the pairwise comparisons between the second marker and the others, i.e., 1 0 1, where again 1s and 0s are matches and mismatches, respectively. One notes that these pairwise comparisons are the same as NOT(markers 3 4 5)= NOT(0 1 0)= 1 0 1.

The general procedure emerges that, for a binary DM, the PM tract with the pairwise comparisons between the $i$-th marker and the others is either equal to the DM column's markers from row $i$+1 to row R if the marker at row $i$ is a 1, or equal to the said DM tract with boolean-negated markers if the $i$-th marker is a 0. In other words, the column with the pairwise comparisons of the markers in a column from a binary DM with R rows is composed of R-1 tracts of pairwise comparisons, where the $i$-th tract is equal to the tract of markers from the ($i$+1)-th row to the R-th row if the $i$-th marker is a 1, or equal to the same tract but with boolean-negated values if the $i$-th marker is a 0.

This procedure requires performing only R-1 pairwise comparisons rather than the usual ½R(R-1). Importantly, this regularity means that the vertical order of the markers in a DM column at stake allows one to predict straightforwardly the row number of every match and mismatch in the corresponding PM column, for any vertical order of said markers.

Panels (b,d) in Figure 16 show, for the PM of a binary DM in which every possible 1-column pattern appears from left to right in the order adopted for the figure, how the just mentioned stacking of successively top-marker-truncated vertical tracts in the PM creates multicolumn blocks of matches and mismatches with a bilaterally symmetrical patterning that repeats the applicable blocks of 1s and 0s in the DM (directly or boolean-toggled). The PM blocks are from top to bottom: two blocks each with $S^R/2$ columns and R-1 rows, four blocks each with R-2 rows and $S^R/4$ columns, four blocks each with R-4 rows and $S^R/8$ columns, etc, until the ½ R(R-1)-th PM row is reached that has $S^R/2$ 1-row 2-column blocks.

The stereotypical left-to-right, top-to-bottom appearance of PM blocks with rigidly predictable row and column positions and within which matches appear at fully predictable positions allows one to address (i.e., index) every match in the every-column-pattern PM. This, together with the straightforward probability of each involved 1-column marker pattern in the DM, may allow a skilled applied-mathematician to address systematically with formulae any group of matches anywhere in this PM. In particular, one could so query the matches in each of the ½R(R-1) rows of the PM in order to derive exact --if cumbersome-- algebraic expressions for $m^1$ and $m^2$ and their two sample variances, at least for values of R and L that are not exceedingly big.

**Likelihoods of the $H_o$ obtained by *brute force* given the *m*'s in the PMs of very small DMs.**
Under the $H_o$ that columns of a DM are independent from each other, it is possible to get by *brute force,* for say a 3-row 3-column DM with binary markers of a given expected frequency at every column, exact algebraic expressions like those that one would obtain using the indexing contemplated above. This includes expressions for the likelihood of said $H_o$ given that one observes a PM's "vector" of *m*'s (the set {$M_m$} above sorted by increasing *m*).

Under this $H_o$ in the said 3x3 case, there are 512 different DMs and seven different vectors, so that it is computationally tractable to generate every DM that delivers a given vector and sum the probabilities of these DMs to compute the likelihood of the $H_o$ given that one observes the vector. The seven likelihoods are in Formula (11) and in them, obviously, the coefficients are the number of ways in which one can rearrange the markers of a DM with as many 1s and 0s as specified in the exponents of the *p*'s and *q*'s, respectively, without altering the vector at stake.

$$Lik(003) = 3p^3q^6 + 9p^4q^5 + 9p^5q^4 + 3p^6q^3$$
$$Lik(012) = 18p^3q^6 + 54p^4q^5 + 54p^5q^4 + 18p^6q^3$$
$$Lik(111) = 6p^3q^6 + 18p^4q^5 + 18p^5q^4 + 6p^6q^3$$
$$Lik(113) = 9p^2q^7 + 18p^3q^6 + 9p^4q^5 + 9p^5q^4 + 18p^6q^3 + 9p^7q^2$$
$$Lik(122) = 18p^2q^7 + 36p^3q^6 + 18p^4q^5 + 18p^5q^4 + 36p^6q^3 + 18p^7q^2$$
$$Lik(223) = 9pq^8 + 9p^2q^7 + 18p^4q^5 + 18p^5q^4 + 9p^7q^2 + 9p^8q$$
$$Lik(333) = q^9 + 3p^3q^6 + 3p^6q^3 + p^9$$

(11)

With these likelihoods and the seven vectors of *m*'s, it is straightforward to calculate the expected $m^1$ and $m^2$ and their sample variances. For a 3x3 binary DM with 0.2 marker frequency, e.g., they are **2.0400**(0.48000) and 0.33280(0.23680), respectively, i.e., nearly identical to the values from numeric and DM-to-PM simulations presented above.

For the 4x3 and 4x4 cases the different Ms are 19 and 38, the different DMs are 4'096 and 65'536, and the likelihoods are in Formulae (12) and (13), respectively (in the Supplementary Materials since unwieldy). Again for the *p* 0.2 case, the 4x3 and 4x4 algebraic and simulated $m^1(m^2)$ and the two sample variances are nearly identical, namely **2.0376**(0.51456);0.22326(0.14005) vs. 2.0399(0.51430);0.22417(0.14001) and **2.7200**(0.68608);0.29867(0.23919) vs. 2.7199(0.68616);0.29866(0.23944), respectively. Working out the formulae by *brute force* becomes impractical when DM columns and rows increase. (As a curiosity, note that the 2x2, 3x3, and 4x3 cases generate the prime numbers 3, 7, and 19).

Obviously, the main challenge for say developing a recursion to generate the formulae of these likelihoods is that many distinct DMs map to the same vector of *m*'s so that one needs a smart algorithm to find every DM that yields the targeted vector and then add up the identified DMs' probabilities to get the likelihood given the vector.

Indeed, the perhaps most immediate empirical use of these likelihoods is calculating the likelihood of the $H_o$ given the observed vector of the PM from the DM at stake, a smaller task that still requires finding all the DMs that generate the vector and summing up their probabilities. However, it appears hard to devise an algorithm to do this, e.g., by modifying systematically the original DM.

As a challenge to the reader, two 4x4 binary DMs are included here that yield the same vector (1 1 2 2 3 3) but for which a procedure to get from one to the other is not obvious. The rows of the two DMs are 0 0 0 0, 0 1 0 0, 0 0 1 1, and 1 0 0 0 vs. 0 0 0 0, 0 1 0 0, 0 0 1 1, and 0 1 1 1. Permuting the vertical and horizontal order of rows and columns, respectively, plus toggling column by column the markers of every distinct subset of columns yields many DMs with identical vector, but in these two DMs it is the bottom rows (not columns) that have toggled markers (1 0 0 0 vs.0 1 1 1), which in most cases triggers a drastic change in the vector.

These likelihoods for very small DMs apparently do not shed light on why PAS and dvPAS have similar and much higher power to detect trinary than binary models of *n*-column association and *n*-IV DV association, respectively (see power section further below). In the trinary 3x3 case in Formula (14), e.g., there are 13 distinct vectors (another prime number), i.e., the probabilities of 19'683 DMs ($3^9$) end up subsumed into 13 likelihoods, whereas in the binary 3x3 case 512 DMs map onto 7 vectors.

It is likely that in general in the trinary case many more DMs map onto only nominally more numerous vectors than in the corresponding binary cases. In other words, in the trinary case the vectors may in general fail to distinguish between many more distinct DMs than in the binary case. This lower resolution would suggest much lower PAS and dvPAS power in the trinary case, contrary to what is observed below. Therefore it seems unlikely that improving how PAS and dvPAS exploit the pairwise matrix --not even through generalized use of exact likelihoods of the $H_o$-- will increase by much the resolution of PAS and dvPAS in binary DMs.

Nonetheless, the likelihoods suggest that PAS and dvPAS resolution in trinary DMs could become even higher if one increased the number of distinct vectors onto which the subsets of distinct DMs map, i.e., if one increased the vectors "complexity" (their number) in a way that lets fewer DMs map onto each vector.

However and as intimated when discussing Fig.6, the use of ratios between the likelihoods of specific $H_1$'s and/or the $H_o$, should both be fully viable computationally and boost power over that of the moment-based PASs and dvPASs. Calculating the likelihood of a model that postulates groups of associated and random columns given an observed distribution of matches at non-focal-column spots in pairwise comparisons should be as straightforward (and tedious) as calculating the likelihood of the $H_o$ given the distribution of matches in the pairwise comparisons in the PM of a null DM, as was done above by *brute force* for very small null DM *sans* the focal-column complication.

Indeed under such an $H_1$ the number of non-focal-column matches in a pairwise comparison is a simple sum of the matches at non-focal random columns and those at non-focal model columns (with column order being ignored). Estimating the maximum-likelihood association parameters for the group of model columns postulated to be in association, the maximum-likelihood number of model and random columns, etc, will be of course necessary and will be facilitated by the i.i.d. nature of the different levels of association under PAS and dvPAS, which will be shown and discussed further below. The ratios of these likelihoods can be expected to behave normally.

Concluding, many vertical orders of matches in PM columns are impossible because this order is straight-jacketed by the partial repetition in every PM column of the vertical order of the markers in the corresponding DM column in the form of vertical tracts of matches that, in the binary case, repeat the order of the markers in the DM column at stake (sometimes boolean-toggled) as they are successively apically truncated by one match/mismatch (Fig.16).

We saw further above that CHIx is not $\chi^2$-distributed but becomes so if one forces matches and mismatches to appear independently and randomly vertically across PM columns. The order destroyed by this randomization is due to said successively shortened repetitions of tracts of matches in real PMs, given that at least in the first tract formed by the comparisons of the row-1 marker vs. the others, the order of matches is only constrained by marker frequencies, since it is identical (in the binary case) to the random vertical order of the markers at rows 2 to R in each DM column at stake. These repetitions in the PM of apically one-marker-decremented DM tracts (sometimes boolean-toggled) must be crucial for PAS and dvPAS power.

## Type I error Study

As intimated above, studying the type I error of the P values of PAS and dvPAS is complex because their main use is to assess for several if not all of the columns of a data matrix DM with many columns, the associations of each examined column with other columns in the DM and the DV association of every IV by itself or in synergy with other IVs, respectively. Therefore multiple tests are a major aspect of the type I error of PAS and dvPAS.

Furthermore, because every column and column subset within a DM can be randomly associated with any other, PAS and dvPAS type I error needs to be studied both in DMs where columns are independent from each other and in DMs that mix independent columns and subsets of associated columns which under the $H_o$ can become randomly associated with others subsets. The latter indeed is approximately the case of chromosomal data in which mutations are said to occur in "blocks" characterized by strong column associations within blocks and random associations between blocks.

Ideally the P values of the PAS and the dvPAS of a given column over multiple null DMs should behave like independent uniform-(0,1)-distributed random numbers. But the PAS and dvPAS P values of several columns from a given null DM are not expected to be independent from each other because same-DM columns must show random levels of association with each other to which PAS and dvPAS react.

Therefore PAS and dvPAS P values from the same null DM must be "randomly" positively correlated with each other, even when individually they are uniform-(0,1)-distributed. Taken together, indeed, the random associations among the columns of a DM are themselves a random variable that is averaged away when PAS and dvPAS P values from multiple DMs are pooled. Therefore the "joint" type I error of the PAS and dvPAS P values of subsets of columns from an individual null DM needs to be studied too.

For the same reason, also the probability of observing any PAS and dvPAS P value no larger than the Sidak cutoffs for a "family of $n$ tests" when $n$ such P values are estimated for a subset of as many columns in a given DM is expected to differ to some extent from the probability of observing such a cutoff among $n$ independent uniform-(0,1)-distributed random numbers. When a family of 100 independent individual tests are carried out, e.g., the Sidak cutoff that delivers say a 0.1 "test-family" type error is equal to $1 - (1 - 0.1)^{1/100}$, i.e.,

~0.001053. Ideally under the $H_o$ therefore one or more columns with PAS or dvPAS P value$\leq$0.001053 should be found in ~10% of say 10'000 100-column "test families" carried out in 10'000 independently generated null DMs when every single-column test generates an independently uniform-(0,1)-distributed P value.

Finally, in a DM whose columns form independent blocks of strongly associated columns, the distribution of any function of several same-DM PAS or dvPAS P values should depart from that of the same function when this is fed independent uniform-(0,1)-distributed random numbers. As an example of such a function, the product of the PAS and dvPAS P values of pairs of columns in a DM is also studied below. The product is referred to as the "2-P value product".

The type I error of the highest-power CHIx's, Mom$^n$'s, dvCHIx's, and dvMom$^n$'s is studied in this section. In the case of PAS these are i) CHIx-M, that can detect any type of association and happens to be the most powerful PAS of this kind when detecting the columns of randomly encountered model DMs of significant (*sensu* Fig.3) 5- and 10-column association; and ii) Mom$^n$i$_Z$, the moment-based PASs that among the PAS considered here have the highest power to detect such model columns.

The studied dvPASs are i) dvCHIx-ijkl, the most powerful PAS when detecting IVs in randomly encountered models of significant (*sensu* Fig.3) $n$-IV association with a DV when marginal effects contribute, and ii) dvMom$^1$ik$_Z$ and dvMom$^2$i$_Z$, that have highest power when the randomly encountered models to be detected lack marginal effects (M&M). Details about these PASs and dvPAS are given in the "Introduction to Method" section above.

The focus is first on the type I error of PAS and dvPAS P values in DMs generated under the $H_o$ that columns are independent from each other (and the DV if present; M&M). To pose a more realistic and possibly harder challenge to PAS and dvPAS, the columns of binary DMs have o12345 minor marker frequencies rather than say binary-marker frequency 0.5 (M&M; see also power section below). Trinary DMs are generated by turning the applicable binary frequencies $p$ and $q$ at each column ($q$=1-$p$) into the trinary H&W frequencies $p^2$ $2pq$ $q^2$. With one exception, the presented PAS and dvPAS P values are estimated by permuting vertically the markers at the focal column and the binary markers ("statuses") at the DV column, respectively.

A second focus is on the type I error of PAS and dvPAS in null DMs in which all columns (except the DV when present) belong to blocks of strongly associated columns that under the $H_o$ are independent from other blocks (and the DV if present). The generation of null DMs with such blocks is described in detail in the M&M and summarily when presenting results below.

**Type I error of the PAS P values of the columns of a null DM.**

A P value for the PAS of a column can be readily obtained by permuting vertically the markers in that column. Such P values quantify how strongly the markers at the column are associated with others at other columns, to the extent that the PAS of interest reacts to such associations (see Form.(1) and (2) and Fig.1a,b). With such P values for a given column from multiple independently generated null DMs, one can estimate the type I error of the PAS P values at that column. And with such P values for many columns from a single null DM, one can estimate their same-DM type I error.

Below these two types of type I error are studied for the P values of CHIx-M and Mom$^1$i$_Z$ in DMs generated under the $H_o$ that the vertical order of the markers at DM columns is independently vertically random across columns. As intimated, these two PAS were chosen because the power assays further below show them to have greatest power to detect randomly encountered models of significant $n$-column association. Details about these PASs are given in the "PAS algorithms in depth" section above. Only two punctual results are included with null DMs whose columns form blocks.

**Type I error of the PAS P values of an individual column across multiple null DMs.** Figure 17 shows the type I error of the CHIx-M and Mom$^1$i$_Z$ P values of individual columns over multiple independently generated null DMs in which columns are independent from each other. In such a DM, the type I error of a focal column's PAS P value is mainly due to the cumulated effect over the entire DM of random associations of said column with subsets of other columns, including with other columns individually. Studied are two sets of 1'000 binary and trinary null DMs, all of which have 2'000 rows and 1'000 columns whose marker frequencies cycle according to the o12345 scheme (M&M).

The upper c.d.f.s in Figure 17 show the type I error of the CHIx-M and Mom$^1$i$_Z$ P values at each of the first ten columns of said null DMs. Both the binary and trinary c.d.f.s tend to match the c.d.f. of a uniform-(0,1)-distributed random number (the diagonal), which is the expectation for P values from independent tests that each generates correct type I error under the $H_o$. The c.d.f.s of Mom$^1$i$_Z$ P values show slightly more noise.

The middle plots show, on the horizontal axis, the user-specified amount of type I error to be tolerated when families of ten independent individual-column tests are carried out at once under the $H_o$. On the vertical axis is the frequency, over the 1'000 binary and trinary null DMs, of null DMs in which at least one of the ten tested columns has a P value no larger than the Sidak cutoff for 10-test-family type I error equal to 0.01, 0.05, 0.1, and 0.2, respectively, i.e., for overall rejection of the $H_o$ at the 99, 95, 90, and 80% significance level. The cutoffs are ~0.0010, 0.0051, 0.010, and 0.022.

The resulting c.d.f.s match the diagonal in both the binary and trinary cases. The departures from the diagonal are like those observed when replacing the PAS P values with as many uniform-(0,1)-distributed random numbers (not shown). The results show that same-DM CHIx-M and Mom$^1$i$_Z$ P values are nearly independent from each other under the $H_o$, at least in the sense of their being accurately corrected for multiple tests with the Sidak method.

The expected non-independence of same-DM PAS values under the $H_o$ is not apparent in the bottom plots of Figure 17 either. These plots show the c.d.f.s of the product of two same-DM P values for every possible pairing of five columns with different marker frequency (15 pairs). The c.d.f.s match those of the products of independent uniform-(0,1)-distributed random numbers that are shown as benchmarks and that discard products involving random numbers no larger than various thresholds to mimic the finiteness of P value permutations. The c.d.f.s of the products of CHIx-M and Mom$^1$i$_Z$ P values are similar to each other and do not appear to react to marker frequencies.

**Type I error when all of the columns of an individual null DM are tested**. As mentioned, the PAS values of the columns of a null DM vary together up and down across null DMs, an effect that is erased when P values from several independently generated null DMs are pooled. Figure 18 showcases these same-DM effects by plotting ten separate same-DM distributions for the CHIx-M and Mom$^1$i$_Z$ P values of all of the columns in ten binary and ten trinary 2'000-row 1'000-column null DMs.

Panels (a,b) and (e,f) in Figure 18 show that the same-DM c.d.f.s of the P values of the ten binary and the ten trinary null DMs tend to match the diagonal but the c.d.f.s of each group of ten sets of P values pooled match it better. Panels (c,d) and (g,h) show that in individual null DMs the P values of the 200 columns with marker frequency 0.1 and 0.5, respectively, tend towards the diagonal too (if more noisily), with the fit being again better when pooling by frequency group. CHIx-M noise is larger in the trinary case.

The results in the bottom panels of Figure 18 show that the distribution of the cartesian (pairwise) product of 1'000 same-DM CHIx-M and Mom$^1$i$_Z$ P values in the ten binary and ten trinary null DMs fits that of the cartesian product of 1'000 independent uniform-(0,1)-distributed random numbers, with CHIx-M noise vs. Mom$^1$i$_Z$'s being again larger in the trinary case. Therefore if one is interested a priori in only the smallest cartesian products of the PAS P values of a DM and one wants to derive their null distribution, one can assume safely that same-DM P values under the $H_o$ behave approximately like independent uniform-(0,1)-distributed random numbers.

The last result is disorienting since if anything it is same-DM PAS P values that should be correlated to some extent rather than PAS P values of given columns across multiple null DMs pooled. Note also that the c.d.f.s of reference products of independently uniform-(0,1)-distributed random numbers in the bottom panels in Figure 18 and Fig.17 are of different nature, since in the products in Fig.17 there are no massive reappearances of random numbers, unlike in the cartesian products of Figure 18.

What is most relevant to experimentalists, however, are the c.d.f.s of same-DM PAS P values. They show that same-DM PAS P values behave under the $H_o$ approximately like independent uniform-(0,1)-distributed random numbers so that when one tests several or all columns of a given many-column null DM the Sidak method should correct the resulting P values quite accurately for multiple tests.

**Type I error of dvPAS P values**

Independent-variable (IV) columns that are strongly associated in any way with a dependent-variable (DV) column in a DM can be flagged if one or more of their dvPASs have low P values. By permuting vertically the markers (categories, statuses) in the DV, a P value can be estimated for the dvPAS of any IV of interest. This P value quantifies how strongly the IV by itself and in synergy with others is associated with the DV (see Form.(4) and (5) and Fig.1c,d).

With such P values for a given IV from multiple independently generated null DMs, one can estimate the type I error of the IV under the $H_o$. And with P values for many IVs from a single null DM, one can estimate their same-DM type I error. However, also the false positives (at known random non-model IVs) in the power assays further below shed additional indirect light on the null behavior of dvPAS scores.

Here the type I error of the P values of dvCHIx-ijkl and dvMom$^1$ik$_Z$ is first studied in DMs generated under the H$_o$ that the vertical order of the markers at each IV is independent from the order at both the other IVs and the DV. This is also done under the H$_o$ that in null DMs all IVs belong to blocks of strongly within-block associated columns that are independent from other blocks and the DV.

As mentioned above, the focus is on the type I error of the P values of these dvPASs because they are among those with greatest power to detect the mentioned randomly encountered models of significant $n$-IV association with a DV with and without marginal effects, respectively (see power section below). Details about the various dvPASs are given in the "PAS algorithms in depth" section above.

**Type I error at individual IVs over multiple null DMs**. Figure 19 presents the type I error of the P values of the dvCHIx-ijkl, dvMom$^1$i, and dvMom$^2$i's of the DV and nine IVs across 1'000 2'000-row 1'000-IV binary and trinary null DMs generated under the H$_o$ of independence of all columns; dvMom$^1$i$_Z$ and dvMom$^2$i$_Z$ are used in the trinary case. The top c.d.f.s tend to match the diagonal regardless of the IVs' marker frequencies.

The middle panels show that the Sidak method corrects quite accurately for multiple tests when the dvPAS P values of nine IVs are estimated in each DM and those DMs are reported where any of the nine IVs' P values is no larger than four standard Sidak cutoffs. The departures from the diagonal are like those observed when dvPAS P values are replaced with independent uniform-(0,1)-distributed random numbers (not shown).

The bottom panels in Figure 19 show null c.d.f.s for each of the products of two same-DM P values from the 15 possible pairings of five IV with different marker frequencies, over the null DMs (i.e., 1'000 2-P value products for each pairing). The 2-P value products behave like the products of uniform-(0,1)-distributed random numbers that are shown as benchmarks.

**Type I error when all of the IVs in a null DM are tested.** As it was the case above for PAS P values, when evaluated as entire sets also same-DM dvPAS P values can be expected to show heterogeneity across null DMs that is erased if one pools the P values of the null DMs. This heterogeneity is explored here by plotting the same-DM c.d.f.s of the P values of every IV in each of ten binary and ten trinary null DMs.

Panels (a-b) and (e-f) in Figure 20 show same-DM c.d.f.s for the P values of dvCHIx-ijkl and dvMom$^1$ik$_Z$ when every column in each of the ten binary and ten trinary null DMs, respectively, is evaluated and c.d.f.s are computed for each DM. The c.d.f.s tend to match the diagonal, with the fit being better when the sets of P values of individual PMs are pooled. The c.d.f.s of dvCHIx-ijkl in (a,e) and (c,g), respectively, are less noisy than those (not shown) of dvMom$^1$i and dvMom$^1$i$_Z$ P values, respectively. Panels (c,d) and (g,h) show that in every null DM, the P values of the 200 IVs with marker frequency 0.1 and 0.5 tend also to match the diagonal (if more noisily) and again better when P values are pooled by frequency group. Noise is more marked for dvMom$^1$ik$_Z$ P values. There are no clear differences between binary and trinary DMs. Therefore these c.d.f.s tend to match the diagonal so that one can use the Sidak correction also when one estimates PAS P values for every IV in a null DM with independent IVs.

The bottom panels in Figure 20 show both that the c.d.f. of the cartesian product of the 1'000 dvCHIx-ijkl and dvMom$^1$ik$_Z$ P values from each of the ten binary and trinary null DMs varies markedly across DMs (e.g., more than for CHIx-M and dvMom$^1$i$_Z$ in Fig.18). This variation increases if the number of permutations used to estimate P values are increased by about one order of magnitude (tried only in the trinary case; not shown), pointing to real across-DM variation in the overall level of DV association shown by the IVs of a DM. Therefore, at least with respect to their cartesian product, same-DM dvPAS P values are not strictly independent from each other. The implications, if any, of this inter-DM variation for data-mining work are unclear.

**Type I error of PAS P values when columns form blocks of associated columns**

DMs can include subsets of columns among which there are strong "background" associations of no interest. However, PAS cannot distinguish such associations from those of interest. In an L-column DM with, e.g., L/2 column pairs in "perfect" 2-column intra-pair association that are randomly associated with other pairs, one could be interested in eventual inter-pair associations rather than the intra-pair ones. PAS, however, would react very strongly to the intra-pair associations, i.e., the PASs of all the columns would be inflated and their P values would be very low, even if one forced the pairs to show no association with each other.

What is then the correct type I error when under the H$_o$ every PAS in a DM is expected to be extreme and every PAS P value is expected to be very small? In the example with L/2 pairs above, all PAS P values must be equally low if the pairs in perfect association show no association with other pairs. But when inter-pair associations are random, the PAS P values of columns from pairs involved by chance in strong inter-pair associations should be additionally reduced.

Type I error in such DMs is therefore still about the lowest P values observed under the $H_o$ that specifies how random associations among pairs originate, even if all P values are very small. One would simulate the null model to generate null data in which one can estimate the null distribution of the PAS P value of interest. And since estimating very low P values is computationally impractical, one would rather estimate directly the empirical P value of the PAS of every column at stake over the simulated null DMs.

An extreme $H_o$ is studied here under which adjacent columns form 100-column blocks with strong intra-block column associations and these blocks are independent from other blocks; the mixed case with independent co-occurrence of individual columns and column blocks is studied towards the end of the power section. As intimated, an empirical P value of a column is used that is estimated by comparing the PAS of the column in the DM of interest to the PAS of the same column in additional null DMs generated under the same $H_o$.

Therefore these P values are not estimated by shuffling only the vertical order of the marker sequences in the block with the tested column (which would follow the logic adopted so far of permuting only the markers in the focal column), since doing this would require a major rewrite of the PAS code's shuffling of focal-column markers. However, most likely these results would be very similar if for P value estimation one shuffled only the vertical order of the sequences in the block with the evaluated column of interest.

To generate an N-row H-block DM with 100-column blocks under said $H_o$, the N sequences of each block are sampled with replacement from a single source set of 116 100-marker sequences of human LPL mutations (Nickerson et al.1998). The rows of every block are shuffled independently vertically before the blocks are joined laterally (M&M). These multi-column blocks are intermediate between independent columns and independent column pairs with complete intra-pair association. For the trials with block-structured trinary "diploid" DMs, one samples from the set of all possible pairings of the LPL sequences after changing the paired 00s, 01(10)s, and 11s into the trinary markers 0, 1, and 2 (M&M).

As an additional challenge, the sequences of one or more blocks are "force-sampled" to let the sampled sequences show a pre-specified marker frequency at the designated "anchor" column in each block (column nr.49). Assessing the effect under the $H_o$ of such forced sampling is relevant to the power assays further below, but the type-I-error results presented here are also confirmed using null DMs generated without forced sampling.

No work is shown about the Sidak-corrected "family of *n* individual tests" type I error when the PAS P values of *n* columns from a block-structured DM are estimated, but below such type I error is presented when a group of several IVs is tested from a DM that has a given type of block structure. No work is presented either about uncorrected same-DM PAS type I error when all the columns are tested in a null DM with blocks, albeit the work on false positives in the dvPAS power section further below is related.

**At chosen columns, over multiple null DMs.** Figure 21 shows the type I error c.d.f.s of the empirical P values introduced above for the CHIx-M of chosen columns in 10'000 binary and 10'000 trinary null DMs. The P values are estimated at every chosen column in each null DM by scoring how often the column's CHIx-M matches or exceeds the CHIx-M of the same column in the other 9'999 null DMs.

Every DM in the set has ten blocks each of which consists of 2'000 100-marker sequences. This gives 10'000 independent null P values for each studied column. P values are estimated for five of the ten (different-block) pos.49 "anchor" columns and five additional "anchor-linked" columns (cols. 1, 22, 50, 63, and 88; M&M) in the first block with a force-sampled anchor column. Identical results were obtained without force-sampling any of the ten blocks.

Unsurprisingly given how P values are calculated, the null c.d.f.s at the top of Figure 21 match the diagonal perfectly. This is so for both the P values of the five (different-block) anchor columns and the P values of the five additional 1st-block columns.

The bottom c.d.f.s in Figure 21 are for the products of two same-DM PAS P values, namely for every possible pairing of the five different-block columns and every possible pairing of the five anchor-linked columns. The former c.d.f.s fit those of the product of two independently uniform-(0,1)-distributed random numbers that are shown for comparison. However, the 2-P value products when pairing the five 1st-block columns are up to one order of magnitude richer in small products than are the products of random numbers.

This excess in small products is likely due to the fact that same-block LPL columns are strongly

(chromosomally) associated with each other so that their PAS P values must be positively correlated to some extent, even if this does not cause the c.d.f.s of the P values of individual columns to depart from the c.d.f of the uniform-(0,1) distribution, as we saw above.

**Type I error of dvPAS P values when IVs form blocks**

When a DM has a DV column and its IVs are associated into IV blocks that are independent from the DV and other blocks, one can nonetheless estimate dvPAS P values and use them to flag IVs associated with the DV. The P value of the dvPAS of an IV is indeed estimated by permuting vertically the binary markers in the DV while keeping everything else unchanged. This makes it impossible for the estimated P values to be affected by background IV associations that do not involve the DV.

Studying the type I error of dvPAS P values is therefore straightforward: Under the $H_o$ of block sequences being randomly associated with both the sequences at other blocks and the (binary) markers at the DV, the c.d.f. of the P value of any individual IV and the DV should match that of a uniform-(0,1)-distributed random number. However, under this $H_o$ the joint distribution of the dvPAS P values of two or more IVs that belong to the same block should be positively correlated due to intra-block IV associations.

Generating an R-row H-block null DM for the assays of type I error presented below requires creating a DV column with R/2 0s and R/2 1s (i.e., marker frequency is 0.5 at the binary DV), sampling independently and randomly (with replacement) the R marker sequences for each block of IVs, and then joining the DV column and the H blocks of sampled sequences laterally. This procedure allows random marginal effects to arise at IVs (M&M). Also here the sampled binary block sequences are those in the aforementioned 100-marker LPL source set and the sampled trinary block sequences are the pairwise comparisons of binary LPL sequences (M&M).

Two procedures are used when sampling block sequences. Under the first one the first four blocks are force-sampled so their pre-specified central "anchor" columns have marker frequencies o1524 (columns 50, 150, 250, and 350 in a DM where the DV is the first column; M&M). Under the second scheme the sequences of the first block are force-sampled in a way that let four (rather than only one; M&M) pre-specified "anchor" columns have markers with frequency o1524 that form random combinations at said columns.

Forced sampling of the first four blocks and specially of the four anchor columns of the first block, respectively, makes the diversity of block sequences different across blocks. This is relevant to the power assays further below where both procedures of forced sampling are tried. Type I error trends are shown below to be identical with one-column forced sampling and random sampling, but with four-column one the trends differ.

Type I error of dvPAS P values is studied first at chosen columns across multiple null DMs with IV blocks and then over all IVs of individual null DMs with blocks. As intimated further above, the focus is on the P values of dvCHIx-ijkl as well as those of dvMom$^1$i and dvMom$^2$i in the binary case and dvMom$^1$i$_Z$ and dvMom$^2$i$_Z$ in the trinary case, i.e., the dvPAS P values that further below show highest power to detect randomly encountered models of significant association of a DV with $n$ IVs with and without marginal effects contributing to the significance.

**At chosen IVs, over multiple null DMs with IV blocks.** Figure 22 shows the distribution of the dvPAS P values of chosen IVs across null DMs whose IVs form blocks. The distributions are over 2'000 1'000-row 100-block DMs generated under the $H_o$ of the DV column and the 100 100-IV LPL blocks being all independent from each other (100 is the number of LPL blocks used in the power assays further below). The focus is on nine IVs in the first LPL block of which four are anchor IVs force-sampled to have random combinations of o1524 binary-markers or the H&W thereof. The other five first-block IVs are the "anchor-IV-linked" IVs described in the M&M.

The top plots in Figure 22 show that the c.d.f.s of said IVs' dvPAS P values tend to match the diagonal. The middle plots show the occurrence of null DMs in which any of the nine IVs has a nominal P value no larger than the Sidak cutoffs for 0.01, 0.05, 0.1, and 0.2 type I error when a "family" of nine independent individual-IV tests is carried out. The observed Sidak family-level type I error sinks from about correct to clearly below expectation as the Sidak cutoff increases, and does this more strongly for both dvMom$^n$i's than dvCHix-ijkl and more in the binary than the trinary case, with the latter being consistent with intra-block associations being stronger in the binary LPL blocks than in the trinary wLPL blocks.

In additional non-shown results when placing every anchor IV in a different block, the observed Sidak type I error when testing four of these IVs as a "test family" is correct (on the diagonal), consistent with the results in Fig.19 with independent IVs. In this case, however and surprisingly, the Sidak correction of the P values of the five first-block (non-anchor) IVs is also very accurate (not shown), suggesting that forced sampling of only one anchor IV per block does not exacerbate same-block marker associations enough to let the Sidak correction of the P values of the assayed same-block IVs become inaccurate.

The bottom plots in Figure 22 show c.d.f.s of the products of two same-DM P values for every pairing of the mentioned 4- and 5-IV subsets. The c.d.f.s of dvMom$^2$i$_Z$ products show too many values compared to the c.d.f.s of the products of two independently uniform-(0,1)-distributed random numbers that are shown for comparison. This effect is much weaker both with only one anchor IV force-sampled per block and without any forced sampling (neither shown).

One can conclude that both the Sidak-corrected type I error when testing families of same-DM IVs and the c.d.f.s of the 2-P value products of same-block dvPAS P values can react to the non-independence of same-block IVs when the non-independence is augmented by force-sampling multiple same-block IVs.

However, neither the Sidak-corrected type I error nor the 2-P value c.d.f.s react markedly to the "native" level of marker non-independence in LPL sequences. This needs confirmation with other types of blocks, but even if confirmed it would not protect users from being misled when unbeknownst to them there is harsh forced sampling of multiple linked IVs like in Figure 22.

The power assays further below will show that only some pairs of the five same-block IVs at stake here are in strong 2-way (intra-block) association dvPAS-speaking. And yet, these assays will also show that dvPAS power to detect 4- and 8-IV DV associations in DMs with 100 100-IV blocks (10'000 IVs total) is about the same as when DMs have somewhere between 100 and 1'000 independent IVs, i.e., the assays will show that the partial non-independence of the markers in the LPL sequences boosts dvPAS power dramatically.

**Over all of the IVs of a null DM with blocks.** The c.d.f.s of same-DM dvPAS P values should vary across null DMs whose IVs form blocks more than when the IVs are independent. In the extreme case of perfect association of the IVs in each block, whole-DM noise becomes quantitatively equal to that of a DM with as many independent IVs as the number of independent perfect-association blocks of the DM.

This effect is assessed here by estimating the P values of dvCHIx-ijkl as well as of the dvMom$^2$i and the dvMom$^2$i$_Z$ (binary and trinary case) of every IV in each of ten binary and ten trinary 100-block null DMs, and then plotting a separate P value c.d.f. for each DM (100-IV blocks; 10'000 IV per DM). IVs have random marginal effects at the DV.

Panels (a,e) in Figure 23 show that dvCHIx-ijkl c.d.f.s vary around the diagonal much less than those of the two dvMom$^2$i's in panels (b,f); this is so also for the c.d.f.s of the two dvMom$^1$ik's (not shown). Non-surprisingly, the c.d.f.s of all the P values pooled are closer to the diagonal. The c.d.f.s of the 4'000 IVs in each DM with lowest and highest marker frequencies in panels (c,g) vs. (d,h) show similar trends but more noisily, specially those of dvMom$^2$i$_Z$ P values in trinary DMs. The c.d.f.s of either set of P values pooled are again closer to the diagonal. The P values of dvMom$^2$i$_Z$ P values in the trinary case are non-conservative (e.g., 20% occurrence of nominal P values no larger than 0.1) due to noise since only ten DMs are studied; false positives are indeed fully correct in the extensively replicated results with dvMom$^2$i$_Z$ P values in the power section further below.

In the bottom of Figure 23 are the c.d.f.s of the cartesian product of the same-DM P values of dvCHIx-ijkl and the two dvMom$^2$i's from the two sets of ten DMs each. Specially the dvMom$^2$i$_Z$ c.d.f.s are much noisier than those of same-DM PAS P values presented in Fig.20 for 2'000-row 1'000 IV DMs without IV blocks. But all c.d.f.s vary around the c.d.f.s of the cartesian products of 10'000 independent uniform-(0,1)-distributed random numbers that are shown as benchmarks.

In the top of Figure 23, the random departures from the diagonal by the same-DM c.d.f.s are clearly stronger than when all IVs are independent (Fig.20), despite the block-structured DMs of Figure 23 having ten times as many IVs. Furthermore, in Figure 23 the departures from the diagonal by individual same-DM c.d.f.s are less marked than in DMs with only 10 100-IV blocks (rather than 100 such blocks; not shown). Therefore in DMs with say more than 400 LPL blocks (40'000 nominal IVs), same-DM c.d.f.s are likely to depart from the diagonal as little as when DMs have 1'000 independent IVs.

Using most powerfully the Sidak correction when one tests every IV in a block-structured DM requires knowing the number of independent tests that are performed, which in DMs with blocks is the number of effectively independent blocks rather than the number of IVs. Empirically, the number of effective blocks should be impossible to estimate without external information (like genetic crosses) or having real replicates of the DM of interest, since it is hard to imagine that a non-parametric trick will ever manage to tell apart within-block IV associations from between-block ones arisen by chance under the $H_o$.

The number of effectively independent blocks is known for the DMs used in the assays of PAS and dvPAS power further below, but nonetheless the type I error (and the power) of naively or block-informedly Sidak-corrected PAS and dvPAS P values was not studied since estimating very small P values with permutations is computationally very onerous.

Above it was shown that across null DMs with 100 or more IVs, PAS and dvPAS type I error are uniform-(0,1)-distributed. Therefore a researcher intending to test several columns in such DMs could simply declare his/her adherence to the rule of accepting say at most 5% of the IVs that he/she will ever test as potential type I error at the 95% level and commit therefore to using a nominal P value of 0.05 as cutoff when identifying a tested IV as "publication significant at the 95% level". This implies embracing the concept of a lifetime per-tested-IV per-researcher type I error rate rather than a rate per researcher per "family of tests" published by a researcher.

Aside of being simpler, the per-test per-researcher approach does away with the arbitrariness of *ad hoc* definitions of the "families of tests" at hand when one publishes several tests, e.g., definitions that invoke natural-world features. A 5% per-test lifetime false-positive rate at the 95% level, e.g., will always result in a 5% lifetime per-test type I error, regardless of whether during one's career the null DMs that one ends up studying are i) null DMs in which all IVs are perfectly 2-way associated (so that same-DM IVs always have the same P value, one that is uniform-(0,1)-distributed across null DMs); ii) null DMs in which all IVs are independent from each other, so that the c.d.f. of every same-DM P value tends to match the c.d.f. of the uniform-(0,1) distribution independently from the other P values; or iii) null DMs with mixtures of the previous two extremes.

These last remarks and proposals are meant to help the reader engage more critically and from more angles the power section below, and in particular to help him/her consider how one's adherence to a lifetime per-tested-IV type I error rate translates into a lifetime per-tested-IV false-positive rate when DMs have mixtures of model and non-model IVs (which makes genuine false positives possible), in particular when DMs have block structure.

## Study of the power and false positives of the P values of PAS and dvPAS.

PAS and dvPAS were conceived and developed as methods for flagging columns that are associated with others in a DM and IVs that individually and in synergy with others are associated with the DV of a DM, respectively. The main DM envisioned has many columns most of which are random "background" columns presumably of no interest. For this reason the power of PAS and dvPAS is studied below for a variety of models of multi-columnar association as a function of the number of non-model background columns.

Furthermore, since PAS and dvPAS can react to associations involving many columns, including random and eventual background associations among non-model columns, the study below of the power of PAS and dvPAS P values has as second main focus the generation of **false positives** as a function of the level of PAS and dvPAS detection of the columns of the models of multi-columnar association whose detection is attempted.

False positives are a cousin of type I error that cannot be studied in null data and thus they were not studied in the preceding section. Indeed the phrase "type I error" was used above only to refer to low-P value columns observed under the $H_o$ that all of the columns or IVs (or blocks thereof) of a DM are independent from each other and the DV when applicable.

Therefore the phrase "false positives" refers below only to low P values of non-model columns in a DM in which other, "model" columns are forced to be strongly reciprocally associated or DV-associated. Additionally, it is stipulated that only columns that are truly independent from model columns under the $H_1$ can show genuine false positives (below).

As it was the case for type I error above, the consequences for power and false positives of the presence of background associations within and between model and non-model columns are of great interest here too, because a main intended use of PAS is to find "disease mutations" in chromosomal data in which causal and non-causal mutations are expected to be associated (linked) with each other when they are close in the DNA sequence, forming blocks that are more or less independent from each other.

In such a case, non-model IVs in blocks with one or more model IVs can show low P values that are not legitimate false positives because they are low due to their being in blocks with model IVs.

Albeit interesting, the power of PAS and dvPAS is rarely compared below to that of exhaustive evaluation of every column subset. Indeed when a DM has more than a few dozen columns, this bench mark has no practical relevance since, as mentioned in the Introduction, it is highly unlikely that computers will ever be able to evaluate every one of say $10^{300}$ distinct column subsets.

Moreover, the power edge of exhaustive evaluation over PAS and dvPAS may change when the number of column subsets to be tested explodes in DMs that have very large numbers of columns. Below, indirect light will be shed on this issue when studying the reaction of PAS and dvPAS power to increasing numbers of background columns.

The exploration of the power of PAS and dvPAS here focusses on estimating the number of DM rows that suffice for 60% of the first two model columns or model IVs in the model(s) of interest to show a PAS or dvPAS P value no larger than 0.1, in the presence of increasing numbers of random columns. Below this number of DM rows is called "the 60%-0.1 P value detection sample size" of the model or model set at stake, or simply its **"60% detection sample"**, and the two model columns are called "**the reference model columns**".

In the aforementioned randomly encountered models introduced again below and in detail in the M&M, the two reference columns always have binary marker frequency 0.1 and 0.5, respectively, or the trinary H&W thereof. The two frequencies can make a model column hardest and easiest to detect, respectively (or vice versa), depending on the kind of models at stake (below). But the first two model columns are studied as mutual controls also when studying the detection of models of pure $n$-column association and pure $n$-IV DV association that only involve binary markers of frequency 0.5.

The concomitant exploration of the generation of false positives at non-model columns is also conditional on 60% detection with PAS or dvPAS P value$\leq$0.1 of the first two columns or IVs of the models at stake.

The power assays try to be illuminating as well as explicit both marker-pattern-wise and frequentistico-probabilistically with respect to the models of $n$-column association and $n$-IV association with a DV that are chosen for detection. This is also done because it is unclear how to carry out a systematic parametric exploration of the "model space" given the complexities adumbrated immediately below.

As intimated, the first class of models that are detected are randomly encountered $r$-row $n$-column model DMs that were selected for studying because every one of their columns was found to be significant *sensu* Fig.3 (M&M). The $r$'s of the chosen model DMs make the models detectable in DMs with genomics-plausible numbers of rows. The trends inferred from studying sets of model DMs with various $n$'s allow one to extrapolate the likely power for model DMs with more rows.

The second class of models have fully specified association structure, e.g., pure $n$-way association and extended "in-phase" 2-IV DV association (Figs.1 and 2). As stated, only the detection of binary IVs in extended 2-way DV association could be studied here because it is unclear to MAA how to generate extended pure $n$-way association when $n>2$ and/or when the markers at so-associated columns are trinary or above.

However, the presented assays of the detection of IVs in extended 2-way DV association include observations about detecting a DV with which many IVs are associated; these observations are *de facto* about the PAS power and the accompanying PAS false positives when detecting a focal column (shown right before the Discussion).

The insight gained from studying the power of PAS and dvPAS when detecting models with specific association structure sheds light on the nature of the associations that make more or less detectable the columns of the randomly encountered model DMs.

Some DMs in genomics have already millions of rows and columns, which is too many for the computational resources available for this study. With that many DM rows, model DMs become PAS- and dvPAS-detectable that have many more rows than do the model DMs considered here. These model DMs may include more cases in which many marker combinations subtly depart from expectation and fewer cases in which only few such combinations (in absolute terms) do so but more crassly, than are found among the 50- to 800-row model DMs studied below.

The latter case of fewer, crasser departures favors apparently PAS power (below). But as intimated, the trends inferred from both the results with the various numbers of model-DM rows and model-DM columns and the results with pure $n$-way association and pure $n$-IV DV association, should allow one to project the power of PAS and dvPAS when detecting model DMs with more rows and columns.

Power assays are first done in DMs in which all non-model columns are independent from each other and the DV when applicable. For dvPAS only, assays of power are also done using DMs in which adjacent IVs are associated into blocks.

The results with models simulated individually in a DM are complemented by results with two models co-occurring independently in the same DM, to look for effects on power and false positives that may be due to interference between co-occurring models of similar or different nature and detectability.

When, e.g., a model is detected very strongly by PAS because it causes strong perturbations of matches patterns, these perturbations may interfere with the reaction by PAS to the subtler matches perturbations that in isolation would allow PAS to detect the columns of a second, less readily detected model.

Interference effects may also be caused by co-occurring lower- and higher-order associations which involve a given column or a given intersect of columns. Only the first possibility is studied here using the randomly encountered model DMs whose nature more or less guarantees coincidence and overlapping of signals at individual columns (below; M&M). Contrasts with situations in which the nature of models and model combinations guarantees no overlapping of association signals are also carried out.

As stated, a first focus of the power assays below is about detecting the columns of randomly generated model DMs selected for the assays because every one of their columns was found to be significantly associated with others or the DV *sensu* Fig.3. Therefore the columns in the selected model DMs tend to have P values that are near the maximum tolerated when selecting their model DM. Indeed random DMs in which one or more columns have P values that are much lower than the tolerated maximum are by definition less frequently encountered.

This bias does not affect the probability of sampling $n$-column DMs with pure $n$-column association, because the P values of such columns tend to be the same. However, other types of column association that would be detectable in a given DM may involve columns with disparate P values, some high and others low.

One could argue that it is reasonable not to include in "a model that is detectable in, and described by, a model DM with $r$ rows and $n$ columns" an additional column that would get a very high P value were it to be added to the model DM at stake and tested. However, this sleight-of-hand argument fails when say a model column's low P value is due to many other high-P value columns that cause cumulatively the low P value of said model column, necessitating so the inclusion of the high-P value columns in the model DM.

The low P value of a DV, e.g., may be due to many independent, weakly 2-way DV-associated IVs. This kind of association is systematically absent from the random model DMs sampled here that have a mere five and ten columns and four and eight IVs, respectively. However and as intimated, results are presented before the Discussion about detecting a DV with which many IVs are weakly 2-way-associated. No further such models are studied.

A parametric exploration of the model space was not attempted here both because of the above complexity and because when one expands enough multinomially the rows of a random DM every column must at some point become significant (except of course when in the "random" initial DM every possible marker sequence is present exactly with its expected count).

Finally, one can note that the associations likely to be present and detectable in a real DM from the natural world need not be a random sample of the associations likely to be found in random DMs, even when these DMs have realistic numbers of rows, columns, and markers per column.

Summarizing, the study of PAS power below uses models of i) randomly encountered 5- and 10-column association with o15241(23451) binary frequencies or the trinary H&W thereof (M&M), and ii) pure 2- to 8-column association of columns with 0.5 binary-marker frequency, i.e., pure associations involving the minimum needed number of columns.

As intimated, neither the patterns of extended pure $n$-way association above 2-way nor the diagrams of pure association when there are more than two markers in each involved column are known to MAA and therefore neither could be studied here. As stated, the effect on power of increasing the number of random columns in the tested DM is examined in both cases.

No study of PAS power and PAS false positives is carried out in DMs with blocks of associated columns and no dedicated attempt is made at studying the power of PAS to detect 2-way associations that extend over many columns. However, the study below of dvPAS power to detect a DV in extended 2-way association with many IVs is essentially a study of PAS power and so is the study of dvPAS power to detect a DV associated with IVs that are located in blocks while the DV is not.

The specific PASs whose power is examined are CHIx-M, Mom$^n$i, and Mom$^n$i$_Z$. They are re-introduced below and were introduced above when their inclusion in the type I error section was justified. Details about these PASs are given in the "Introduction to the Method" section.

When studying dvPAS power a first focus is on a) the detection of randomly encountered binary and trinary models of 4- and 8-IV association with a DV when model-IV markers have binary o15241 frequencies or the trinary H&W thereof (M&M; sampling random 16-IV models was too slow computationally); a second focus is b) the detection of models of pure 2- to 8-IV association with the DV, e.g., three IVs that together with the DV form a pure 4-way association (Fig.1b,d); and a third focus is c) the detection of "in-phase" (Fig.2) and "off-phase" 2-way IV associations with the DV that involve up to hundreds of IVs.

The study of (a,b) is done in DMs with increasing numbers of random IVs while that of (c) is carried out in DMs with 1'000 total IVs (problematized in that section). As intimated, the work in (a,b) is partially repeated in DMs whose IVs form blocks (M&M; see also type-I-error section).

In all cases the DV has two binary affected-or-control "status" markers of frequency 0.5, i.e., the marker rows from affecteds and controls are equally numerous in the simulated DMs.

The specific dvPASs whose power is examined are dvCHIx-ijkl and dvKSi as well as dvMom$^1$i, dvMom$^1$ik, and their Z-valued versions. Details about these dvPASs are in the "Introduction to the Method" section. They too are re-introduced below and were introduced above when their type I error was studied.

The reader is warned that until shortly before the end of the Discussion, all results are about the power and false positives of the P values of various PASs and dvPASs under examination, P values that are estimated by permuting the markers at each column at stake and the binary markers at the DV, respectively, albeit for conciseness the text often talks about the power and false positives of the PASs and dvPASs at stake, e.g., as "the power of Mom$^1$i" and "the power of dvMom$^1$ik".

**Power and false positives of PAS P values when detecting columns associated with others in a DM.**

This section presents simulations of the power and false positives of PAS P values. The first sets of models studied are randomly encountered 100-row 5- and 10-column model DMs in which every column is significantly associated with at least another column *sensu* DV test in Fig.3 used successively at each column (M&M and above).

The model DMs are expanded multinomially to a desired number of rows and then PAS-tested in DMs that have increasing numbers of independent random columns. Model-DM columns and added random columns have o15241(23451) and o12345 binary-marker frequencies, respectively, or the trinary H&W thereof (M&M). The three PASs whose P values are studied are CHIx-M, Mom$^1$i, and Mom$^1$i$_Z$. P values are estimated using at least 100 permutations.

The 60%-0.1 P value detection samples for the two aforementioned **reference model columns** pooled and the accompanying false positives at random columns are studied as functions of i) the number of random columns in the DMs in which the detection occurs, ii) the number of model-DM columns, and iii) the number of model-DM rows. As intimated, the latter number may be negatively correlated with the extent to which the overall departure from expectation of a model DM is dominated by big departures of only few of the marker sequences in the model DMs.

This work is repeated by letting the marker sequences of two independently expanded and vertically shuffled model DMs co-occur in tested DMs, to look for eventual interference between co-occurring models on both the PAS power to detect the model columns and the accompanying false positives, compared to when only one model is simulated per DM.

Then follows work about detecting columns involved in pure *n*-way *n*-column associations of binary markers (of frequency 0.5) in DMs with increasing numbers of random columns. Also here power and false positives are studied for selected cases in which a pure *m*-column model and a pure *n*-column model co-occur independently in a DM, to look for eventual interference between co-occurring models of different purity.

Finally, the study of model space is expanded to the detection of randomly encountered 5- and 10-column binary and trinary model DMs with 100, 200, 400, and 800 rows, but only in DMs with 995 and 990 random columns, respectively (1'000 total columns). These results together with those about detecting pure *n*-column associations shed light on both the nature of the column associations in the randomly encountered model DMs and the likely power of the P values of the selected PASs in the general model space.

No attempt is made to assess the power of PAS P values when DMs have blocks of columns nor to study the possibility of flagging columns directly that have extreme PAS values (rather than low PAS P values), but direct flagging of extreme dvPAS values (not P values) is explored further below.

**Detecting columns in randomly encountered models of significant inter-columnar association.** The models whose detection is studied are the aforeintroduced randomly encountered binary and trinary 100-row 5- and 10-column model DMs. To be kept as model DM every one of the columns in the randomly generated DM under consideration had to have P value$\leq$0.01 under the DV test in Fig.3 (M&M).

The model rows are expanded multinomially as stated and then PAS-detected in DMs having increasing numbers of random columns. Note that when the models DMs are "expanded" to the same number of rows of the original model DM (i.e., when model-DM rows are bootstrapped), only ~90% of their columns' P values *sensu* Fig.3 continues being no larger than 0.01.

The studied PASs are: $Mom^1i_Z$ that has highest power to detect these models, CHIx-M that is the all-signals PAS with highest power to detect these models, and $Mom^1i$ because it helps one make sense of why power is boosted when $Mom^1i_Z$ sums up every individual score that is conditional on a match for a given marker, after Z-valueing the scores (see Introduction).

Figure 24 shows for two 1'000-model sets of binary and trinary 100-row 5-column model DMs  i) the 60%-0.1 P value detection samples and false positives of the three PASs when DMs have up to 9'995 random columns;  ii) P value c.d.f.s for individual model and random columns of different marker frequency and groups thereof when there are 1'000 total columns; and  iii) the corresponding c.d.f.s when pairs of these models co-occur independently in 1'000-column DMs.

Panels (a,b) in Figure 24 show  i) that the log of the detection samples for the two pooled reference model columns (of binary frequency 0.1 and 0.5 or the trinary H&W thereof) increases near-linearly with the log of the number of random columns and  ii) that false positives are correct. They also show that the power of CHIx-M, $Mom^1i$, and $Mom^1i_Z$ increases markedly in that order, specially that of $Mom^1i_Z$ in the trinary case. When, e.g., the binary models are detected in DMs with 995 random columns, the detection samples are 12'000, 3'500, and 1'170 rows for the three dvPASs vs. ~30'000 (extrapolated), ~16'000 (extrapolated), and 1'590 rows in the trinary case, i.e., the power boost delivered by $Mom^1i_Z$ is most substantial in the trinary case. Furthermore the circles in the two plots show that when two models co-occur $Mom^1i_Z$ power is unchanged and false positives remain correct.

Panels (c,d) show that the $Mom^1i_Z$ c.d.f.s of individual model columns with different marker frequencies can differ from each other in the binary and trinary case, with the differences persisting in panels (e,f) with two models co-occurring. However, the corresponding c.d.f.s of $Mom^1i$ P values differ much more markedly across marker frequencies (not shown), e.g., in the binary case model columns with marker frequency 0.5 are detected much better by $Mom^1i$ than frequency-0.1 ones. $Mom^1i_Z$ detects the latter very well instead, owing to it a big part of its edge over $Mom^1i$. In the trinary case, $Mom^1i_Z$ detects better all of the model columns albeit the detection of the frequency-0.1 and frequency-0.5 ones is specially boosted. The latter is puzzling but is also observed for the second frequency-0.1 column in the 5-column model DMs at stake (o1524<u>1</u>).

When detecting the sets of randomly encountered binary and trinary 10-column model DMs, the power of $Mom^1i_Z$ P values is much higher than for the 5-column DMs, and it shows little reaction to marker frequencies (neither is shown). In a 1'000-column DM, e.g., the $Mom^1i_Z$ detection samples for binary and trinary 10-column model DMs are 600 and 650 rows (vs. 1'170 and 1'670 rows for the 5-column models). This increase in power is welcome but its reasons are unclear. CHIx-M, on the other hand, has much lower power with the 10-column models than $Mom^1i$ and $Mom^1i_Z$ have.

Figure 25 compares the $Mom^1i_Z$ detection samples of sets of 200 5- and 10-column randomly encountered model DMs with 100, 200, 400, and 800 rows in DMs with 1'000 total columns (but the 100-row 5- and 10-column model DMs comprise 1'000 model DMs each). False positives are correct throughout and not shown. Detection samples increase roughly linearly with the number of model-DM rows for a given number of model-DM columns (requires double-log plot in the trinary case), are roughly one order of magnitude larger than the rows of the model DMs being detected in the binary and the 10-column trinary case, and are smaller for the 10- than the 5-column models for any given number of model-DM rows, be the models binary or trinary.

In the case of the trinary 5-column model DMs, however, model-DM columns become faster much harder to detect than those of the corresponding binary models, with the detection sample reaching ~80'000 rows (extrapolated) in the trinary 800-row 5-column model DMs (but 100- and 200-row 10-column trinary models are easier to detect than their haploid counterparts). The rough linearity with which detection samples increase in reaction to increasing numbers of model-DM rows appears solid but more simulations are desirable.

Since MAA has no intuition about the extent to which pure *n*-way effects contribute to the power of the P values of moment-based PASs when detecting columns of randomly encountered trinary model DMs, no explanation is given here for why $Mom^1i_Z$ loses power faster when the trinary model DMs to be detected have many rows, compared to when detecting matched binary models.

Fig.24 showed that ~1'200 rows are needed by $Mom^1i_Z$ for 60% detection with P value$\leq$0.1 of the two reference columns of 1'000 100-row 5-column binary model DMs when these are simulated in DMs with

1'000 total columns. These are clearly fewer rows than the 4'000 rows needed by meePAS in Fig.10 to detect four columns in pure 4-way binary association in the presence of 996 random columns of marker frequency 0.5.

Furthermore, the two reference columns are detected by $Mom^2i_Z$ (that reacts to pure 3-way signal) only ~12% of the time with P value$\leq$0.1 in said 1'200-row 1'000-column DMs (not shown), i.e., barely above false positives; the percent for $Mom^3i_Z$ that reacts to pure 4-way signal, is 10%. Indeed ~6'100 DM rows are needed for 60% detection of the two reference model columns given only 100 random columns, in which case furthermore the false positives of $Mom^2i_Z$ and $Mom^1i_Z$ remain correct and $Mom^1i_Z$ detects the two columns 99.5% of the time with P value$\leq$0.001.

The normal behavior of $Mom^2i_Z$ P values in the last case indicates that in general a focal column's strong lower-order associations both may not compromise the reaction of $Mom^ni_Z$ to higher-order signal involving the focal column and do not inflate the accompanying false positives of higher-order $Mom^ni_Z$'s either. Note in passing that when DM rows are many, the P value of the $Mom^2i$ (non-Z-valued) of the frequency-0.1 reference model column almost does not detect the column albeit the accompanying false positives are fully correct (neither is shown).

The observations above, together with that of $Mom^1i_Z$ having both maximal power to detect the sets of 5- and 10-column binary (and trinary) models with 100, 200, 400, or 800 rows as well as maximal power to detect 2-way associations, indicate that the randomly encountered model DMs used here are much richer in PAS-detectable 2-way associations than in similarly PAS-detectable higher-order associations. However, the latter does not mean that exhaustive 2-column scans can detect these 2-column sub-associations as effectively as would an exhaustive scan of the cumulated 2-way signal of every distinct *n*-column subset up to *n* equal 5 and 10, respectively (see Introduction).

Finally, the results in Figs.24,25 use the all-signals CHIx-M rather than its CHIx-ij counterpart because CHIx-M's conditioning on any match at the focal column delivers higher power with these model DMs. However, the study of dvPAS power below will show that conditioning on a generic match generates too many false positives when some IVs are strongly associated with the DV, whereas the false positives of various dvPASs of type "i", "$i_Z$", "ik", "$ik_Z$", and "ijkl" behave much better, including the false positives of the P values of $dvMom^nik_Z$ and dvCHIx-ijkl.

**Detecting columns in pure *n*-column association.** The indications above about randomly encountered 5- and 10-column models of significant inter-columnar association being richer in more readily PAS-detectable 2-way associations, are consistent with the results in Figure 26 on $Mom^ni$ power when detecting columns in pure 2- to 8-way association as a function of the number of random columns of a DM. These results use $Mom^nM$ P values that here have the same power as $Mom^ni_Z$ P values because pure *n*-marker associations involve binary markers of frequency 0.5.

Panel (a) in Figure 26 shows that the log of the 60%-0.1 P value detection sample increases also here near-linearly with that of the number of random columns, for any order of pure *n*-column association (linearity requires double-log plot). Panel (b) shows that the log of the detection sample increases near-linearly with the order of pure *n*-column association, for any given number of total columns of the DMs in which the models are embedded and detected, e.g., in the case with 300 random columns every increase by one unit in the order of pure association necessitates increasing the detection sample by about one order of magnitude. False positives are correct throughout and not shown.

Panels (c,d,e,f) in Figure 26 show that co-occurrence in a DM of a strongly PAS-detected pure *m*-column model and a much more weakly PAS-detected pure *n*-way model affects neither the power to detect the second model nor the accompanying false positives when (*m,n*) is (3,2), (4,3), (2,3), and (3,4), respectively. Additional simulations show that the presence of a pure higher-order association affects neither the detection of a pure co-occurring lower-order one nor the generation of any of the false positives of concern (not shown).

These mixed-order results complement the facts i) that the top plots of Figure 26 remain unchanged when two identical pure *n*-column models co-occur independently in a DM (not shown) and ii) that when two randomly encountered models co-occur in Fig.24 they affect neither each other's detection nor the accompanying false positives. Taken together these observations suggest strongly that the power and false positives of $Mom^ni$ and $Mom^ni_Z$ (and $Mom^1M$'s when binary markers have frequency 0.5) are not altered when multiple groups of columns that are internally associated in the same or different ways co-occur independently in a DM.

Above it was shown that the 3-way and higher associations of individual columns in the randomly encountered model DMs do not affect the detection by $Mom^1i_Z$ P values of these columns' 2-way associations, but one could argue that said higher-order signals are perhaps too weak to create problems. However, no problem was observed above when $Mom^2i_Z$ P values were used to detect a column's 3-way associations when the column was at the same time in very strong 2-way association with other columns, i.e., when a column's 2-way signal is much more strongly PAS-detectable than its 3-way signal.  See also further below several examples of how further co-occurrences of associations with similar and disparate dvPAS detectabilities affect or not the power and false positives of relevant $dvMom^n i$'s and $dvMom^n ik$'s and their Z-valued versions.

Soberingly, Figure 26 shows that ~24'500 rows are needed by $Mom^3 M$ P values to detect a column involved in a pure 4-column association when a DM has 996 random columns.  These are many more rows than the 4'000 rows that meePAS needed in Fig.10 and shows that meePAS has much higher power to detect a pure $n$-column association than do $Mom^{n-1} i$ P values obtained through permutation of the focal column.  No systematic study of the power and false positives of meePAS is carried out here, a decision not only due to meePAS's much higher computational cost.

Indeed, it will be shown further below that the $dvMom^3 i$ detection sample (and $dvMom^1 M$'s one) for IVs in pure 3-IV association with a DV (equivalent of a pure 4-column association, see Fig.1d) is only ~2'000 rows in a DM with 1'000 random IVs, despite $dvMom^n i$ being orders of magnitude less computationally demanding than meePAS.  Furthermore and as intimated in the "Introduction to the Method", the power of meePAS should fade away when detecting models in which $c$ columns are in extended pure $n$-way association and $c>>n$ (Fig.2), since with increasing $c$ the individual exclusion of any of the $c$ columns begins affecting the level of pure $n$-way association of the other $c$-1 columns by (relatively) increasingly smaller $1/c$ amounts.

Readers interested in pure $n$-way associations involving only $n$ columns in DMs without blocks of columns (see below) are warmly invited to study in depth the power of meePAS and hybrid approaches like that of excluding only the DV and flagging IVs whose $Mom^n$ reacts most strongly to the exclusion.  However, the combination of high power and computational thrift offered by the P values of $dvMom^n i$ and $dvMom^n ik$ (below) is already very favorable when detecting DV-associated IVs.  It is the focus of the power assays below.

**Power of dvPAS when detecting individual columns associated with a dependent variable DV.**

This part is about the power and false positives of dvPAS P values when detecting IVs associated with a binary DV.  A shorter section is about flagging IVs directly when they have a large dvPAS rather than a small dvPAS P value.  Direct detection may be the only computationally tractable option when a DM has say $10^{12}$ rows and $10^{12}$ IVs, albeit doing say 100 permutations of the DV of a $10^6$-row $10^6$-IV DM, each followed by $10^6$ dvPAS recalculations for as many IVs, is fully tractable in a large parallel supercomputer.

As above, power is shown by plotting the number of rows that DMs need to have for 60% of the dvPAS P values of two **reference model IVs** in the tested set of model DMs (pooled) to be no larger than 0.1.  These IVs are the first two from the left in each model DM and in the case of the randomly encountered model DMs to be tested here they have binary (minor)marker frequency 0.1 and 0.5, respectively, or the H&W thereof. Power and accompanying false positives are studied as functions of the number of IVs and rows of the model DMs at stake as well as as functions of the number of rows and random IVs of the DMs in which model DMs are embedded and their IVs detected.

The first models studied are randomly encountered model DMs with significant 4- and 8-IV association with a DV.  These models are of two types:  i) model DMs that when generated, tested, and kept as models (with most others being discarded) are allowed to have IVs whose significant DV association *sensu* Fig.3 may be partly due to random marginal effects of said IVs at the DV; and  ii) similarly significant model DMs whose IVs' marker frequencies are identical in the two DV categories so that none of their IVs can have a marginal effect at the DV (M&M).

The reaction of power and false positives to the number of random IVs is first studied in DMs in which all non-model IVs are independent from each other and have binary markers with o12345 frequencies or the trinary H&W thereof (M&M). This work is repeated in DMs in which two models co-occur independently, to establish whether dvPAS power and accompanying false positives change compared to when the models are detected in isolation, i.e., to look for model co-occurrence effects that may compromise power and false positives and, when so, to try and find ways to neutralize such effects.  This is done by pairing models that both have or lack marginal effects, by mixing models of the two types, and by weakening (diluting) one model but not the other.

As intimated, the power assays focus on the P values of the $dvCHIx$-$ijkl$, $dvMom^1 i$, $dvMom^2 i$, and $dvMom^1 ik$ of the two reference model IVs and the P values of applicable Z-valued versions.  These dvPASs have greatest power to detect the aforementioned randomly encountered binary and trinary models with and without marginal effects. But the power and false positives of dvKS are also studied when dvKS power is close to the others'.

The work with randomly encountered model DMs is followed by work about detecting IVs in pure $n$-IV association with a DV, using dvMom$^n$i P values (as intimated the Z-valued versions deliver no extra power when focal-column marker frequency is 0.5). This two-pronged approach sheds light on both dvPAS power in the general model space and the nature of the DV associations in the randomly encountered model DMs to which the dvPASs reacts. Interference effects on dvPAS power and false positives due to co-occurrence of two pure models of different order are studied only in DMs whose IVs form blocks (below).

Indeed much attention is given to the detection of the IVs of the randomly encountered model DMs in DMs whose IVs form blocks of intra-block-associated IVs and where blocks without model IVs are independent from other blocks and the DV. These studies include placing each model IV in a different block as well as placing four model IVs per block, with the latter case being studied only for the 8-IV models.

The work with blocks is expanded to the detection of additional randomly encountered binary and trinary 50-, 100-, 200-, 400-, and 800-row 8-IV model DMs with and without marginal effects in DMs whose IVs form blocks. This is done only for DMs with blocks both because block structure is empirically relevant in genomics and because blocks boost dvPAS power markedly, making the study faster computationally. Also here the two cases are studied of placing model IVs one and four per block.

The detection of model IVs when pairs of models co-occur independently in DMs with blocks is also studied to look for eventual interactions with block structure. This includes pairing randomly encountered model DMs with and without marginal effects as well as pairing a pure $m$-IV model and a pure $n$-IV model for various values of ($m$,$n$).

Then, as intimated, the case is considered of a single DM with so many IVs that calculating a P value for every IV appears impractical and the flagging of IVs directly by the magnitude of their dvPAS becomes desirable. This possibility is quickly shown to be unviable when a DM has IV blocks.

A case study follows of how an experimentalist who wishes to analyze a single empirical DM (**eDM**) can use dvPAS P values to detect IVs associated with the DV in said DM. Two of the randomly encountered model DMs of 8-IV no-marginal-effect association with a DV are chosen and each model is simulated together with its matched, nominally corresponding marginal-effects counterpart in five independent eDMs with blocks.

Only in the latter section are DMs studied in which IVs can belong to two types of column blocks arranged in tandem along the DM, which allows for several controls. Each tandem includes a block of the usual 100-marker LPL sequences and a block of "CHJP" sequences sampled from a set of 1'000-marker Chinese and Japanese haplotypes (M&M), either in that order (LPCJ tandems) or reversed (CJLP tandems).

Finally the last section before the Discussion addresses the lacuna in model space that is left unexplored by the power assays adumbrated above which rely on model DMs with no more than eight DV-associated IVs. The section studies the detection of models with up to 495 pairs of IVs (i.e., 990 model IVs) that are in extended (Fig.2) and independent weak 2-way DV association in binary and trinary 1'000-IV DMs with no block structure. These results shed bright light on the marker combinations that make model-DM columns and model-DM IVs more readily detectable by PAS and dvPAS, respectively.

**Detecting IVs belonging to randomly encountered 4- and 8-IV models of DV association.** Figure 27 shows the power and false positives of the P values of dvCHIx-ijkl, dvKSi, dvMom$^1$i, dvMom$^2$i, dvMom$^3$i, dvMom$^1$ik, and selected Z-valued versions when detecting IVs that alone and/or in synergy with others are associated with a binary DV. Detected are the reference columns in sets of 1'000 randomly encountered 200-row 4- and 8-IV binary and trinary model DMs with or without marginal effects at the DV (i.e., eight sets of 1'000 model DMs each). As stated, in these models every IV is DV-associated with standard P value$\leq$0.05 *sensu* Fig.3 (M&M).

The IVs of each model DM are tested when the model is embedded in DMs with increasing numbers of additional random IVs. The 200 4- and 8-marker sequences (rows) of each model DM at stake are expanded multinomially up to the number of rows (the "sample size") desired for the DM in which detection is to be tried. However, when a no-marginal-effect model DM is expanded, the two subsets of 100 4- or 8-marker model-DM sequences each that are associated with either DV category are expanded separately in order to keep the two resulting groups of sequences equally numerous.

Therefore when testing the no-marginal-effect models the only random marginal effects of the model IVs in the tested DM are those created when expanding multinomially the 100 4- or 8-marker model-DM sequences associated with "affecteds" separately from the 100 sequences associated with "controls" (M&M).

Random IV columns with o12345 binary-marker frequencies or the trinary H&W thereof are generated independently and added laterally. Each model DM is expanded and tested at least once, except when the size of the final DM exceeds 6'000 columns and 10'000 rows, in which case at least 200 model DMs are studied. P values are estimated using at least 100 permutations of the DV markers.

Panels (a,b) in Figure 27i,ii show that as random IVs increase the 60% 0.1-P value detection samples for the models with marginal effects increase monotonously and the false positives with P value$\leq$0.1 sink steadily towards expectation in both the binary and trinary case, albeit the detection samples of dvCHIx-ijkl, dvMom$^1$i, and dvMom$^1$i$_Z$ plateau already with few random columns.

For any number of random IVs the dvCHIx-ijkl false positives no larger than 0.1 are about 11% (rather than 10%), those of both dvMom$^2$i's are ~10%, but those of dvMom$^2$i$_Z$, dvMom$^3$i$_Z$, and dvMom$^1$ik$_Z$ are much higher, hovering in the case of 100 random IVs, e.g., at about 40, 55, and 29%, respectively, for the 4-IV models and about 36, 46, and 27% for the 8-IV models.

When in panels (a,b) of Figure 27i the binary 4- and 8-IV model DMs with marginal effects are tested in the presence of at least 20 and 40 random IVs, respectively, dvCHIx-ijkl has the smallest detection samples (the greatest power), followed closely by dvMom$^1$i. In the trinary results in Figure 27ii, this is observed already with 10 random IVs. The plateaued power edge of dvCHIx-ijkl vs. dvMom$^1$i and dvMom$^1$i$_Z$ is ~10% in the binary case but it is ~0% and ~20% vs. dvMom$^1$i$_Z$ and dvMom$^1$i in the trinary case.

Unexpectedly, panels (a,b) in Figure 27i show the P values of the higher-order dvMom$^2$i, dvMom$^3$i$_Z$, and dvMom$^1$ik as having seemingly superior power when detecting the IVs of the 4- and 8-IV marginal-effects binary model DMs in the presence of fewer than 100 and 600 random IVs, respectively, while in the trinary cases in Figure 27ii this happens only below 40 and 200 random IVs.

The higher power is partly a reaction to real higher-order DV associations in the model DMs (below), but these plots do not demonstrate power cleanly because concomitantly the corresponding false positives are in marked excess, e.g., the detection samples of dvMom$^3$i$_Z$ are larger but are accompanied by less of an excess in false positives.

On the other hand, the highest power shown by dvCHIx-ijkl and both dvMom$^1$i's in panels (a,b) of Figure 27i,ii should be safe to exploit when random IVs are numerous and one is detecting IVs with substantial marginal effects, since the corresponding false positives are uniform-(0,1)-distributed. In Figure 27i,ii(a,b), e.g., the false positives of dvCHIx-ijkl are ~11% at the 0.1 P value level when random IVs are 1'000 or more, respectively, while those of dvMom$^1$i and dvMom$^1$i$_Z$ are ~10% already with 10 random IVs.

As intimated and possibly related and welcome (if puzzling), in panels (a,b) of Figure 27i,ii the detection samples of dvCHIx-ijkl and both dvMom$^1$i's *de facto* stop increasing above 100 and 30 random IVs, respectively, being barely larger with 10'000 random IVs than say with 100 (while false positives with P value$\leq$0.1 remain at 10%). The plotted dvCHIx-ijkl detection samples for, e.g., binary 8-IV marginal-effects models with 10, 100, 1'000, and 10'000 random IVs are 744, 1'160, 1'200, and 1'200 rows, respectively (1'000 model DMs each expanded, embedded, and tested 10 times).

When random IVs are 100 or more, binary and trinary 4-IV models with marginal effects are only slightly easier to detect by the P values of dvCHIx-ijkl and both dvMom$^1$i's than the corresponding 8-IV models are. And the panels (e) in Figure 27i,ii show that the power of dvCHIx-ijkl barely reacts to marker frequencies when detecting the binary and trinary models with marginal effects in DMs with 1'000 random IVs while false positives show little reaction as well. The same applies when using dvMom$^1$i$_Z$ (not shown).

Regarding the detection of the no-marginal-effect models, panels (c,d) in Figure 27i,ii show that the most powerful dvPASs for these models are dvMom$^2$i and dvMom$^1$ik in the binary case, and their Z-valued versions in the trinary case, with all of them generating correct amounts of false positives. Their detection sizes for, e.g., the binary and trinary 8-IV no-marginal-effect models in the presence of 1'000 random IVs are about fivefold smaller than those of dvKSi. But because of the aforementioned excessive false positives of these dvMom's in the presence of even very few IVs with strong marginal effects (four and eight in panels (a,b) of Figure 27i,ii), their superior power cannot be exploited naively when studying empirical DMs that may include IVs with marginal effects. One needs additional manipulations of the DM to make sure that their false positives are not inflated by such effects (below).

When detecting binary no-marginal-effect models in the presence of 1'000 random IVs, dvMom$^1$ik$_Z$ and dvMom$^2$i$_Z$ perform equally and worse than their non-Z-valued versions, respectively. The 60%-0.1 P value detection samples for dvMom$^1$ik$_Z$(dvMom$^1$ik) and dvMom$^2$i$_Z$(dvMom$^2$i) are 3'800(3'900) and 4'300(3'850) for the 4-IV models, e.g., and 2'790(2'910) and 3'300(2'960) for the 8-IV models (1'000 models each replicated five times). The dvMom$^2$i$_Z$ vs. dvMom$^2$i difference is due to dvMom$^2$i detecting the model IVs with marker frequency 0.1 much better than the 0.5-frequency ones, e.g., ~14% better in the case of the 8-IV models, whereas dvMom$^2$i$_Z$ detects the frequency-0.5 model IVs as well as dvMom$^2$i does it, but detects the 0.1-frequency ones like the 0.5-frequency ones.

In the trinary case the detection samples are, in the same order, 2'850(3'040) 2'800(3'100) and 1'680(1'880) 1'570(1'720), respectively, i.e., the Z-ed versions have higher power, e.g., the dvMom$^2$i$_Z$ detection sample is ~10% smaller. Also here the model IVs with trinary frequency 0.1 (H&W thereof) are detected much better by dvMom$^2$i$_Z$ but the detection of the others is improved too, unlike in the binary case.

Although dvKSi has very low power with no-marginal-effect models, its detection sample hints at some plateauing above 300 random IVs when detecting the IVs of the 4-IV no-marginal-effect models binary and trinary. Therefore dvKSi may turn out to be the most powerful option for such models when random IVs are 100'000 or more, that is if dvMom$^2$i and dvMom$^1$ik do not plateau earlier below dvKSi. This could not be studied here because of computer-power limitations.

Panel (e) in Figure 27i shows that the marker frequencies of model IVs do not affect markedly the detection by dvCHIx-ijkl of the IVs of the binary 200-row 8-IV model DMs with marginal effects in DMs with 1'000 random IVs. But the few subtle differences are puzzling since they do not track marker frequencies.

However, the corresponding results in panel (f) of Figure 27i when detecting binary model IVs without marginal effects show that the detection of these IVs by dvMom$^2$i (i.e., not Z-valued) worsens noticeably as marker frequency increases from 0.1 to 0.5. In contrast, the effect of marker frequencies is much subtler in the trinary cases in panels (e,f) of Figure 27ii. The steadier behavior of dvMom$^2$i$_Z$ in trinary DMs suggests that similar may be achievable for binary DMs through more careful applied-mathematical work, without lowering power.

Figure 27ii shows that the detection samples for trinary models with and without marginal effects are about half as big as those of the binary models, which is good news for human genetics where trinary data is most important. Given 1'000 random IVs, e.g., the smallest detection samples for 8-IV models with and without marginal effects binary(trinary) are 1'540(590) and 2'800(1'570) rows when using dvCHIx-ijkl(dvCHIx-ijkl) and dvMom$^2$i(dvMom$^1$i$_Z$), respectively, i.e., the trinary detection samples are 38 and 56% of the binary ones.

The plots in Figure 27i,ii show indirectly how to erase the excess in dvMom$^2$i and dvMom$^1$ik false positives caused by IVs with strong marginal effects. Indeed both of these false positives are correct in panels (c,d) in Figure 27i,ii where the tested model DMs have no marginal effects and the only marginal effects that are present arise from the multinomial expansion of model-DM rows (M&M and above). This shows that the excessive false positives of dvMom$^2$i and dvMom$^1$ik in the (a,b) panels of Figure 27I,ii are mainly due to the strong marginal effects of the IVs in the model DMs studied there and that therefore the two excesses must disappear if one erases every strong marginal effect in the DM at stake.

The power and false positives of dvMom$^2$i and dvMom$^1$ik after erasing strong marginal effects are studied immediately below for the case of a DM in which an 8-IV no-marginal-effect model co-occurs with a more readily PAS-detectable 8-IV marginal-effects model. This work shows that a careful such erasure lowers the false positives of dvMom$^2$i and dvMom$^1$ik (and Z-valued versions) down to expectation and increases their detection samples up to those for isolated models, without discernible disadvantages.

However, the above excess in the false positives of dvMom$^2$i, dvMom$^1$ik, and dvMom$^3$i$_Z$ P values that are caused by strong marginal (1-way) effects points to the possibility of similar excesses arising when a pure *m*-IV and a pure *n*-IV DV association co-occur and (*m,n*), unlike in the (a,b) panels of Figure 27i,ii, is not (1,2). Indeed it is very likely that the strong 2-way DV associations of the IVs in the tested model DMs are the only cause of the excess in false positives of dvMom$^3$i (not dvMom$^2$i) that is apparent in panels (c,d) of Figure 27i,ii in the absence of IVs with strong marginal effects.

The above suggests the general rule that a pure *m*-IV DV association that is strongly detected by say dvMom$^m$i$_Z$ given the rows of the DM at hand, should inflate the false positives of dvMom$^n$i, dvMom$^{n-1}$ik, etc, whenever $m \leq n-1$. The rule is confirmed further below for additional (*m,n*)'s but also substantially refined. Also pertinent, immediately below are results on dvPAS power and false positives when the model DMs studied in Figure 27 co-occur independently as pairs in DMs, and further below are results with pairs of pure *m*- and *n*-IV models of DV association co-occurring in DMs with IV-blocks, for several values of (*m,n*).

**Interference between two co-occurring models of *n*-IV DV association, one with and one without marginal effects.** Figure 28 shows the power and false positives of selected dvPAS when detecting two binary or trinary models of 8-IV DV association, one with and one without marginal effects, that co-occur independently in the same DM; their model IVs below are often called first- and second-model IVs. The paired models are studied in Fig.27 in isolation with the same dvPASs.

Figure 28 is exclusively about the detection of the IVs of the pairs' second models (that lack marginal effects) as well as about the concomitant generation of false positives at five random IVs. Indeed i) in both the binary and trinary case the detection by dvCHIx-ijkl, dvMom$^1$i, and dvMom$^1$i$_Z$ of the reference IVs of the pairs' marginal-effects models is identical to that in isolation and ii) the generated false positives fit expectation, suggesting that the presence of IVs involved in moderately dvPAS-detected higher-order DV-associations affects neither. Further below, however, this lack of interference is confirmed for cases of very strongly detected pure 2- and 3-IV models co-occurring with a weakly detected 1-IV model.

Figure 28 shows both that the detection samples for the model IVs without marginal effects are smaller than in isolation and that, as expected, this is accompanied by an excess in the concomitant false positives at non-model IVs. The excess generated by dvCHIx-ijkl and dvKSi sinks as random IVs increase but that of dvMom$^2$i and dvMom$^1$ik (and Z-valued versions) does not. In the trinary case with ~10'000 independent random IVs, e.g., the false positives of dvMom$^2$i$_Z$ and dvMom$^1$ik$_Z$ with P value$\leq$0.1 are still about 26 and 20% (down from 30 and 26% with 10 random IVs).

Again as expected, the excess in false positives of dvMom$^2$i and dvMom$^1$ik (and Z-valued versions) disappears when one erases the marker-frequency differences behind the strong marginal effects that are present in the simulated DMs. The erasure is done here by toggling in both affecteds and controls minimal numbers of randomly chosen excess markers at every IV with a standard marginal-effect $\chi^2$ P value no larger than a threshold value chosen through trial and error.

For instance, in the trinary case when DMs have 10, 100, 1'000, and 10'000 random IVs, the excess in false positives when the number of DM rows yields 60% detection with dvMom$^1$ik$_Z$ and dvMom$^2$i$_Z$ P value$\leq$0.1 of the reference model IVs in the 8-IV no-marginal-effect models, disappears after toggling excess markers at every IV showing a standard marginal-effect P value$\leq$ 0.019, 0.0065, 0.001, and 0.0002, respectively (2 d.f). As intimated, detection samples sink too after the erasure and become like those in isolation.

The correction might be done more elegantly by letting the formulae of dvMom$^2$i and dvMom$^1$ik react in some way to the marginal effects of the entire DM at hand, but this was not attempted here. Note that apparently when one toggles only randomly chosen excess markers one does not erase higher-order effects to a great extent (below).

The above erasure, however, does not reduce the large excess in the false positives of dvMom$^3$i$_Z$ P values (and dvMom$^2$ik ones; not shown) nor raises noticeably its detection samples, suggesting that the 2-way DV associations in the two sets of model DMs (that apparently survive to a great extent the toggling of excess markers) are the main cause of the inflation of dvMom$^3$i$_Z$ false positives, rather than the marginal effects of first-model IVs.

More light on this is shed by looking at dvMom$^3$i$_Z$ detection samples and false positives after randomizing the higher-order DV associations of the model IVs with marginal effects but without erasing these IVs' marginal effects, i.e., by shuffling vertically the markers in first-model IVs separately in affecteds and controls. In the binary case with 100 random IVs when the IVs are: unchanged, with erased marginal effects, and with randomized higher-order DV associations, the dvMom$^3$i$_Z$ detection sample(false positives) values are 970(0.49), 980(0.49), and 1'650(0.40), respectively, while in the trinary case they are 652(0.43), 674(0.42), and 1'020(0.34), respectively. These values show that in the binary and the trinary case both the dvMom$^3$i$_Z$ detection samples of second-model IVs (that lack marginal effects) increase more and the concomitant excess in dvMom$^3$i$_Z$ false positives is more reduced if one randomizes the higher-order DV associations of the (marginal-effects) first models than if one erases the latter models' above-threshold marginal effects.

Note that the effect of erasing the marginal effects of first-model IVs on the P values of the dvMom$^2$i$_Z$ (not dvMom$^3$i$_Z$) of second-model IVs, is fully consistent with the effect of randomizing the first models' higher-order DV associations on the dvMom$^3$i$_Z$ P values of second-model IVs (while keeping first-model marginal effects unchanged). Indeed and again for the said three treatments, the dvMom$^2$i$_Z$ detection sample(false positives) values in the binary case are 602(0.20), 1000(0.10), and 594(0.20) while in the trinary case they are 334(0.27), 560(0.10), and 340(0.29), i.e., the plain and randomized-higher-orders values are nearly identical.

The above therefore shows that it is mainly the marginal effects of first-model IVs what inflates dvMom$^2$i false positives at non-model IVs and it also suggests that the 2- and 3-way effects of first-model IVs do not interfere with the detection by dvMom$^2$i (and dvMom$^2$i$_Z$) of second-model IVs that lack marginal effects.

The last conclusion is consistent with both the fact that dvMom$^n$i detection samples and false positives do not change when two models of similar order co-occur independently and the fact that higher-order DV associations do not affect the power of dvMom$^n$i P values when detecting IVs involved in pure lower-order $n$-IV DV associations. Both was preliminarily shown above for *de facto* $n$=1 and 2 and is confirmed in several additional cases below.

Panel (c) in Figure 28 shows for DMs with 1'000 random IVs, the individual dvMom$^2$i c.d.f.s of the eight model IVs of each model type and five random IVs with binary frequencies o15241234 and o12345, respectively, while in panel (d) are the dvMom$^2$i$_Z$ results for the corresponding trinary case. The DMs have 2'950 and 1'600 rows, respectively, that in both cases yield ~60% detection with dvMom$^2$i and dvMom$^2$i$_Z$ P value$\leq$0.1, respectively, of the two reference IVs of the no-marginal-effect second models in the pairs, after marginal effects with standard P value$\leq$0.001 (1 and 2 d.f., resp.) are erased by toggling excess markers. The c.d.f.s of individual IVs with different marker frequencies are like those observed in isolation (Fig.27).

Interestingly, in the binary and trinary cases in panels (c,d), first-model IVs (with erased marginal effects) are detected about 50% and 45% of the time with dvMom$^2$i and dvMom$^2$i$_Z$ P value$\leq$0.1, respectively (despite their marginal effects' erasure), showing that these model IVs' higher-order DV associations survive the toggling of excess markers so well that the IVs are detected by dvMom$^2$i and dvMom$^2$i$_Z$ almost as strongly as the IVs of the second-model DMs which were selected as model DMs only because of their significant higher-order DV associations.

The perhaps unexpected higher-order effects of the models with marginal effects must be what inflates the false positives of dvMom$^3$i$_Z$ in panels (a,b) of Fig27i,ii where these models are simulated in isolation. (Remember that IVs from both kinds of model DMs are equally significant *sensu* Fig.3.)

As intimated, however, the fact that moderately strong first-order "1-way" marginal effects inflate the false positives of dvMom$^2$i and dvMom$^1$ik (and Z-valued versions) and also, but more weakly, those of dvMom$^3$i and dvMom$^2$ik, raises the possibility of similar excesses in other cases too, e.g., when two co-occurring models have strong 3- and 5-way effects, respectively. This is studied further below after presenting the results about detecting both the IVs in pure *n*-IV DV association and the IVs of the two sets of randomly encountered model DMs when the IVs of DMs form blocks.

**Detecting IVs that are in pure *n*-IV association with the DV.** Figure 29 shows the power of dvMom$^n$i when detecting individual IVs in pure binary *n*-IV association with a DV and the concomitant dvMom$^n$i false positives at random IVs with o12345 binary marker frequencies, for $2\leq n\leq 8$. As intimated, no work is shown about detecting trinary models of pure DV association, because the diagrams of such associations are unknown to MAA. The power and false positives of the P values of dvMom$^n$M, dvMom$^n$i$_Z$, and dvMom$^n$ik$_Z$ are like those shown of dvMom$^n$i.

Two binary cases are considered. In the first case, *n* IVs are in pure *n*-way association with "affecteds" (the 0 markers at the DV) and are contrasted to "randoms" while in the second case said "affecteds" are contrasted to "controls" that have a 1 at the DV and *n* IVs that show the complementary pattern of pure *n*-way association (see pure 2- and 3-IV DV associations in Fig.1).

The plots on the left in Figure 29 show both that the log of the 60%-0.1 P value detection sample increases near-linearly with that of the number of random IVs and that false positives with P value$\leq$0.1 are correct throughout. Detection samples are several times smaller in the "vs. controls" case.

The right-hand plots show that given a number of random IVs the detection samples increase exponentially as the order of pure *n*-IV DV association increases linearly, e.g., in the case with 30 random IVs every increase by one unit in the order of pure association necessitates increasing the detection sample by about a factor of ten. Note that detecting say a pure 6-IV DV association in the more favorable vs.-controls situation requires, if one extrapolates, about 60'000 rows when random IVs are 100, which is sobering data-gathering-wise but far from intractable computationally.

In panel (c) of Figure 29, the black dots over the pure 2-, 3-, and 4-IV lines are for the vs.-controls power of dvMom$^n$ik and its false positives with P value$\leq$0.1. They show that dvMom$^n$ik offers no advantage over dvMom$^n$i when detecting pure models of *n*-IV DV association involving IVs with binary markers of frequency 0.5, at least under these testing conditions.

**Power of dvPAS when IV columns form blocks with strong internal associations**

The above results on the power of selected dvPAS and their generation of false positives need confirmation, at least for workers in genomics, in DMs where all IVs form blocks within which IV markers are intra-block associated say like actual mutations in human chromosomes allegedly are.

This section attempts this confirmation using DMs in which the blocks harboring no model IVs are randomly associated with each other and the DV. The main block types that are used are the 100-column binary and trinary LPL and wLPL blocks introduced above and in the M&M.

In such a DM one can still permute the DV to obtain a dvPAS P value for an IV of interest and quantify so how strongly the IV, by itself and/or in synergy with others, is associated with the DV. Of interest here too are the false positives and the detection of the IVs of the randomly encountered model DMs and the IVs in the pure $n$-IV associations (vs. controls) that were studied above in DMs without blocks, both when each model is simulated and tested in isolation and when two models co-occur independently in the same DM.

The type I error of dvPAS P values in DMs with blocks was explored above and found to be correct, i.e., uniform-(0,1)-distributed, albeit P values of same-block IVs were at times strongly correlated. Therefore when in such a DM a group of IVs is forced to be DV-associated, one expects the non-model IVs that are in blocks with one or more strongly DV-associated model IVs to have P values that are lower than those of the IVs from blocks without model IVs. This is confirmed below.

Furthermore, intra-block IV associations should increase dvPAS power in a DM with a given number H of $n$-IV blocks over that in a DM with H*$n$ independent IVs, because such intra-block associations reduce the number of effectively independent random IVs in the DM, a number whose increase was shown above to reduce dvPAS power in most cases.

We also saw that related block effects on type I error cannot be demonstrated clearly by studying how they perturb the Sidak correction of P values obtained from testing same-block "families of IVs" in DMs generated under the $H_o$ of blocks being independent from each other. It is therefore of interest to study whether IV blocks boost dvPAS power and how they affect the generation of false positives.

The focus here is again on i) the number of DM rows that are needed for 60% of the first two ("reference") model IVs of the model DMs at stake to have dvPAS P value$\leq$0.1 and ii) the concomitant generation of false positives at non-model IVs, including the **faux false positives** at non-model IVs from blocks with one model IV or more (called "**model-linked IVs**" below).

The section starts with simulations of randomly encountered 50-, 100-, 200-, 400-, and 800-row 4- and 8-IV binary and trinary model DMs with and without marginal effects of which the 200-row ones were studied in detail in Figs.27,28 in DMs without blocks. Two placements of model IVs across IV blocks are tried, namely i) placing every model IV at the 49$^{th}$ column (from the left) of a different block and ii) placing the $n$ model IVs four at a time at columns 5, 35, 49, and 66 of each of the $n$/4 blocks that are needed to allocate the $n$ model IVs when $n$ is a multiple of 4 as is the case here.

Results are also shown for the first four model IVs individually and all four pooled, as well as for the faux false positives at five model-linked IVs at columns 2, 23, 51, 64, and 89 of the first block (from the left) that harbors one or more model IVs (M&M). Comparing the detection of the five model-linked IVs to that of their same-block model IVs is an assay of "**coarse vs. fine mapping**" resolution (below).

To study genuine false positives, results are shown for five position-49 IVs from five blocks without model IVs, for each IV individually and the five IVs pooled. Forced sampling of LPL sequences is used to let the markers at these five IVs have o12345 binary frequencies or the trinary H&W thereof (M&M).

The chosen dvPASs are those that above were shown to detect best the IVs in the model DMs at stake when random IVs do not form blocks, except when block structure lets another dvPAS be most powerful, in which case the latter is used. As intimated, results are shown for both the case of each model DM being simulated alone in a DM and the case of two models of equal or different type co-occurring independently in the same DM.

**Power and false positives as a function of IV-marker frequencies in DMs with blocks.** Figures 30 and 31 show dvPAS power and false positives for model, model-linked, and random-block IVs with different marker frequencies in DMs with 100 100-IV blocks when four or eight pos.49 IVs in as many blocks are forced to be DV-associated like the IVs of the randomly encountered 200-row 4- and 8-IV model DMs studied in Fig.27.

As usual, the results are conditional on the rows of the DMs allowing the showcased dvPAS to detect the first two (reference) model IVs ~60% of the time with P value$\leq$0.1. The five model-linked IVs are from the block with the first model IV.

Figure 30 shows the results with the model DMs with marginal effects. In the four cases shown, dvCHIx-ijkl is used because it is most powerful for these models when both IVs form blocks and one model IV is placed per block, but dvMom$^1$i and dvMom$^1$i$_z$ have nearly identical power here (like above with independent random IVs).

The detection samples in panels (a,b,c,d) are 1'160, 1'450, 574, and 670 rows for the 4- and 8-IV binary and the 4- and 8-IV trinary models, respectively. Like in Fig.27, the samples are both bigger for the 8- than the 4-IV binary models and about twice as big for binary than trinary models. In all panels the c.d.f.s of the two reference model IVs pooled behave like those of the other model IVs albeit there is more variation in the binary case with the 8-IV models.

Model-linked IVs at pos.89 are detected like model IVs and those at pos.64 are detected intermediately, but with pos.89 being 21 and 15% less detected in the trinary cases in panels (c,d) and pos.64 40 and 32% less. The c.d.f.s of the other three model-linked IVs and the DV are on the diagonal much like the c.d.f.s of the false positives at five IVs from random blocks.

The results with model-linked IVs indicate that dvPAS coarse-mapping is more effective when detecting the binary models, whereas fine-mapping has higher resolution in the trinary cases. Albeit confirmation with additional types of blocks is needed, these two result make sense when one considers that the associations of binary markers in the LPL sequences become "averaged" (diluted) in the pairwise comparisons of the LPL sequences that are used to generate the associations of trinary markers in the wLPL sequences (M&M).

Figure 31 shows the results with the model DMs whose IVs lack marginal effects. In panels (a,b) and (c,d) $dvMom^2i$ and $dvMom^2i_Z$ are used, respectively. In (a,b) $dvMom^1ik$ and $dvMom^1ik_Z$ have similar power but in panel (c,d) $dvMom^1ik$ has ~7% lower power than the others (not shown; case with four IVs per block is below).

The detection samples in panels (a,b,c,d) are 3'000, 2'150, 2'100, and 1'190 rows for the 4- and 8-IV binary and the 4- and 8-IV trinary models, respectively, i.e., the 8-IV trinary models are detected much better. Only in the two binary cases the no-marginal-effect model IVs with different marker frequency are detected differently by the higher-power $dvMom^1i$ (non-Z-valued), with low- and intermediate-marker-frequency IVs being best and worst detected, respectively. In the 4-IV binary case, e.g., the frequency-0.1 model IVs are easiest to detect, the 0.4 ones hardest to detect, and the 0.5 and 0.2 ones second-hardest and average to detect, respectively, i.e., there is no linear effect of marker frequency.

Like in Fig.30, the model-linked IVs at pos.89 and pos.64 are detected best and second-best, respectively, and both IVs are more detected in the binary case, with pos.89 being 33 and 24% less detected in the trinary cases in panels (c,d) and pos.64 53 and 17% less. In other words, when detecting models lacking marginal effects the resolution of fine mapping appears to be higher in the trinary case. The c.d.f.s of the other three model-linked IVs and those of the false positives are on the diagonal. Therefore also the results in Figure 31 point to coarse- and fine-mapping being more effective in binary and trinary data, respectively. The conclusion is confirmed in Figure 32 below where the IVs of the 8-IV models of either type are placed four per block.

Fortunately for geneticists of diploid organisms, the detection samples for the trinary 4- and 8-IV no-marginal-effect models are smaller and much smaller, respectively, than for the corresponding binary models. Also welcome, power increases when going from 4- to 8-IV models in both the binary and trinary no-marginal-effect cases, like in Fig.27 without blocks. As mentioned and most felicitously, the two boosting effects are compounded when detecting the trinary 8-IV no-marginal-effect models.

If one compares Figs.30,31 with Figs.27,28, one can conclude that block structure boosts power by reducing the "effective number" of independent IVs of a DM. When the DMs in Figs.27,28 have, e.g., 1'000 random IVs the smallest detection samples for the 4- and 8-IV binary and trinary no-marginal-effect models are 3'800, 2'800, 2'600, and 1'570 rows, vs. 3'000, 2'150, 2'100, and 1'190 in Fig.31 with 100 100-IV-block DMs. Both sets of detection samples are in the same order of magnitude, albeit in Figure 31 the tested DMs have nominally 10'000 total IVs rather than 1'000. In DMs with 10'000 independent IVs, the detection samples in Figure 31 would yield almost no detection of the no-marginal-effect model IVs at stake.

Since dvPAS power to detect DV-associated IVs in a DM with $n$ independent random blocks each consisting of any number of IVs in "perfect" extended 2-way-association is the same as when there is a total of $n$ independent random IVs, one can conclude that the 100 IVs in each LPL block are about equivalent to between five and ten random independent IVs, at least as far as dvPAS power is concerned.

The same effect must also apply when detecting randomly encountered models with marginal effects, but less clearly so since when detecting such models the detection sample plateaus very quickly as random IVs or blocks increase in number (Figs.27,28; results with increasing numbers of blocks further below).

**Power and false positives when model IVs appear four in each block.** Figure 32 repeats the results with 8-IV models in Figs.30,31 but with the eight model IVs being placed in two blocks, four IVs per block, rather than one per block in eight different blocks as in Figs.30,31. Other details are like in Figs.30,31, including the use of dvCHIx-ijkl and $dvMom^2i_Z$ when detecting the marginal- and non-marginal-effect model IVs, respectively. Note, however, that $dvMom^1i_Z$ has ~7% smaller detection sample and fully correct false positives (not shown) in the panel-(b) case of Figure 32, but dvCHIx-ijkl was preferred to allow comparisons with Fig.30.

The c.d.f.s of false positives in Figure 32 are slightly above the diagonal likely because of the subtle marginal effects caused by the harsher sampling of LPL and wLPL sequences that have four model-IV markers per sequence. Indeed the excess in false positives disappears after above-threshold marginal effects are erased like in Fig.28 (see also next figures). It is not further discussed here.

The detection samples in panels (a,b) of Figure 32 for model IVs with marginal-effects models in the binary and trinary case are 1'500 and 670 vs. 3'500 and 1'800 in panels (c,d) for model IVs without marginal effects. In the given order these are 3 and 0% vs. 63 and 51% more rows than in Figs.30,31 with one model IV per block, i.e., only the detection of the IVs of the no-marginal-effect models is reduced when model IVs are placed four per block. However and like in Figs.30,31, power is clearly lower in the binary than the trinary case also when model IVs are placed four per block.

Figure 32 also shows that the individual model IVs in the trinary no-marginal-effects case are detected differently, possibly due to interactions with the wLPL marker combinations near each of the four model positions, unlike in Figs.30,31 where the same model IVs are placed at different-block pos.49s but all are almost identically detected regardless of model type. The strongest difference is between the trinary no-marginal-effect model IVs with H&W trinary-marker frequency 0.1 and 0.5 that are detected ~53 and 65% in panel (d) of Figure 32 vs. both ~60% in panel (d) of Fig.31. This observation is disorienting since in the binary case with four model IVs per block the detection of model IVs is more similar across marker frequencies.

Among the model-linked IVs those at positions 2, 89, and 64 are detected best in all cases in Figure 32, with that at pos.2 being detected ~60% of the time with P value$\leq$0.1 (i.e., like the best detected model IV) in all cases except for the 51% in panel (d) when the models being detected are trinary and have no marginal effects. In the latter case all model-linked IVs are less detected than in the other cases, favoring fine-mapping, whereas coarse-mapping appears to be most effective in the binary case with marginal effects in panel (a) where all model-linked IVs are detected 50% of the time or more with P value$\leq$0.1.

**Power as a function of the number of model-DM columns and rows when IVs form blocks.** Figure 33 shows the 60%-0.1 P value detection samples when sets of 1'000 randomly encountered *r*-row binary and trinary model DMs of 8-IV DV association either with or without marginal effects, respectively, are simulated in DMs with 100 100-IV LPL blocks. The P values of dvCHIx-ijkl are used for marginal-effects models while those of dvMom$^1$i and dvMom$^1$iz are used for no-marginal-effect models binary and trinary, respectively. This is done for *r* equal 50, 100, 200, 400, and 800 rows and with one and four model IVs being placed per block (see also Figs.30,31,32 and M&M). False positives are uniform-(0,1)-distributed throughout and not shown.

Figure 33 shows that regardless of the model-DM type and the placement of model IVs, detection samples increase roughly linearly with the number of model-DM rows. In all cases, except in panel (c) when detecting the binary no-marginal-effect model IVs placed four per block, the detection sample is clearly less than one order of magnitude bigger than the number of rows of the model DM at stake, i.e., less than ten times the number of rows that exhaustive evaluation would need to detect ~90% of the model IVs of interest with P value$\leq$0.05 (*sensu* Fig.3) directly in their model DMs, i.e., in the absence of random IVs (and without correcting for multiple-testing all the columns of the model DM). However, even in panel (c) the detection samples are barely above one and a half orders of magnitude higher, e.g., for the 400-row models they are 4'250 and 6'650 rows with one and four model IVs per block, respectively.

Consistent with Figs.30,31, Figure 33 shows that, given a number of model-DM rows, the IVs of the binary and trinary models with marginal effects are easier to detect than the model IVs that lack such effects. Furthermore the detection samples of trinary model DMs with and without marginal effects are ~40% and ~60% of those of the corresponding binary models, respectively. And the detection samples are ~45% smaller when binary and trinary no-marginal-effect model IVs are placed one per block vs. four per block, whereas the samples are very similar when model IVs have marginal effects.

Furthermore and as intimated, dvMom$^1$ik has about the same power as dvMom$^2$i and the Z-valued versions, except when detecting the trinary non-marginal-effects model IVs placed one per block. In this case dvMom$^1$ik has lower power than the other dvMom's, requiring, e.g., 1'500 DM rows to detect the two reference IVs of the applicable 200-row model DMs (vs. the ~1'200 rows required by say dvMom$^2$i$_Z$ in panel (d) of Fig.31; 25% more rows). This is consistent with the detection samples of dvMom$^1$ik and dvMom$^1$ik$_Z$ in Fig.27ii for the same trinary 200-row model DMs in the presence of say 1'000 independent random IVs, which are 1'880 and 1'680 rows, respectively, i.e., ~12% more rows are needed by dvMom$^1$ik in Fig27ii (dvMom$^2$i and dvMom$^2$i$_Z$ require 1'720 and 1'570 rows in Fig.27ii).

Finally and remarkably, for every given class of model DMs in Figure 33 the ratio remains about constant between the detection sample of a set of model DMs and the rows of said model DMs, over a number of model-DM rows spanning about one order of magnitude in Figure 33 (50-800 rows and 100-800 rows, resp.).

These model-class-typical ratios suggest that the DV associations in the randomly encountered model DMs remain qualitatively similar when the rows of the model DMs increase, to the extent that association quality is reflected by the dvPAS-detectability of the model IVs that were tested.

**Detection samples as a function of the number of blocks.**

The last aspect considered in this section on detecting the IVs of models that are embedded individually in larger DMs with IV blocks, is how power and false positives react when said DMs have increasing numbers of random blocks. This was studied for the P values of dvCHIx-ijkl and dvMom$^2$i$_Z$ by simulating 500 200- and 100-row 8-IV binary model DMs with and without marginal effects, respectively, and placing model IVs one and four per block in DMs having 25, 50, 100, 200, 400, and 800 total blocks of which the blocks without model IVs have o12345 marker frequencies at their pos.49 IVs.

The observed trends (not shown) are consistent with those in Fig.27, with the 60%-0.1 P value detection samples for models with marginal effects plateauing above 100 blocks, those for models without marginal effects increasing log-linearly with the log of the number of blocks, the false positives of dvMom$^2$i$_Z$ being uniform-(0,1)-distributed throughout, and those of dvCHIx-ijkl becoming so with 50 random blocks or more.

**No interference when a 4- and an 8-IV model of DV association, both with or without marginal effects, co-occur in a DM with blocks.**

Figure 34 shows the power and false positives of the P values of selected dvPASs when detecting the sets of 1'000 model DMs studied in Figs.30,31 but with the models co-occurring independently as pairs of a 4- and an 8-IV model, both either with or without marginal effects. The results are based on simulating and detecting 1'000 such pairs and are conditional on the rows of the tested DMs sufficing for 60%-0.1 P value detection of the two reference IVs of the 4-IV models in the pairings. Model IVs are placed one per block; the case with four model IVs per block is not studied (but see below).

The figure shows no interference between the independently co-occurring models, i.e., the detection of both the IVs of the 4-IV models and their model-linked IVs as well as the generation of false positives are the same as when the models are simulated and detected in isolation. Indeed i) in all panels of Figure 34 the detection samples of the IVs of the 4-IV models and their five model-linked IVs are like in isolation, ii) the c.d.f.s of the binary-model IVs without marginal effects vary most across marker frequencies, also like in isolation, and iii) the c.d.f.s of false positives remain on the diagonal. Further, the c.d.f.s of the 8-IV models without marginal effects are clearly above 60% at P value 0.1, consistent with their detection samples in isolation being smaller than those of the corresponding 4-IV models.

**Using the P values of dvMom$^2$i and dvMom$^1$ik (and Z-valued versions) requires erasing strong marginal effects.**

Figure 35 studies, again in DMs whose IVs form blocks, whether the independent co-occurrence of a marginal- and a no-marginal-effect model of 8-IV DV association affects the power of the moment-based dvPASs that detect the two types of model IVs, the detection of model-linked IVs, and the generation of dvPAS false positives at IVs in blocks with no model IVs. Results are based on pairing the binary and trinary 8-IV marginal-effects models studied in isolation in Fig.30 to the corresponding no-marginal-effect models studied in isolation in Fig.31.

Figure 35i shows results when model IVs are placed one per block in 16 blocks. Panels (a,c) show that the IVs of the models with marginal effects and those of their model-linked IVs are detected by dvCHIx-ijkl exactly as they are in isolation in panels (b,d) of Figs.30,31. The results, e.g., show both that the c.d.f.s of individual model IVs with different marker frequency differ from each other as they do in Figs.30,31 and that the c.d.f.s of false positives remain on the diagonal.

Since there is no detection by dvCHIx-ijkl of the IVs of the co-occurring no-marginal-effect model IVs, their dvCHIx-ijkl c.d.f.s and those of the corresponding model-linked IVs are on the diagonal like those of the false positives. The same conclusion can be drawn from panels (a,c) in Figure 35ii where the eight plus eight model IVs are placed four per block in two blocks and patterns are very much like in panels (a,b) of Fig.32.

Panels (b,d) in Figures 35i,ii show the detection of both types of model IVs by dvMom$^2$i$_Z$. To let the inflated false positives of dvMom$^2$i$_Z$ become uniform(0,1)-distributed, every marginal effect is erased that has standard marginal-effect P value≤ 0.002 and 0.003 in Fig.35i and 0.0023 and 0.0015 in Fig.35ii, in the binary and the trinary cases with 1 and 2 d.f., respectively. The same was done in Fig.28 to the same end. Otherwise, as was also shown in Fig.28, the presence of IVs with strong marginal effects inflates the P values of dvMom$^2$i$_Z$ (as well as of dvMom$^2$i and the two dvMom$^1$ik's), making it impossible to take them at face value.

Panels (b,d) in Figures 35i,ii show both that, after erasing above-threshold marginal effects, the IVs of the no-marginal-effect model DMs and their model-linked IVs are detected like in isolation and that the c.d.f.s of the five IVs from blocks with no model IV(s) are on the diagonal as in isolation. And like in panels (c,d) of Fig.28, the detection of the IVs of model DMs with marginal effects and their model-linked IVs (both IV types freed of strong marginal effects) is comparable to that of the corresponding IVs of, or related to, the models without marginal effects.

Indeed in Figure 35i the two reference model IVs of the models with erased marginal effects are detected 54 and 49% of the time with P value$\leq$0.1 in the binary and trinary cases, respectively, vs. 55 and 51% in Figure 35ii, i.e., much like the 60% detection of the two no-marginal-effects reference IVs focal to each plot. This shows again that the randomly encountered model DMs with marginal effects contain genuine higher-order DV associations that survive to a great extent the erasure of their marginal effects.

The above together with the observations in Figs.27,28 about dvMom$^3$i$_Z$ in DMs without blocks, indicate strongly that in general higher-order effects in randomly encountered model DMs make their IVs almost as dvMom-detectable as their lower-order effects do and that these higher-order effects must contribute to these IV's significance *sensu* Fig.3 (below). This makes it hard to defend the traditional to-test-first status of marginal effects from an heuristic-frequentistic perspective.

**Interference when two randomly encountered models co-occur in a DM whose IVs form blocks: Remaining cases.** Figure 36 shows results with additional binary and trinary pairs of randomly encountered 100- and 800-row 8-IV model DMs of DV association, namely when two types of model DMs of interest co-occur independently in a DM with IV blocks and DM rows allow the IVs of the 100-row "first" model type to be 85%+ detected with P value$\leq$0.1 and those of the plot-focal 800-row "second" set to be detected ~60% of the time, by applicable dvMom$^n$i's.

Results are based on 500 pairings of a no-marginal-effect model with a marginal-effects model, 500 pairings of two no-marginal-effect models, and 500 pairings of two marginal-effects models. As intimated, in all three cases the first and the second model DMs have 100 and 800 rows, respectively. The results complement those above with pairs of randomly encountered 200-row model DMs of the same or different type and equal detectability *sensu* Fig.3.

Figure 36 shows that in these three additional cases the IVs of the less detected models are detected as they are in isolation and the accompanying false positives remain uniform-(0.1)-distributed. Furthermore the results show that the presence of a strongly dvPAS-detected model with or without marginal effects affects neither the power of dvMom$^1$i P values to detect the IVs of a co-occurring weakly detected marginal-effects model nor the false positives of the P values of dvMom$^1$i and dvMom$^1$i$_Z$.

Note that when the 60%-detected second model is an *ad hoc* weakened marginal-effects model, the plots use dvMom$^1$i's rather than the higher-power dvCHIx-ijkl, because the latter shows too many false positives as DM rows are increased in order to detect 85%+ of the time the higher-order first model. The excess in false positives of dvCHIx-ijkl detracts from the score's empirical usability, but dvCHIx-ijkl keeps its slight power edge over the dvMom$^1$i's, e.g., when the dvCHIx-ijkl value for 0.1 false positives is inferred from a strongly non-conservative c.d.f. of simulated dvCHIx-ijkl false positives and used as empirical 0.1 dvCHIx-ijkl P value (not shown).

Figure 36 and Fig.35 above suggest the general rule that IVs in lower-order DV association interfere with the detection of IVs in higher-order DV association by inflating the false positives of the dvMom$^n$ P values that detect the higher-order IVs. However, the figures also show that lower-order IVs can do this only if the rows of the DM in which both types of IVs co-occur are numerous enough for the lower-order IVs to be strongly dvPAS detected. This preliminary rule is revisited and refined immediately below.

**Strongly dvPAS-detected pure *m*-IV DV associations inflate dvMom$^n$i false positives whenever *m*=*n*-1 or when both *m*=1 and *m*<*n*.**

Figure 37 shows dvMom$^n$i detection power and false positives when an *m*- and an *n*-IV binary model of pure DV association co-occur independently in a DM whose IVs form blocks and whose rows suffice for both 95%+ detection with dvMom$^m$i P value$\leq$0.1 of the IVs of the *m*-IV "first" model and ~60% detection with dvMom$^n$i P value$\leq$0.1 of the IVs of the *n*-IV "second" model. Results with dvMom$^n$M and Z-valued dvMom$^n$i's are identical here because the binary markers at both the DV and the targeted IVs have frequency 0.5.

The results are for ($m,n$) pairings (1,3), (1,4), (2,3), and (3,4). Also studied are (3,2) and (4,2) pairings of pure 3- and 4-IV first models with pure 2-IV second models that are *ad hoc* "weakened" to lower to ~60% the detection of their IVs with dvMom$^2$i P value$\leq$0.1 given the desired DM rows, while dvMom$^3$i and dvMom$^4$i detect 100% and ~96% of the time with P value$\leq$0.1 the IVs of the pure 3- and 4-way first models in either pairing, respectively.

The ($m,n$) case (1,2) is examined in detail in Fig.35 (1-way: marginal effect) and the (2,1) case is indirectly treated in panels (a,b) of Fig.36 where the strongly detected randomly encountered no-marginal-effect first models and the weakened-detection marginal-effects second models are, to a first approximation, models of mainly 2- and 1-way DV association, respectively, at least regarding their empirical dvPAS-detectability.

In the (a,b) panels of Figure 37 for (1,3) and (1,4) pairings, respectively, marginal effects with standard P value$\leq$0.0009 and 0.001 (1 and 2 d.f., resp.) are erased to let the false positives of the P values of dvMom$^3$i and dvMom$^4$i, respectively, become uniform-(0,1)-distributed. These results, together with the afore-documented inflated false positives of dvMom$^2$i in the presence of IVs with strong marginal effects, show i) that strong enough marginal effects can deflate the P values of dvMom$^n$ when $n=2$, 3, and 4, i.e., possibly whenever $n>1$, resulting in excessive false positives, and ii) that erasing marginal effects that are stronger than an *ad hoc* chosen threshold succeeds in forcing the c.d.f. of dvMom$^n$ false positives down to the diagonal in these cases too.

Panels (c,d) show excessive false positives for two cases with $n=m+1$, namely for ($m,n$) equal (2,3) and (3,4) and dvMom$^3$i and dvMom$^4$i, respectively. This is consistent with the uncorrected results (not shown) of the corrected results in panels (a,b) for the ($m,n$) pairings (1,3) and (1,4). In any case, since the excesses in panel (c,d) are not observed when detecting pure 3- and 4-IV models of DV association in isolation (with and without IV blocks), these excesses too must disappear after erasing from the DM at stake every strong pure 2- and 3-way effect, respectively.

In a non-shown additional case of $n=m+2$ for ($m,n$) equal (2,4) there is no inflation of dvMom$^4$i false positives despite the two IVs in "perfect" 2-way DV association being detected 100% of the time with dvMom$^2$i P value$\leq$0.00. This is welcome but not consistent with how in panels (a,b) of Figure 37 1-way effects deflate the P values of dvMom$^n$i when ($m,n$) is (1,3) and (1,4), i.e., when $n=m+2$ and $n=m+3$, respectively.

Finally, in panels (e,f) the ($m,n$) values are (3,2) and (4,2), i.e., $m>n$, but the c.d.f.s of the false positives of dvMom$^2$ match the diagonal without additional manipulations. This indicates that strongly detected $m$-way effects cannot deflate the P values of dvMom$^n$ when $m>n$. This conclusion too needs mathematical proof or confirmation by simulation of additional cases with $m>n$ and $n>2$.

The above observations confirm that when models of pure $m$- and $n$-IV DV association co-occur in a DM, the false positives of dvMom$^n$i P values (and dvMom$^{n-1}$ik's) can become inflated, compromising thereby the use of these P values in detecting pure $n$-IV DV associations. The inflation happens when both $n=m+1$ and the pure $m$-IV DV associations of interest are strongly detected by say dvMom$^m$i given the rows of the DM at stake. Additionally, strong enough 1-way (marginal) effects can inflate the false positives of any dvMom$^n$ when $n>1$. Also these two conclusions need confirmation with additional simulations and mathematical proof.

Furthermore, note in panels (c,d) of Figure 37 for the ($m,n$) pairings (2,3) and (3,4) the remarkably dissimilar behavior of the c.d.f.s of the dvMom$^n$i P values ($n$, not $m$) of first-model $m$-type IVs and those of random-block IVs: The dvMom$^n$i c.d.f.s of said first-model P values tend towards the diagonal while those of random-block IVs and those of, or related-to, second-model $n$-type IVs are clearly above the diagonal.

Indeed also the dvMom$^n$i P values of the model-linked IVs from the block with the first $m$-model IV tend towards the diagonal and in the order suggested by the extent to which these model-linked IVs behave like model IVs in Figs.30,31, e.g., those at pos.89 are closer to the diagonal than those at pos.64.

Therefore, given these ($m,n$) pairings the P values of the dvMom$^n$i's ($n$, not $m$) of the IVs in pure $m$-IV DV association are immune to the deflation that affects the dvMom$^n$i P values of both random-block IVs and the IVs in pure $n$-IV DV association, albeit the naive expectation would have been that the dvMom$^n$i P values of $m$-model and $m$-model-linked IVs would be deflated as well.

Be the latter oddities as they may and despite the desirability of confirming the results above using pure $n$-IV DV associations involving trinary IV markers and above, it appears fully safe for a user who wants to analyze an empirical DM with dvMom$^n$i and dvMom$^n$ik (or Z-valued versions) to start by first finding all IVs in strong 1-way DV association, erase these associations, move on to finding all IVs in 2-way DV association, erase them, etc., until reaching the $n$-way level.

Indeed no problems were observed when detecting one or more independently occurring pure $n$-way DV associations of any strength neither in random DMs with and without blocks nor when additionally a higher-order $m$-way DV association ($m>n$) of any strength co-occurs in the same DM. Obviously, a mathematical demonstration of this conclusion too is highly desirable.

**Mining an empirical DM by flagging extreme dvPAS values directly.**

The *modus operandi* is studied here that a user may wish to follow in analyzing an empirical DM (eDM) with the goal of flagging DV-associated IVs in the eDM if they have low dvMom$^1$ik P values. First, however, the possibility is explored of flagging directly such IVs when they have extreme dvMom$^1$ik scores, which would eliminate the computational burden of having to estimate P values and/or Z values by permuting the markers in the DV.

In the chosen case study, two 8-IV binary models co-occur in a binary eDM whose IVs form blocks, one model with and one without marginal effects at the DV. Whenever possible the two cases are studied of placing model IVs one per block in 16 blocks and four per block in four blocks, respectively, as was done in Fig.35. For the first time here, however, DMs with two types of blocks are also studied: In addition to the usual 100-IV LPL blocks, some DMs below consist of tandems of LPL blocks and "CHJP" blocks of 1'000-marker chinese and japanese haplotypes (M&M).

The results show that direct flagging of extreme dvMom$^1$ik values in DMs with IV blocks is not effective, making the afore-demonstrated power and resolution of the P values of these scores in the presence of block structure even more remarkable and felicitous. Even less effective is direct use of dvMom$^2$i, which is also shown for completeness.

Direct use of dvMom$^1$ik values is likely viable and powerful in DMs without IV blocks, but this is not studied here because background associations are pervasive in genomics.

The focus is on the detection of the IVs of the no-marginal-effect model in each pair. The marginal-effects model is included in the simulations to exemplify how a user can deal with strong marginal effects that inflate the false positives of dvMom$^2$i and dvMom$^1$ik (and Z-valued versions). Indeed IVs with strong marginal effects at a DV are best detected with standard methods, unless they are also involved in statistically stronger higher-order DV associations.

Therefore throughout the sections below, the user is assumed to find him/herself forced to bring the c.d.f. of dvMom$^1$ik P values at non-model IVs down to the diagonal by erasing eventually present marginal effects that exceed a threshold which the user determines empirically for each eDM at hand before estimating dvMom$^1$ik's and dvMom$^1$ik P values (below).

Since the user is assumed to know neither the block structure nor the blocks with model IVs, the study examines whether one can tune the false positives at the IVs of the unknown blocks without model IVs by both adding 500 random independent IVs to the eDM of interest and determining for these IVs the maximal P value of the marginal effects to be erased. This addition and erasure are also performed when studying the flagging of IVs with extreme dvMom$^1$ik scores (rather than small dvMom$^1$ik P values), an *a priori* imposition that is left unexplored.

Three types of DMs with different block configuration are considered as intimated. The first type of DM uses only LPL blocks. In the second, "LPCJ type" the odd blocks (from the left) are LPLs and the even blocks are CHJPs. In DMs of the third, CJLP type the order is reversed. By comparing results with these DM types one can assay how interactions of block sequences with subsets of model IVs affect dvPAS power and false positives. If in an LPCJ DM, e.g., the sixteen model IVs occur four per block in four successive blocks, then the first and second tetrads of each type of model IV fall in a LPL and a CHJP block, respectively, whereas if the model IVs occur one per block then the two block types harbor odd and even model IVs, respectively. Obviously the allocations and thus the comparisons are toggled in CJLP DMs.

**Flagging directly IVs having extreme dvMom$^1$ik and/or dvMom$^2$i values is not powerful in DMs with IV blocks.** Figure 38 shows that the resolving power is very low if one tries to flag directly the IVs that have extreme dvMom$^1$ik and/or dvMom$^2$i values relative to the values of random-block IVs, despite the DMs having enough rows for 95%+ detection with dvMom$^1$ik P value$\leq$0.1 of the eight no-marginal-effect model IVs. As intimated, these results are conditional on strong marginal effects above an empirically determined threshold having been erased through toggling of excess markers at involved IVs.

Figure 38 presents distributions based on simulations of the first 100 model pairs from the sets of 100-row 8-IV model DMs with and without marginal effects from Fig.33. These pairs are akin to the pairs of 200-row 8-

IV model DMs used in Figs.35i,ii but were preferred over the 200-IV ones in order to speed up simulations. Shown are c.d.f.s for the dvMom$^2$i and dvMom$^1$ik scores (not P values) of the 8+8 model IVs, 5+5 model-linked IVs from the first block with one or more model IVs of either type, 16 IVs from as many blocks without model IVs, and five IVs from the 500 aforementioned user-added random IVs. Dedicated c.d.f.s are also shown for the no-marginal-effect model IVs that were placed in even vs. odd blocks.

In both LPCJ and CJLP DMs the total number of blocks is 20 and 32 when model IVs occur four and one per block, respectively, in order for DMs to have 16 random blocks with no model IVs. Therefore these DMs have 11'000 and 17'600 block-forming IVs; their rows are 4'000 and 6'000, respectively. Also LPL DMs have 4'000 and 6'000 rows in said two cases (and 10'000 block-forming IVs in 100 100-IV blocks). As stated, independently every DM receives 500 "user-added" random IVs.

Figure 38 shows that the dvMom$^2$i's of marginal and no-marginal-effect model IVs are similar whereas the dvMom$^1$ik's of no-marginal-effect model IVs are clearly larger; this is so across all types of DMs. With one model IV per block ~90%+ of the dvMom$^1$ik's of the no-marginal effect model IVs are larger than the 10% largest dvMom$^1$ik among the user-added random IVs (top, bottom), nearing so the 95%+ detection with dvMom$^1$ik P value$\leq$0.1 in LPL DMs. However, the difference is not nearly as marked vs. random-block IVs.

Noticeable block-type effects are seen with one model IV per block, pointing to interactions of LPL and CHJP marker sequences with the sampling of block sequences that have desired model-IV markers at their pos.49s and pos.502s, respectively (M&M). Note, in passing, that the dvMom$^1$ik's of model-linked IVs are at times larger than those of model IVs, specially in LPCJ DMs.

In general the dvMom$^1$ik's of model IVs tend to be different from those of the independent random IVs whereas the corresponding dvMom$^2$i's show no such tendency. The dvMom$^2$i result is not surprising given that when one compares the dvMom$^n$i's of different focal columns (as opposed to comparing their P values obtained through DV permutation) one is really comparing values that quantify those focal columns' ($n$-1)-order associations with all other columns, i.e., not specifically their ($n$-1)-order associations with the DV (see Fig.1).

The P values of dvMom$^n$i are instead obtained by permuting the DV and therefore they reflect very narrowly the ($n$-1)-order associations of those focal columns with the DV. Indeed and not surprisingly, flagging directly the IVs with large dvMom$^2$i$_Z$ delivers approximately the same detection power as does using dvMom$^2$i$_Z$ P values (not shown), because both dvMom$^2$i$_Z$ and its P value are estimated by permuting the DV.

The better performance of dvMom$^1$ik is also expected since this score assesses directly both the pairwise state at the DV and that at the focal column at stake. If anything, dvMom$^1$ik was expected to compete much better with its P value obtained through permutation of the DV than it does in Figure 38. This mixed performance, however, suggests that there may be an opportunity to refine dvMom$^1$ik to turn it into a score that can be used directly.

The strong differences in Figure 38 between the c.d.f.s of the dvMom$^n$ik's of the added random IVs vs. those of the IVs in random blocks, together in most cases with i) the latter's disorientingly weak differentiation from those of model IVs embedded in blocks and ii) the very strong reaction of the dvMom$^2$i and dvMom$^1$ik c.d.f.s of model IVs when individual marker sequences with applicable model-IV markers or model-marker combinations are sampled from either source of IV blocks, make direct flagging of model IVs with extreme dvMom$^2$i and dvMom$^1$ik values very ineffective, compared to flagging those IVs that have small dvMom$^1$ik P values and/or small dvMom$^2$i P values.

All in all, the dismal lack of resolving power of dvMom$^1$ik and dvMom$^2$i when these scores are used directly in DMs with IV blocks highlights one more time the felicitously high resolving power of both the P values of the scores and the scores' Z-valued versions used directly, which both are obtained by permuting the DV of the DM at stake.

**Mining a single empirical DM using permutation-obtained P values of dvPAS: Background Distributions.**

Figure 39 shows the background distributions pertinent to the analyses of individual eDMs further below. It shows the expected power and false positives when testing the two model pairs chosen for the eDM study and, as benchmark, the first 100 model pairs used in Fig.35 in DMs with 100 LPL blocks total. The shown c.d.f.s are based on 100 simulations of each of the two model pair and 100 simulations of the 100 model pairs (one simulation per pair). The two chosen 100-row 8-IV model-DM pairs are numbers 265 and 865 in Fig.33.

C.d.f.s are shown for the P values of the dvMom$^1$ik's and dvMom$^1$ik$_Z$'s (as comparison) of the 8+8 model IVs with and without marginal effects, 5+5 model-linked IVs from the first block with one or more model IVs of either kind, five IVs from five different blocks without model IVs, and five of the 500 user-added random IVs. DM rows are 1'300 and 1'700 for the case with one and four model IVs per block, respectively. These numbers of rows yield ~80 and 60% detection with dvMom$^1$ik P value$\leq$0.1 of the no-marginal-effect model IVs of pair 265 when model IVs are placed one and four per block, respectively vs. 87 and 70% for pair 865.

Figure 39 shows that erasing marginal effects that are as strong or stronger than the threshold lets the false positives at both the 500 user-added random IVs and the IVs in blocks without model IVs have c.d.f.s that overlap the diagonal. Further, the figure shows that model IVs are clearly better detected than model-linked IVs when one model IV is simulated per block, whereas when the eight model IVs of model pair 865 are placed four per block said model IVs are detected like their model-linked IVs. When pair 265 is simulated, both types of model-linked IVs are detected like the model IVs from the marginal-effects model if four model IVs are placed per block (vs. almost not detected when only one model IV is placed per block).

The results with one model IV per block show that the P values of the odd and even model IVs of model pair 865 behave differently but not those of pair 265, while with four model IVs per block the P values of the first- and second-tetrad model IVs of pair 265 behave differently but not those of pair 865.

Not shown in Figure 39 is that when simulating model pairs 265 and 865, the forced sampling of the sequences of the random LPL blocks (without model IVs; M&M), that lets the sampled blocks' pos.49 markers show the o12345 frequency scheme, causes noticeable, if much weaker, block-type effects that reverse when going from dvMom$^2$i to dvMom$^1$ik, all of which is puzzling. However, no such effects are noticeable in the pooled results of the 100 model pairs.

**Studies of individual eDMs.**

The above analyses of individual eDMs can be expected to be hindered by the noise from adding random IVs and that from toggling excess markers to erase above-threshold marginal effects. This noise is jointly demonstrated here by showing five separate analyses ("runs") of every eDM at stake, during each of which an independently generated batch of 500 random IVs is added to each eDM of interest before the modified eDM undergoes an independent erasure of marginal effects. Two eDMs are generated and studied for each of the two model pairs introduced above (i.e., four total).

In Figure 40a,b only LPL blocks are used and the c.d.f.s of the dvMom$^1$ik P values of the various types of IVs in the first 30 blocks (100 blocks total) are presented, together with the c.d.f. of the 500 user-added random IVs pooled. The cases of model IVs being placed one and four per block, respectively, are studied. In each figure the maximal P value for erasure of marginal effects is determined by tuning the false positives at the 500 added random IVs of the first pair-265 eDM over 100 independently generated sets of DMs. The erasure P value is also used for other eDMs to showcase the need for a dedicated tuning of the erasure P value for each eDM at stake.

Note i) in each run, the similarity between the c.d.f.s of the random-block IVs and the added random IVs, even when they depart from the diagonal and the empirical "overall null c.d.f." (that pools the ten runs with each model pair); and ii) how model-IV c.d.f.s rise and sink across eDMs in concert with those of random-block IVs and user-added random IVs, keeping so the c.d.f.s' relative departure from both the diagonal and the two null c.d.f.s of the eDM at stake, even when a run's empirical null c.d.f.s are clearly above or below the diagonal because of noise across eDMs.

For model pair 265 in the top five plots in the left half of Figure 40a, it would seem that the erasure of marginal effects with standard P value$\leq$0.001 (1 d.f.) lets the individual-run c.d.f.s of the 500 user-added random IVs both be close to the diagonal and show almost no noise across the five plots. However, this is not guaranteed since these c.d.f.s depart much more strongly from the diagonal in Figure 40b (where the erasure P value turns out to be the same by coincidence).

The lack of noise across the runs with each eDM is not unexpected either, since the type-I-error results further above showed much less noise across DMs with 1'000 independent random IVs than across DMs with 100 100-IV LPL blocks (i.e., 10'000 IVs; Fig.20 and Fig.23, see also below). Here only the 500 used-added IVs create random noise across same-eDM runs (if one neglects the noise from toggling randomly chosen excess markers to erase strong marginal effects) and the large number of these IVs reduces the noise in the actually evaluated DM.

Figure 40a,b also suggests that the extra noise from independently generating and adding a batch of random IVs as well as from tuning false positives through a single erasure of marginal effects, does not reduce by much the power of dvMom[1]ik P values. Indeed the c.d.f.s of model-IV P values are always clearly above both the diagonal and the c.d.f.s of the two types of random IVs (i.e., model-IV P values tend to be smaller). Indeed in Fig.33 when model IVs are placed one per block and no random IVs are user-added, one needs ~1'100 rows to detect with P value$\leq$0.1 60% of the IVs of the 1'000 200-row 8-IV no-marginal-effect model DMs, whereas one needs ~1'300 rows when 500 random IVs are added (~21% more rows).

As intimated, almost correct false positives of both kinds are observed in the top right halves of Fig.40a,b that each shows five runs with the first pair-865 eDM. However, in the left and right bottom halves of Figure 40a,b with the five runs of the second eDMs of pair 265 and 865, respectively, the c.d.f.s of random-block IVs and added random IVs are clearly above or below the diagonal, respectively. This is not surprising since, as mentioned, the threshold P value for marginal-effect erasure was determined over 100 runs with the first pair-265 eDM of each figure, to each of which independently 500 random IVs were added. Therefore the nearly correct false positives observed with the first quintet of pair-865 eDMs in Fig.40a,b are a coincidence. The conclusion is that tuning the false positives of any given eDM requires a dedicated estimation of the threshold marginal-effects P value to be used.

The results in Figure 40 are consistent with the fact that across null DMs whose IVs form 100 independent 100-IV LPL blocks, the c.d.f.s of same-DM dvPAS type I error show stronger departures from the diagonal than do null DMs with 1'000 independent IVs. As intimated, adding the 500 independent random IVs (for a total of 10'500 IVs) should result in false positives with less across-DMs noise than without the addition.

Figure 40a,b also shows that the P values of non-model IVs from blocks with one or four model IV behave at times almost like those of model IVs, with marked variation across eDMs and much weaker variation across the analysis-by-user runs with a given eDM. This dual behavior can be shown by both or only one of the two kinds of model-linked IVs examined here.

The main hope of adding random IVs was that by tuning their false positives one would also tune those of the IVs in unknown random blocks (with no model IVs). And it works: Figure 40 shows that i) such tuning lets the c.d.f.s of dvMom[1]ik P values of IVs in blocks with no model IVs tend towards the diagonal and that ii) the c.d.f.s of both types of "random" IVs across the runs with a given eDM swing together up or down the diagonal (which is welcome and in hindsight should have been expected). This shows that tuning the false positives of added random IVs is an effective way of normalizing the false positives at both the added random IVs and the IVs in the blocks without model IVs, blocks normally unknown to the user.

Figure 41 presents additional results when two different types of blocks occur in the eDM at stake. The results shed light on how block-type interacts with both power and the generation of false positives at user-added random IVs. The two blocks are the aforementioned 100- and 1'000-marker LPL and CHJP blocks, respectively. The eDMs studied have 20 blocks in order to keep the number of total IVs around 10'000 and the computational expense similar (but certainly not the noise combinatorics). Only the case with four model IVs per block is studied (with one model IV per block and 20 total blocks one would have only four blocks without model IVs, two of each type).

Figure 41 confirms that by forcing the c.d.f. of the false positives at added random IVs to match the diagonal, one also tunes the false positives at the IVs in random blocks of either type. Indeed in every one of the five runs in the top rows of the four figures, the c.d.f. of the run's false positives at random blocks matches very closely that of the run's added random IVs, with both curves departing from the diagonal in the same direction and to a similar extent in a given run and the only clear exception being in Figure 41a (i.e., one out of 20 shown runs).

However, the similarity of departure is most marked for the IVs that belong to random CHJP blocks, i.e., the random blocks with most IVs in the eDM, whereas for the IVs in the random LPL blocks the similarity is less marked. This suggests that extreme dvPAS P values that are caused by noise are due to random associations among the types of random IVs that are more numerous in the DM at hand.

Regarding power, however, there is in Figure 41 a possible tendency for the no-marginal-effect model IVs in LPL blocks (thickest solid black lines; mid-row plots) to be detected better than those in CHJP blocks, which if confirmed would be disorienting. More important and like in Fig.40, Figure 41 shows that generally model-IV c.d.f.s tend to be above those of both added random IVs and random-block IVs, even when noise pushes the latter's c.d.f.s to be markedly above or below the diagonal. This includes cases in which the c.d.f.s. of the IVs in random LPL and CHJP blocks depart from the diagonal in opposite directions. In these cases the latter c.d.f.s are pushed by noise in the same direction as those of model IVs in LPL and CHJP blocks, respectively, so that the difference between model and random-block IVs that share a block type is more or less maintained. This result too if confirmed would be disorienting.

The latter observation suggests that the power to detect model IVs may be higher and the generation of false positives may be more correct, if one compares model and non-model IVs from blocks of the same type. However and of course, whenever possible one should prefer the *brute-force* approach of increasing DM rows until background noise becomes irrelevant. As a curiosity, note that all of the observations about random-block c.d.f.s in Figure 41 can already be made by assaying only two blocks of each type (not shown) rather than eight blocks of each type as was done here.

**Detecting IVs in 2-way DV association when many such IVs occur in a DM.**

Without exception dvPAS power and false positives were studied above using models comprising at most eight DV-associated IVs. When two models co-occurred independently in the same DM, a maximum of 16 model IVs was ever present in a DM. However, DV associations in a DM may involve many more IVs, including all IVs of a DM. One could imagine that problems may arise for PAS and dvPAS power and false positives when many columns and IVs in a DM are reciprocally associated and DV-associated, respectively.

In this last section before the Discussion, dvPAS power and false positives are studied when the models to be detected involve up to 495 pairs of IVs (i.e., 990 IVs) in extended or independent weak 2-way DV association. This is done in 1'000-IV binary and trinary DMs without block structure. The alternative of conditioning all assays on there being always 1'000 random columns in the DM at stake is not studied.

In the binary case, the DV associations of the desired IV pairs are generated by adding to "affecteds" additional rows with 00 and 11 2-marker combinations at each involved IV pair, whereas in the considered trinary H&W case additional 00, 11, and 22 2-marker combinations are added. In both cases the 2-marker combinations of "controls" are random. Also here the additions of 2-marker combinations are made in a way that no marginal effects are generated (M&M). Two cases are studied in which the minor markers of the IV pairs have either binary-marker frequency 0.5 or 0.1, or the trinary H&W thereof.

To generate extended "in phase" 2-way DV association at say three IV pairs in the affecteds half of a binary DM (Fig.2), one adds additional 00'00'00 and 11'11'11 6-marker combinations at the six corresponding model IVs. Instead to generate three "off phase" independently DV-associated pairs one adds, again in affecteds, additional 00s and 11s combinations independently at each of three IV pairs such that all the 2-marker combinations at any of the three IV pairs co-occur randomly with those at the other two pairs. Obviously the markers at each of the remaining random IVs are simulated to be independent from those at the other IVs.

In Figure 42 specifically, when a 2-way DV association of IVs involves $n$ in-phase binary markers of frequency 0.5 and 0.1 then 10% and 20% of the applicable $n$-marker combinations at stake, respectively, are additional $n$-marker runs; or 10% when the model IVs have trinary markers with the H&W of frequency 0.5 (see text). The same percent additions happen independently at every IV pair in the corresponding cases of off-phase DV association.

As usual the focus is on the DV-associated IVs' 60%-0.1 P value detection samples and the concomitant false positives at random IVs. An additional focus is how the detection sample of the DV sinks with increasing numbers of IV pairs in in- or off-phase 2-way DV association. As stated, studying the detection of the DV in this case sheds light on PAS power and false positives when a focal column is in weak in- or off-phase 2-way association with many other columns. False positives turned out to be uniform-(0,1)-distributed in all simulations and are not further discussed.

Figure 42 shows that when IVs are in-phase 2-way DV-associated, without exception the log of the detection sample of these IVs decreases monotonously with the log of the number of DV-associated IVs, until the sample plateaus about 100 and 600 DM rows for binary frequency 0.5 and 0.1, respectively, when the associated IVs are ~100 or more vs. about 200 rows with 600 in-phase IVs or more in the trinary 0.5 H&W case. The in-phase curves in the binary and trinary H&W case for both the model IVs and the DV, are similarly shaped.

The similarity of plateauing in the binary and trinary in-phase cases most likely reflects the very discrete number of copies of the two or three extended multi-marker combinations (i.e., 000... and 111... vs. 000…, 111…, and 222...) that the in-phase models generate in the affecteds of these few-row DMs. Indeed in all cases a mere increase by two in the rows of the tested DM lets the detection jump from say 50 to 80%, a clear sign that discrete jumps in the counts of crucial combinations of same-row markers in reaction to increasing DM rows are involved, a situation in which exhaustive evaluation combined with non-parametric tricks like bootstrapping of crucial patterns should deliver their maximal power (Antezana and Hudson, 1999; in the case of PAS and dvPAS there is also the constraint that at least two DM rows be present with the same marker run of interest for the PM to register the run).

One can note that in general the sites of the markers of any crucial, sufficiently long DV-associated marker run disjunct or not and present in three or more copies in the DM should be as readily detected by dvMom$^2$i's and dvMom$^2$ik's as are the runs in Figure 42 that lack marginal effects, i.e., even when the runs' markers "cause" weak marginal effects at the DV (say because the effects are erased if the markers are toggled).

In the binary and trinary off-phase cases, however, the detection samples of the DV-associated IVs are one to three orders of magnitude larger and as expected they do not become smaller as the numbers of off-phase IV pairs increses. Detecting 60% of the time with dvMom$^2$i P value$\leq$0.1 the IV pairs in off-phase DV association requires, e.g., ~35'000 DM rows when binary-model-IV marker frequency is 0.5, regardless of the number of pairs.

However, the detection sample of the DV sinks rapidly with increasing numbers of off-phase IV pairs (as expected given Fig.12), albeit for any given number of off-phase IV pairs the sample is many times larger than when as many IV pairs are in-phase DV-associated. 60% detection of the DV in the off-phase case when binary-IV marker frequency is 0.5 requires, e.g., ~11'000 and ~1'500 rows with 50 and 445 off-phase pairs, respectively, vs. ~150 and ~80 rows when 100 and 990 IVs, respectively, are in-phase DV-associated.

As intimated in the introduction to the power section, when many IVs pairs are in off-phase DV association the situation may arise of a model that involves a single, readily PAS- or dvPAS-detected column, given the rows of the DM at hand, and many additional columns that only become PAS-detectable if one increases in a major way the number of rows of the DM, because each of the latter columns contributes only weakly to the total association that causes the readily detected column to be such.

This model with only one very PAS-detectable column and numerous additional column pairs that are independently 2-way-associated in a much less PAS-detectable way with the very detectable column, could possibly be found in the affected rows of an affected-vs.-controls DM whose total rows only suffice for strongly detecting the more PAS-detectable column (formerly the DV, now an IV). The question would then arise of how to erase the excess in false positives that the presence of such a model in affecteds may cause when one scans say for IVs in 3-way DV association.

In the specific off-phase cases in Figure 42, it is obvious that if one randomizes the one column that is more dvPAS-detectable then the column pairs previously equally independently weakly 2-way associated with said column cannot anymore cause an excess in false positives, since it was through their association with the now-randomized column that these pairs could trigger the to-be-erased affecteds-vs.-controls excess in false positives.

To avoid that the randomization of the one readily detected IV erase all higher-order associations of that column, one can choose a DM in which the permutation of the markers of the IV minimizes both the excess in false positives of concern and the number of toggled markers. This, however, needs to be confirmed with simulations and there is no guarantee that similar erasures can be targeted and performed as easily when dealing with excesses in false positives that are caused by other patterns of association.

## Discussion

PAS and dvPAS are unprecedented algorithmically and datamining-wise. Their robustness, specially dvPAS', with respect to their generation of type I error and false positives under the $H_o$ and various $H_1$'s, respectively, was tested above by exposing the methods to almost every possible challenge that MAA could think of. However, examining their robustness to additional complications is highly desirable.

The main intended use of the various PASs and dvPASs is to test in a many-column DM every column of interest for its associations with others or a DV, respectively. The study of type I error and false positives showed that when testing columns from a variety of DM types the resulting same-DM PAS and dvPAS P values tend to be uniform-(0,1)-distributed. Perhaps non-surprisingly therefore, the Sidak method was found to correct accurately same-DM PAS and dvPAS P values for multiple tests of columns and IVs, respectively, in null DMs with many independent random columns.

Surprisingly, the Sidak correction was also quite accurate when correcting dvPAS P values from a block of columns in which intra-block marker associations come from a sample of chromosome fragments. The correction became conservative only after columns within blocks were made more strongly non-independent by force-sampling block sequences. Both observations have to be replicated using blocks from other chromosomal regions.

if contrary to the above indication, the non-independence of neighboring mutations is mostly strong enough to make the standard Sidak correction markedly conservative, researchers in genomics may consider helping the correction with estimates of the number of effectively independent blocks. Studying this should be specially interesting and challenging when block boundaries are as fuzzy as they must be in real chromosomes away from hot spots of recombination.

The power assays in this paper did not use Sidak-corrected P values because estimating with permutations very small Sidak cutoff P values was too taxing for the available computational resources. Instead the assays used uncorrected permutation-obtained P values, i.e., adopted the criterion that over all the tests that one publishes over a lifetime one should be allowed to generate say a 5% rate of "individual-test error at the 95% significance level", regardless of how one allocates tests across publications and "families of tests".

The criterion is simple enough but does not tell us what to do when whole-DM noise pushes the c.d.f. of all of the P values of a DM away from the diagonal. Heuristically in this case, one may choose to report say every P value$\leq 0.05$ regardless of the whole-DM c.d.f., which would generate said 5% individual-test lifetime type I error at the 95% level. The alternative of taking every PAS or dvPAS P value no larger than the $^L/_{20}$-th smallest P value observed among the L columns of the DM at stake, would not allow the user to report all highly significant columns when these are more than 5% of the columns in the DM at hand.

The power and false positives of the P values of PAS and specially dvPAS were assayed above using nearly every model of association that came to mind and could be simulated. However, models harder to simulate should be also tried, e.g., i) the afore-discussed models of extended DV association in which some IVs are weakly off-phase connected to a pivotal focal IV which is the only IV readily detected by dvPAS in the DM at hand; ii) models of pure 3-way and higher association involving two or more markers per column and extending over many columns, iii) models in which individual columns are involved at once in two or more types of $n$-way association (other than the mixes of mainly 1- and 2-way effects in the randomly encountered model DMs studied above), and iv) trinary models of minimal and extended pure $n$-way association and DV association.

PAS P values were found to have slightly higher power when detecting randomly encountered binary model DMs of column association than when detecting matched trinary model DMs, specially when model DMs have numerous rows. On the other hand, in trinary DMs where IVs form blocks dvPAS P values had much higher power when detecting randomly encountered model DMs of $n$-IV DV association and when discriminating model IVs from model-linked IVs ("fine-mapping"), both of which is good news for geneticists of diploid organisms. The second, however, also means that model-linked IVs are flagged more strongly in the binary case, favoring "coarse-mapping". Another trend is that the power to detect pure $n$-way associations and pure $n$-IV DV associations in binary DMs decreases logarithmically as $n$ increases.

The community is invited to confirm these trends by studying randomly encountered model DMs with four or more markers in every column, as well as pure associations involving such columns. However, 2-way DV associations that extend "in-phase" (Fig.2) over many IVs were shown above to be detected with near-ultimate power regardless of them involving binary or trinary markers. This will hardly change for similar models with four or more markers per column.

The community is also invited to propose and study modifications of PAS and dvPAS that may boost power. Ironically, the likelihoods of the $H_o$ that were worked out by *brute force* above suggest that true-likelihood versions of PAS and dvPAS may boost power more in the trinary than the binary case, but only implementing these versions can tell. The community can also consider the intuition, shared in the likelihoods section, that boosting power may require changing the nature of the items in the rows of the PM while still exploiting the summing of selected same-row items.

There should be much room for improvement of the power of $Mom^n i$, $dvMom^n$, and $dvMom^n ik$ and their Z-valued versions when detecting higher-order associations that involve focal-markers of frequency other than 0.5 and H&W thereof. These PASs and dvPASs indeed were chosen because they detect best the columns of randomly encountered model DMs and columns in pure associations involving binary markers of frequency 0.5.

No attempt was made, however, at optimizing in the trinary case the moment-based dvPASs' summing of marker-conditional scores in order to increase their power when detecting pure 3-way and higher column associations and pure 3-IV and higher DV associations when focal column markers have frequency other than 1/3. But it seems logic that with three or more markers per column one could, e.g., weigh more heavily the marker-conditional scores that are least correlated across the permutations to get P values and/or individual-marker Z scores, before summing the scores up, etc.

It may be also worth exploring what was proposed above when discussing the substracted distributions of Mom[1] and Mom[2] in Fig.6 for focal columns in pure 2- and 3-way association, respectively. As stated there, the power vs. the various Mom$^n$s and dvMom$^n$s should increase if one uses ratios of the likelihoods of models in which some columns are in $n$-column association and the others are random, over the likelihoods of simpler such models or the H$_o$ with only independent random columns (see the likelihoods of said H$_o$ given very small null DMs worked out above by *brute force*).

The just mentioned models ignore indeed column order, so that using such likelihoods and likelihood ratios will be computationally tractable at least in the sense that it will not require exhaustive evaluation of subsets of columns at fully specified positions. Indeed, dealing with background associations may turn out to be a bigger obstacle when calculating the desired likelihoods.

Likelihood-based dvPASs are also worth pursuing because one would expect them to be unaffected by interference of a lower-order $m$-way DV association with the detection of IVs in higher-order $n$-way DV association, an interference shown and neutralized above when detecting with dvMom$^n$i P values the IVs of model pairs with ($m,n$) equal to (1,$n>1$) and shown but not neutralized for several cases of (m>1,n=m+1) but not for (m>1,n>m+1).

Note, however, i) that no interference was observed when using moment-based PASs to detect the columns of two independently co-occurring models of $m$- and $n$-way association (not DV association), albeit several ($m,n$) combinations were tried and ii) that in the only case examined of a column involved in two types of $n$-way association at once the false positives of the two applicable Mom$^n$i$_Z$'s were both correct.

Regarding developing an intuition about the reaction, or lack thereof, of PAS and dvPAS P values to marker patterns in the DM, the study above of power and false positives suggests that PAS and dvPAS power are higher when the $n$-column model DM to be detected contains only a few, in absolute terms, $n$-marker sequences with marker motifs that both depart strongly from expectation and are each present three or more times, the number needed for ~95% significance when bootstrapping the counts of an inference-deciding marker sequence in a DM (or the affecteds of a DM), if one disregards that in the PAS context a motif must be in at least two DM sequences for the PM to register it.

For instance, i) only two and three $n$-marker sequences are forced to be over-represented when a binary and a trinary 2-way DV association, respectively, extend in-phase over $n$ IVs and $n >>2$, two cases in which the power of dvPAS P values is similarly dramatically high; and ii) a pure binary 6-way association is detected by PAS and dvPAS with much lower power than is say a pure binary 3-column association, i.e., power is much lower when $2^5$ 6-marker combinations out of $2^6$ possible ones are over-represented than when this happens to $2^2$ 3-marker combinations out of $2^3$ possible ones.

Fully unclear to MAA is instead why the detection by moment-based PASs of the columns of binary and trinary models of $n$-column association ($n$=5 and 10) is similar but the detection by moment-based dvPASs of the IVs of $n$-IV trinary models of DV association ($n$=4 and 8) is much better than for the corresponding binary models.

For the time being, users wishing to scan a multi-column DM should use the P values of the moment-based PASs and dvPASs that detect with highest power, but separately, each different level of pure $n$-way association and pure $n$-way DV association, respectively. Indeed among the P values of the various all-signals PASs and dvPASs studied above, only those of dvCHIx-ijkl were found to have high power and only when detecting randomly encountered model DMs with substantial marginal effects. It was not studied whether when IVs form blocks the high power of dvCHIx-ijkl (and dvLKx-IJ) for IVs with marginal effects can be exploited directly without estimating P values by permuting the DV.

Using the moment-based PASs and dvPASs requires, however, assaying in an orderly manner the different levels of $n$-way association in a DM, and then combining at each column the P values of the assayed levels. Indeed the above procedure may result in both different columns showing low P values for different orders of $n$-way association and individual columns showing multiple low P values, one for each order of $n$-way association in which the column happens to be strongly involved.

The power simulations above showed indeed that the IVs in the randomly encountered models of DV association with marginal effects tend to show, after their marginal effects are erased, levels of 2-way DV association that are nearly as strong (*sensu* the IVs being detected by dvMom$^2$i and dvMom$^1$ik P values) as those of the IVs of comparably significant random models generated without marginal effects to begin with.

This is observable also standard-frequentistically, e.g., under the tests in Fig.3: When one simulates a randomly encountered 8-IV marginal-effects model with a no-marginal-effects one and the number of rows is enough for 75% of the marginal-effects model IVs to be detected with standard P value$\leq$0.1(*sensu* Fig.3), such detection sinks only to 60% after erasing the first model's marginal effects by toggling randomly chosen excess markers.

This means that also *sensu* Fig.3 the IVs that are 1-way DV-associated in these marginal-effects models are involved at the same time in comparably strong 2-way and higher DV associations. This observation and the exploratory results on partitioning the $\chi^2$ of a DM (Fig.12) confirm the point advanced tentatively in the Introduction that no level of *n*-way association is special frequentistically.

The order to be followed when assaying various levels of *n*-way and *n*-IV DV association in a DM depends on the presence and the intricacies of the causation of eventual interference that affects PAS and dvPAS power and false positives and may arise between co-occurring associations of the same or different orders which can involve overlapping and disjunct subsets of columns in a DM.

On the one hand and as intimated shortly above, no interference of any kind was observed between two co-occurring models of pure *n*- and *m*-way association when several (*m,n*) combinations were tested using the P values of moment-based PASs (Fig.26) nor when both the same column participates in strongly and weakly PAS-detectable 2- and 3-way associations, respectively, and the column's 3-way signal is detected 60% of the time with $Mom^2 i$ P value$\leq 0.1$ (paragraphs after Fig.24), albeit the second conclusion is only about false positives being correct at random columns since detection power was not compared to the one in isolation.

On the other hand, IVs in strong lower-order DV association were found to inflate the false positives of the P values of the moment-based dvPASs that detect IVs involved in one-order-higher DV associations, with the exception of strong enough 1-way (marginal) effects that inflated the false positives of the P values of $dvMom^n i$ and $dvMom^{n-1} ik$ (and Z-valued versions) for *n*=2, 3, and 4, i.e., possibly they can do so for any *n*>1 (from Fig.27 on). Therefore when detecting IVs in pure *n*-order DV association with correct false positives, one must first erase every strong one-order-lower DV association as well as every strong marginal effect at the DV, by toggling suitable randomly picked markers at the involved IVs.

Fig.37, e.g., showed that the presence in a DM of three IVs in pure 3-way DV association inflates the false positives of $dvMom^4 i$ P values, hindering so the detection by $dvMom^4 i$ P values of four IVs in pure 4-way DV association that co-occurred in the same DM. Since there was no such inflation when the latter four IVs were both simulated and $dvMom^4 i$-detected in isolation, it is clear that also in this case one can eliminate the excess in the false positives of $dvMom^4 i$ P values (and of $dvMom^3 ik$ ones) by toggling suitable markers at the three IVs in pure 3-way DV association. When instead *m>n*, it was observed that not even very strong higher-order *m*-way signal manages to inflate the false positives of the P values of $dvMom^n i$ and $dvMom^{n-1} ik$ that in isolation detect lower-order signal without trouble.

To flag most effectively the IVs in pure *n*-way DV association using the P values of moment-based dvPASs, one must therefore scan first for the IVs with strong marginal effects (i.e., in 1-way DV association), erase these IVs' marginal effects, scan for IVs in strong 2-way DV association, erase their 2-way signal, scan for the IVs in pure 3-way DV association, etc. This is also the order in which the power of moment-based dvPAS P values decreases when detecting pure *n*-way DV associations involving the minimal number of IVs that can show the pure association of interest (Fig.1). Out of prudence it appears advisable to proceed in the same order when using $Mom^n i_Z$ to detect columns associated with others.

Inflated *m*-way false positives, however, should also be caused by strong pure *n*-way DV associations that extend over many more than *n* IVs when *n<m* (but this inflation was not simulated above). Even in this case, however, it should suffice to toggle minimal numbers of markers at such *n*-way DV-associated IVs in order to suppress the excess in *m*-way false positives that these IVs would otherwise trigger.

Indeed the work above with 2-way DV associations extending in-phase over hundreds of IVs showed that these IVs can be readily flagged by their low dvPAS P values. Knowing said DV-associated IVs allows one to populate a contingency table in which the to-be-erased over- and under-represented marker combinations become apparent. And if one has to fend with effects caused by low-count marker combinations, one can choose randomly among the configurations with toggled markers that happen to best erase the targeted association, e.g., by using a randomly chosen such configuration in every permutation of the DV during Z or P value estimation.

Note finally that numerous but individually weak independent ("off-frame") *n*-column DV associations can let the DV be strongly PAS-detected even if every one of the *n*-column associations is too weak to be PAS-detected. However, these 2-column associations should be too weak to inflate the false positives of higher-order dvMom's.

As a *brute force* last resort, one can generate DMs in which markers at the biasing IVs of interest are shuffled vertically and choose those DMs in which both the excess in false positives of concern and the number of toggled IV markers are minimized. This may be the only way to erase an excess that is caused by models in which only one column is detectable given the rows of the DM at hand, e.g., because the column is involved in hundreds of weak independent associations with columns that are at unknown positions in the DM at hand and only become dvPAS-detectable when DM rows are increased by orders of magnitude (see paragraphs immediately above the Discussion).

There is a major additional advantage in the proposed general procedure that one must first identify the IVs with strong 1-way DV association and erase their signal before estimating the P values of the moment-based dvPASs that react to the 2-way DV associations of individual IVs, erase their signal, etc. This erasure guarantees indeed that the obtained 2-way P value of each IV will be independent from the 1-way P values of any other IVs, including that of the IV itself.

By extension, it seems reasonable to assume that when one follows this procedure under the $H_o$ and one assays say $n$-way DV associations up to $n=10$, the $n$ P values generated for each IV are i.i.d. under the $H_o$ and as such can be easily condensed into a single P value say by turning each of them into $-2ln(P\ value_i)$, summing the $n$ resulting logs, and getting the P value of this sum (that is $\chi^2$-distributed with $2n$ d.f.; Hedges & Holkin, 1985, Wardrop, 2015).

**Acknowledgements.** I thank most warmly Jean-Claude Walser of the *Zentrum für die Erforschung der genetischen Vielfalt* (GDC) of ETH-Zürich and the people of Switzerland for too many, decisive years of super-computer access and six months of financial aid; Eric Müller for expert, heart-warmingly forthcoming help with using ETH supercomputers; R.R.Hudson for insightful feedback and thanklessly sacrificed grant writing during the early exploration of the method; Stuart A.Kurtz for expert help in understanding C's weaknesses; Pablo Weber-Cornejo and Juan C.Antezana for expertly choosing the components of the project's small server; Francisco Pina-Martins for expertly assembling the server and sys-administering it since 2012, as well as for invaluable help with LINUX, scripts, firewalls, etc; M.Kreitman for hosting me during the first four years of the project; G.Kotoulas for server access in 2015-16; D.R.Weinberg for long-shot financial support early in the project; P.Fernandes of IGC for server access upon my arrival in Portugal; the Portuguese people for 13 years of kind daily contacts and five years of financial support; A.Palsson for his early enthusiasm; the Icelandic people for one year of financial support; and Lee and Gloria Stephens for selfless financial support. Last but not least, I want to thank in eternity my parents Hector and Gladys for their unwavering, most-crucially decisive emotional and financial support since 2013, without which everything would have been lost.

## Materials and Methods

**Implementation of the PAS and dvPAS algorithms.** The PAS code is written in plain C using long doubles and long ints to reduce accuracy problems, albeit working with very large numbers that differ only several places after the decimal point required *ad hoc* programming. The LKx and CHIx statistics are logarithmized but the moment-based PASs and the Kolmogorov-Smirnov PASs are not. The factorials used by LKx and CHIx are calculated *brute force* up to $2 \times 10^4!$ after which the program uses Gospers' improvement of the De Moivre-Stirling approximation for $ln(n!)$, discussed online at the *Mathematica* website.

**Details about the PAS algorithms.** See dedicated section "PAS Algorithms in Depth".

**Generation of a random data matrix DM with sequences of markers.** When simulating the random DMs used to confirm the results in the mathematico-analytical section, the counts of the various markers at each individual DM column are generated through multinomial sampling (Devroye 1996) using as multinomial probabilities the marker frequencies desired for the column at stake.

In all other cases, random DMs with markers are generated column by column by forcing the markers at every column to have frequencies as close as possible to those desired for the column. When the number of DM rows times a desired marker frequency does not yield an integer, the product is rounded down and at the end all the non-assigned counts are assigned a marker multinomially according to individual marker probabilities equal to the marker frequencies pre-specified for the column, minus the frequency of the already assigned markers of each type, divided by the sum of these differences. After generating the marker pool for a column, the markers in the pool are picked randomly (without replacement) one by one and placed vertically down the column in the order in which they are picked. Therefore, repeating the described generation of a DM is very close and at times identical to permuting vertically the markers at every column of a single DM with the same marker frequencies.

The DMs studied below include "haploid" binary-marker DMs in which minor binary-marker frequency ranges from 0.1 to 0.5 with 0.1 increments, to cover cases possibly increasingly favorable to PAS. Throughout the paper this is referred to as a **o12345 marker-frequency scheme**, under which five columns have minor binary-marker frequency 0.1 0.2 0.3 0.4 0.5 from left to right in the DM, with the series cycling when additional columns are needed. Also "diploid" trinary-marker DMs are used in which the frequencies of three markers at every column are the **HW** (Hardy&Weinberg) trinary proportions $p^2$ $2pq$ $q^2$ ($q= 1-p$), where $p$ cycles according to the o12345 scheme.

When detecting columns associated with others according to various models of column association, random columns are added laterally to the model columns in association. Also these random columns are generated as described above so that in this case too marker frequencies are as close as possible to those pre-specified.

In the assays of PAS power when detecting IV columns that are DV-associated according to various models of $n$-IV association with a DV, random columns are added laterally to the right of the columns of the DV and the DV-associated IVs. The random columns are generated at once for all DV categories together, as it was described above for the o12345 frequency scheme, so that also here their marker frequencies are as close as possible to the pre-specified ones (see above). This bulk generation for all DV categories pooled allows marker frequencies at random IVs to vary randomly across DV categories, i.e., it allows individual random columns to both have marginal effects and participate in combination effects at the DV, which may result in interactions with the signals from model IVs. Therefore random columns are always allowed to have marginal effects regardless of whether the models of IV association with the DV studied in a power assay have such effects or not (see below), which is both more realistic and creates more noise.

**Generation of DM columns according to a pre-specified association model.** To generate an L-column DM where the numbers of specific same-row $n$-marker marker combinations at $n$ "model" columns reflect the frequencies of same-row $n$-marker sequences in an $r$-row $n$-column model DM (e.g., the binary-marker combinations 0_0 0_1 1_0 1_1 may have counts 20 30 40 10 in a 100-row 2-column model DM), so that the detection of the $n$ model columns by PAS or dvPAS can be studied in the L-column DM, one expands the model DM's $n$-marker sequences multinomially according to their counts, up to the total number of rows desired for the L-column DM. L-$n$ random sites are then generated as described and added laterally to the generated block of model columns.

**Generation of IV columns according to a pre-specified model of $n$-IV association with a DV.** To generate marker combinations at $n$ IV columns in a DM in accordance to the frequencies of the $n$-marker sequences in a model DM of $n$-IV DV association (e.g., the binary-marker combinations 0_0 0_1 1_0 1_1 may have counts 20 30 40 10 in the "affecteds" half of a 200-row 2-IV model DM vs. 25 25 25 25 in "controls"), one expands multinomially, but separately for each DV category in order to keep affecteds and controls equally numerous, the model DM's marker sequences according to their frequencies within each DV category of the model DV, up to the total number of rows desired for that DV category in the final DM in which model-IV detection will be attempted. Random sites are generated as described above and added laterally.

**Generation of DMs with blocks of columns associated like mutations in a real set of DNA sequences.** To generate **blocks** of columns in intra-block association, a "source set" is used that comprises 116 distinct 100-marker **"LPL"** sequences of binarized "mutations" that Nickerson et al (1998) characterized in the LPL gene region of a set of human chromosomes. In some simulations a set of 178 1'000-marker binarized human-chromosome **"CHJP"** sequences is also used that was kindly provided by Sidhar Sudarvali. Both sets are available from MAA upon request.

To generate a random binary N-row H-block DM, the N sequences of each block are sampled, with replacement, from the source set. The rows of each block are independently shuffled vertically before the H generated blocks are joined laterally.

To generate an N-row H-block null DM with desired marker frequencies at a chosen "**anchor**" column (below) in each of its H LPL blocks, one first generates a N-row guiding model with H independent columns with randomly vertically ordered markers of the frequencies desired for said "anchor" columns. Then one samples H LPL sequences whose anchor-column markers match those in the first H-marker sequence in the guiding model, one repeats for the second sequence, etc, until the N H-block LPL sequences with the H desired anchor-column marker frequencies are sampled. Such guided sampling is also used when sampling N LPL sequences that have pre-specified marker combinations at four same-block "anchor" positions (below).

To sample LPL sequences according to a model of $n$-column association or according to a model of $n$-IV association with a DV, a quartet of four columns in the LPL source sequences were chosen as potential model anchor columns, namely columns 5, 35, 49, and 66, but only column 49 is used when a single model column is allowed per block. For CHJP sequences the lone model column is nr.501 and the quartet are columns nrs.11, 501, 830, and 848. Therefore, if one wants to sample sequences according to say a 8-column association model and one wants the eight model columns to fall on as few blocks as possible, two blocks with four model columns each are necessary regardless of the chosen block type.

The columns in the quartets were chosen because all of the 16 possible 4-marker combinations are found at those columns, allowing so the sampling of any possible 4-marker combination that may be present in a model DM. For this reason, the columns in the quartets are less "linked" than most random column tetrads in the LPL sequences.

Five non-model columns that are "**model-column-linked**" or "**model-linked**" to a model column or model IV, respectively, in a block with one or more model columns, are often also studied. They are always at positions 1, 22, 50, 63, and 88 of the first LPL block with a model IV or more; in CHJP blocks they are at positions 60, 108, 579, 727, and 902.

To generate similarly **block-structured trinary "diploid" DMs**, one samples from a source set of trinary LPL-based 100-marker sequences that was generated by performing every possible pairwise pairing of the 116 LPL haplotypes and then turning the paired 00s, 01(10)s, and 11s in the pairwise comparisons into the trinary markers 0, 1, and 2. The resulting trinary wLPL source set consists of 6'670 100-trinary-marker sequences.

Since eight of the 81 possible 4-marker combinations of trinary markers were not found at the four anchor positions 5, 35, 49, and 66 of the mentioned trinary sequences (mainly because no pairwise comparison of an LPL sequence with itself was performed) and in order to allow simulations of four model IVs per block, the missing trinary sequences were created by altering eight randomly chosen wLPL sequences that differed by one trinary marker from one of the missing sequences; the eight altered wLPL sequences were added to the those directly generated through pairwise comparisons. No simulation uses diploid CHJP sequences.

**Encountering random "model DMs" with statistically significantly associated columns.** To find random $r$-row $n$-column DMs with a desired marker-frequency scheme and whose every column is statistically significantly associated with others in the DM, one generates a random $r$-row $n$-column DM as described above and then tests every column with the DV-column test in Fig.3. If every column of the random DM has a DV P value$\leq$0.01 (200 permutations) the random DM is retained as model DM. This guarantees that every column of a retained model DM is statistically significantly associated with at least another column in the DM. A set of such model DMs is therefore characterized by the number of rows and columns of the DMs, the columns' marker frequency scheme, and the retention P value that the columns of the DM did not exceed. Note that when say 100-row 5-column binary and trinary such model DMs are individually multinomially "expanded" (resampled) again to 100 rows, their columns have again P value$\leq$0.01 (*sensu* Fig.3) only ~90% of the time (vs. 100% when the model DMs are tested directly). Note also that encountering the 800-row model DMs consumed a lot of computer time, specially when the desired model DMs were trinary and/or had ten columns. For this reason no model DMs with more than 800 rows and 10 columns are studied here.

**Encountering random model DMs whose IVs are significantly DV-associated without marginal effects.** To find random $r$-row $n$-IV model DMs in which both every IV is part of an IV subset that is significantly DV-associated and no IV has a marginal effect at the DV (i.e., no IV shows a marker-frequency difference in say affected vs. controls), one generates two random ($r$/2)-row $n$-column DMs (see above) using the same marker-frequency scheme and then creates a DV column with "affecteds" and "controls" labels by adding a 0 to the left of every row in the first DM and a 1 to the left of the rows of the second DM. The two DMS are stacked (vertically) to generate a candidate $r$-row $n$-IV model DM and then the P values *sensu* Fig.3 of the DV and the IVs of this DM are estimated. If every IV and the DV have P value$\leq$0.05 (100 permutations), the model DM is retained for the power assays.

Therefore in every model DM belonging to such a set of model DMs  i) the DV is statistically significantly associated with each IV but in synergy with at least another IV and  ii) every IV is statistically significantly associated with the DV but in synergy with at least another IV (since no marker-frequency differences are allowed between DV categories so that no IV can be individually associated with the DV). Also here is every group of model DMs characterized by the number of rows and IVs of the DMs, the IVs' marker-frequency scheme, and the cutoff that the P values of the DV and the IVs did not exceed.  Multinomial "expansion" of these model DMs to their original number of rows reduces also here the detection of their IVs, e.g., the detection of the IVs in the 8-IV models sinks to ~85% with P value$\leq$0.05 (from 100%) be the models binary or trinary.

**Encountering a random model DM whose IVs are significantly DV-associated with eventual marginal effects.**  To find random $r$-row $n$-IV model DMs whose every IV by itself and/or in synergy with others is associated with the DV (i.e., IVs that are allowed to show a marginal effect at the DV due to random marker-frequency differences in say "affected vs. controls"), one generates an $r$-row $n$-column DM according to the desired marker-frequency scheme and one creates a DV column with "affecteds" and "controls" labels by adding a 0 to the left of the top $r/2$ rows and a 1 to the left of the bottom $r/2$ rows.  Then DM columns are tested sensu Fig.3 and the DM is retained as model DM if the P values of the DV and the IVs are all no larger than 0.05 (100 permutations). Therefore in the saved DMs  i) the DV is statistically significantly associated with every IV individually or because the IV contributes non-additively to the DV-association of a subset of the IVs, and  ii) every IV is statistically significantly associated with the DV by itself (since IVs with marker-frequency differences across DV categories are allowed) and/or through synergies with other IVs.  And also here  i) every set of model DMs is characterized by the DMs' rows, columns, and marker frequencies as well as by the P value used for model-DM retention, and  ii) multinomially "expanding" the binary and trinary 200-row model DMs to the original 200 rows lowers the IVs' detection (e.g., in the case of the 4-IV model DMs, to 93% with P value$\leq$0.05).

# Supplementary Materials

$Lik(0\,0\,3) = 3\,p^3 q^6 + 9\,p^4 q^5 + 9\,p^5 q^4 + 3\,p^6 q^3$
$Lik(0\,1\,2) = 18\,p^3 q^6 + 54\,p^4 q^5 + 54\,p^5 q^4 + 18\,p^6 q^3$
$Lik(1\,1\,1) = 6\,p^3 q^6 + 18\,p^4 q^5 + 18\,p^5 q^4 + 6\,p^6 q^3$
$Lik(1\,1\,3) = 9\,p^2 q^7 + 18\,p^3 q^6 + 9\,p^4 q^5 + 9\,p^5 q^4 + 18\,p^6 q^3 + 9\,p^7 q^2$
$Lik(1\,2\,2) = 18\,p^2 q^7 + 36\,p^3 q^6 + 18\,p^4 q^5 + 18\,p^5 q^4 + 36\,p^6 q^3 + 18\,p^7 q^2$
$Lik(2\,2\,3) = 9\,p q^8 + 9\,p^2 q^7 + 18\,p^4 q^5 + 18\,p^5 q^4 + 9\,p^7 q^2 + 9\,p^8 q$
$Lik(3\,3\,3) = q^9 + 3\,p^3 q^6 + 3\,p^6 q^3 + p^9$

(11) 3x3

$Lik(0\,0\,0\,0\,3\,3) = 24\,p^6 q^6$
$Lik(0\,0\,0\,3\,3\,3) = 4\,p^3 q^9 + 12\,p^5 q^7 + 12\,p^7 q^5 + 4\,p^9 q^3$
$Lik(0\,0\,1\,1\,2\,2) = 144\,p^6 q^6$
$Lik(0\,0\,1\,1\,2\,3) = 144\,p^5 q^7 + 144\,p^7 q^5$
$Lik(0\,0\,1\,2\,2\,3) = 72\,p^4 q^8 + 144\,p^6 q^6 + 72\,p^8 q^4$
$Lik(0\,1\,1\,1\,2\,2) = 288\,p^5 q^7 + 288\,p^7 q^5$
$Lik(0\,1\,1\,2\,2\,2) = 144\,p^4 q^8 + 288\,p^6 q^6 + 144\,p^8 q^4$
$Lik(0\,1\,1\,2\,2\,3) = 36\,p^3 q^9 + 108\,p^5 q^7 + 108\,p^7 q^5 + 36\,p^9 q^3$
$Lik(1\,1\,1\,1\,1\,1) = 48\,p^6 q^6$
$Lik(1\,1\,1\,1\,1\,3) = 72\,p^4 q^8 + 144\,p^6 q^6 + 72\,p^8 q^4$
$Lik(1\,1\,1\,1\,3\,3) = 36\,p^4 q^8 + 36\,p^8 q^4$
$Lik(1\,1\,1\,2\,2\,2) = 24\,p^3 q^9 + 72\,p^5 q^7 + 72\,p^7 q^5 + 24\,p^9 q^3$
$Lik(1\,1\,1\,3\,3\,3) = 12\,p^{10} q^2 + 12\,p^2 q^{10} + 24\,p^4 q^8 + 24\,p^6 q^6 + 24\,p^8 q^4$
$Lik(1\,1\,2\,2\,2\,2) = 72\,p^4 q^8 + 72\,p^8 q^4$
$Lik(1\,1\,2\,2\,2\,3) = 144\,p^3 q^9 + 144\,p^5 q^7 + 144\,p^7 q^5 + 144\,p^9 q^3$
$Lik(1\,2\,2\,2\,2\,3) = 36\,p^{10} q^2 + 36\,p^2 q^{10} + 72\,p^4 q^8 + 72\,p^6 q^6 + 72\,p^8 q^4$
$Lik(2\,2\,2\,2\,3\,3) = 18\,p^{10} q^2 + 18\,p^2 q^{10} + 36\,p^6 q^6$
$Lik(2\,2\,2\,3\,3\,3) = 12\,p^{11} q + 12\,p q^{11} + 12\,p^3 q^9 + 24\,p^5 q^7 + 24\,p^7 q^5 + 12\,p^9 q^3$
$Lik(3\,3\,3\,3\,3\,3) = q^{12} + p^{12} + 3\,p^4 q^8 + 3\,p^8 q^4$

(12) 4x3

$Lik(0\,0\,0\,0\,4\,4)=48\,p^8 q^8$
$Lik(0\,0\,0\,4\,4\,4)=16\,p^{10}q^6+4\,p^{12}q^4+4\,p^4 q^{12}+16\,p^6 q^{10}+24\,p^8 q^8$
$Lik(0\,0\,1\,1\,3\,3)=384\,p^8 q^8$
$Lik(0\,0\,1\,1\,3\,4)=384\,p^7 q^9+384\,p^9 q^7$
$Lik(0\,0\,1\,3\,3\,4)=96\,p^{11}q^5+96\,p^5 q^{11}+288\,p^7 q^9+288\,p^9 q^7$
$Lik(0\,0\,2\,2\,2\,2)=288\,p^8 q^8$
$Lik(0\,0\,2\,2\,2\,4)=288\,p^{10}q^6+288\,p^6 q^{10}+576\,p^8 q^8$
$Lik(0\,1\,1\,2\,2\,3)=2304\,p^7 q^9+2304\,p^9 q^7$
$Lik(0\,1\,1\,2\,3\,3)=1152\,p^{10}q^6+1152\,p^6 q^{10}+2304\,p^8 q^8$
$Lik(0\,1\,1\,3\,3\,4)=192\,p^{10}q^6+48\,p^{12}q^4+48\,p^4 q^{12}+192\,p^6 q^{10}+288\,p^8 q^8$
$Lik(0\,1\,2\,2\,3\,3)=576\,p^{11}q^5+576\,p^5 q^{11}+1728\,p^7 q^9+1728\,p^9 q^7$
$Lik(0\,2\,2\,2\,2\,2)=576\,p^{10}q^6+576\,p^6 q^{10}+1152\,p^8 q^8$
$Lik(0\,2\,2\,2\,2\,4)=144\,p^{10}q^6+36\,p^{12}q^4+36\,p^4 q^{12}+144\,p^6 q^{10}+216\,p^8 q^8$
$Lik(1\,1\,1\,1\,2\,2)=576\,p^8 q^8$
$Lik(1\,1\,1\,1\,2\,4)=288\,p^{10}q^6+288\,p^6 q^{10}+576\,p^8 q^8$
$Lik(1\,1\,1\,1\,4\,4)=96\,p^{10}q^6+96\,p^6 q^{10}$
$Lik(1\,1\,1\,2\,2\,2)=768\,p^7 q^9+768\,p^9 q^7$
$Lik(1\,1\,1\,2\,2\,3)=1152\,p^{10}q^6+1152\,p^6 q^{10}+2304\,p^8 q^8$
$Lik(1\,1\,1\,2\,2\,4)=288\,p^{11}q^5+288\,p^5 q^{11}+864\,p^7 q^9+864\,p^9 q^7$
$Lik(1\,1\,1\,4\,4\,4)=48\,p^{11}q^5+16\,p^{13}q^3+16\,p^3 q^{13}+48\,p^5 q^{11}+64\,p^7 q^9+64\,p^9 q^7$
$Lik(1\,1\,2\,2\,2\,3)=576\,p^{11}q^5+576\,p^5 q^{11}+1728\,p^7 q^9+1728\,p^9 q^7$
$Lik(1\,1\,2\,2\,3\,3)=1152\,p^{10}q^6+144\,p^{12}q^4+144\,p^4 q^{12}+1152\,p^6 q^{10}+864\,p^8 q^8$
$Lik(1\,1\,2\,2\,3\,4)=576\,p^{11}q^5+576\,p^5 q^{11}+576\,p^7 q^9+576\,p^9 q^7$
$Lik(1\,1\,2\,3\,3\,4)=576\,p^{10}q^6+288\,p^{12}q^4+288\,p^4 q^{12}+576\,p^6 q^{10}+576\,p^8 q^8$
$Lik(1\,2\,2\,2\,3\,3)=1152\,p^{11}q^5+1152\,p^5 q^{11}+1152\,p^7 q^9+1152\,p^9 q^7$
$Lik(1\,2\,2\,3\,3\,3)=1152\,p^{10}q^6+576\,p^{12}q^4+576\,p^4 q^{12}+1152\,p^6 q^{10}+1152\,p^8 q^8$
$Lik(1\,2\,2\,3\,3\,4)=432\,p^{11}q^5+144\,p^{13}q^3+144\,p^3 q^{13}+432\,p^5 q^{11}+576\,p^7 q^9+576\,p^9 q^7$
$Lik(2\,2\,2\,2\,2\,2)=288\,p^{10}q^6+24\,p^{12}q^4+24\,p^4 q^{12}+288\,p^6 q^{10}+144\,p^8 q^8$
$Lik(2\,2\,2\,2\,2\,4)=576\,p^{10}q^6+288\,p^{12}q^4+288\,p^4 q^{12}+576\,p^6 q^{10}+576\,p^8 q^8$
$Lik(2\,2\,2\,2\,4\,4)=72\,p^{12}q^4+72\,p^4 q^{12}+144\,p^8 q^8$
$Lik(2\,2\,2\,3\,3\,3)=288\,p^{11}q^5+96\,p^{13}q^3+96\,p^3 q^{13}+288\,p^5 q^{11}+384\,p^7 q^9+384\,p^9 q^7$
$Lik(2\,2\,2\,4\,4\,4)=72\,p^{10}q^6+48\,p^{12}q^4+24\,p^{14}q^2+24\,p^2 q^{14}+48\,p^4 q^{12}+72\,p^6 q^{10}+96\,p^8 q^8$
$Lik(2\,2\,3\,3\,3\,3)=144\,p^{12}q^4+144\,p^4 q^{12}+288\,p^8 q^8$
$Lik(2\,2\,3\,3\,3\,4)=288\,p^{11}q^5+288\,p^{13}q^3+288\,p^3 q^{13}+288\,p^5 q^{11}+576\,p^7 q^9+576\,p^9 q^7$
$Lik(2\,3\,3\,3\,3\,4)=216\,p^{10}q^6+144\,p^{12}q^4+72\,p^{14}q^2+72\,p^2 q^{14}+144\,p^4 q^{12}+216\,p^6 q^{10}+288\,p^8 q^8$
$Lik(3\,3\,3\,3\,4\,4)=72\,p^{10}q^6+24\,p^{14}q^2+24\,p^2 q^{14}+72\,p^6 q^{10}$
$Lik(3\,3\,3\,4\,4\,4)=48\,p^{11}q^5+16\,p^{13}q^3+16\,p^{15}q+16\,p^3 q^{13}+48\,p^5 q^{11}+48\,p^7 q^9+48\,p^9 q^7+16\,pq^{15}$
$Lik(4\,4\,4\,4\,4\,4)=4\,p^{12}q^4+p^{16}+4\,p^4 q^{12}+6\,p^8 q^8+q^{16}$

(13)

4x4

$Lik(0\,0\,0) = 216\,p^3q^3r^3$

$Lik(0\,0\,1) = 324\,p^2q^3r^4 + 324\,p^2q^4r^3 + 324\,p^3q^2r^4 + 324\,p^3q^4r^2 + 324\,p^4q^2r^3 + 324\,p^4q^3r^2$

$Lik(0\,0\,2) = 54\,pq^3r^5 + 108\,pq^4r^4 + 54\,pq^5r^3 + 108\,p^2q^2r^5 + 108\,p^2q^3r^4 + 108\,p^2q^4r^3 + 108\,p^2q^5r^2 + 54\,p^3qr^5 + 108\,p^3q^2r^4 + 324\,p^3q^3r^3 + 108\,p^3q^4r^2 + 54\,p^3q^5r$
$+ 108\,p^4qr^4 + 108\,p^4q^2r^3 + 108\,p^4q^3r^2 + 108\,p^4q^4r$
$+ 54\,p^5qr^3 + 108\,p^5q^2r^2 + 54\,p^5q^3r$

$Lik(0\,0\,3) = 3\,q^3r^6 + 9\,q^4r^5 + 9\,q^5r^4 + 3q^6r^3 + 9\,pq^2r^6 + 18\,pq^3r^5 + 18\,pq^4r^4 + 18\,pq^5r^3 + 9\,pq^6r^2 + 9\,p^2qr^6 + 18\,p^2q^2r^5 + 45\,p^2q^3r^4 + 45\,p^2q^4r^3 + 18\,p^2q^5r^2 + 9\,p^2q^6r$
$+ 3\,p^3r^6 + 18\,p^3qr^5 + 45\,p^3q^2r^4 + 36\,p^3q^3r^3 + 45\,p^3q^4r^2 + 18\,p^3q^5r$
$+ 3\,p^3q^6 + 9\,p^4r^5 + 18\,p^4qr^4 + 45\,p^4q^2r^3 + 45\,p^4q^3r^2 + 18\,p^4q^4r + 9\,p^4q^5 + 9\,p^5r^4 + 18\,p^5qr^3 + 18\,p^5q^2r^2 + 18\,p^5q^3r$
$+ 9\,p^5q^4 + 3\,p^6r^3 + 9\,p^6qr^2 + 9\,p^6q^2r + 3\,p^6q^3$

$Lik(0\,1\,1) = 108\,pq^3r^5 + 216\,pq^4r^4 + 108\,pq^5r^3 + 216\,p^2q^2r^5 + 216\,p^2q^3r^4 + 216\,p^2q^4r^3 + 216\,p^2q^5r^2 + 108\,p^3qr^5 + 216\,p^3q^2r^4$
$+ 648\,p^3q^3r^3 + 216\,p^3q^4r^2 + 108\,p^3q^5r + 216\,p^4qr^4 + 216\,p^4q^2r^3 + 216\,p^4q^3r^2 + 216\,p^4q^4r + 108\,p^5qr^3 + 216\,p^5q^2r^2 + 108\,p^5q^3r$

$Lik(0\,1\,2) = 18\,q^3r^6 + 54\,q^4r^5 + 54\,q^5r^4 + 18\,q^6r^3 + 54\,pq^2r^6 + 108\,pq^3r^5 + 108\,pq^4r^4 + 108\,pq^5r^3 + 54\,pq^6r^2 + 54\,p^2qr^6 + 108\,p^2q^2r^5$
$+ 270\,p^2q^3r^4 + 270\,p^2q^4r^3 + 108\,p^2q^5r^2 + 54\,p^2q^6r + 18\,p^3r^6 + 108\,p^3qr^5$
$+ 270\,p^3q^2r^4 + 216\,p^3q^3r^3 + 270\,p^3q^4r^2 + 108\,p^3q^5r + 18\,p^3q^6 + 54\,p^4r^5 + 108\,p^4qr^4$
$+ 270\,p^4q^2r^3 + 270\,p^4q^3r^2 + 108\,p^4q^4r + 54\,p^4q^5 + 54\,p^5r^4 + 108\,p^5qr^3 + 108\,p^5q^2r^2 + 108\,p^5q^3r + 54\,p^5q^4 + 18\,p^6r^3 + 54\,p^6qr^2 + 54\,p^6q^2r + 18\,p^6q^3$

$Lik(1\,1\,1) = 6\,q^3r^6 + 18\,q^4r^5 + 18\,q^5r^4 + 6\,q^6r^3 + 18\,pq^2r^6 + 36\,pq^3r^5 + 36\,pq^4r^4 + 36\,pq^5r^3 + 18\,pq^6r^2 + 18\,p^2qr^6 + 144\,p^2q^2r^5 + 90\,p^2q^3r^4 + 90\,p^2q^4r^3 + 144\,p^2q^5r^2 + 18\,p^2q^6r$
$+ 6\,p^3r^6 + 36\,p^3qr^5 + 90\,p^3q^2r^4 + 72\,p^3q^3r^3 + 90\,p^3q^4r^2 + 36\,p^3q^5r$
$+ 6\,p^3q^6 + 18\,p^4r^5 + 36\,p^4qr^4 + 90\,p^4q^2r^3 + 90\,p^4q^3r^2 + 36\,p^4q^4r + 18\,p^4q^5 + 18\,p^5r^4 + 36\,p^5qr^3 + 144\,p^5q^2r^2 + 36\,p^5q^3r + 18\,p^5q^4$
$+ 6\,p^6r^3 + 18\,p^6qr^2 + 18\,p^6q^2r + 6\,p^6q^3$

$Lik(1\,1\,2) = 108\,pq^2r^6 + 108\,pq^3r^5 + 108\,pq^5r^3 + 108\,pq^6r^2 + 108\,p^2qr^6 + 108\,p^2q^3r^4 + 108\,p^2q^4r^3 + 108\,p^2q^6r$
$+ 108\,p^3qr^5 + 108\,p^3q^2r^4 + 108\,p^3q^4r^2 + 108\,p^3q^5r + 108\,p^4q^2r^3 + 108\,p^4q^3r^2 + 108\,p^5qr^3 + 108\,p^5q^3r + 108\,p^6qr^2 + 108\,p^6q^2r$

$Lik(1\,1\,3) = 9\,q^2r^7 + 18\,q^3r^6 + 9\,q^4r^5 + 9\,q^5r^4 + 18\,q^6r^3 + 9\,q^7r^2 + 18\,pqr^7 + 18\,pq^2r^6 + 18\,pq^3r^5 + 36\,pq^4r^4 + 18\,pq^5r^3 + 18\,pq^6r^2 + 18\,pq^7r$
$+ 9\,p^2r^7 + 18\,p^2qr^6 + 54\,p^2q^2r^5 + 27\,p^2q^3r^4 + 27\,p^2q^4r^3 + 54\,p^2q^5r^2 + 18\,p^2q^6r$
$+ 9\,p^2q^7 + 18\,p^3r^6 + 18\,p^3qr^5 + 27\,p^3q^2r^4 + 54\,p^3q^3r^3 + 27\,p^3q^4r^2 + 18\,p^3q^5r$
$+ 18\,p^3q^6 + 9\,p^4r^5 + 36\,p^4qr^4 + 27\,p^4q^2r^3 + 27\,p^4q^3r^2 + 36\,p^4q^4r$
$+ 9\,p^4q^5 + 9\,p^5r^4 + 18\,p^5qr^3 + 54\,p^5q^2r^2 + 18\,p^5q^3r$
$+ 9\,p^5q^4 + 18\,p^6r^3 + 18\,p^6qr^2 + 18\,p^6q^2r$
$+ 18\,p^6q^3 + 9\,p^7r^2 + 18\,p^7qr + 9\,p^7q^2$

$Lik(1\,2\,2) = 18\,q^2r^7 + 36\,q^3r^6 + 18\,q^4r^5 + 18\,q^5r^4 + 36\,q^6r^3 + 18\,q^7r^2 + 36\,pqr^7 + 36\,pq^2r^6 + 36\,pq^3r^5 + 72\,pq^4r^4 + 36\,pq^5r^3 + 36\,pq^6r^2 + 36\,pq^7r$
$+ 18\,p^2r^7 + 36\,p^2qr^6 + 108\,p^2q^2r^5 + 54\,p^2q^3r^4 + 54\,p^2q^4r^3 + 108\,p^2q^5r^2 + 36\,p^2q^6r + 18\,p^2q^7 + 36\,p^3r^6 + 36\,p^3qr^5 + 54\,p^3q^2r^4 + 108\,p^3q^3r^3 + 54\,p^3q^4r^2 + 36\,p^3q^5r$
$+ 36\,p^3q^6 + 18\,p^4r^5 + 72\,p^4qr^4 + 54\,p^4q^2r^3 + 54\,p^4q^3r^2 + 72\,p^4q^4r$
$+ 18\,p^4q^5 + 18\,p^5r^4 + 36\,p^5qr^3 + 108\,p^5q^2r^2 + 36\,p^5q^3r + 18\,p^5q^4 + 36\,p^6r^3 + 36\,p^6qr^2 + 36\,p^6q^2r + 36\,p^6q^3 + 18\,p^7r^2 + 36\,p^7qr + 18\,p^7q^2$

$Lik(2\,2\,2) = 18\,pqr^7 + 36\,pq^4r^4 + 18\,pq^7r + 36\,p^4qr^4 + 36\,p^4q^4r + 18\,p^7qr$

$Lik(2\,2\,3) = 9\,qr^8 + 9\,q^2r^7 + 18\,q^4r^5 + 18\,q^5r^4 + 9\,q^7r^2 + 9\,q^8r$
$+ 9\,pr^8 + 9\,pq^2r^6 + 18\,pq^3r^5 + 18\,pq^5r^3 + 9\,pq^6r^2 + 9\,pq^8 + 9\,p^2r^7 + 9\,p^2qr^6 + 18\,p^2q^3r^4 + 18\,p^2q^4r^3 + 9\,p^2q^6r$
$+ 9\,p^2q^7 + 18\,p^3qr^5 + 18\,p^3q^2r^4 + 18\,p^3q^4r^2 + 18\,p^3q^5r$
$+ 18\,p^4r^5 + 18\,p^4q^2r^3 + 18\,p^4q^3r^2 + 18\,p^4q^5 + 18\,p^5r^4 + 18\,p^5qr^3 + 18\,p^5q^3r + 18\,p^5q^4 + 9\,p^6qr^2 + 9\,p^6q^2r + 9\,p^7r^2 + 9\,p^7q^2 + 9\,p^8r + 9\,p^8q$

$Lik(3\,3\,3) = r^9 + 3\,q^3r^6 + 3\,q^6r^3 + q^9 + 3\,p^3r^6 + 6\,p^3q^3r^3 + 3\,p^3q^6 + 3\,p^6r^3 + 3\,p^6q^3 + p^9$

(14)

3-marker
3x3

**Table 1. Observed and predicted counts of the pairwise comparisons with *m* matches when every possible marker combination in the DM occurs exactly *n* times.**

a) L=5 columns; S=2 markers; n=1

| m | pw-count | $_LC_i$ | $_LC_i\, 2^L/2$ | $2^L/2$ |
|---|---|---|---|---|
| 0 | 16 | 1 | 16 | 16 |
| 1 | 80 | 5 | 80 | 16 |
| 2 | 160 | 10 | 160 | 16 |
| 3 | 160 | 10 | 160 | 16 |
| 4 | 80 | 5 | 80 | 16 |
| 5 | 0 | 1 | 16 | 16 |
| total | 496 | | | |

b) L=7 columns; S=2 markers; n=3

| m | pw-count | $_LC_i$ | $2^L/2$ | $n^2$ | $n^2{_LC_i}\,2^L/2$ | $_LC_i/S^L$ | observ. | binomial |
|---|---|---|---|---|---|---|---|---|
| 0 | 576 | 1 | 64 | 9 | 576 | 0.00781 | 0.00783 | 0.00796 |
| 1 | 4'032 | 7 | 64 | 9 | 4'032 | 0.05469 | 0.05483 | 0.05540 |
| 2 | 12'096 | 21 | 64 | 9 | 12'096 | 0.16406 | 0.16449 | 0.16535 |
| 3 | 20'160 | 35 | 64 | 9 | 20'160 | 0.27344 | 0.27415 | 0.27415 |
| 4 | 20'160 | 35 | 64 | 9 | 20'160 | 0.27344 | 0.27415 | 0.27272 |
| 5 | 12'096 | 21 | 64 | 9 | 12'096 | 0.16406 | 0.16449 | 0.16278 |
| 6 | 4'032 | 7 | 64 | 9 | 4'032 | 0.05469 | 0.05483 | 0.05398 |
| 7 | 384 | 1 | 64 | 6 | 384 | 0.00781 | 0.00522 | 0.00767 |
| total | 73'536 | | | n(n-1) | $m^1$ | **3.5000** | **3.4909** | **3.4909** |
| | | | | | $m^2$ | **1.7500** | **1.7225** | **1.7500** |

c) L=6 columns; S=4 markers; n=2

| m | pw-count | $_LC_i$ | $(S-1)^{L-i}$ | $S^L/2$ | $n^2$ | $1/2\, n^2{_LC_i}\, S^L(S-1)^{L-i}$ | approx. | observ. | binomial |
|---|---|---|---|---|---|---|---|---|---|
| 0 | 5'971'968 | 1 | 729 | 2048 | 4 | 5'971'968 | 0.17798 | 0.17800 | 0.17798 |
| 1 | 11'943'936 | 6 | 243 | 2048 | 4 | 11'943'936 | 0.35596 | 0.35600 | 0.35596 |
| 2 | 9'953'280 | 15 | 81 | 2048 | 4 | 9'953'280 | 0.29663 | 0.29667 | 0.29663 |
| 3 | 4'423'680 | 20 | 27 | 2048 | 4 | 4'423'680 | 0.13184 | 0.13185 | 0.13184 |
| 4 | 1'105'920 | 15 | 9 | 2048 | 4 | 1'105'920 | 0.03296 | 0.03296 | 0.03296 |
| 5 | 147'456 | 6 | 3 | 2048 | 4 | 147'456 | 0.00439 | 0.00440 | 0.00439 |
| 6 | 4'096 | 1 | 1 | 2048 | 2 | 4'096 | 0.00024 | 0.00012 | 0.00024 |
| total | 33550336 | | | | n(n-1) | $1/2\, n(n-1){_LC_i}\, S^L(S-1)^{L-i}$ | | | |

**Table 2. Observed and numerically evaluated exact frequencies of the possible $m$'s, $m^1$, and $m^2$ vs. binomial predictions; $10^5$ 10-row 40-column DMs; $p=0.2$**

| m's | $10^5$ simuls. | prob(m) | binom | pred -sim | binom -sim | pred -binom | sim.var(#) | var: sim -binom |
|---|---|---|---|---|---|---|---|---|
| 0 | 0.000E+00 | 1.6069E-20 | 1.6069E-20 | 2.E-20 | 2.E-20 | -4.E-29 | 0.00000 | - |
| 1 | 0.000E+00 | 1.3659E-18 | 1.3659E-18 | 1.E-18 | 1.E-18 | 2.E-27 | 0.00000 | - |
| 2 | 0.000E+00 | 5.6599E-17 | 5.6599E-17 | 6.E-17 | 6.E-17 | -3.E-26 | 0.00000 | - |
| 3 | 0.000E+00 | 1.5235E-15 | 1.5235E-15 | 2.E-15 | 2.E-15 | -8.E-25 | 0.00000 | - |
| 4 | 0.000E+00 | 2.9946E-14 | 2.9946E-14 | 3.E-14 | 3.E-14 | -4.E-23 | 0.00000 | - |
| 5 | 0.000E+00 | 4.5817E-13 | 4.5817E-13 | 5.E-13 | 5.E-13 | 5.E-22 | 0.00000 | - |
| 6 | 0.000E+00 | 5.6794E-12 | 5.6794E-12 | 6.E-12 | 6.E-12 | -1.E-21 | 0.00000 | - |
| 7 | 0.000E+00 | 5.8619E-11 | 5.8619E-11 | 6.E-11 | 6.E-11 | 2.E-20 | 0.00000 | - |
| 8 | 0.000E+00 | 5.1383E-10 | 5.1383E-10 | 5.E-10 | 5.E-10 | 3.E-19 | 0.00000 | - |
| 9 | 0.000E+00 | 3.8823E-09 | 3.8823E-09 | 4.E-09 | 4.E-09 | 2.E-18 | 0.00000 | - |
| 10 | 0.000E+00 | 2.5575E-08 | 2.5575E-08 | 3.E-08 | 3.E-08 | 4.E-17 | 0.00000 | - |
| 11 | 4.445E-07 | 1.4822E-07 | 1.4822E-07 | -3.E-07 | -3.E-07 | -5.E-16 | 0.00000 | - |
| 12 | 2.222E-06 | 7.6116E-07 | 7.6116E-07 | -1.E-06 | -1.E-06 | 6.E-17 | 0.00001 | -71% |
| 13 | 3.111E-06 | 3.4838E-06 | 3.4838E-06 | 4.E-07 | 4.E-07 | 3.E-15 | 0.00006 | -62% |
| 14 | 1.667E-05 | 1.4277E-05 | 1.4277E-05 | -2.E-06 | -2.E-06 | -3.E-14 | 0.00058 | -10% |
| 15 | 4.933E-05 | 5.2588E-05 | 5.2588E-05 | 3.E-06 | 3.E-06 | -5.E-14 | 0.00220 | -7% |
| 16 | 1.653E-04 | 1.7461E-04 | 1.7461E-04 | 9.E-06 | 9.E-06 | -1.E-13 | 0.00655 | -17% |
| 17 | 5.202E-04 | 5.2382E-04 | 5.2382E-04 | 4.E-06 | 4.E-06 | -4.E-13 | 0.02355 | 0% |
| 18 | 1.428E-03 | 1.4223E-03 | 1.4223E-03 | -6.E-06 | -6.E-06 | 4.E-12 | 0.07018 | 10% |
| 19 | 3.588E-03 | 3.4997E-03 | 3.4997E-03 | -9.E-05 | -9.E-05 | 3.E-12 | 0.17927 | 14% |
| 20 | 7.841E-03 | 7.8086E-03 | 7.8086E-03 | -3.E-05 | -3.E-05 | -4.E-12 | 0.41767 | 20% |
| 21 | 1.589E-02 | 1.5803E-02 | 1.5803E-02 | -8.E-05 | -8.E-05 | 4.E-11 | 0.90996 | 30% |
| 22 | 2.910E-02 | 2.9002E-02 | 2.9002E-02 | -1.E-04 | -1.E-04 | -3.E-11 | 1.77559 | 40% |
| 23 | 4.788E-02 | 4.8232E-02 | 4.8232E-02 | 3.E-04 | 3.E-04 | 3.E-11 | 2.90676 | 41% |
| 24 | 7.239E-02 | 7.2599E-02 | 7.2599E-02 | 2.E-04 | 2.E-04 | 2.E-13 | 4.19517 | 38% |
| 25 | 9.858E-02 | 9.8735E-02 | 9.8735E-02 | 2.E-04 | 2.E-04 | -8.E-12 | 5.16875 | 29% |
| 26 | 1.212E-01 | 1.2105E-01 | 1.2105E-01 | -1.E-04 | -1.E-04 | -3.E-10 | 5.62737 | 18% |
| 27 | 1.335E-01 | 1.3337E-01 | 1.3337E-01 | -1.E-04 | -1.E-04 | 1.E-10 | 5.67553 | 9% |
| 28 | 1.316E-01 | 1.3159E-01 | 1.3159E-01 | -2.E-05 | -2.E-05 | -2.E-10 | 5.71295 | 11% |
| 29 | 1.154E-01 | 1.1571E-01 | 1.1571E-01 | 3.E-04 | 3.E-04 | -2.E-10 | 5.77788 | 25% |
| 30 | 9.028E-02 | 9.0155E-02 | 9.0155E-02 | -1.E-04 | -1.E-04 | -5.E-11 | 5.28478 | 43% |
| 31 | 6.213E-02 | 6.1800E-02 | 6.1800E-02 | -3.E-04 | -3.E-04 | 2.E-11 | 4.00030 | 53% |
| 32 | 3.693E-02 | 3.6935E-02 | 3.6935E-02 | 6.E-06 | 6.E-06 | -2.E-11 | 2.52928 | 58% |
| 33 | 1.909E-02 | 1.9027E-02 | 1.9027E-02 | -6.E-05 | -6.E-05 | 3.E-11 | 1.27008 | 51% |
| 34 | 8.228E-03 | 8.3244E-03 | 8.3244E-03 | 1.E-04 | 1.E-04 | 4.E-12 | 0.52796 | 42% |
| 35 | 3.013E-03 | 3.0324E-03 | 3.0324E-03 | 2.E-05 | 2.E-05 | -4.E-14 | 0.17117 | 26% |
| 36 | 9.180E-04 | 8.9499E-04 | 8.9499E-04 | -2.E-05 | -2.E-05 | -3.E-13 | 0.04863 | 21% |
| 37 | 1.980E-04 | 2.0561E-04 | 2.0561E-04 | 8.E-06 | 8.E-06 | 2.E-13 | 0.01112 | 20% |
| 38 | 3.578E-05 | 3.4493E-05 | 3.4493E-05 | -1.E-06 | -1.E-06 | 1.E-14 | 0.00149 | -4% |
| 39 | 5.778E-06 | 3.7589E-06 | 3.7589E-06 | -2.E-06 | -2.E-06 | -4.E-15 | 0.00008 | -53% |
| 40 | 2.089E-05 | 1.9969E-07 | 1.9969E-07 | -2.E-05 | -2.E-05 | 5.E-17 | 0.00000 | - |
| **$m^1$** | **27.2014** | **27.2000** | **27.2000** | | | | | |
| **$m^2$** | **8.714** | **8.704** | **8.704** | | | | | |
| **numric.$m^2$** | **7.878** | | | | | | | |
| **obsrvd.$m^2$** | **7.869** | | | | | | | |

a) pure 3-way association

| | | |
|---|---|---|
| 0 | 0 | 0 |
| 0 | 1 | 1 |
| 1 | 0 | 1 |
| 1 | 1 | 0 |

b) pure 4-way association

| | | | |
|---|---|---|---|
| 0 | 0 | 0 | 0 |
| 0 | 0 | 1 | 1 |
| 0 | 1 | 0 | 1 |
| 0 | 1 | 1 | 0 |
| 1 | 0 | 0 | 1 |
| 1 | 0 | 1 | 0 |
| 1 | 1 | 0 | 0 |
| 1 | 1 | 1 | 1 |

c) pure 2-way disease model

| | | |
|---|---|---|
| affecteds | 0 | 0 |
| | 1 | 1 |
| controls | 0 | 1 |
| | 1 | 0 |

d) pure 3-way disease model

| | | | |
|---|---|---|---|
| affecteds | 0 | 0 | 0 |
| | 0 | 1 | 1 |
| | 1 | 0 | 1 |
| | 1 | 1 | 0 |
| controls | 0 | 0 | 1 |
| | 0 | 1 | 0 |
| | 1 | 0 | 0 |
| | 1 | 1 | 1 |

**Figure 1. Pure 3- and 4-way associations among binary markers and related pure 2- and 3-way binary disease models**. Panels (a) and (b) show schemes of "pure" 3- and 4-way association of binary markers, respectively, that characteristically miss half of the eight and sixteen possible 3- and 4-marker same-row combinations, despite showing all possible lower-order same-row marker combinations. If the 1s and 0s in the leftmost columns in (a,b) are relabelled as "affecteds" and "controls," resp., one obtains the models in (c,d) of "perfect" 2- and pure 3-way disease "causation" in which a set of markers in pure association causes disease and the other set protects against it (i.e., both sets have 100% "penetrance"), without any differences in the lower-order associations of affecteds and controls, i.e., no marker-frequency differences in (c) and neither marker-frequency nor 2-marker-combination differences in (d). Similarly, a pure 5-way pattern becomes a disease model with pure 4-way causation, etc (not shown).

combination counts

| affecteds | randoms | 6-marker combinations | | | | | |
|---|---|---|---|---|---|---|---|
| **3** | 1 | **0** | **0** | **0** | **0** | **0** | **0** |
| 1 | 1 | 0 | 0 | 0 | 0 | 0 | 1 |
| 1 | 1 | 0 | 0 | 0 | 0 | 1 | 0 |
| 1 | 1 | 0 | 0 | 0 | 0 | 1 | 1 |
| 1 | 1 | 0 | 0 | 0 | 1 | 0 | |
| 1 | 1 | 0 | 0 | 0 | 1 | 0 | |
| 1 | 1 | 0 | 0 | 0 | | | |
| 1 | 1 | 0 | 0 | | | | |
| 1 | 1 | 0 | | | | | 1 |
| 1 | 1 | 0 | | | | 0 | 0 |
| | | | | | 1 | 0 | 1 |
| | | | 1 | 0 | 1 | 1 | 0 |
| | | | 1 | 0 | 1 | 1 | 1 |
| | | | 1 | 1 | 0 | 0 | 0 |
| 1 | 1 | 1 | 1 | 1 | 0 | 0 | 1 |
| 1 | 1 | 1 | 1 | 1 | 0 | 1 | 0 |
| 1 | 1 | 1 | 1 | 1 | 0 | 1 | 1 |
| 1 | 1 | 1 | 1 | 1 | 1 | 0 | 0 |
| 1 | 1 | 1 | 1 | 1 | 1 | 0 | 1 |
| 1 | 1 | 1 | 1 | 1 | 1 | 1 | 0 |
| **3** | 1 | **1** | **1** | **1** | **1** | **1** | **1** |

**Figure 2. In-phase extended 2-way causation of affecteds vs. randoms**. Example of an association of "affected" status with a weak 2-way association of binary markers that extends "in-phase" over six IV columns, in the absence of other differences between affecteds and randoms. Idealized counts are used with 68 affecteds and 64 randoms. In such cases, the 64x2 6-column-vs.-status contingency table can be statistically significant even if none of the "6 choose 2" 2-column-vs.-status 4x2 tables is significant. The mid-plot 6-marker combinations and counts are excised to both shorten the figure vertically and still show the two "extended" 6-marker combinations 000000 and 111111 at the top and bottom, respectively (boldface), that are both equally over-represented in affecteds.

| Comb. nr. | dependent variable (DV) Affected or Control? | | independent-variable IDs (IV nr.) | | | | P(comb): prod(freqs) | observeds | expecteds | Chi$^2$ |
|---|---|---|---|---|---|---|---|---|---|---|
| | | | 1 | 2 | 3 | 4 | | | | |
| 1 | affected | 0 | 0 | 0 | 0 | 0 | 0.002 | 0 | 0.4 | 0.40 |
| 2 | affected | 0 | 0 | 0 | 0 | 1 | 0.003 | 0 | 0.6 | 0.60 |
| 3 | affected | 0 | 0 | 0 | 1 | 0 | 0.008 | 0 | 1.6 | 1.60 |
| 4 | affected | 0 | 0 | 0 | 1 | 1 | 0.012 | 6 | 2.4 | 5.40 |
| 5 | affected | 0 | 0 | 1 | 0 | 0 | 0.002 | 0 | 0.4 | 0.40 |
| 6 | affected | 0 | 0 | 1 | 0 | 1 | 0.003 | 0 | 0.6 | 0.60 |
| 7 | affected | 0 | 0 | 1 | 1 | 0 | 0.008 | 3 | 1.6 | 1.23 |
| 8 | affected | 0 | 0 | 1 | 1 | 1 | 0.012 | 1 | 2.4 | 0.82 |
| 9 | affected | 0 | 1 | 0 | 0 | 0 | 0.018 | 4 | 3.6 | 0.04 |
| 10 | affected | 0 | 1 | 0 | 0 | 1 | 0.027 | 9 | 5.4 | 2.40 |
| 11 | affected | 0 | 1 | 0 | 1 | 0 | 0.072 | 18 | 14.4 | 0.90 |
| 12 | affected | 0 | 1 | 0 | 1 | 1 | 0.108 | 13 | 21.6 | 3.42 |
| 13 | affected | 0 | 1 | 1 | 0 | 0 | 0.018 | 4 | 3.6 | 0.04 |
| 14 | affected | 0 | 1 | 1 | 0 | 1 | 0.027 | 3 | 5.4 | 1.07 |
| 15 | affected | 0 | 1 | 1 | 1 | 0 | 0.072 | 11 | 14.4 | 0.80 |
| 16 | affected | 0 | 1 | 1 | 1 | 1 | 0.108 | 28 | 21.6 | 1.90 |
| 17 | control | 1 | 0 | 0 | 0 | 0 | 0.002 | 0 | 0.4 | 0.40 |
| 18 | control | 1 | 0 | 0 | 0 | 1 | 0.003 | 2 | 0.6 | 3.27 |
| 19 | control | 1 | 0 | 0 | 1 | 0 | 0.008 | 2 | 1.6 | 0.10 |
| 20 | control | 1 | 0 | 0 | 1 | 1 | 0.012 | 0 | 2.4 | 2.40 |
| 21 | control | 1 | 0 | 1 | 0 | 0 | 0.002 | 1 | 0.4 | 0.90 |
| 22 | control | 1 | 0 | 1 | 0 | 1 | 0.003 | 0 | 0.6 | 0.60 |
| 23 | control | 1 | 0 | 1 | 1 | 0 | 0.008 | 0 | 1.6 | 1.60 |
| 24 | control | 1 | 0 | 1 | 1 | 1 | 0.012 | 5 | 2.4 | 2.82 |
| 25 | control | 1 | 1 | 0 | 0 | 0 | 0.018 | 1 | 3.6 | 1.88 |
| 26 | control | 1 | 1 | 0 | 0 | 1 | 0.027 | 7 | 5.4 | 0.47 |
| 27 | control | 1 | 1 | 0 | 1 | 0 | 0.072 | 21 | 14.4 | 3.03 |
| 28 | control | 1 | 1 | 0 | 1 | 1 | 0.108 | 17 | 21.6 | 0.98 |
| 29 | control | 1 | 1 | 1 | 0 | 0 | 0.018 | 2 | 3.6 | 0.71 |
| 30 | control | 1 | 1 | 1 | 0 | 1 | 0.027 | 7 | 5.4 | 0.47 |
| 31 | control | 1 | 1 | 1 | 1 | 0 | 0.072 | 13 | 14.4 | 0.14 |
| 32 | control | 1 | 1 | 1 | 1 | 1 | 0.108 | 22 | 21.6 | 0.01 |
| | freq(0) | 0.5 | 0.1 | 0.5 | 0.2 | 0.4 | 1.0 | 200 | 200 | **41.39** |

| DV P value H$_o$: random DV-marker order | IV-dependent P value of the DV P value H$_o$: random marker order in the tested IV | Table P value **0.02835** (26 d.f.) |
|---|---|---|
| **0.028** | 0.002  0.006  0.010  0.012 | |

**Figure 3. Permuting the markers at one or two columns allows one to flag significant columns and bypass model selection.** The shown 5-column binary-marker data matrix (DM) has much fewer columns (variables) than rows ("observeds"). This allows one to permute vertically the markers in the **Affected or Control** dependent-variable column DV on the left to so estimate non-parametrically the **DV P value** (bottom left) that a random vertical order of DV markers generate a Chi$^2$ at least as large as the overall Chi$^2$ (41.39, bottom right), while keeping everything else unchanged. Permuting the markers of an independent-variable IV and re-estimating each time the DV P value (as just described), allows one to estimate an **IV-dependent P value** that a random vertical order of said IV's markers let the DV P value become no larger than the original DV P value, while keeping everything else unchanged (but requires squaring the permutations). The shown **P(comb)** expectations of the 32 5-marker sequences are products of five marker frequencies and deliver greater power in cases like Fig.2's than using the frequencies of the combinations of non-tested-column markers pooled (e.g., IV-marker combinations from affecteds and controls; not shown). The **Table P value** for the overall Chi$^2$ given 26 d.f. (32 -5 -1; bottom right) coincides here with the permutation-obtained DV P value because the departures from expectation in this particular DM are mainly of affecteds-vs.-controls nature. In random DMs the two P values tend to differ because of IV associations that do not involve the DV. The shown P values of the DV and the IVs are estimated using 10'000 and 10'000$^2$ permutations, respectively.

|   | site nr. | 1 | 2 | 3 | 4 | 5 | 6 | 7 | 8 | 9 | 10 | 11 | 12 | 13 | 14 | 15 | 16 | 17 |
|---|---|---|---|---|---|---|---|---|---|---|---|---|---|---|---|---|---|---|
|   |   |   |   |   | *i* |   |   |   |   |   | *j* |   |   |   |   |   |   |   |
| a) | sequence 1: | 0 | 1 | 1 | 0 | 1 | 1 | 0 | 0 | 1 | 0 | 0 | 0 | 0 | 1 | 0 | 0 | 0 |
|   | (1;2) pairwise comparison | n | M | M | **M** | n | M | n | M | n | **M** | n | n | n | n | M | n | M |
|   | sequence 2: | 1 | 1 | 1 | 0 | 0 | 1 | 1 | 0 | 0 | 0 | 1 | 1 | 1 | 0 | 0 | 1 | 0 |
| b) | sequence 3: | 0 | 0 | 0 | 0 | 0 | 0 | 0 | 0 | 0 | 0 | 1 | 1 | 1 | 0 | 1 | 1 | 0 |
|   | (3;10) pairwise comparison | M | n | n | **n** | n | M | M | n | n | **n** | n | M | n | M | n | n | n |
|   | sequence 10: | 0 | 1 | 1 | 1 | 1 | 0 | 0 | 1 | 1 | 1 | 0 | 1 | 0 | 0 | 0 | 0 | 1 |

**Figure 4. Markers co-occurring in the rows of a data matrix compel matches to co-occur in the pairwise comparisons of the rows.** If in the rows of a data matrix DM with binary markers of frequency 0.5 the markers of type 0 at column *i* co-occur only with markers of type 0 at column *j* (and therefore also 1s with 1s) then, as shown in (a), matches must co-occur with matches at those two columns in the pairwise comparisons of the rows and so must mismatches with mismatches, as shown in (b) for two other rows. The same happens when 0s co-occur with 1s and 1s with 0s at the two columns.

a)

|   | 1 | 2 | 3 | 4 | 5 | 6 | 7 | 8 | 9 |
|---|---|---|---|---|---|---|---|---|---|
| 1 | 0 | 1 | 0 | 1 | 0 | 1 | 0 | 0 | 1 |
| 2 | 0 | 0 | 0 | 0 | 1 | 0 | 1 | 1 | 0 |
| 3 | 1 | 0 | 0 | 0 | 1 | 1 | 0 | 0 | 1 |
| 4 | 1 | 0 | 1 | 1 | 1 | 0 | 0 | 0 | 1 |
| 5 | 1 | 1 | 1 | 1 | 0 | 1 | 1 | 1 | 0 |
| 6 | 0 | 1 | 1 | 0 | 0 | 0 | 1 | 1 | 0 |

**Figure 5. The data matrix DM, the pairwise matrix PM, and the scoring of conditional matches.** Panel (a) shows a binary-marker data matrix DM with six random 9-marker rows ("marker sequences").

b)

|    |       | 1 | 2 | 3 | 4 | 5 | 6 | 7 | 8 | 9 |
|----|-------|---|---|---|---|---|---|---|---|---|
| 1  | (1,2) | M | - | M | - | - | - | - | - | - |
| 2  | (1,3) | - | - | M | - | - | M | M | M | M |
| 3  | (1,4) | - | - | - | M | - | - | M | M | M |
| 4  | (1,5) | - | M | - | M | M | M | - | - | - |
| 5  | (1,6) | M | M | - | - | M | - | - | - | - |
| 6  | (2,3) | - | M | M | M | M | - | - | - | - |
| 7  | (2,4) | - | M | - | - | M | M | - | - | - |
| 8  | (2,5) | - | - | - | - | - | - | M | M | M |
| 9  | (2,6) | M | - | - | M | - | M | M | M | M |
| 10 | (3,4) | M | M | - | - | M | - | M | M | M |
| 11 | (3,5) | M | - | - | - | - | M | - | - | - |
| 12 | (3,6) | - | - | - | M | - | - | - | - | - |
| 13 | (4,5) | M | - | M | M | - | - | - | - | - |
| 14 | (4,6) | - | - | M | - | - | M | - | - | - |
| 15 | (5,6) | - | M | M | - | M | - | M | M | M |

c)

| pair  | focal column S | \_ | \_ | \_ | \_ | \_ | \_ | \_ | \_ | total $M_m$ |
|-------|----------------|---|---|---|---|---|---|---|---|---|
| (1,2) | match 0/0      | - | M | - | - | - | - | - | - | 1 |
| (1,6) |                | M | - | - | M | - | - | - | - | 2 |
| (2,6) |                | - | - | M | - | M | M | M | M | 5 |
| (3,5) | match 1/1      | - | - | - | - | M | - | - | - | 1 |
| (4,5) |                | - | M | M | - | - | - | - | - | 2 |
| (3,4) |                | M | - | - | M | - | M | M | M | 5 |
| (3,6) | mismatch 0/1 (1/0) | - | - | M | - | - | - | - | - | 1 |
| (4,6) |                | - | M | - | - | M | - | - | - | 2 |
| (2,4) |                | M | - | - | M | M | - | - | - | 3 |
| (2,5) |                | - | - | - | - | - | M | M | M | 3 |
| (1,4) |                | - | - | M | - | - | M | M | M | 4 |
| (1,5) |                | M | - | M | M | M | - | - | - | 4 |
| (2,3) |                | M | M | M | M | - | - | - | - | 4 |
| (1,3) |                | - | M | - | - | M | M | M | M | 5 |
| (5,6) |                | M | M | - | M | - | M | M | M | 6 |

**(Figure 5.Cont.)** Panel (b) shows the pairwise matrix PM with the fifteeen pairwise comparisons of the six 9-marker sequences in (a). Panel (c) shows how after choosing a "focal column" S in the PM (here the leftmost column), one can group the pairwise comparisons by their $S_i..M_m$ combinations, i.e., by the "pairwise state" $S_i$ at their focal-column site and the number of matches $M_m$ at their non-focal-column sites.

d)

| $S_i..M_m$ count | $S_i$ | $M_m$ |
|---|---|---|
| 1 | 0/0 | 1 |
| 1 |  | 2 |
| 1 |  | 5 |
| 1 | 1/1 | 1 |
| 1 |  | 2 |
| 1 |  | 5 |
| 1 | 0/1(1/0) | 1 |
| 1 |  | 2 |
| 2 |  | 3 |
| 3 |  | 4 |
| 1 |  | 5 |
| 1 |  | 6 |

e)

| $S_i..M_m$ count | $S_i$ | $M_m$ |
|---|---|---|
| 2 | match | 1 |
| 2 |  | 2 |
| 2 |  | 5 |
| 1 | mismatch | 1 |
| 1 |  | 2 |
| 2 |  | 3 |
| 3 |  | 4 |
| 1 |  | 5 |
| 1 |  | 6 |

**(Figure 5.Cont.)** When scoring the counts of the distinct $S_i..M_m$ combinations, the distinguished $S_i$ pairwise states can be the fully specified 2-marker pairings 0/0, 1/1, and 0/1(1/0) in (d), or the plain matches and mismatches in (e).

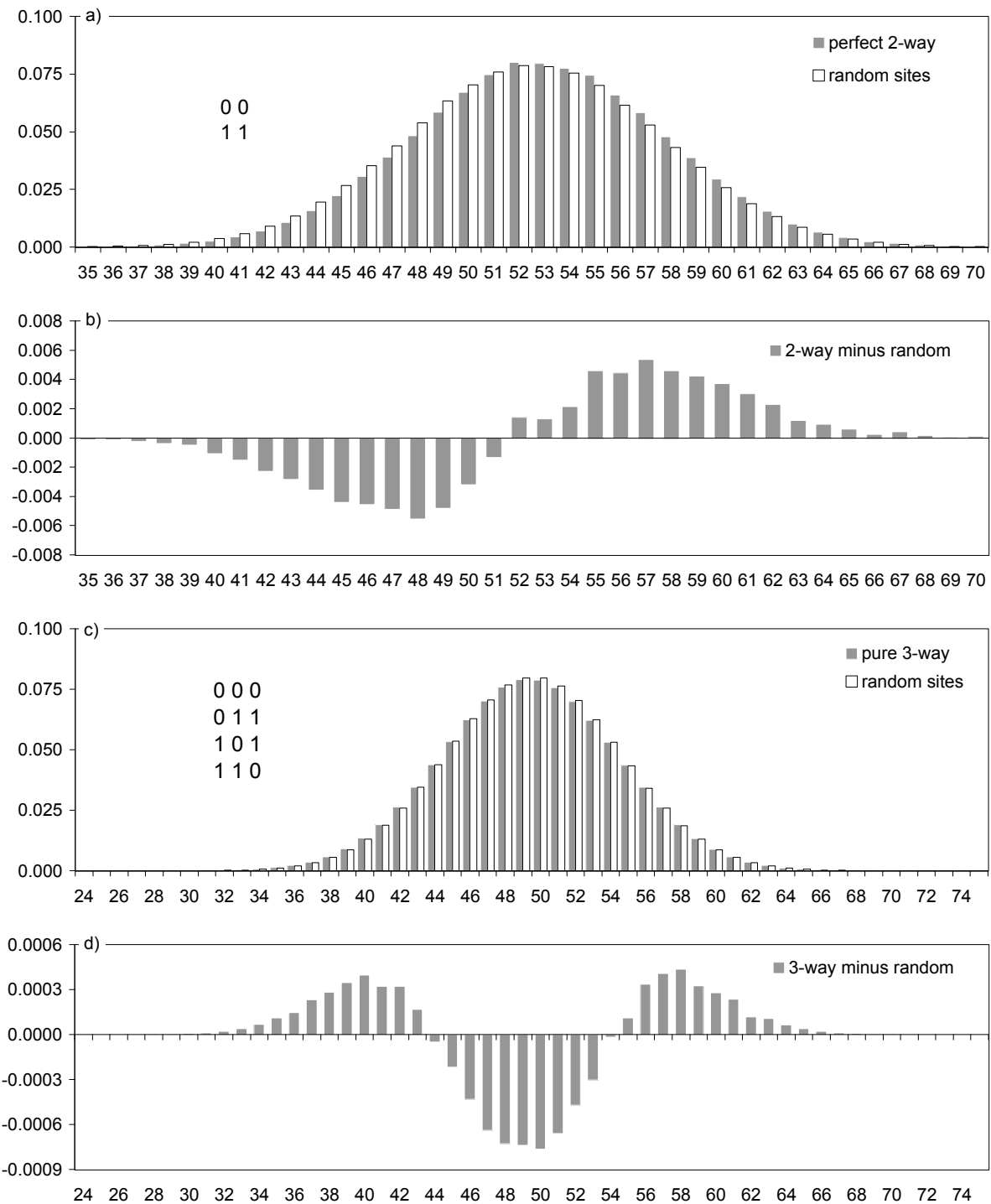

**Figure 6i. Perturbations of the distribution of the number of conditional matches when the focal column is involved in pure 2- or 3-way association with other columns.** Panels (a,d) show the distribution of the number of non-focal-column matches conditional on a match at a focal column when the focal column is pure 2- or 3-way associated with other columns, resp., vs.when the focal column is a random column. This is clearer in (b,d) where the distribution of random columns is substracted from that of the associated columns. In all cases the DM has 100 columns with binary-marker frequency 0.5 at every column. "Perfect" 2-way case: 1'000 100-marker sequences; pure 3-way case: 7'000 100-marker sequences.

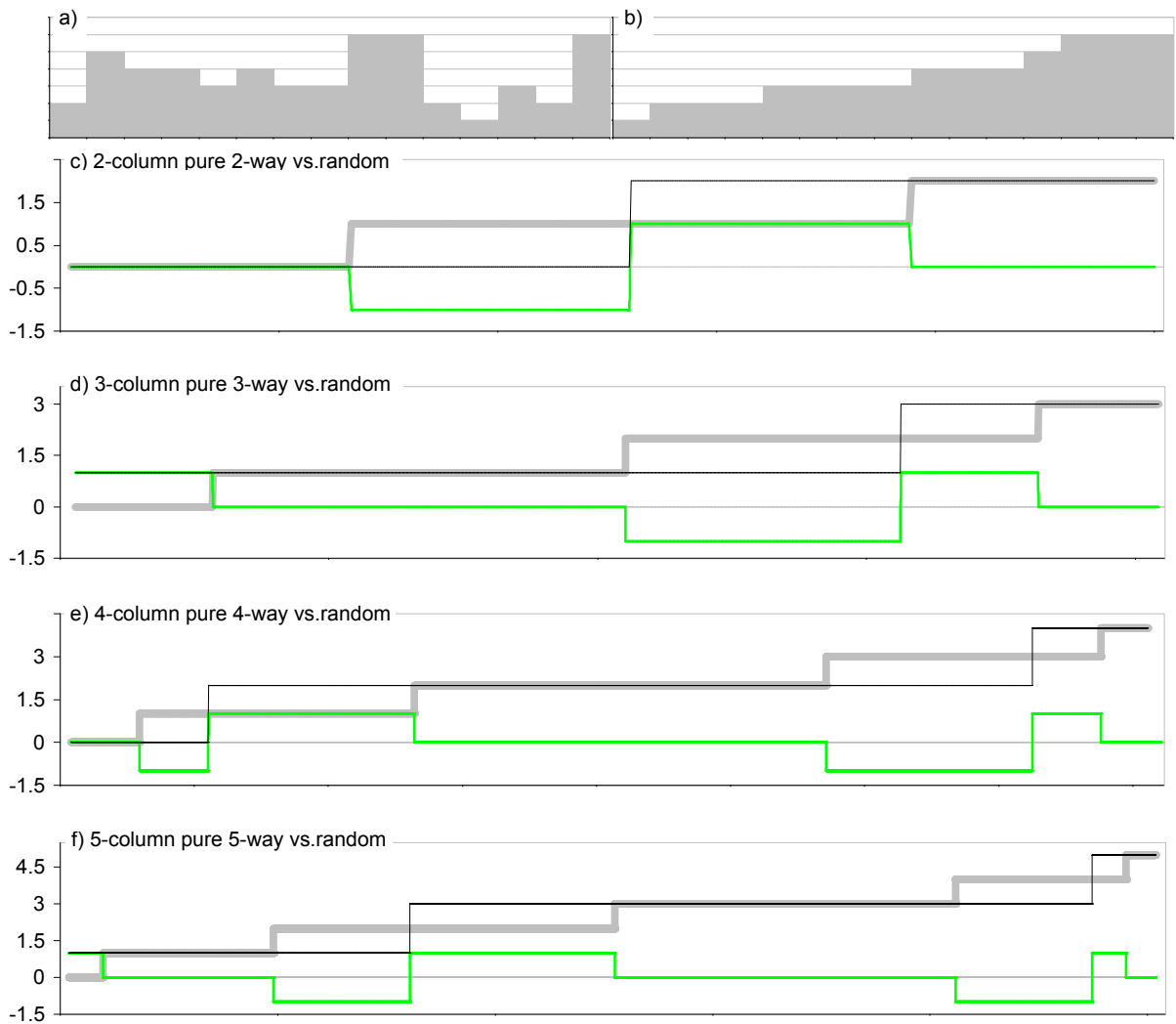

**Figure 6ii. The matches in the pairwise comparisons of *n*-marker sequences in pure *n*-way association vs. when every possible sequence is equally present, sorted by increasing value.** Panel (a) shows the 15 sums of the row matches in Fig.5c (vert.axis) and in (b) the sums are sorted by increasing value from left to right. Panels (c-f) are akin to (b) and show the row sums when there is pure 2- to 5-way association in 2- to 5-column DMs, resp.(thin black), contrasted to the sums when all possible 2- to 5-marker sequences are equally represented in 2- to 5-column DMs (thick grey). The PMs used in (c,d,e,f) have 496, 2'016, 8'128, and 24'976 rows and in all cases the compared 2-, 3-, 4-, and 5-marker sequences, pure and not, are equally represented (i.e., marker frequency is 0.5 at every column). The relative (horizontal) widths of the "ladders" approximate the frequencies of the sums of matches obtained from corresponding infinite-row *n*-column DMs. In the "pure 2-way" case in (c), e.g., there are two equally frequent 2-marker sequences (00 and 11) that yield zero and two matches when compared to each other and themselves, resp., resulting in only two about equally wide plateaus for 0s and 2s, resp.(thin black). The sums when all possible sequences are equally present are in thick grey and in mid-thick green is the latter curve substracted from the *n*-column pure *n*-way curve. Fig.1 shows that the ladders for the sums given a match at a focal column should be in the panel immediately above the pure association at stake in this figure. In (c), e.g., the thin-black ladder coincides with that for the sums which are conditional on a match at a column that is in pure 3-way association with two other columns. Therefore the pattern of conditional matches is missing here for the pure 2-way case (it would show one conditional match given a match at either focal column in pure 2-way association, and nothing else).

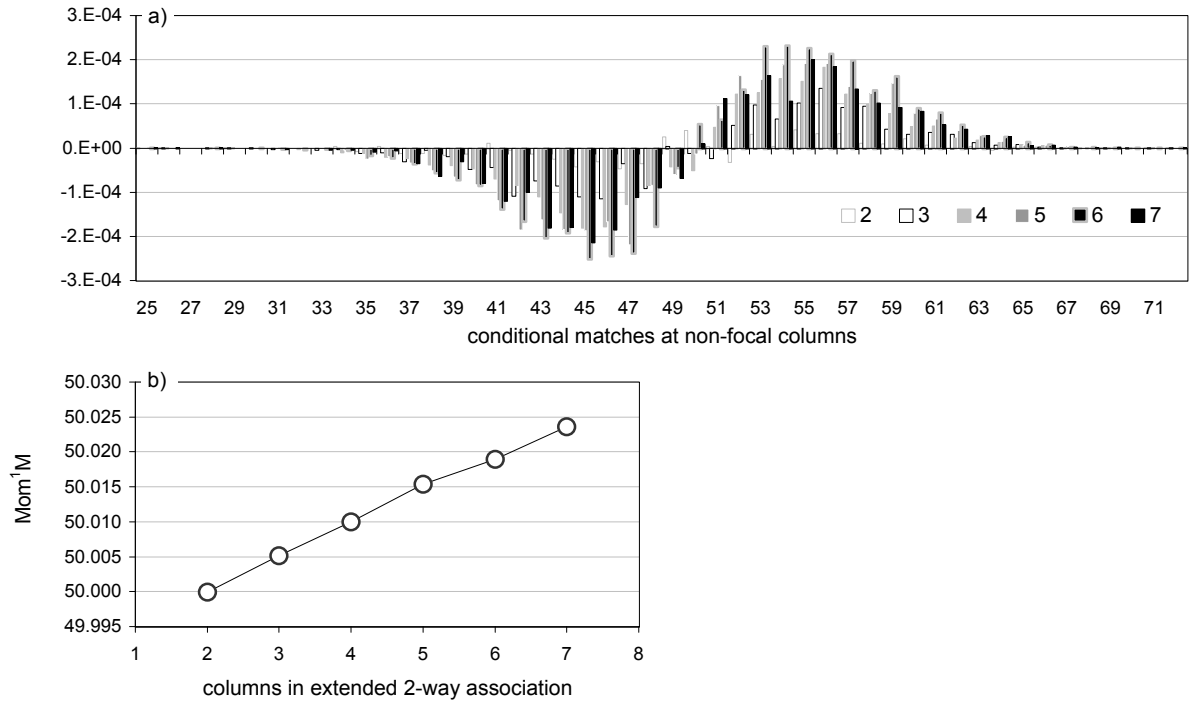

**Figure 7. An extended "in-phase" 2-way association perturbs the distribution of the number of conditional matches.** The distribution of the number of non-focal-column matches given a match at the focal column is shown as it reacts to increasing numbers of columns that are in extended "in-phase" 2-way association with the focal column (the pattern in Fig.2 sans DV). In panel (a) the p.d.f. of the number of matches given a match at a random column is substracted from that given a match at a column that is in in-phase $n$-column 2-way association with others. Every bin of each shown distribution for a column in extended 2-way association is actually an average over two distributions whereas for the columns in random association the average is over three distributions. In (b) is the average number of matches given a match at a column involved in in-phase $n$-column 2-way association. All DMs have 8'592 100-marker sequences where in the first $n$ columns there are 200 copies of each of the two $n$-marker runs (e.g., 1111111 and 0000000 for $n=7$) plus 8'192 copies of the distinct $n$-site combinations in equal number (including the two $n$-marker runs, i.e., in total there are say 64+200 counts of each 7-marker run). The remaining (100-$n$) columns are independent with random vertical marker order. Binary-marker frequency is 0.5 at every column.

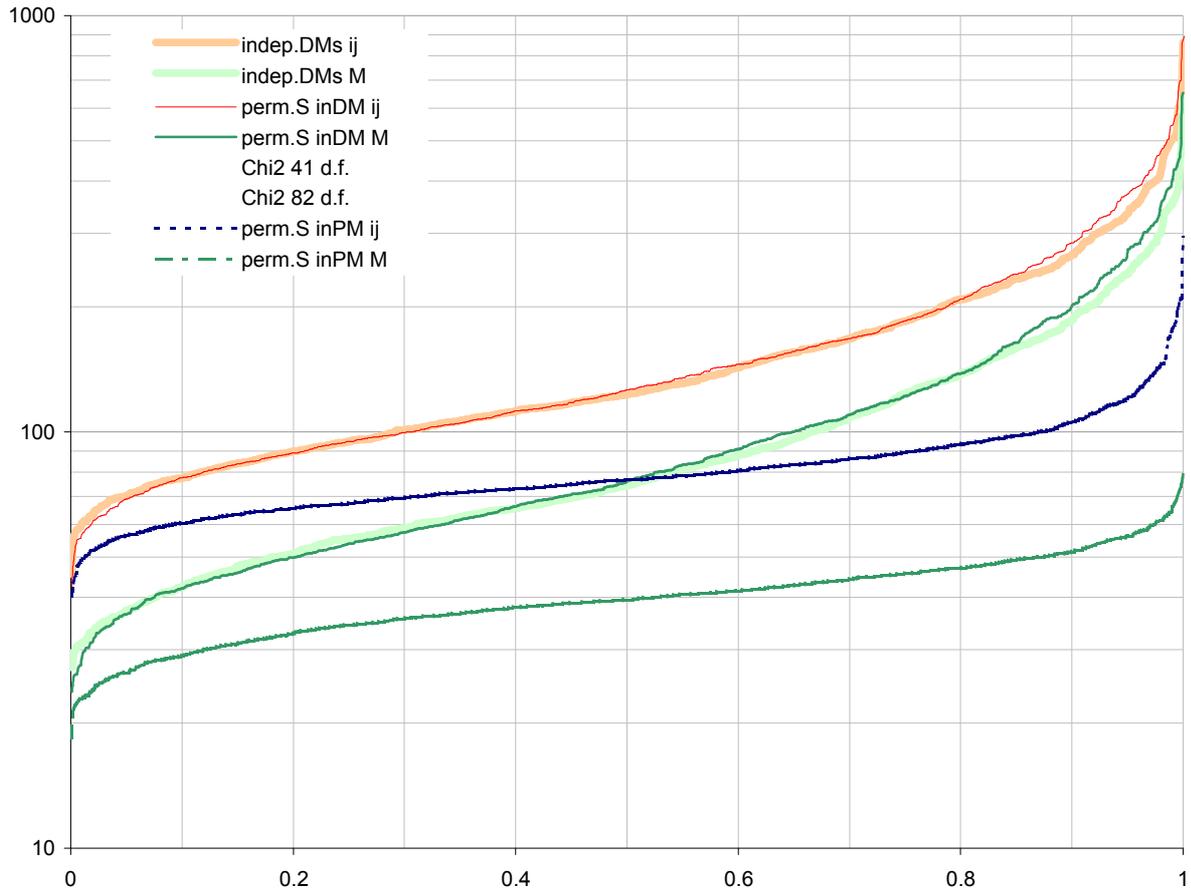

**Figure 8. CHIx-M and CHIx-ij are not Chi$^2$-distributed under the $H_o$.** The distributions of the CHIx-M and CHIx-ij of a column S from a 500-row 100-column binary null DM are estimated by permuting 1'000 times the markers at S while keeping everything else unchanged (labels "perm.S inDM M" and "perm.S inDM ij"). The two distributions do not match any standard Chi$^2$-distribution but match those for the same column across 1'000 independently generated null DMs (label "indep.DMs"). The various sets of 1'000 CHIx-M and CHIx-ij values (vert.axis) are sorted by increasing magnitude from left to right. The original null DM has independent columns where binary markers are randomly vertically ordered with minor-marker frequency cycling in the order 0.1, 0.2, 0.3, 0.4, and 0.5 (o12345 marker-frequency scheme, see M&M). When one permutes the matches and mismatches at column S in the pairwise matrix PM of the null DM, CHIx-M and CHIx-ij are Chi$^2$-distributed, as expected (with 42 and 82 d.f., resp.; label "perm.S inPM"). Minor-marker frequency at S is 0.1. The slight departure by "perm.S in PM ij" from the Chi$^2$ distribution disappears when DMs have more rows (not shown).

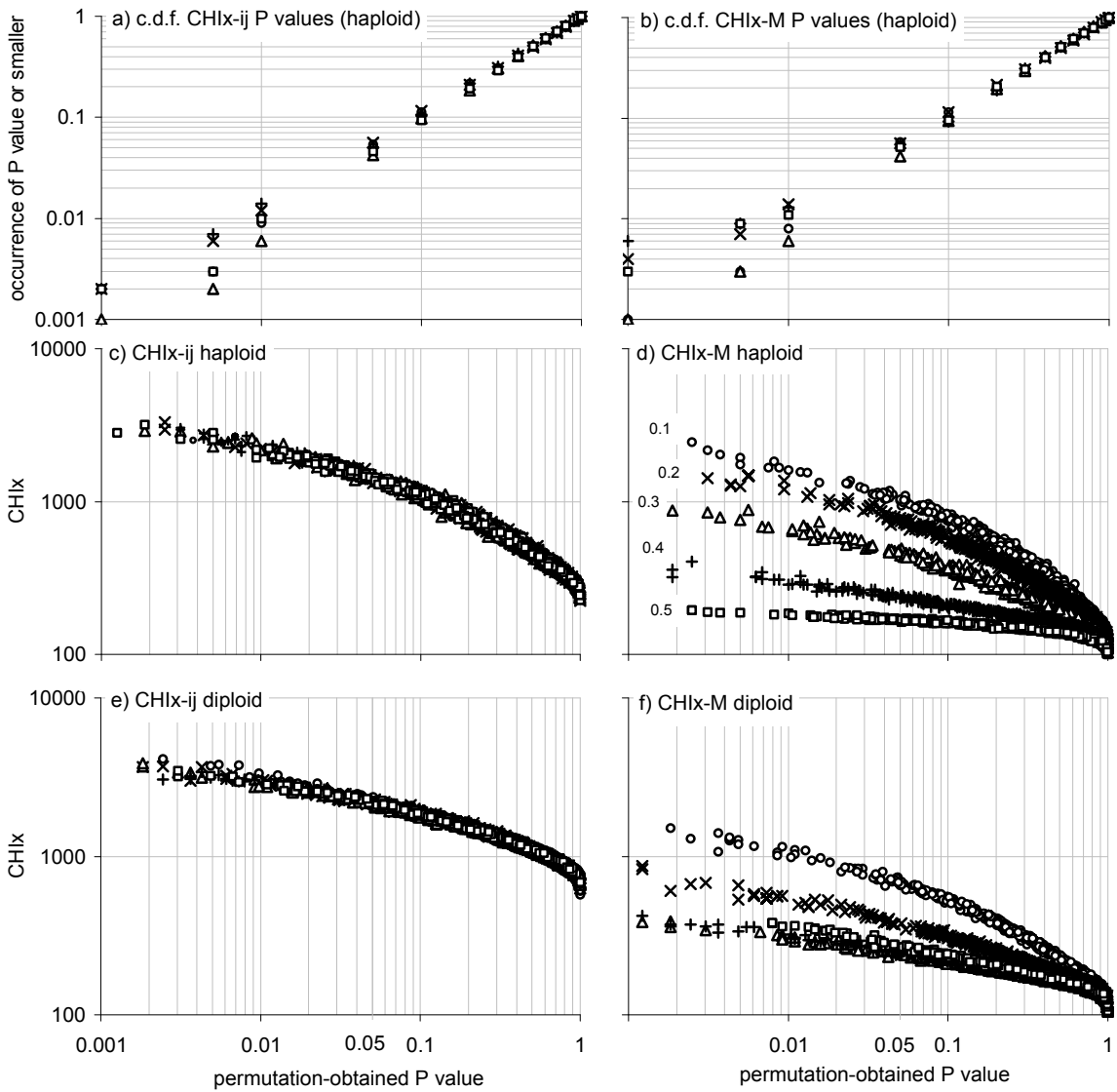

**Figure 9. The values of CHIx-M and CHIx-ij vs. their permutation-obtained P values in null DMs.** P values for the CHIx-ij and the CHIx-M of a column can be estimated by permuting the markers of the column (in the DM). Panels (a,b) show that the c.d.f.s of such P values match that of the uniform-(0,1) distribution, in DMs with independent columns where markers have random vertical order. Panels (c,d) and (e,f) show for the binary and trinary (haploid,diploid) case, resp., that ranking columns by their CHIx is nearly equivalent to ranking them by their permutation-obtained P values, albeit CHIx-M is affected strongly by marker frequency. Therefore, the largest CHIx-ij values have the lowest permutation-obtained P values and can be used directly to flag columns. The DMs in (a,b,c,d) are those described in Fig.8; the 1'000 DMs used in (e,f) have 2'000 rows and 1'000 columns. Binary and trinary marker frequencies follow the o12345 scheme and the trinary H&W thereof, resp.(see M&M). Five columns are scored per DM, one for each o12345 marker frequency. P values are estimated with 1'600 permutations of the markers at the evaluated column in the DM.

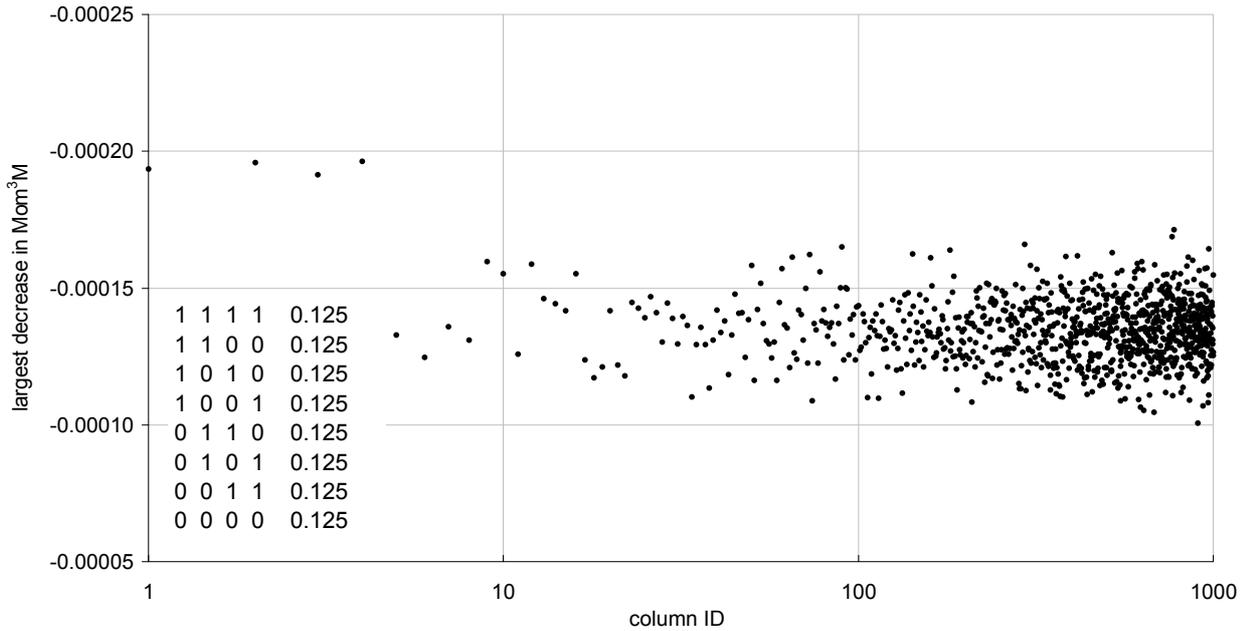

**Figure 10. Removing a column involved in a pure 4-way association reduces the $Mom^3M$ of its three associated columns**. The 3rd-moment PAS conditional on any match, $Mom^3M$, of each of four columns in pure 4-way association sinks strongly when any of the other three associated columns is removed from the DM at stake. Plotted is the largest $Mom^3M$ change (vertical axis) that is scored as one removes, one at a time, each of the 1'000 columns in a DM in which 996 columns are random and the four leftmost ones are in pure 4-way association (i.e., show the 4-marker sequences in the plot's bottom left that lack any 2- and 3-way associations). The DM has 4'000 1'000-marker sequences (rows) and binary marker frequency is 0.5 at every column. Performing the 1'000 individual-column exclusions and recalculating each time the $Mom^3M$ of the 999 remaining columns squares the computational expense. Direct detection with $Mom^3M$ (without column exclusions) requires a DM with 100'000s of rows (extrapolating the trend from results with DMs with fewer random colums).

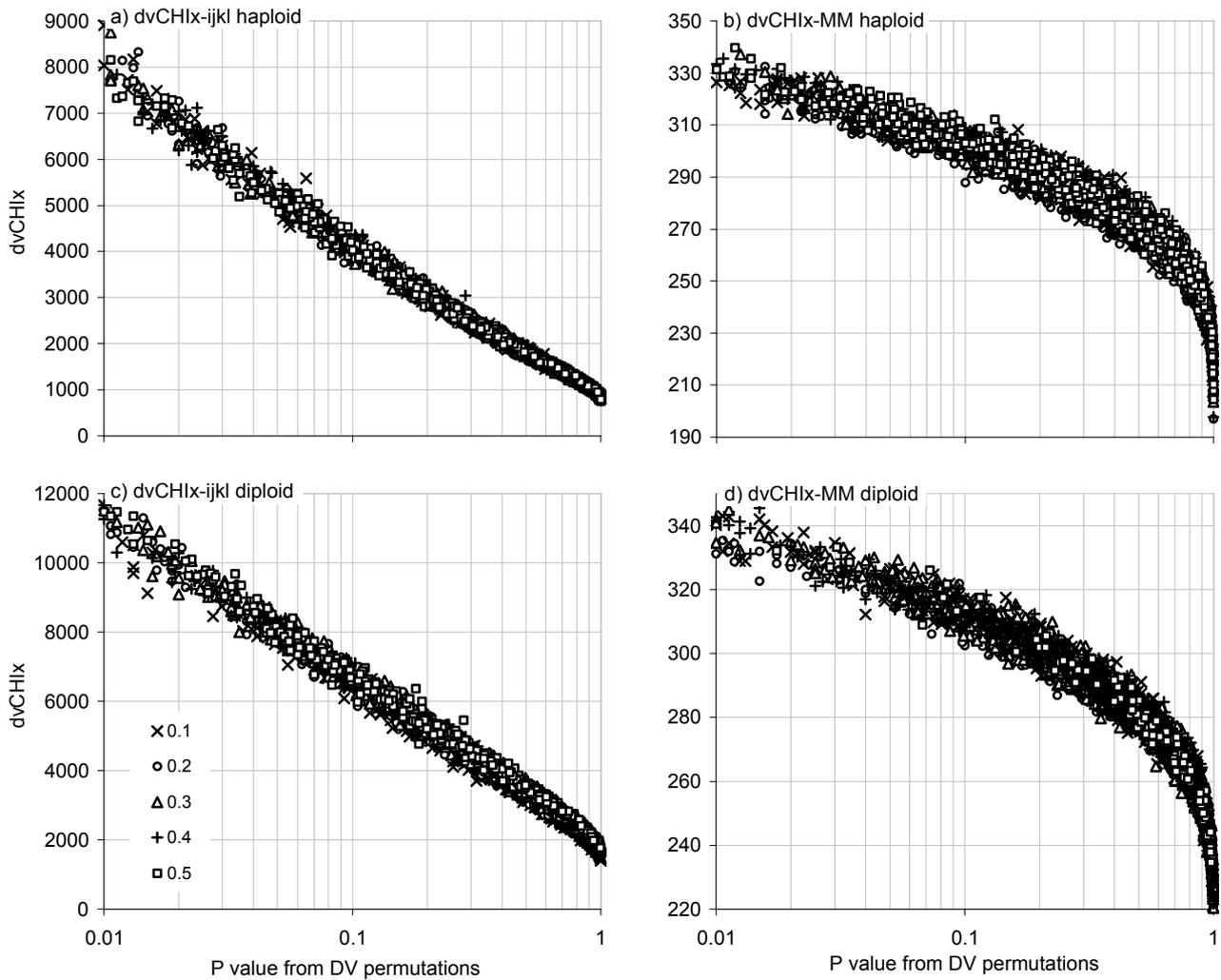

**Figure 11. The values of dvCHIx-ijkl and dvCHIx-MM vs. their P values from shuffling DV markers.** The two double-conditional dvCHIx's of an independent-variable (IV) column increase when the IV by itself or in synergy with others is associated with the dependent-variable DV of a DM (e.g., an "affecteds vs.controls" column). A P value for the dvCHIx of an IV can be estimated by permuting DV markers and scoring how often the dvCHIx of the IV is at least as large as the original dvCHIx. Panels (a,b) and (c,d) show that in both the binary and trinary case, resp., ranking the IVs by their dvCHIx-ijkl and dvCHIx-MM (vert.axes) approximates ranking them by P value (horiz.axes), regardless of IV marker frequencies (here given by the o12345 scheme, see M&M). Flagging the IVs with the largest dvCHIx's is therefore like choosing the lowest-P value IVs and obviates permuting the markers at the DV. Plotted are values for a binary and a trinary set of 1'000 2'000-row 1'000-IV DMs. All IVs are independent with random vertical order of markers. The DV's binary-marker frequency is 0.5, i.e., "affecteds" and "controls" are equally numerous. Random marginal effects are allowed at every IV.

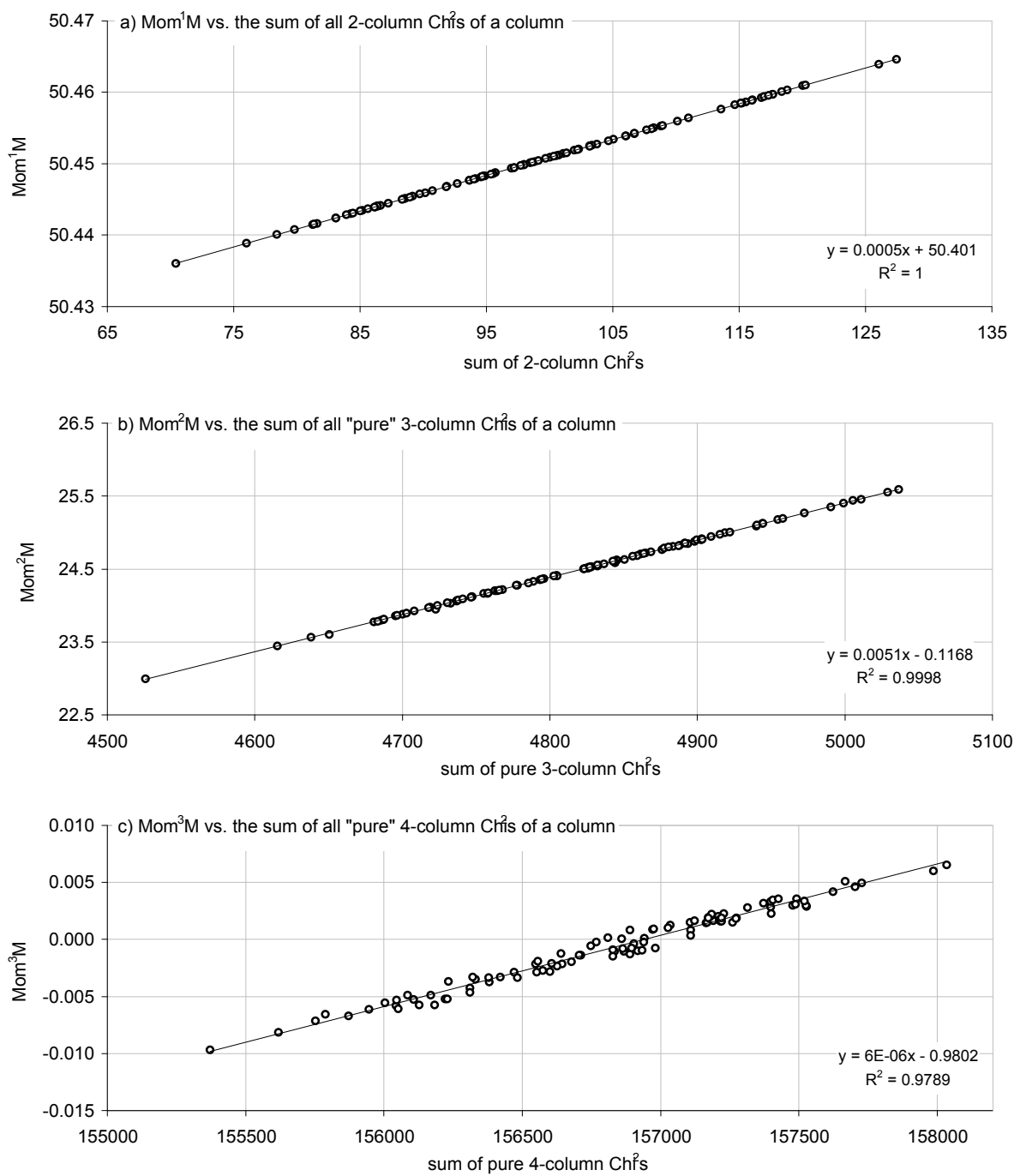

**Figure 12. The first three Mom$^n$M's correlate with conditional sums of standard Chi$^2$s.** The first three PAS moments of a focal column, Mom$^1$M, Mom$^2$M, and Mom$^3$M ("-M"=conditional on any match), correlate tightly with the sums of all 2-column, all "pure" 3-column, and all "pure" 4-column DM-level Chi$^2$s that involve the focal column, resp. Plotted are, for a single 1'000-row 100-column DM where columns are independent with binary marker frequency $p$ 0.5 at every column, in panel (a) the Mom$^1$M of each column vs.the sum of every 2-column Chi$^2$ involving the column; in (b) the Mom$^2$M of each column vs.the sum of every pure 3-column Chi$^2$ involving the column; and in (c) the Mom$^3$M of each column vs.the sum of every pure 4-column Chi$^2$ involving the column. Each pure 3-column Chi$^2$ equals the corresponding plain 3-column Chi$^2$ minus the three "internal" 2-column Chi$^2$s. Each pure 4-column Chi$^2$ equals the plain 4-column Chi$^2$, minus the three internal pure 3-column Chi$^2$s, minus the six internal 2-column Chi$^2$s. The R$^2$s sink with $p$ <0.5, e.g., Mom$^1$M R$^2$s are 81.3 and 15.9% when $p$ is 0.45 and 0.4, resp., in a DM with as many rows and columns (not shown). In general the R$^2$s are ~100% when DM columns have S markers of identical frequency 1/S (1<$n$<3; not shown).

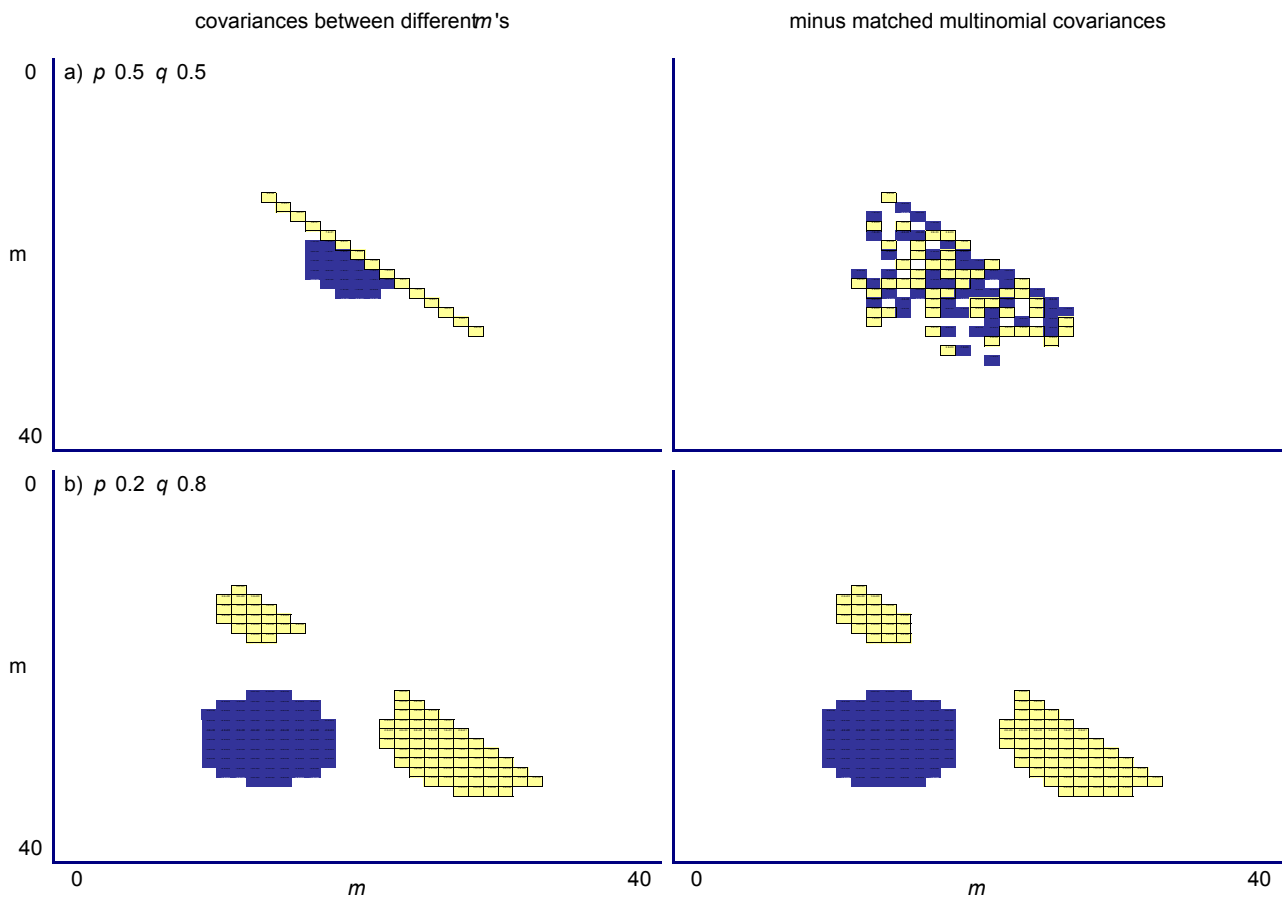

**Figure 13. Expected covariances between any two possible *m*'s in the PM of a 10-row 40-column binary DM.** The value *m* is here the number of matches in an entire row of a PM (see text). On the left is the single-DM covariance of every possible pair of *m*'s averaged over $10^5$ PMs derived from as many independently simulated DMs (average variances on the diagonal). Top and bottom plots are for DMs with minor-marker frequency *p* 0.5 and 0.2 at every DM column, resp. On the left one substracts the corresponding average covariances coming from $10^5$ simulations of a multinomial trial that uses as category probabilities the expected frequencies of the *m*'s in the PM at stake (see text). Values smaller or larger than the mean of the values plotted in each matrix $\pm$ one standard deviation are light- and dark-highlighted, resp. Results with 100-row 40-column DMs are qualitatively identical.

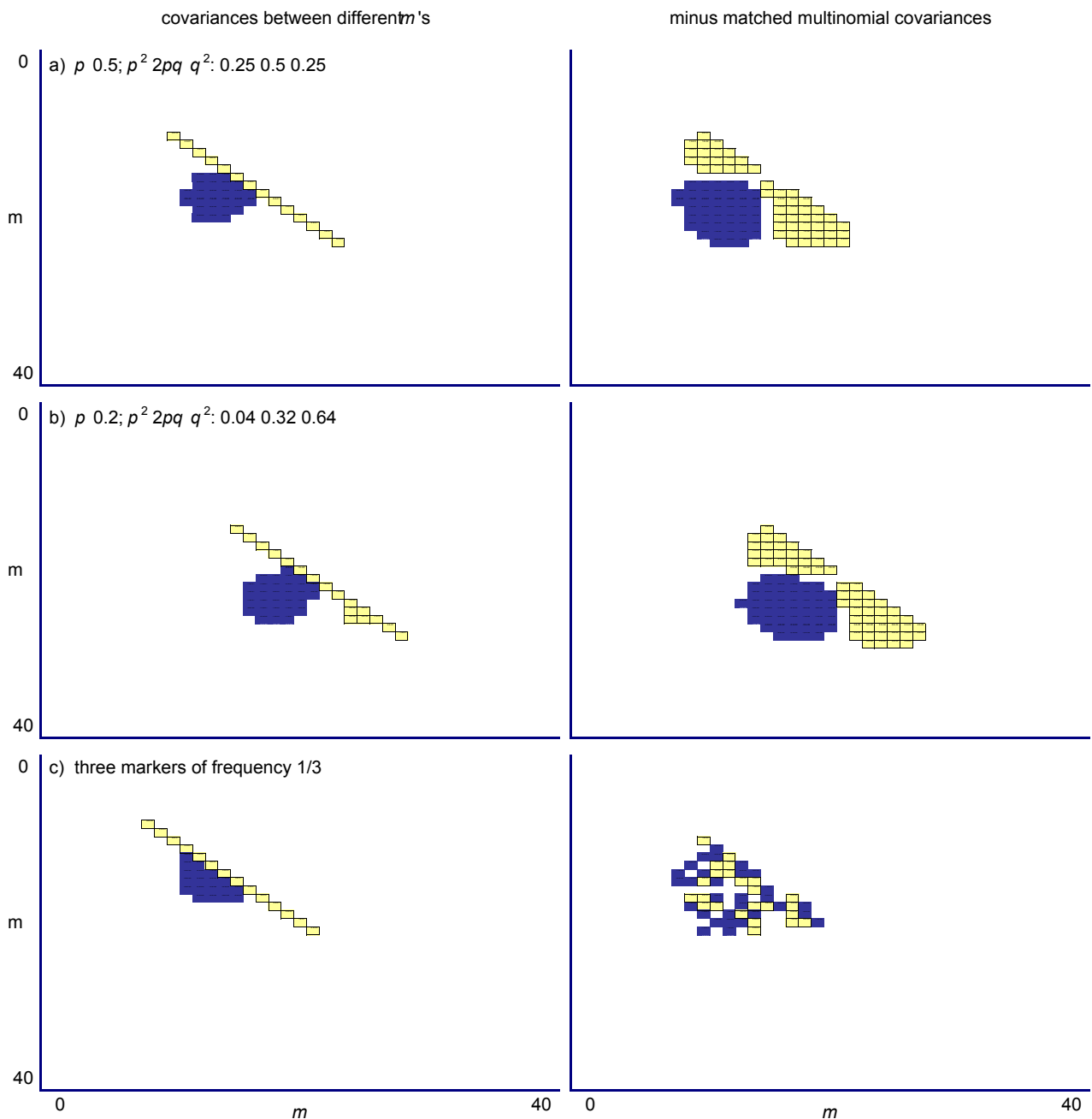

**Figure 14. Expected covariances between any two possible *m*'s in the PM of a 10-row 40-column trinary DM.** Top and middle plots are for DMs with three "diploid" markers of frequency 0.25 0.50 0.25 and 0.04 0.32 0.64, respectively, at every column (i.e., the H&W frequencies $p^2$ $2pq$ $q^2$ for $p$ 0.5 and 0.2). Bottom plots are for trinary DMs with marker frequency 1/3 at every column. Results with 100-row 40-column DMs are qualitatively identical. Labels and other details like in Fig.13.

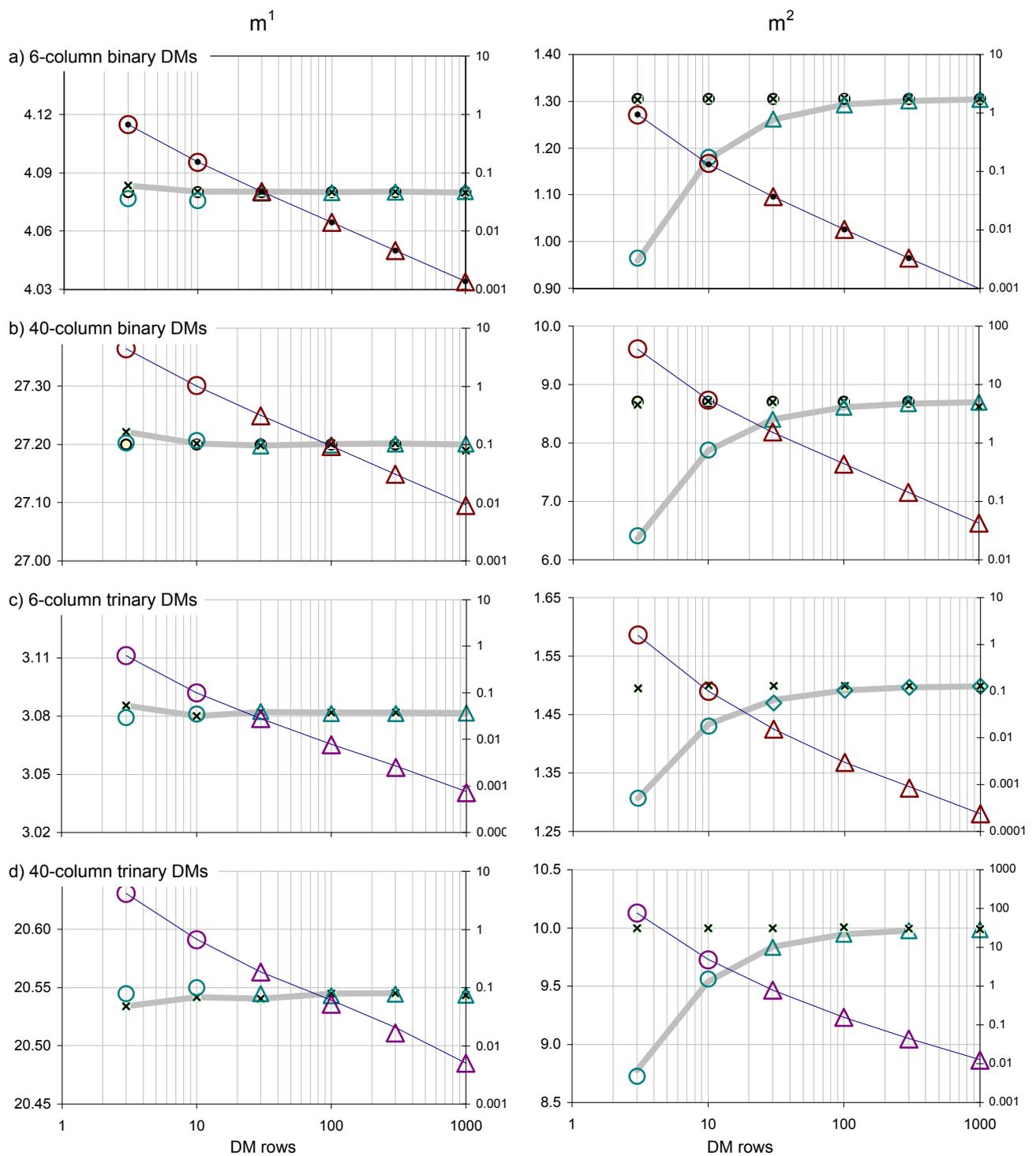

**Figure 15. Predicted and simulated averages and variances of the number of matches in the pairwise comparisons of a DM's rows, plus sample variances.** The average and variance ($m^1$, $m^2$: left, right; left axes) of the number of matches in the pairwise comparisons of the rows of an individual binary or trinary DM (panels a,b vs.c,d) with 6 and 40 columns (a,c vs.b,d) as a function of DM rows, when DMs have minor-marker frequency 0.2 or the trinary H&W thereof. The empirical average $m^1$ and $m^2$ (thick grey lines) and their two across-PMs sample variances (thin black lines) are based on $10^5$ PMs from as many simulated DMs (sample variances on the right axis). Binary expectations (small circles; not for trinary case) are from Formula (10). The x's label the $m^1$ and $m^2$ when pooling the counts of the various $m$'s from $10^5$ PMs. Numeric sample variances of $m^1$ and $m^2$ are the big circles/triangles (till 10 rows, above 10 rows; see text). The binomial $m^1$s (not shown) and pooled-PMs $m^1$s match the empirical average $m^1$s but the $m^2$s match only when DMs have 100 rows or more. In 3-row 40-column binary and trinary DMs (b,d), e.g., the empirical $m^2$s are only 74 and 87% of the pooled-PMs $m^2$s, resp.(left plots). Numeric sample variances match the empirical across-PMs ones.

**Figure 16. The 1-column patterns in a few-row many-column binary-marker DM and its PM.** Every distinct 1-column R-marker pattern should be observed when in an R-row L-column binary-marker DM both $L \gg 2^R$ and $R \gg p^{-1}$ ($p$: marker frequency). Panels (a,b,c,d) show these patterns when DMs have R equal (2;3), 4, 5, and 6, resp. In (a,b) all 1-column patterns are shown. Under the left-to-right pattern order in (a,b), right DM halves are identical to left halves with toggled markers, so in (c,d) only left halves are shown in (a,b)'s order while on the right the 1-column DM patterns are ordered by increasing number of 1s. The plots' left halves show how blocks of DM rows become blocks in the PM as if markers were being turned into Ms and mismatches. Panel (c.2) shows the turning procedure column by column. To generate, e.g., the "row-1 marker vs.other markers" PM tract of a DM column, one removes the top (first) marker of the DM column and turns the 1s into Ms (bold in c.1) and the 0s into mismatches, if the column's row-1 marker is a "1"; otherwise one turns the 0s into Ms and the 1s into mismatches. For the "row-2 vs.others" tract, one removes both the row-1 and row-2 markers of the column and again one turns either the remaining 1s into Ms if the row-2 marker is a "1" or the 0s into Ms if the row-2 marker is a "0", etc. Generating this way the R(R-1)/2 pairwise comparisons of the markers in a DM column requires only R-1 pairwise comparisons of markers, which ideally should speed up the generation of the PM. More importantly, the turning procedure makes Ms easily indexable given any DM column.

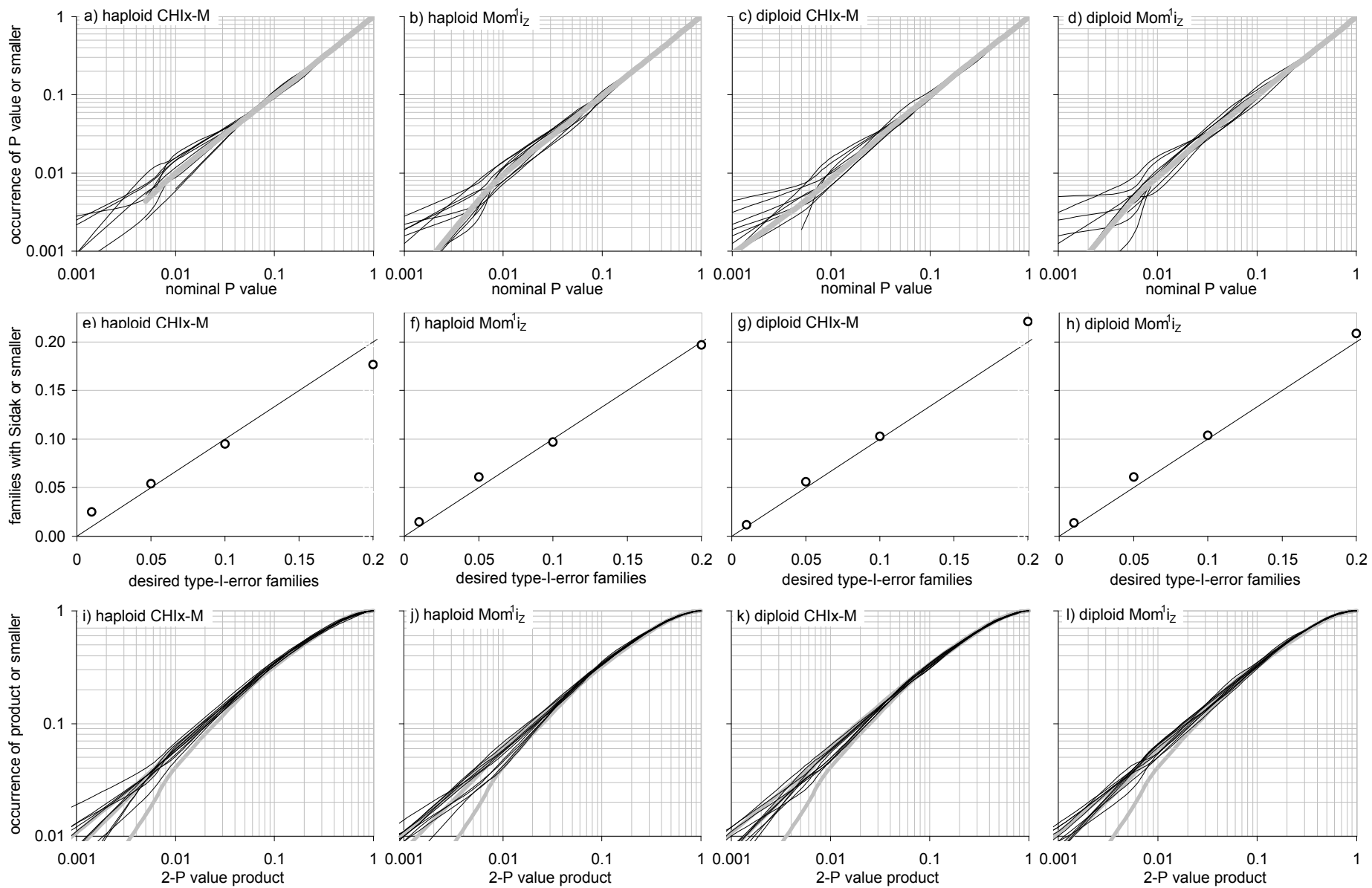

**Figure 17. Type I error of the CHIx-M and Mom $^1i_Z$ P values of individual columns across null DMs.** A P value for the CHIx-M or the Mom$^1i_Z$ of a column can be estimated by permuting the markers at the column (in the DM). With such P values for chosen columns across null DMs one can compute null distributions, including for say the product of two same-DM P values. **Panels (a-d)** show c.d.f.s of such P values for each of ten columns over 1'000 2'000-row 1'000-column binary and trinary null DMs generated under the H$_o$ of columns being independent with marker frequencies that follow the o12345 scheme or the trinary H&W thereof (see M&M). The c.d.f.s tend to match the c.d.f. of the uniform-(0,1)-distribution without discernible marker-frequency effects. **In (e-h)** is the c.d.f. of the uniform-(0,1) distribution (diagonal) and, as circles, the occurrence of null DMs where any of the ten columns has a PAS P value no larger than the Sidak cutoffs for 0.01, 0.05, 0.1, and 0.2 occurrence of "10-test family" type I error. **In (i-l)** the c.d.f.s of the ten pairwise products of five same-DM PAS P values (thin lines) tend to match those of the products of two independently uniform-(0,1)-distributed random numbers that from the bottom up exclude products of numbers below 0.01, 0.001, and 0.00 (thick grey) to mimick finite P value permutations. 3'200 P value permutations of each studied column.

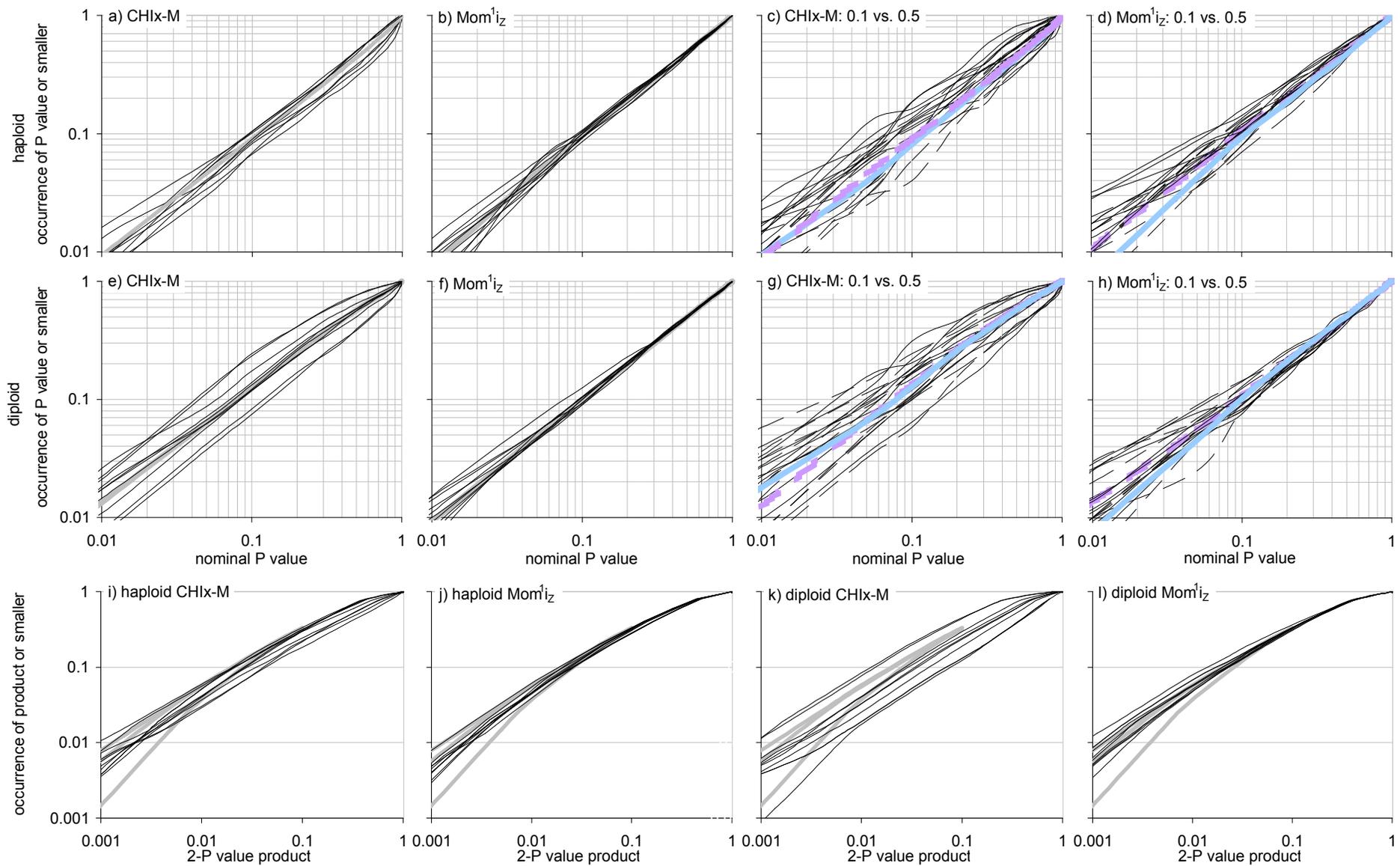

**Figure 18. Type I error of the P values of the CHIx-M and Mom$^1$i$_Z$ of all of the columns in a null DM.** By shuffling the markers at a column (in the DM) one can estimate the P value for the column's PAS. With such P values for all of the columns in a null DM generated under the H$_o$ of columns being independent from each other, one can compute same-DM null distributions. **Panels (a,b)** and **(e,f)** show the c.d.f.s of such P values for each of ten binary and trinary independently generated 2'000-row 1'000-column null DMs, resp.(pooled: thick grey) whose columns have o12345 binary-marker frequencies or the trinary H&W thereof. **Panels (c,d)** and **(g,h)** show the corresponding c.d.f.s for the 200 columns in each null DM with frequency 0.1 and 0.5, resp.(solid and segmented lines; pooled: thick blue and violet). C.d.f.s tend to match the diagonal, with the trinary CHIx-M c.d.f.s varying most across DMs. **Panels (i,j)** and **(k,l)** show the c.d.f.s of the 499'500-value cartesian (pairwise) product of the P values of each PAS for every DM as well as c.d.f.s of matched cartesian products of independent uniform-(0,1)-distributed random numbers with product exclusions as in Fig.17. 3'200 permutations of each studied column.

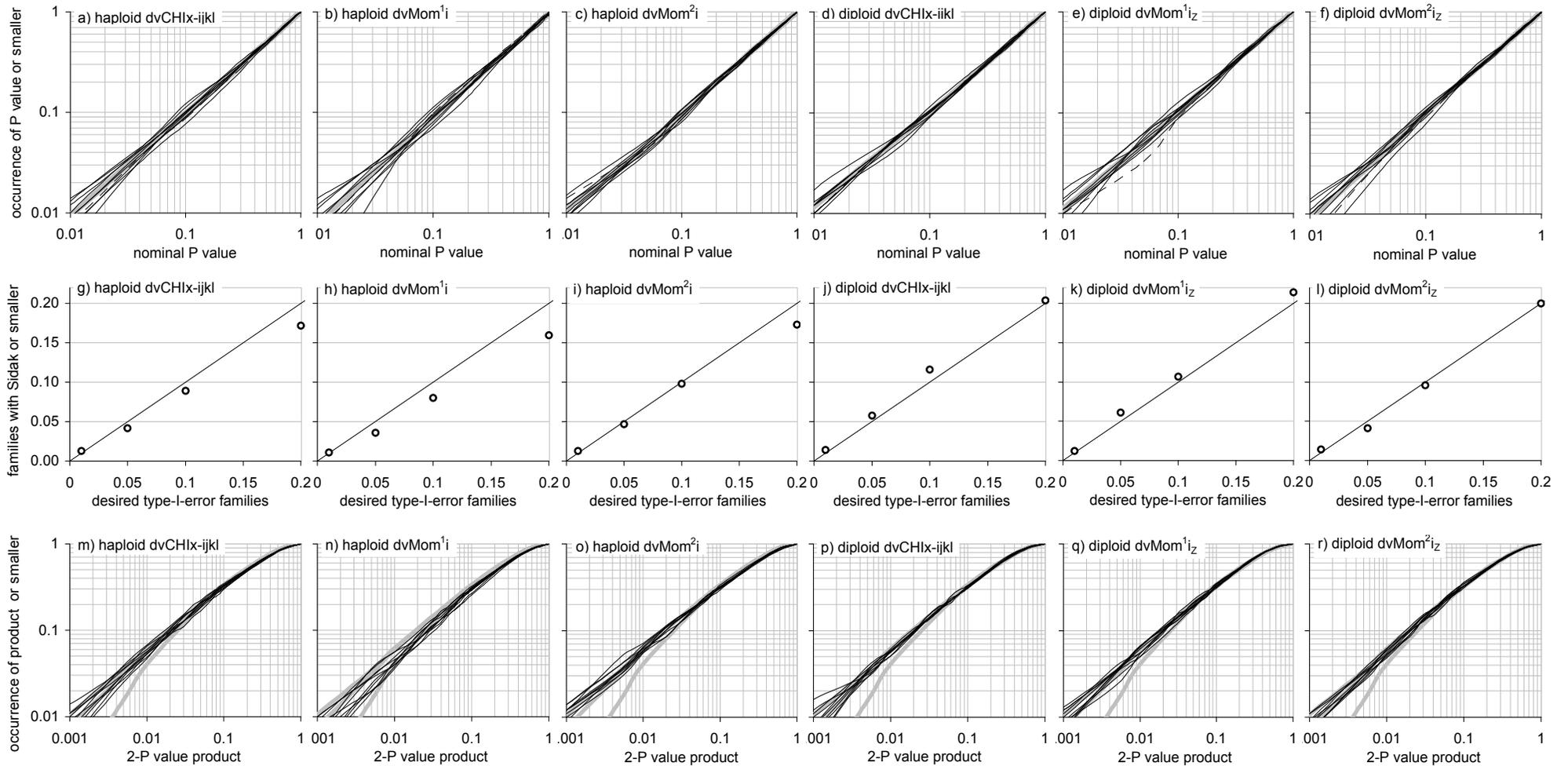

**Figure 19. Type I error of the P values of dvCHIx-ijkl, dvMom$^1$i, and dvMom$^2$i (or Z-valued) of chosen IVs across multiple null DMs.** By permuting the DV markers in a DM generated under the $H_o$ of IVs being independent from others and the DV, one can estimate dvPAS P values for any IVs in the DM and compute same-DM null distributions. The P values quantify the associations of the IVs with the DV. Highest power dvMom$^n$i's and dvMom$^n$i$_Z$'s are used in the binary and trinary cases, resp.(left,right halves). **Panels (a-f)** show that the c.d.f.s of the P values of individual IVs over 1'000 binary and 1'000 trinary 2'000-row 1'000-column null DMs tend towards the diagonal. C.d.f.s are shown for the DV (dashed) and four plus five IVs with o1524 and o12345 binary-marker frequencies, resp.(or the trinary H&W thereof) as thin black lines. DvMom$^2$i$_Z$ c.d.fs vary slightly more across IVs. **Panels (g-l)** show the c.d.f. of the uniform-(0,1) distribution on the diagonal and, as circles, the occurrence of null DMs in which at least one of the nine IVs at stake has a P value no larger than the Sidak cutoffs for 0.01, 0.05, 0.1, and 0.2 occurrence of "9-test family" type I error under said $H_o$. The observed Sidak-corrected family-of-tests type I error is on the diagonal. **Panels (m-r)** show c.d.f.s of the product of two same-DM P values for each of the 15 possible pairings of the five o12345 IVs as well as the c.d.f.s of 1'000 products of two uniform-(0,1)-distributed random numbers that from the bottom up exclude products like in Fig.17. The 2-P value products behave like those of random numbers. P values are estimated with 3'200 DV permutations.

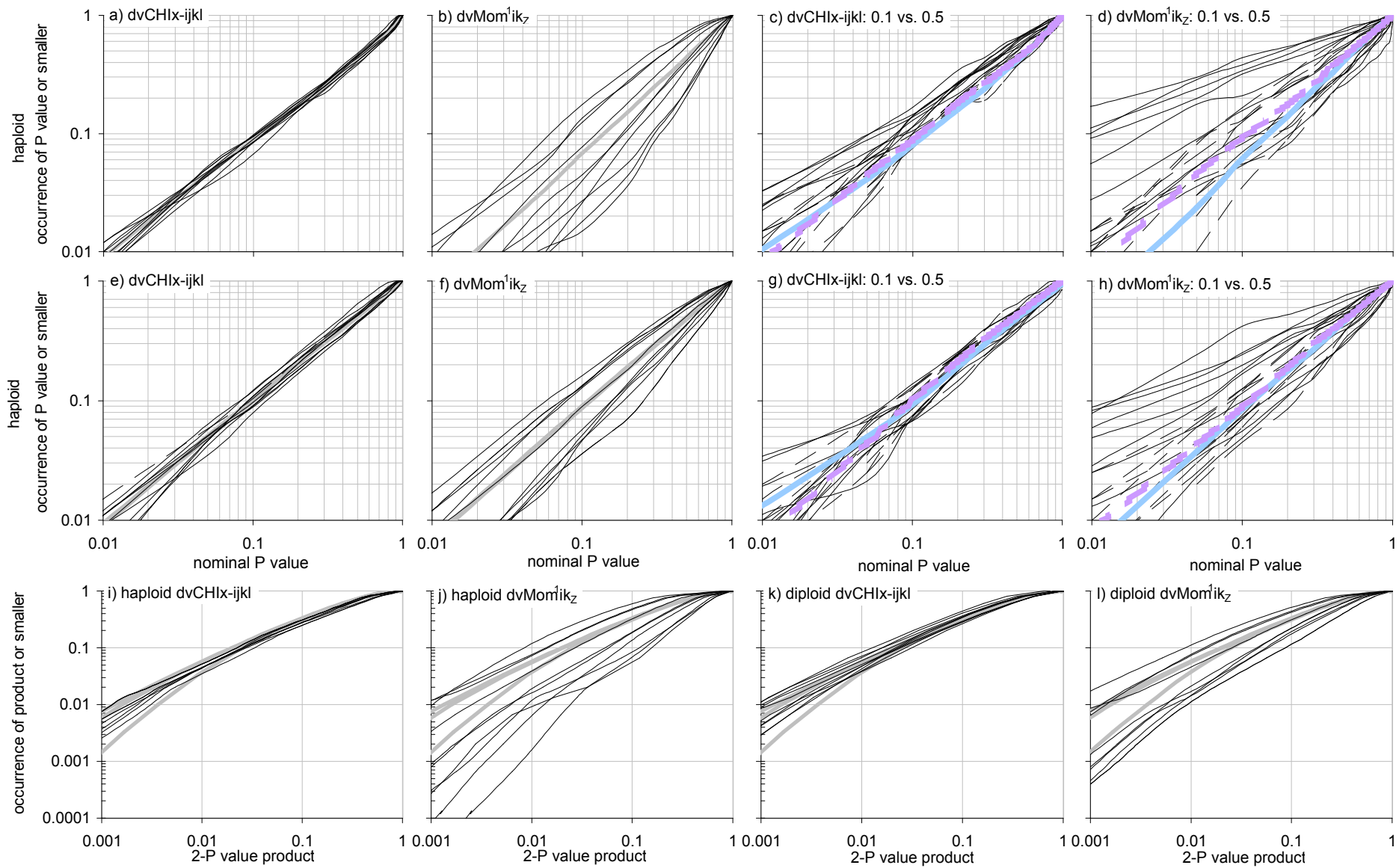

**Figure 20. Type I error of the P values of the dvCHIx-ijkl and dvMom$^1$ik$_Z$ of all of the IVs in a null DM.** By permuting the DV markers in a DM generated under the H$_o$ of IVs being independent from each other and the DV, one can estimate the dvPAS P values of all of the IVs in the DM and compute same-DM null distributions. Here the DV has binary markers of frequency 0.5 and the IVs have o12345 marker frequencies or the trinary H&W thereof. **Panels (a,b)** and **(e,f)** show the c.d.f.s of such P values for each of ten binary and trinary independently generated 2'000-row 1'000-IV DMs, resp. **Panels (c,d)** and **(g,h)** show the corresponding c.d.f.s for the 200 IVs with frequency 01 and 0.5, resp.(solid and segmented thin lines; pooled: thick blue and violet). C.d.f.s tend to match the diagonal, with dvMom$^1$ik$_Z$ c.d.f.s varying more across DMs. Panels **(i,j)** and **(k,l)** show the corresponding individual-DM c.d.f.s of the cartesian (pairwise) products plus the usual reference c.d.f.s (grey; see Fig.19 and text). IVs are allowed to have random marginal effects at the DV; 1'600 and 12'800 DV permutations for binary and trinary dvCHIx-ijkl results, resp., and 1'000 permutations for dvMom$^1$ik$_Z$ results; patterns do not change in (a,c) when using 12'800 permutations. Patterns of dvMom$^1$i and dvMom$^1$i$_Z$ are like dCHIx-ikjl's (not shown).

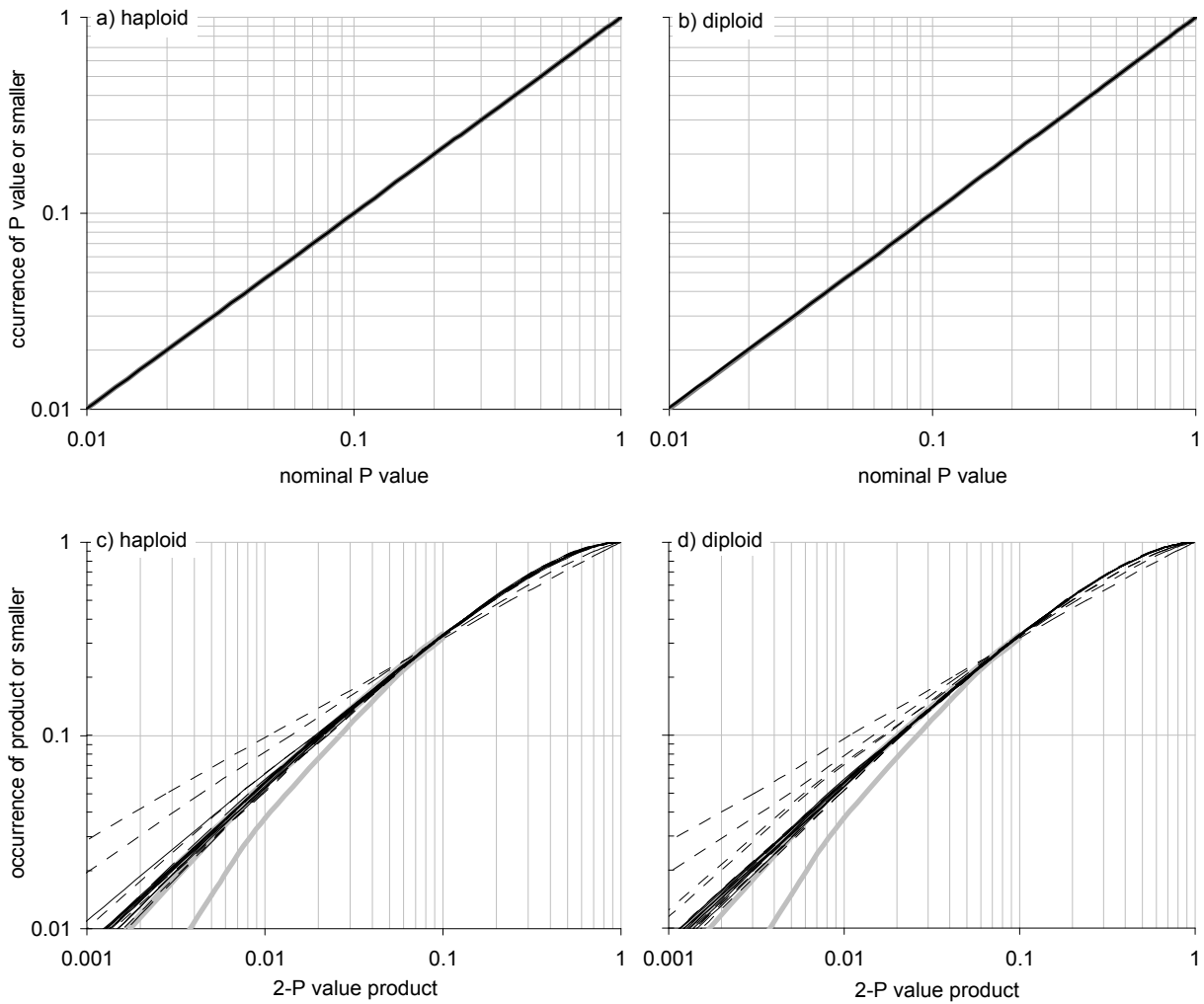

**Figure 21. Type I error of CHIx-M P values over null DMs whose columns are associated into blocks.** In a DM whose columns are associated into blocks that are independent from other blocks, one can shuffle the vertical order of the "sequences" (rows) of every block independently at once to estimate a P value for a column's CHIx-M. This P value reflects how the column's markers are associated to sequences in other blocks and is unaffected by marker associations within blocks. With P values for chosen columns across many null DMs one can compute c.d.f.s also for statistics involving several same-DM P values. **Panels (a,b)** show such c.d.f.s for chosen columns over 10'000 binary and 10'000 trinary null DMs, resp. Each DM has ten independently generated and then laterally joined blocks, and each block has 2'000 100-marker sequences sampled randomly (with replacement) from a set of 100-binary-marker LPL or 100-trinary-marker wLPL sequences (M&M). Forced sampling of the ten blocks lets their pos.49 "anchor" columns have o12345 binary frequencies or the trinary H&W thereof. P values are estimated by comparing the CHIx-M of each column in the DM at hand to those of the same column in the other 9'999 null DMs. C.d.f.s are shown for the five pos.49 columns in the first five blocks, five additional 1st-block columns ("model-linked"; M&Ms), and the columns of each category pooled in thin-black, thin-broken, and thick-grey lines, resp. The c.d.f.s match the diagonal. **Panels (c,d)** show the c.d.f.s of every possible product of two same-DM P values from the five model-linked 1st-block columns and the five pos.49 columns, resp.(thin broken and black lines), compared to c.d.f.s of 10'000 products of two independent uniform-(0,1) random numbers that from the bottom up exclude products involving random numbers below 0.01, 0.001, or 0.00 (grey lines), to mimic finite P value permutations. P values are estimated through 1'600 permutations of the DV.

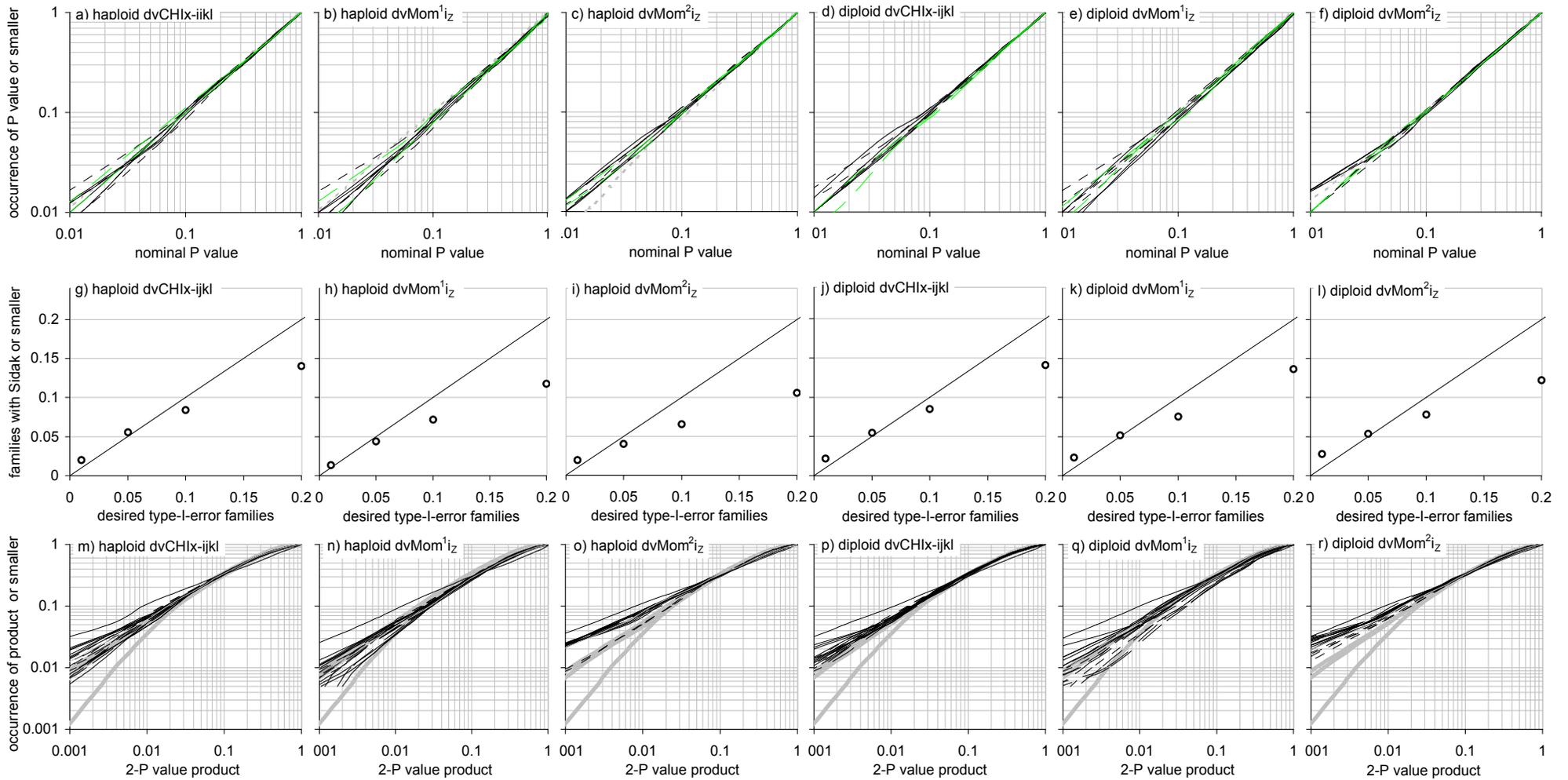

**Figure 22. Type I error of the dvCHIx-ijkl, dvMom $^1i_Z$, and dvMom $^2i_Z$ P values of chosen IVs over multiple null DMs with IV blocks.** By permuting the DV markers of a DM with IV blocks, one can estimate the dvPAS P values of chosen IVs and study them and functions thereof. Here distributions are shown of such P values across 2'000 binary and 2'000 trinary 1'000-row 100-block DMs (100-IV blocks; 10'000 total IVs) generated under the $H_o$ of blocks being independent from each other and the DV. The focus are nine IVs from the first block of which four are "anchor" IVs (M&M) force-sampled to have o1524 frequencies (or the H&W thereof) and the other five are "model-linked" IVs (M&M). **Panels (a-f)** show that individual-IV c.d.f.s tend to match the diagonal (DV dashed, nine IVs thin, two independent-block IVs segmented green). **Panels (g-l)** show the c.d.f. of the uniform-(0,1) distribution (diagonal) and, as circles, the occurrence of null DMs in which any of the nine IVs has a nominal P value no larger than the Sidak cutoffs for 0.01, 0.05, 0.1, and 0.2 occurrence of "family of nine tests" type I error, which here tends to be below expectation (conservative) in both the binary and trinary cases. **Panels (m-r)** show the c.d.f.s of the product of two same-DM P values for the pairings between anchor IVs and those between model-linked IVs, as well as c.d.f.s of 10'000 products of two uniform-(0,1)-distributed random numbers (grey) with the usual exclusions of products with below-threshold random numbers (Fig.17). The products of dvCHIx-ijkl and dvMom $^1i_Z$ P values behave like those of random numbers but dvMom $^2i_Z$ products tend to be above the exclusion-0.01 random products. IVs are allowed to have random marginal effects at the DV; 3'200 permutations of DV-column markers.

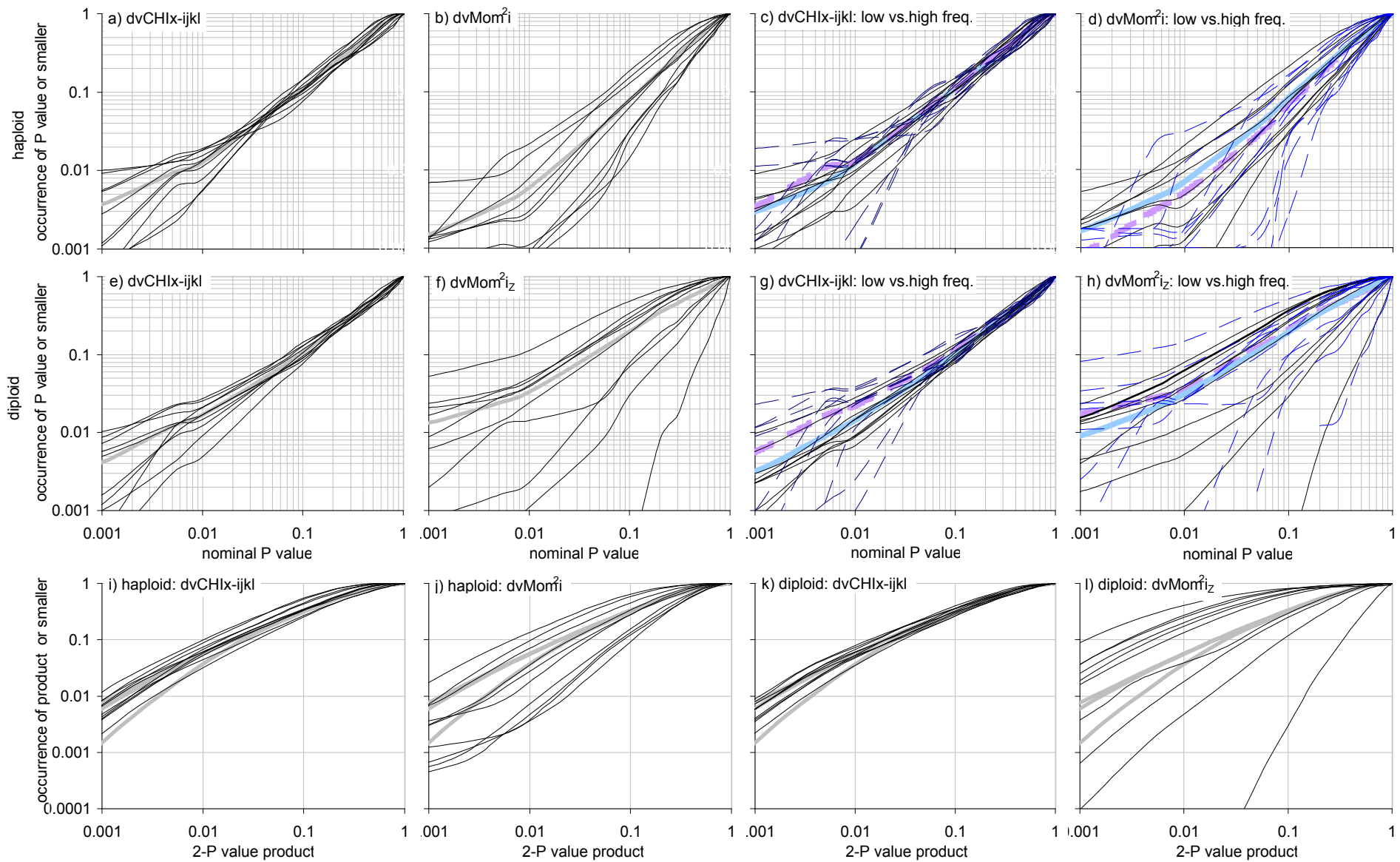

**Figure 23. Type I error of the P values of dvCHIx-ijkl and dvMom$^2$i of all of the IVs in a null DM with blocks.** By permuting the DV markers of a DM generated under the $H_o$ of blocks being independent from other blocks and the DV, one can estimate the dvPAS P values of all of the IVs in the DM and compute same-DM null distributions. Unlike in Fig.22, here five blocks are force-sampled so their pos.49 "anchor" IVs have o12345 binary-marker frequencies or the trinary H&W thereof. **Panels (a,b)** and **(e,f)** show P value c.d.f.s over all IVs in each of ten binary and trinary independently generated 1'000-row 100-block binary DMs, resp.(10'000 IVs per DM; dvMom$^2$i$_Z$ in the trinary case). **Panels (c,d)** and **(g,h)** show the corresponding c.d.f.s of the 4'000 IVs with lowest and highest frequency, resp.(0.27 to 0.35 and 0.4 to 0.5, resp. or H&W thereof; solid and segmented thin blue and black; pooled: thicker blue and violet). C.d.f.s tend to match the c.d.f. of the uniform-(0,1) distribution, with dvMom$^2$i c.d.f.s varying markedly across DMs. **Panels (i,j)** and **(k,l)** show ten c.d.f.s of cartesian (pairwise) products of binary and trinary single-DM P values, resp., and the usual reference c.d.f.s (see Fig.18 and text). Blocks have random marginal effects at the DV; 3'200 permutations of DV-column markers. The null c.d.f.s of dvMom$^1$i and dvMom$^1$i$_Z$ are like dvLKx-ikjl's (not shown).

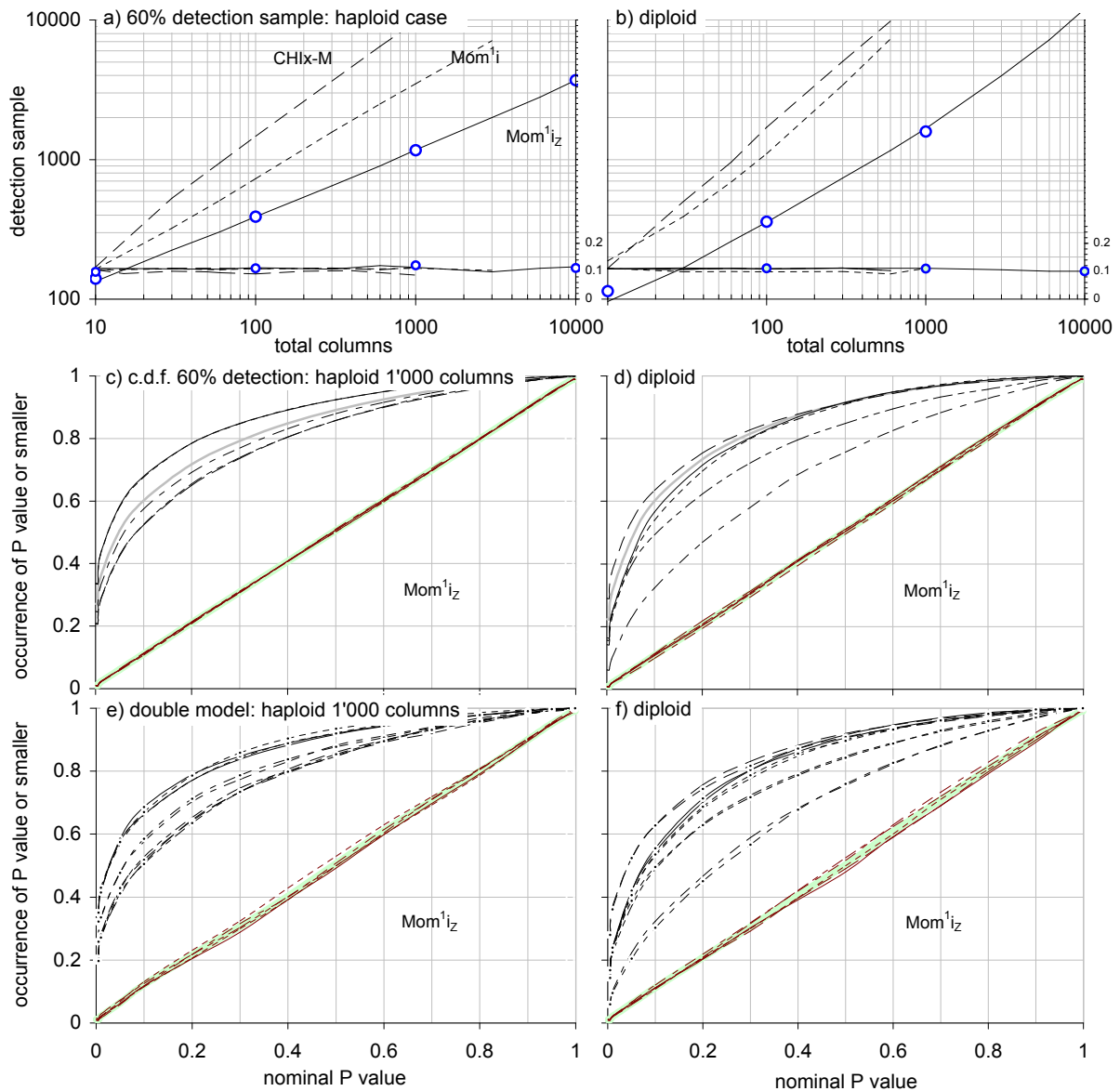

**Figure 24. PAS power and false positives vs. random columns, including with two models co-occurring.** Panels (a,b) show the power and false positives of selected PAS P values as a function of the columns of a DM, when detecting two sets of 1'000 randomly encountered 100-row 5-column binary and trinary model DMs (haploid,diploid) with o15241 binary-marker frequencies or the trinary H&W thereof (M&M), including when two models co-occur independently. Model DMs are expanded multinomially up to a given "sample size" (DM rows) and random columns with o12345 binary or trinary frequencies are added laterally. The plotted DM rows (left vert.axes) yield ~60% detection with PAS P value$\leq$0.1 of the first two "reference" columns of the 1'000 model DMs of interest. False positives with PAS P value$\leq$0.1 at five random columns pooled are on the right vertical axis. **Labels (a,b):** CHIx-M, Mom$^1$i, and Mom$^1$i$_Z$ in segmented, dashed, and solid black, resp.; false positives in thin black. Detection samples increase about double-log-linearly with random columns and false positives occur with 0.1 frequency, with no change when two models co-occur (circles). **Panels (c,d):** individual c.d.f.s in 1'000-column DMs for the five model columns and five random columns in black and red, resp., both pentads as thin solid, segmented, dot-segmented, bidash-segmented, and dashed lines (grey: each pentad pooled). **Panels (e,f):** like (c,d) but with two models co-occurring and black dots for the c.d.f.s of second-model columns (no pooled results). In (c,e) and (d,f) DM rows are 1'170 and (1'590,1'650), resp. In every panel each model or model pair is replicated ten times when the DM is smaller than 3'000 rows x 3'000 columns, and twice otherwise. P values are estimated with 100 permutations of each column at stake.

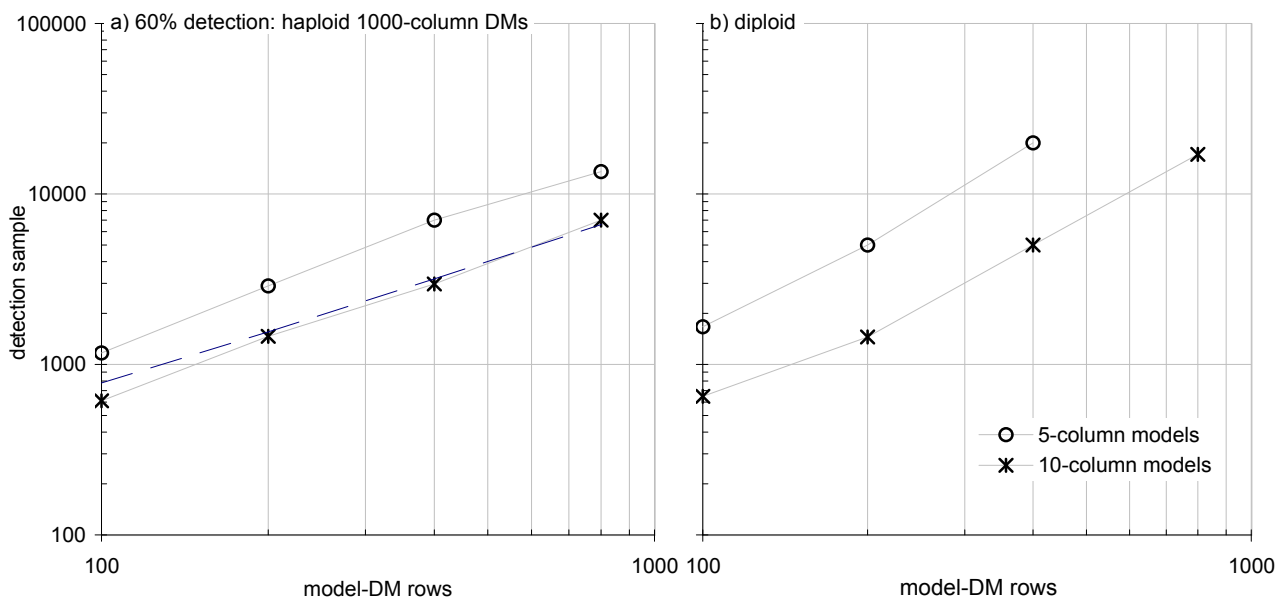

**Figure 25. Power of Mom$^1$i$_Z$ P values as a function of the number of rows and columns of the model DMs to be detected.** The power of Mom$^1$i$_Z$ P values when detecting sets of randomly encountered binary and trinary model DMs (left,right) of significant (*sensu* Fig.3) 5- and 10-column association, as a function of the number of model-DM rows and columns. Model DMs are expanded and tested after embedding in DMs with 1'000 total columns. Model sets have 200 model DMs each, with 5 and 10 model-DM columns (circles and stars) and 100, 200, 400, and 800 model-DM rows (horiz.axis). The 100-row sets comprise 1'000 model DMs each. Plotted is the number of rows needed for 60% detection with Mom$^1$i$_Z$ P value$\leq$0.1 of the first two "reference" columns of the model DMs in the set at stake (vert.axis). Detection samples are lower for 10- than 5-column model DMs given a number of model-DM rows and increase roughly linearly with the number of model-DM rows. For the binary model DMs and the 5-column trinary ones detection samples range from one to one and a half orders of magnitude larger than the number of rows of the model DMs being detected, whereas those of the 5-column trinary models become faster larger as model-DM rows increase, reaching about 80'000 rows (extrapolated) for the trinary 800-row 5-column model DMs. Every detection sample is estimated with at least five replications of the model DMs in the set at stake. P values are estimated with at least 100 permutations of each studied column. False positives are uniform-(0,1)-distributed throughout (not shown). The thin segmented line in (a) is for four sets of 500 10-column DMs each, whose every column has standard P value$\leq$0.05, and the line is included as additional control.

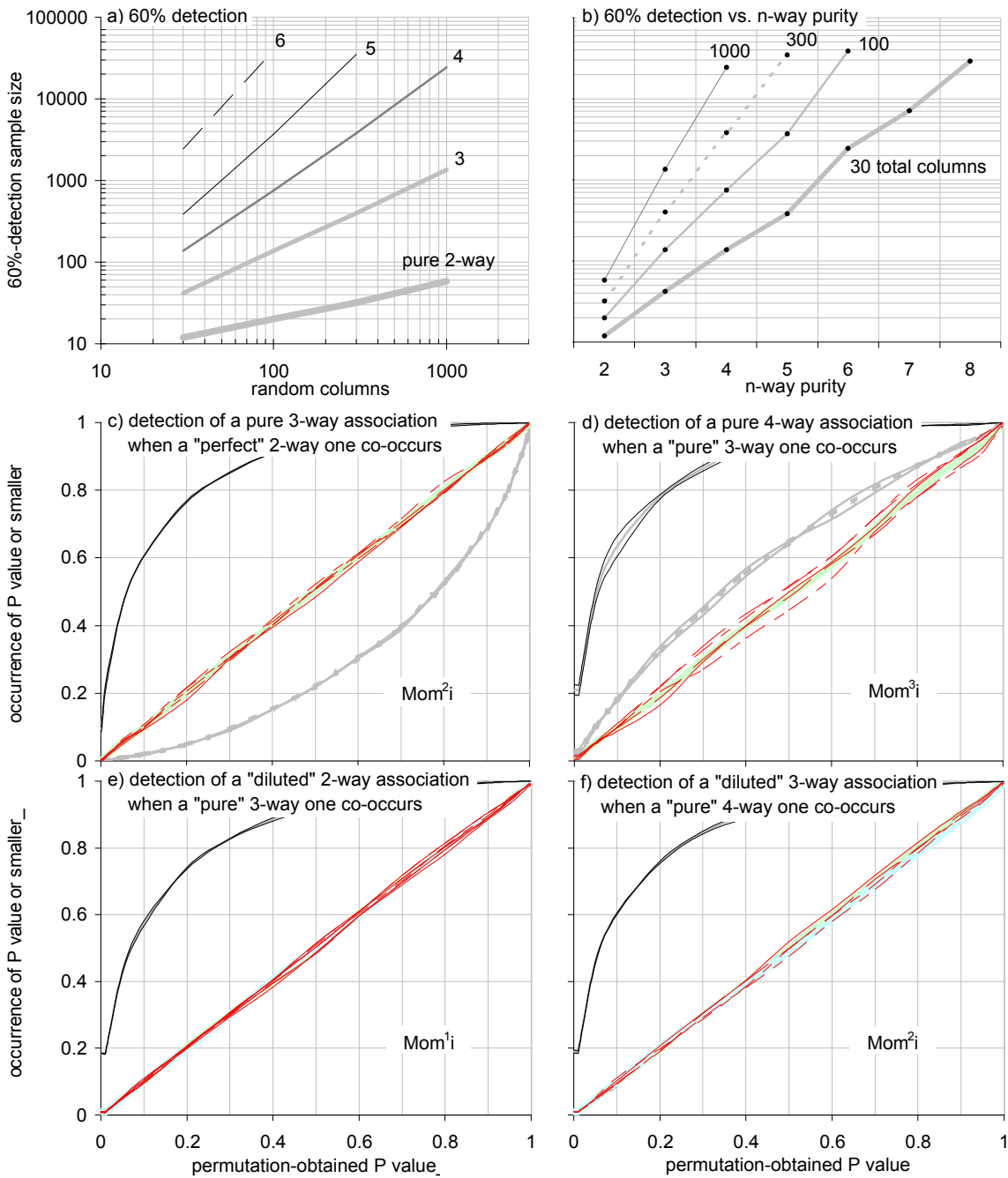

**Figure 26. Power and false positives of $Mom^{n-1}i$ P values when detecting pure $n$-column associations, including two associations co-occurring.** The power of $Mom^{n-1}i$ P values to detect $n$ columns in "pure" binary $n$-way association as random columns increase. The $n$-marker sequences of the pure $n$-column association at stake (Fig.1) are expanded multinomially to the desired rows; independent random columns with o12345 binary markers are added laterally. **Panel (a)** shows that the 60%-0.1 P value detection samples increase linearly with random columns, for any $n$. False-positives are uniform-(0,1)-distributed (not shown). **In (b)** the detection samples increase log linearly as $n$-column $n$-purity increases linearly from 2- to 8-way given DMs with 30, 100, 300, and 1'000 total columns. The results in (a,b) are identical when two pure $n$-column models co-occur. The c.d.f.s **in (c,d,e,f)** are for two models co-occurring in DMs with 1'390, 24'500, 2'500, and 6'000 rows, resp., rows that suffice for 60%-0.1 P value detection of the first models and 100, 100, 90, and 85% detection with P value $\leq 0.1$, resp., of the second models, using applicable $dvMom^n i$'s. DMs have 1'000 and 300 total columns (c,d,e) and (f), resp. The first and second models (black and grey; pooled: thicker grey and thick acqua) are **in (c)** a pure 3-column association and a "perfect" 2-column one and **in (d)** a pure 4-column association and a pure 3-column one, whereas **in (e,f)** the first models are an 84%- and a 74%-diluted pure 2-column and pure 3-column association, resp., and the second models are a pure 3-column and a pure 4-column association, resp. False positives at five o12345 columns are on the diagonal (red; pooled: thick light green). Detection samples until and above 1'000 rows are estimated with at least 1'000 and 200 replications, resp. P values are estimated with at least 100 permutations of each evaluated column.

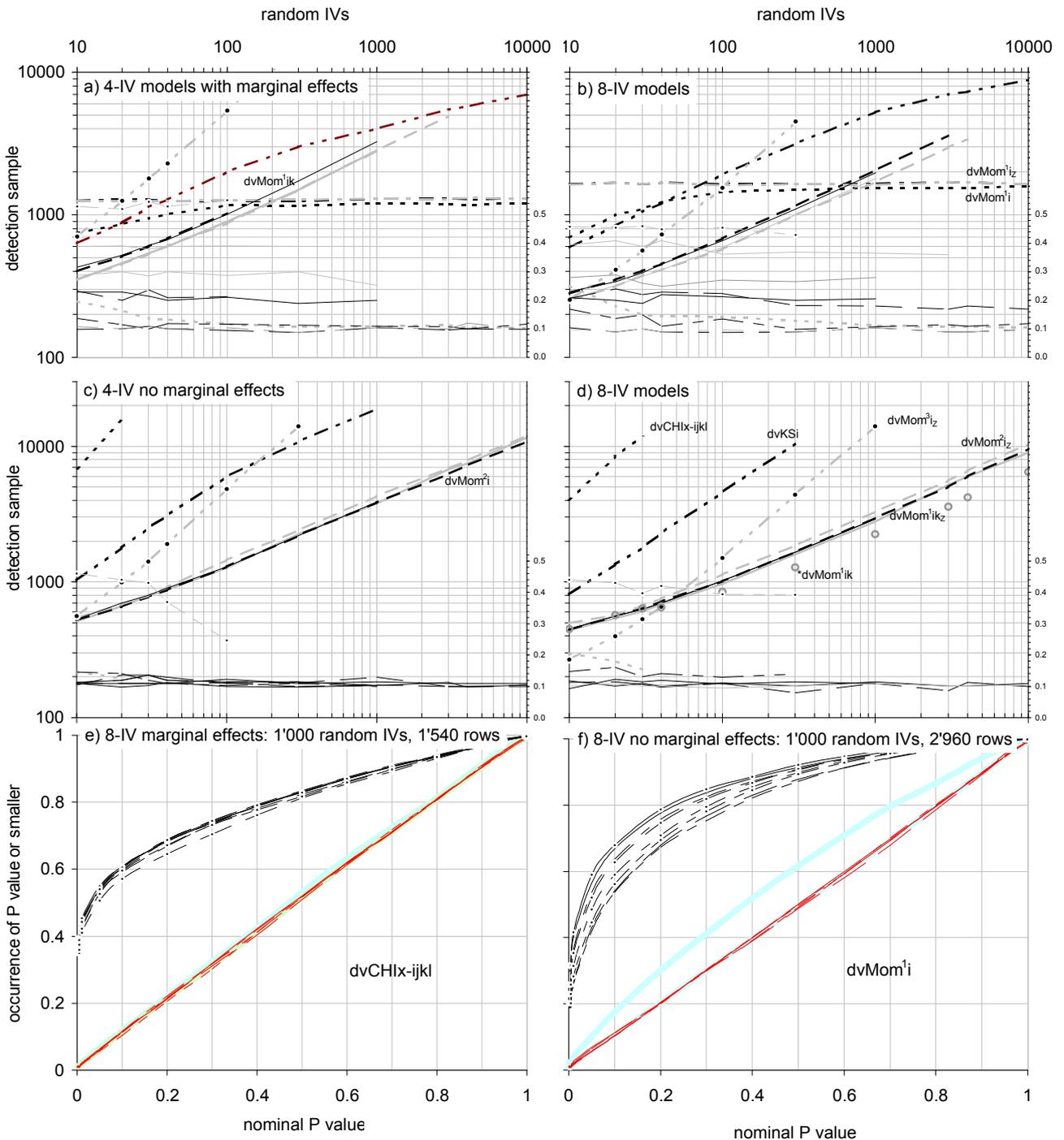

**Figure 27i. Power and false positives of the P values of selected dvPAS when detecting binary 4- and 8-IV models of DV association.** Panels (a,b) and (c,d) show the DM rows (right axis) needed for 60% detection with dvPAS P value$\leq$0.1 of the first two "reference" IVs of 4- and 8-IV binary models of significant DV association (*sensu* Fig.3) with and without marginal effects, resp. and accompanying false positives with P value$\leq$0.1 (right axes). Four sets of 1'000 200-row 4- and 8-IV binary model DMs are studied (left,right). Each model DM is expanded multinomially to a desired number of rows; random IVs with o12345 binary-marker frequencies are added laterally. In (e,f) are the individual-IV c.d.f.s of the highest-power dvPAS P values for the 8-IV cases with 1'000 random IVs. **Labels (a,b,c,d):** Detection samples and false positives are shown as thick and thin lines. Dashed, bidash-segmented, dash-segmented, segmented, and solid lines are for dvCHIx-ijkl, dvKSi, dvMom$^1$i, dvMom$^2$i, and dvMom$^1$ik, resp.(grey: Z-valued); dvMom$^3$i$_Z$ is dot-bidash-segmented grey. Above 100 random IVs in (a-d), detection samples increase near-linearly and false positives are uniform-(0,1)-distributed, except for CHIx-IJ and dvMom$^1$ that in (a,b) plateau quickly as random IVs increase. Note i) the absence in (c,d) of dvCHIx-ijkl and both dvMom$^1$i's that do not detect IVs lacking marginal effects; ii) dvKSi's intermediate power throughout; iii) the excess in false positives of dvMom$^2$i's only in (a,b) vs. dvMom$^3$i's everywhere; and iv) that due to (iii), the high power of dvMom$^2$i and dvMom$^1$ik in (c,d) cannot be exploited naively when IVs have strong marginal effects. In (c,d) dvMom$^n$i$_Z$ and dvMom$^n$ik$_Z$ (grey) reduce and add little power, resp.(see text, Fig.27ii). Circles in (d) are a lost *dvMom$^1$ik$_Z$ with highest power. **Labels (e,f):** The eight model IVs are shown as two IV tetrads (o1524,1234) each with plain, segmented, dash-segmented, and bidash-segmented lines, the second tetrad with dots. In (f) model IVs with marker frequency 0.1 and 0.5 are best and worst detected, resp., whereas in (e) differences are small; DV and pooled random IVs: thickest grey lines. P values are estimated with 100 DV permutations.

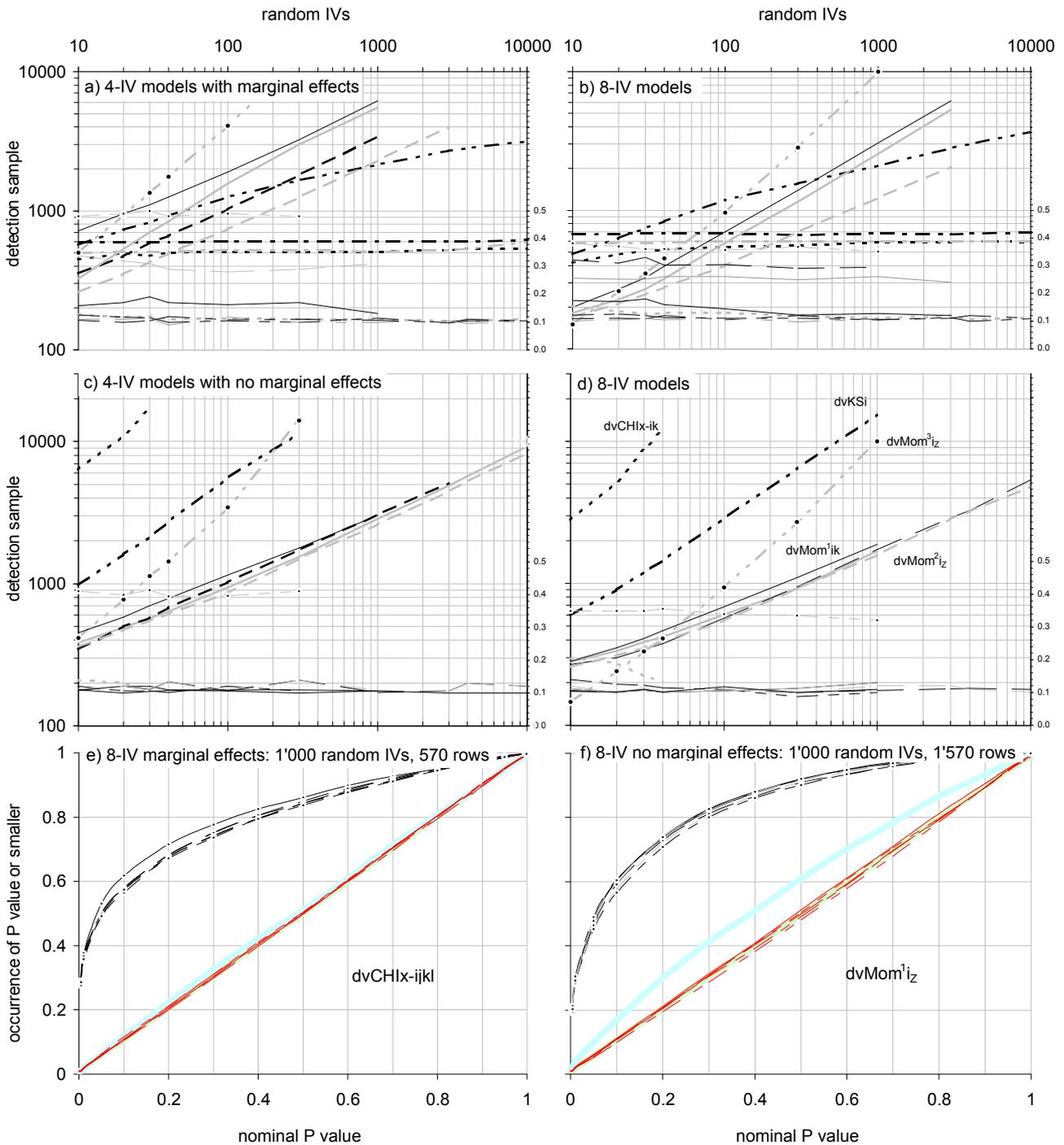

**Figure 27ii. Power and false positives of selected dvPASs when detecting trinary 4- and 8-IV models of DV association.** Repetition of Fig.27i but with model and random IVs having trinary markers. DV associations come from four sets of 1'000 200-row 4- and 8-IV trinary model DMs (left,right). Random IVs with o12345 trinary H&W marker frequencies are added laterally. **Labels** like Fig.27i's but the highest-power dvMom$^2$i$_Z$ is used in (f). No results are shown with the lost *dvMom$^1$ij$_Z$ version. Same trends as in Fig.27i but in (a,b) CHIx-IJ plateaus almost immediately as random IVs increase and in (f) the behavior of the c.d.f.s of model IVs with different trinary-marker frequencies is similar, unlike in panel (f) of Fig.27i.

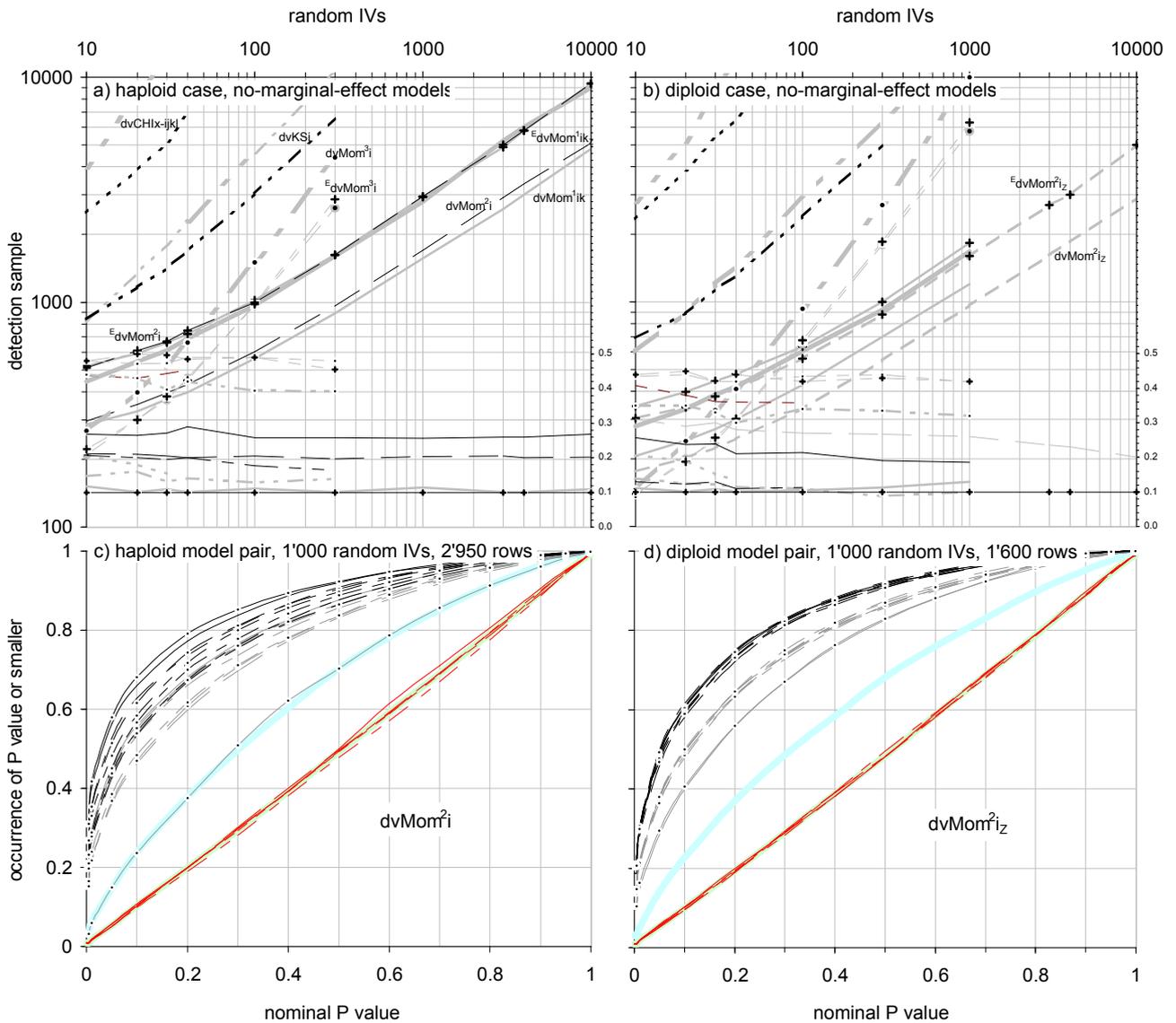

**Figure 28. Power and false positives of selected dvPASs when two 8-IV models co-occur, the first with and the second without marginal effects.** The paired models are studied in Fig.27 in isolation. Panels (a,b) show detection samples and false positives for dvCHIx-ijkl, dvKSi, and the highest-power dvMom$^2$i's and dvMom$^1$ik's, when detecting the IVs of the no-marginal-effect model DMs of each pairing at stake, as random IVs increase. In (c,d) are the c.d.f.s of the dvMom$^2$i and dvMom$^2$i$_Z$ P values of the eight individual model IVs of each co-occurring model type as well as of five random IVs with o12345 binary and trinary H&W markers in DMs with 1'000 random IVs and 2'940 and 1'810 rows, resp, that allow ~60% of the two "reference" IVs of the no-marginal-effect second models to have dvMom$^2$i and dvMom$^2$i$_Z$ P value$\leq$0.1, resp., after strong marginal effects are erased (below). **Labels (a,b)** are like in Fig.27, with +'s when marginal effects are erased ($^E$superscripts) and thick grey lines for Fig.27 results. **Labels (c,d):** The c.d.f.s of the eight model IVs of each model type are shown as two tetrads (o1524,1234) whose IVs are solid, segmented, dash-segmented, and bidash-segmented, the first four IVs in grey, the other four in black, with dots labelling second-model tetrads; DV and pooled random IVs: thickest grey. Detection samples in (a,b) are smaller than in isolation but there is an excess in false positives. For both dvMom$^1$ik and dvMom$^2$i in (a) the excess disappears and detection samples increase to match those in isolation if one toggles all excess markers at every IV with standard marginal-effects P value$\leq$0.019, 0.005, 0.001, and 0.00015 (1 d.f.) when random IVs are 10, 100, 1'000, and 10'000, resp.; in (b) for both dvMom$^1$ik$_Z$ and dvMom$^2$i$_Z$, those with P values$\leq$0.019, 0.0065, 0.001, and 0.0002 (2 d.f.); and in (c,d) those with P values 0.001 and 0.0065, resp.(1 and 2 d.f.). Note in (c,d) that the IVs from model DMs with erased marginal effects are ~50%- and ~45%-detected too, i.e., that their higher-order DV associations despite said erasure are nearly as strong as those of the IVs of the no-marginal-effect model DMs. The detection samples and false positives of the non-shown dvMom$^2$i's and dvMom$^1$ik's behave qualitatively identically. 1'000 pairings and 100 DV permutations.

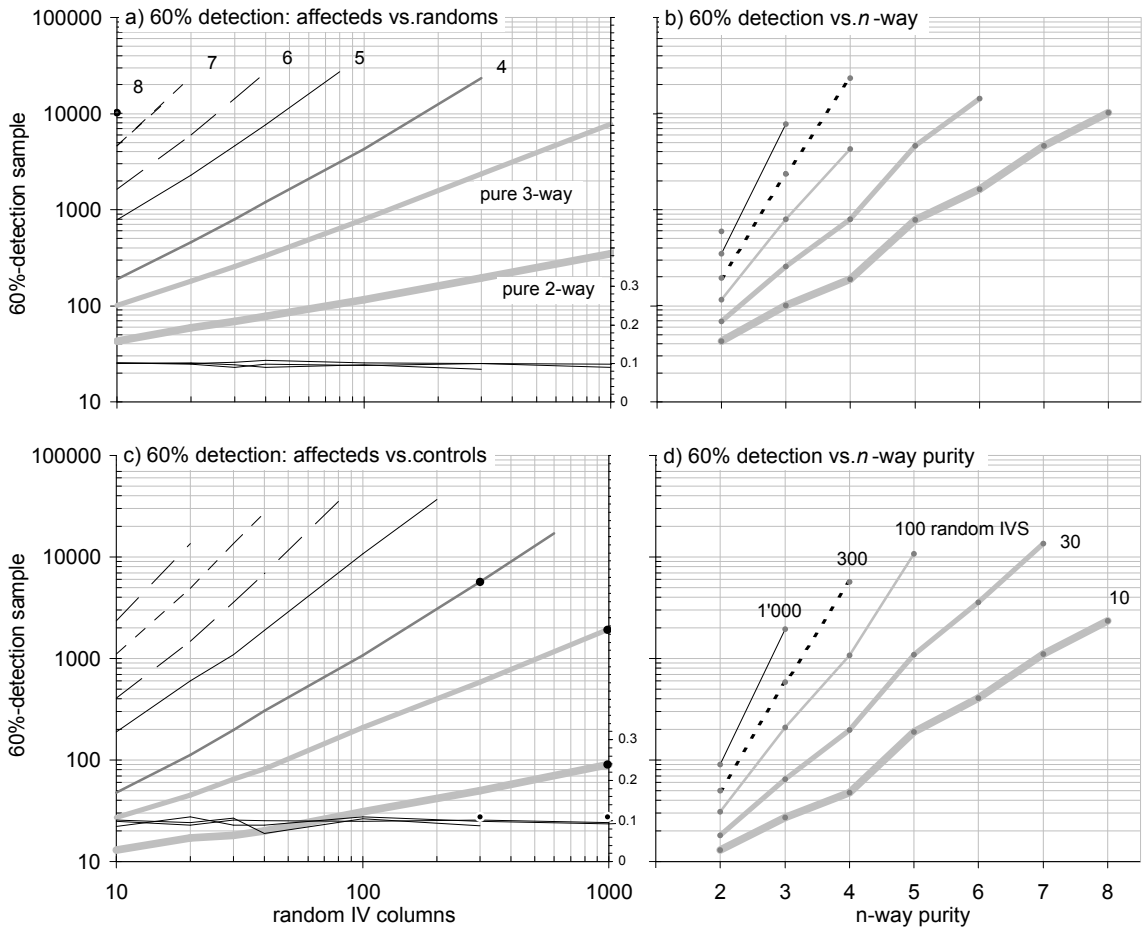

**Figure 29. Power of dvMom$^{n-1}$i when detecting IVs in pure $n$-IV association to a DV**. Power and false positives of dvMom$^{n-1}$i when detecting IVs in "pure" $n$-IV association with a DV, as random IVs increase. **In (a,c)** are the 60%-0.1 P value detection samples when "affecteds" with $n$ IVs in pure $n$-way DV association are contrasted to "randoms" and real controls, resp.(see pure 2- and 3-IV DV associations vs.controls in Fig.1). **In (b,d)** the detection samples react to $n$-way purity increasing from 2- to 8-way in DMs with up to 1'000 random IVs. Pure $n$-IV DV association is simulated by taking one of the two sets of pure $n$-way $n$-marker combinations and expanding its rows multinomially to R rows. Underneath one adds either an R-row $n$-column block of random binary markers of frequency 0.5 or the complementary set of pure $n$-marker combinations expanded to R rows, which means full pure $n$-way "affection" determination with infinitesimal incidence (i.e., ~0% "penetrance") and affection and protection caused fully (100% penetrance) by the two sets of pure $n$-marker combinations, resp. Random IVs have o12345 binary markers. The R rows are increased until 60% of the first two IVs in pure DV-association pooled have dvMom$^n$ P value$\leq$0.1. In (a,c) the detection samples increase about double-log-linearly with the number of random IVs and the false positives with P value$\leq$0.1 at two random IVs pooled hover around 10% (right axes). Detection samples are estimated using at least 10'000 replications when 2R$\leq$ 1'000 rows, and at least 5'000, 1'000, and 500 replications when 2R$\leq$ 3'000, 10'000, and 10'000+ rows, resp. Black dots in (c) are for dvMom$^n$ik and show no advantage over dvMom$^{n-1}$i. DvMom$^n$i$_Z$ and dvMom$^n$ik$_Z$ as well as dvMom$^n$M have the same power and false positives as dvMom$^n$i (not shown). The three values in (b,d) near the inflection point at 4-way purity in DMs with 10, 30, and 100 random IVs are estimated using at least 10'000 and 800 replications and DV permutations, resp.

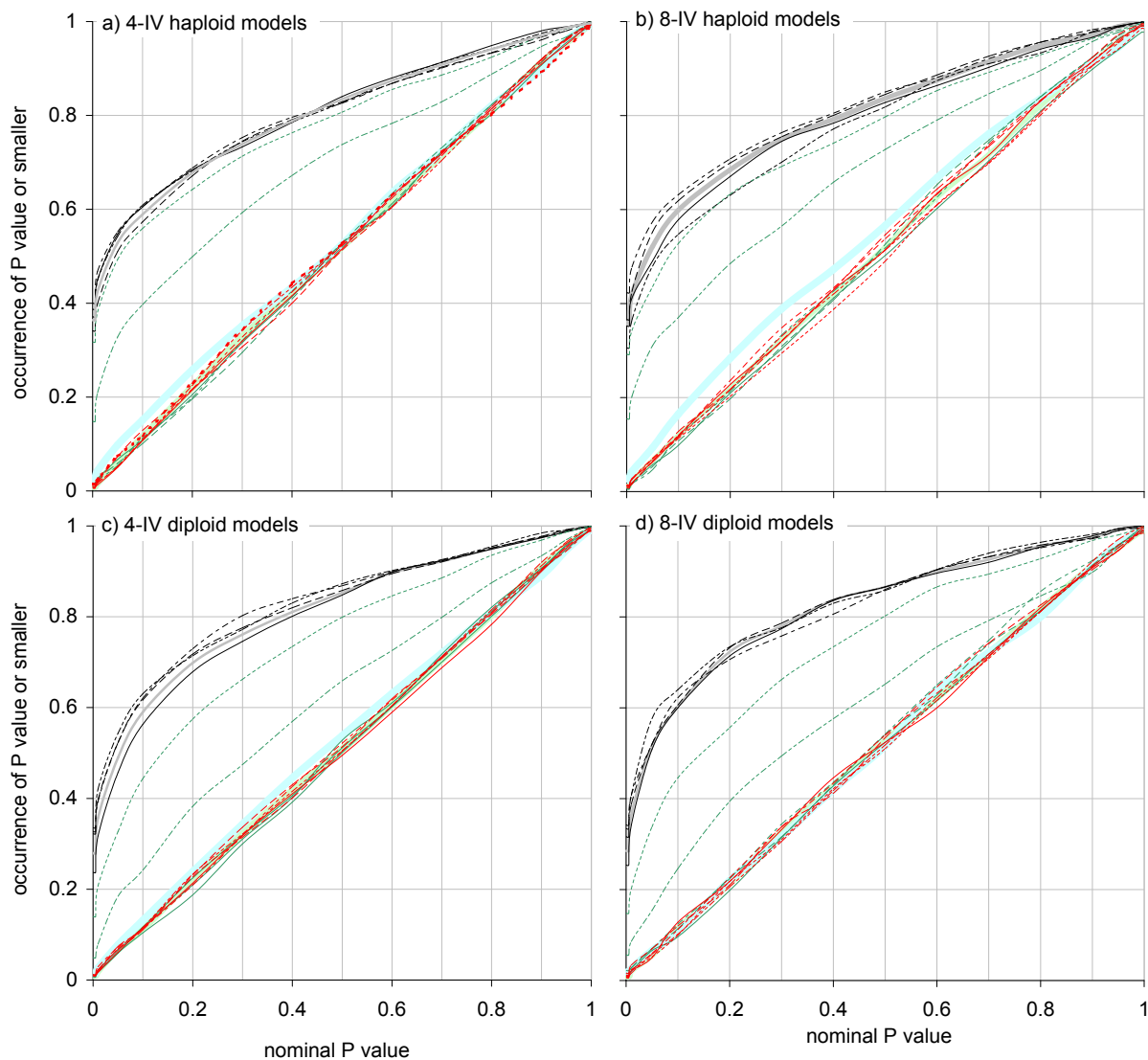

**Figure 30. Power and false positives of dvCHIx-ijkl when detecting IVs of randomly encountered models of significant DV association with marginal effects, in DMs where all IVs form blocks.** The 60%-0.1 P value detection samples and false positives of dvCHIx-ijkl P values are studied in DMs whose IVs form 100 100-IV blocks and where groups of four or eight IVs are forced to be DV-associated, as a function of IV-marker frequency. The four sets of 1'000 randomly encountered 200-row 4- and 8-IV binary and trinary model DMs with marginal effects being detected are those of Fig.27. The rows of the 100 100-IV blocks are sampled with replacement from a set of 116 real 100-mutation LPL chromosomal sequences, or from the 116(115)/2 wLPL pairwise comparisons thereof (M&M), as follows. A model DM's 4- or 8-marker sequences are expanded multinomially up to the number of rows of the DM to be tested. The expanded sequences guide the sampling of four or eight blocks of LPL or wLPL sequences whose pos.49 markers are identical to those in the expanded sequences (M&M); this places every model IV in a different block. Five additional blocks are independently force-sampled to have o12345 marker frequencies (or H&W thereof) at their pos.49s, shuffled vertically, and joined laterally to enough additional freely sampled random blocks to reach 100 total blocks (10'000 total IVs). Every one of the 1'000 model DMs in each set is simulated and tested once. The rows of the DMs used for the shown results suffice for 60% detection with dvCHIx-ijkl P value$\leq$0.1 of the first two "reference" model IVs pooled (see text, M&M). **Labels:** i) two reference model IVs pooled in thick grey; ii) DV in thicker grey; iii) first four pos.49 model IVs in solid, segmented, dash-segmented, and bidash-segmented thin black; iv) five "model-linked" IVs from the block with the first model IV in solid, segmented, dash-segmented, bidash-segmented, and dashed thin grey (pooled: thick grey); and v) five pos.49 IVs from the five force-sampled random blocks, like the model-linked IVs but in red (pooled: thicker grey). The DMs in (a,b,c,d) are 1'160, 1'450, 574, and 670 rows, resp., i.e., detection samples are smaller for trinary than binary models as well as for 4- than 8-IV models. Model-IV c.d.f.s are similar across marker frequencies. Pos.89 model-linked IVs are detected strongly and pos.64s intermediately specially in (a,b), with the others behaving like the false positives (whose c.d.f.s match the diagonal throughout). 100 DV permutations.

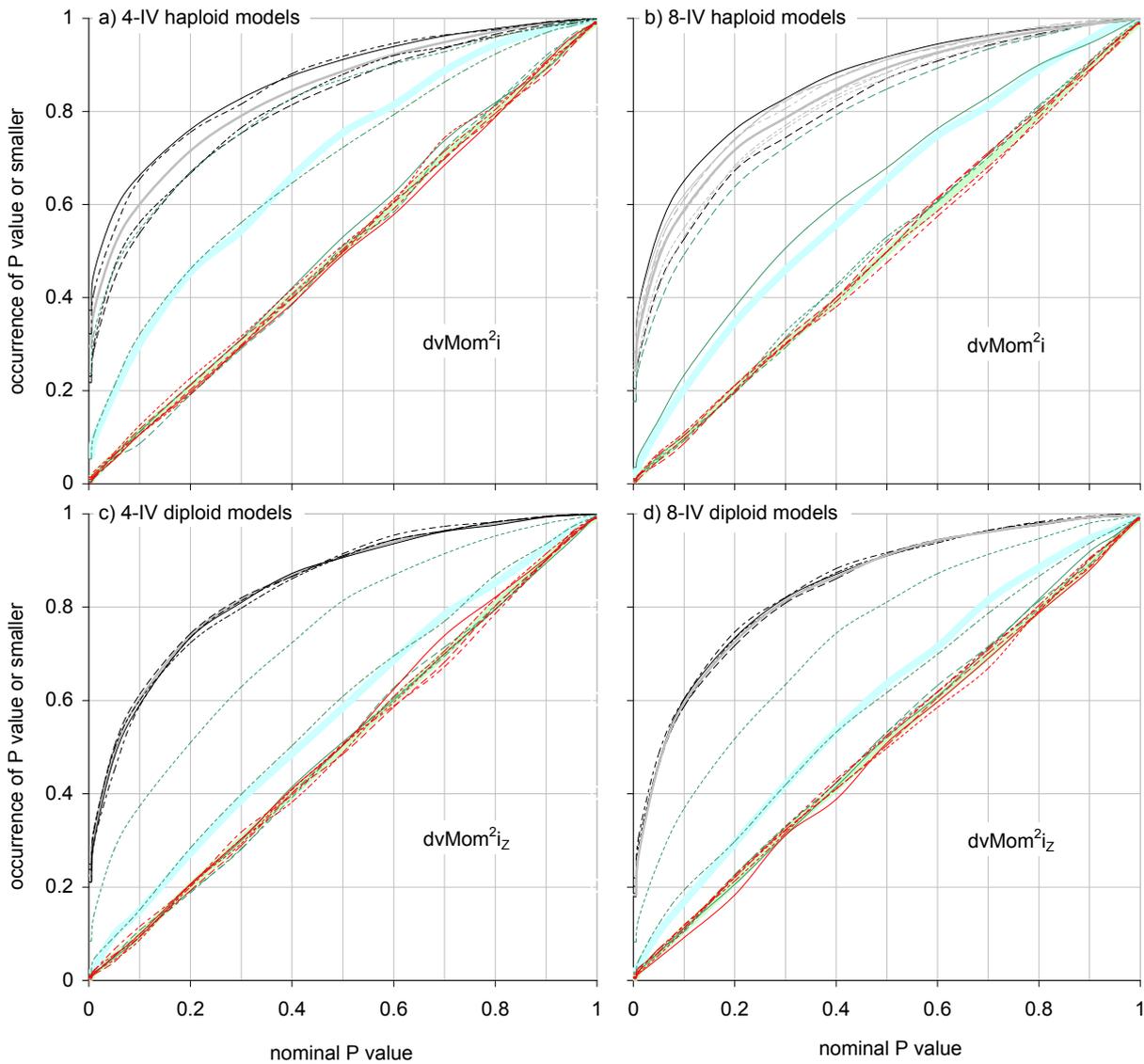

**Figure 31. Power and false positives of selected dvMom's when detecting IVs of randomly encountered models of significant DV association that lack marginal effects, in DMs with blocks.** Repetition of Fig.30 but using model DMs whose IVs have no marginal effects at the DV and the plot-specified dvPAS P values. To generate a tested DM, the marker sequences of affecteds and controls in a model DM are expanded multinomially as separate groups, each up to half the rows of the desired DM. The two resulting groups of 4- or 8-marker sequences guide the sampling of four or eight blocks of LPL or wLPL sequences with the same four and eight same-row markers at the blocks' pos.49s (M&M). Like in Fig.30, additional force- and non-force-sampled random blocks are added to reach 100 total blocks. The four sets of 1'000 model DMs each are those of Fig.27. **Labels** and other details are like in Fig.30 but in (b) the c.d.f.s of the second tetrad of model IVs (freqs. o1234) are added in grey with iterated labelling. The detection samples in (a,b,c,d) are 3'000, 2'150, 2'100, and 1'190 rows, resp., i.e., they are smaller for trinary than binary models and much smaller for 8- than 4-IV models, notably in the trinary case. In the binary case the highest-power $dvMom^2i$ (not Z-valued) detects better the low-marker-frequency model IVs than the intermediate-frequency ones. Pos.89 model-linked IVs are detected best and pos.64 ones intermediately and this more strongly in the binary than the trinary cases, favoring coarse- and fine-mapping, respectively; the others behave like the false positives (that here too are on the diagonal throughout).

Fig.31 hd NOmarg4-8

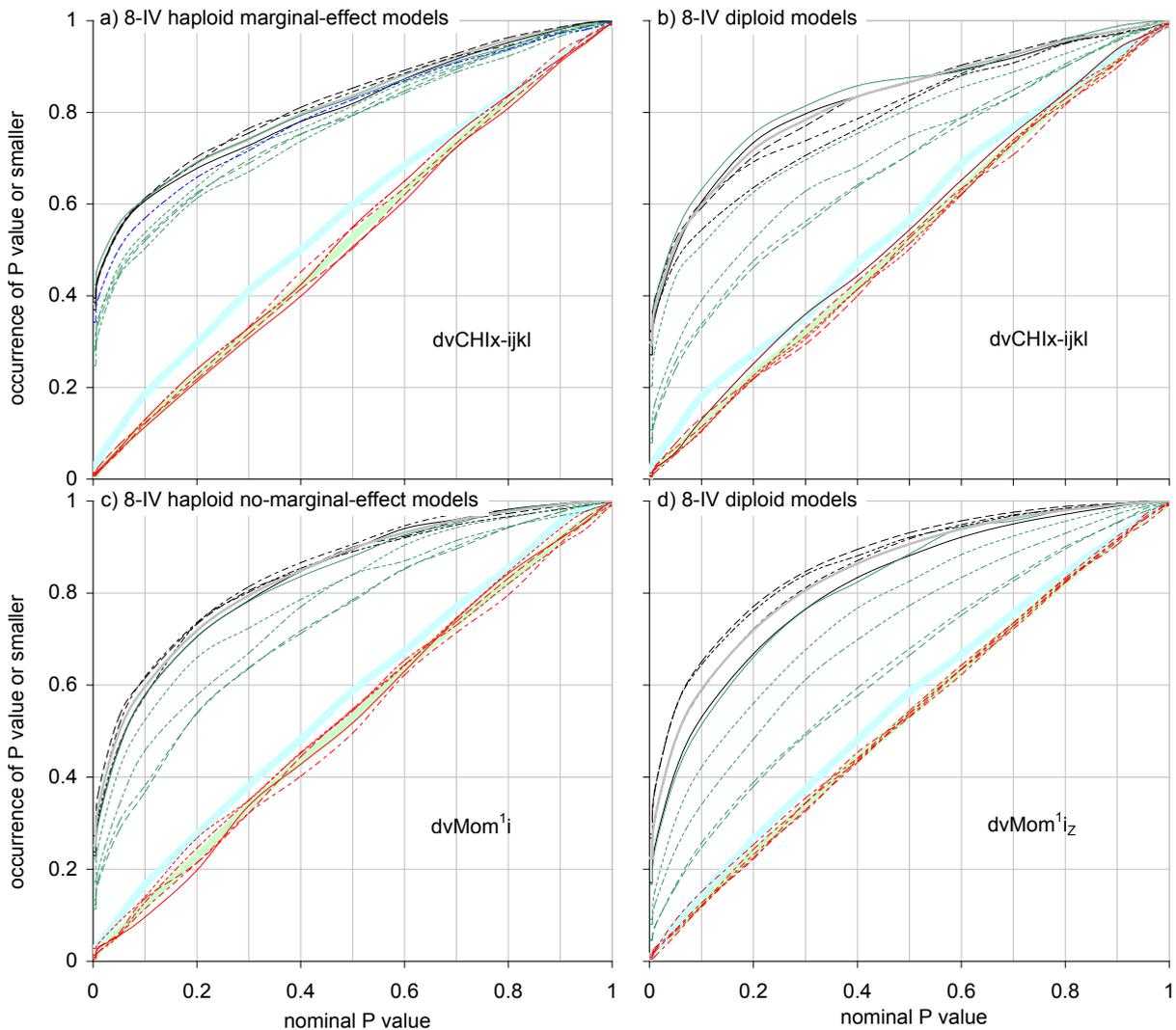

**Figure 32. Power and false positives of selected dvPASs when detecting the IVs of randomly encountered models of 8-IV DV association in DMs with blocks: Four model IVs per block.** Repetition of the results with 8-IV models in the (b,d) panels of Figs.30,31 but with model IVs placed four per block in two blocks rather than one per block. **Labels** and other details are like in Figs.30,31 unless noted. The 60% detection samples in (a,b,c,d) are 1'500, 670, 3'500, and 1'800 rows, resp. In panels (a,c) and (b) the IVs of both kinds of binary models and those of marginal-effects trinary models, resp., are detected similarly regardless of marker frequencies, unlike in (d) where the IVs of the no-marginal-effect trinary models show stronger reactions to marker frequencies (also vs.Fig.31), with those with marker frequency 0.1, 0.2, 0.4, and 0.5 being detected 53, 59, 64, and 65%, resp. Model-linked IVs are strongly detected say with P value$\leq$0.1 specially in the binary cases, namely pos.2s like model IVs, pos.89s and pos.64s like with one model IV per block, and pos.51s and pos.23s both about 53 and 23% in (a) and (d), resp.(vs.~10% in Fig.31). Therefore coarse- and fine-mapping are favored in (a) and (d), resp. False positives are on the diagonal throughout.

Fig.32 8-IV 4ccXblck

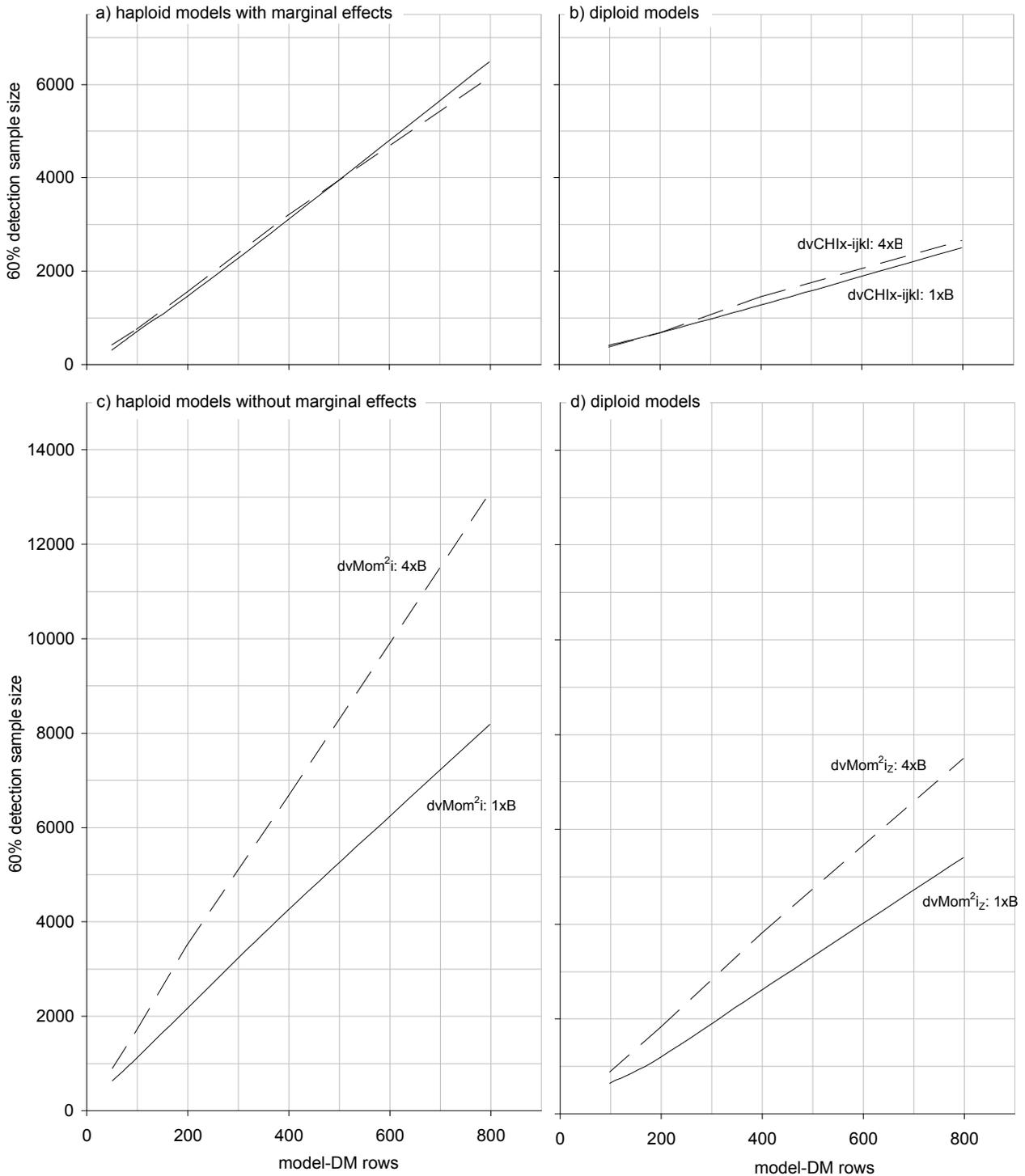

**Figure 33. Power and false positives of highest-power dvPASs when detecting the IVs of $r$-row 8-IV models of DV association in DMs with 100 100-IV blocks, for increasing $r$.** The number of rows (vert.axis) needed for 60% detection with dvPAS P value$\leq$0.1 when detecting the first two IVs of an $r$-row 8-IV model DM of DV association in a DM whose IVs form 100 100-IV blocks (Figs.30,31), as a function of increasing $r$ (horiz.axis). Results for binary and trinary models are in panels (a,c) and (b,d), resp., while for models with and without marginal effects they are in (a,b) and (c,d), resp. Continuous and segmented lines are for results with one and four model IVs per block, resp. In (a,b) only dvCHIx-ijkl is shown but dvMom$^1 i_Z$ power is comparable. In (c) and (d) only dvMom$^2$i and dvMom$^2 i_Z$ are shown, resp., because most powerful but dvMom$^1$ik and dvMom$^1 ik_Z$ have similar power. Detected are the two reference model IVs in 1'000 different multinomially expanded model DMs (see Fig.30, M&M), each model DM studied once. Detection samples increase linearly with model-DM rows, with the detection of the marginal-effects models requiring fewer rows than that of the no-marginal-effect models, except in the trinary case with one model IV per block in (d) that is detected better than both binary marginal-effects cases in (a). No results for 50-row trinary models are shown in (b,d) because no such models were found after months of testing random DMs (see M&M). False positives are uniform-(0,1)-distributed throughout and not shown. P values are estimated with 100 permutations of the DV.

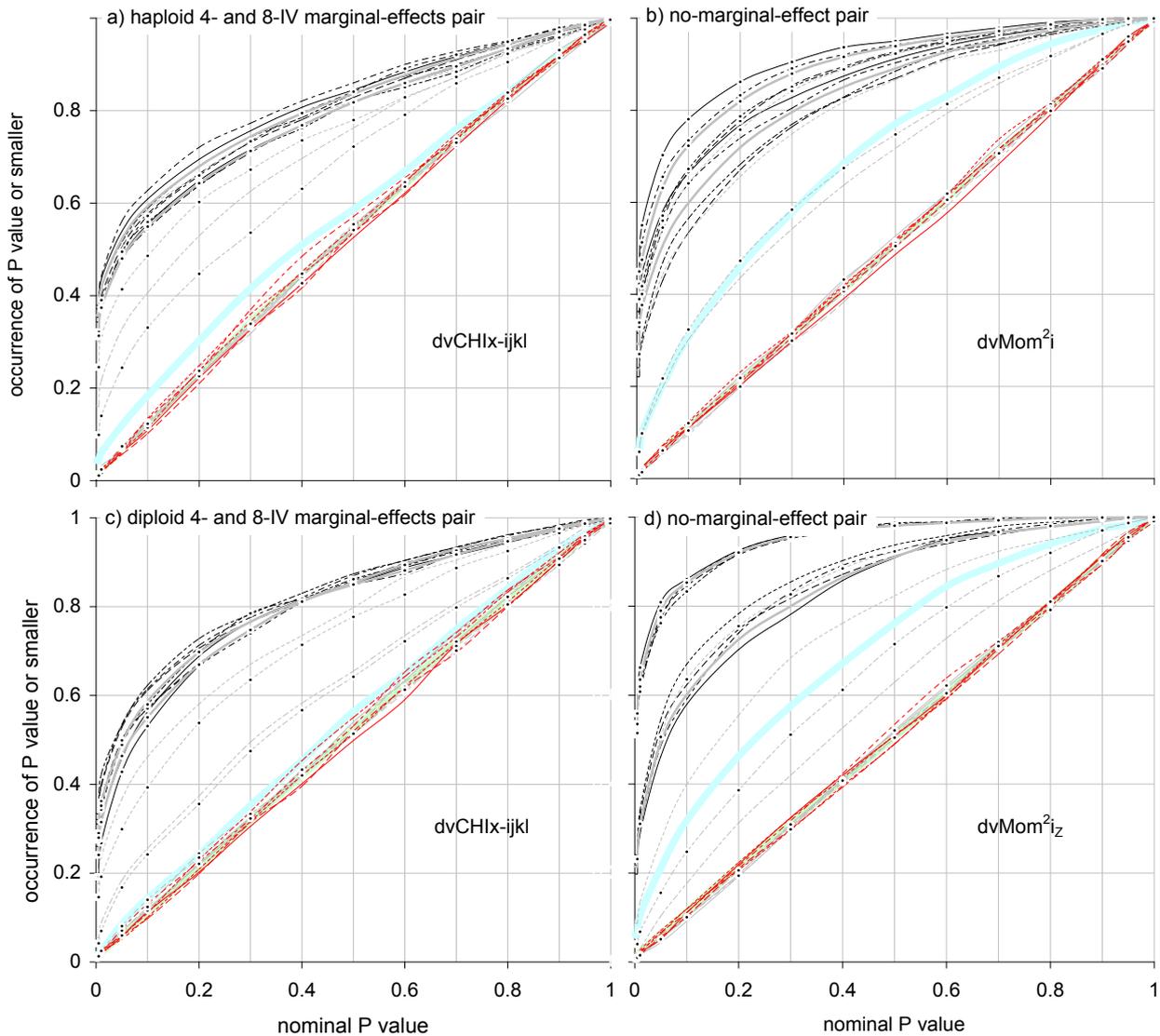

**Figure 34. Power and false positives of selected dvPASs when a 4- and an 8-IV model, both with or without marginal effects, co-occur independently in a DM with blocks**. Each panel presents results with 1'000 pairings of the model DMs used for Figs.30,31, in (a,b) for binary pairs with and without marginal effects and in (c,d) for the corresponding trinary pairings. Model IVs are placed one per block in 12 blocks (100 100-IV blocks total). C.d.f.s are conditional on 60% detection with plot-specified dvPAS P value$\leq$0.1 of the two reference IVs of the 4-IV models in the pairings. **Labels** are like in Fig.30, with dots marking the c.d.f.s related to the 8-IV models. In (a,c) both c.d.f.s are for dvLKx-ijkl P values whereas in (b,d) they are for $dvMom^2i$ and $dvMom^2i_Z$ P values, resp. Model IVs and model-linked IVs are detected like in isolation in Figs.30,31, e.g.: i) in (a,b,c,d) DM rows are 1'160, 2'900, 574, and 2'100 rows, resp., i.e., only the 2'900 differs slightly from the rows in Figs.30,31 (3'000); ii) in (b) the c.d.f.s of binary no-marginal-effect model IVs react most to the IVs' marker frequencies; iii) of the 4-IV-model-linked IVs in (a,b) pos.89s are best detected with P value<0.1 (58%,54%) and pos.64s intermediately (40%,31%), much like the (56%,55%) and (41%,30%) in isolation; and iv) model-linked IVs are less detected in the trinary case, specially when model IVs lack marginal effects. In (b,d) as expected, the easier-to-detect 8-IV models are detected well above 60% with P value 0.1. False positives are on the diagonal, like in isolation. P values are estimated with 100 permutations of DV markers.

Fig.34 double

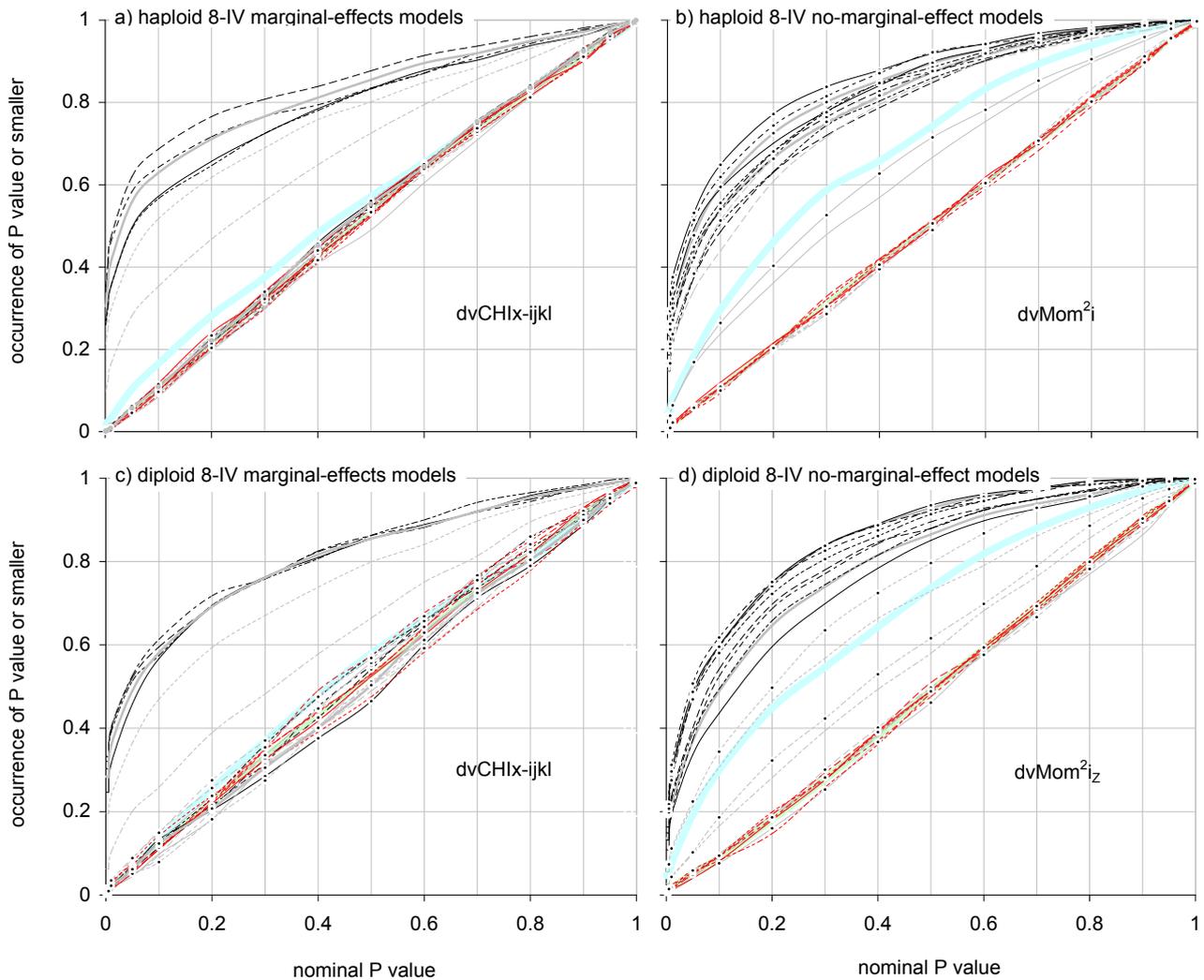

**Figure 35i. Power and false positives of selected dvPASs when a marginal- and a no-marginal-effect model of 8-IV DV association co-occur independently in a DM with blocks: One model IV per block**. Detection of pairs of binary or trinary 8-IV models of DV-association, one with and one without marginal effects, when both models co-occur independently in a DM whose IVs form blocks. Each panel shows results with 1'000 pairings of the model DMs studied in isolation in Figs.30,31, each pair simulated once. Every plot is conditional on DM rows yielding 60% detection with panel-specified P value$\leq$ 0.1 of the first two reference IVs (pooled) of the panel's focal model; 100 100-IV blocks and **labels as in** Figs.30,31, with dots marking the c.d.f.s related to the paired no-marginal-effect models. In (a,c) and (b,d) the c.d.f.s are for dvCHIx-ijkl and dvMom$^2$i, dvMom$^2$i$_Z$, resp., that best detect focal-model IVs. The detection samples in (a,b,c,d) for the two reference model IVs pooled are 1'500, 2'150, 650 and 1'300 rows, i.e., like the 1'450, 2'150, 670, 1'190 rows in the (b,d) panels of Figs.30,31 for the same models in isolation. Marginal effects with standard P value$\leq$ 0.002 and 0.003, resp.(1 and 2 d.f., see Fig.28) are erased in (b,d) to bring false positives down to the diagonal. In (a,b,c,d) the focal-model IVs with (best,worst) detection have binary-marker frequencies (0.5 0.1), (0.1 0.4), (0.2 0.1), and (0.2 0.4), or the trinary H&W thereof, without discernible trend. Note in (b,d,) that model IVs with erased marginal effects are detected 54 and 49% with dvMom$^2$i and dvMom$^2$i$_Z$ P value$\leq$0.1, resp., i.e., almost like the IVs of the no-marginal-effect models. Model-linked IVs at pos.64 and 89 are more detected in (a,b) than (c,d), suggesting again that fine- and coarse-mapping are favored in the binary and trinary case, resp. P values are estimated using 100 permutations of DV markers.

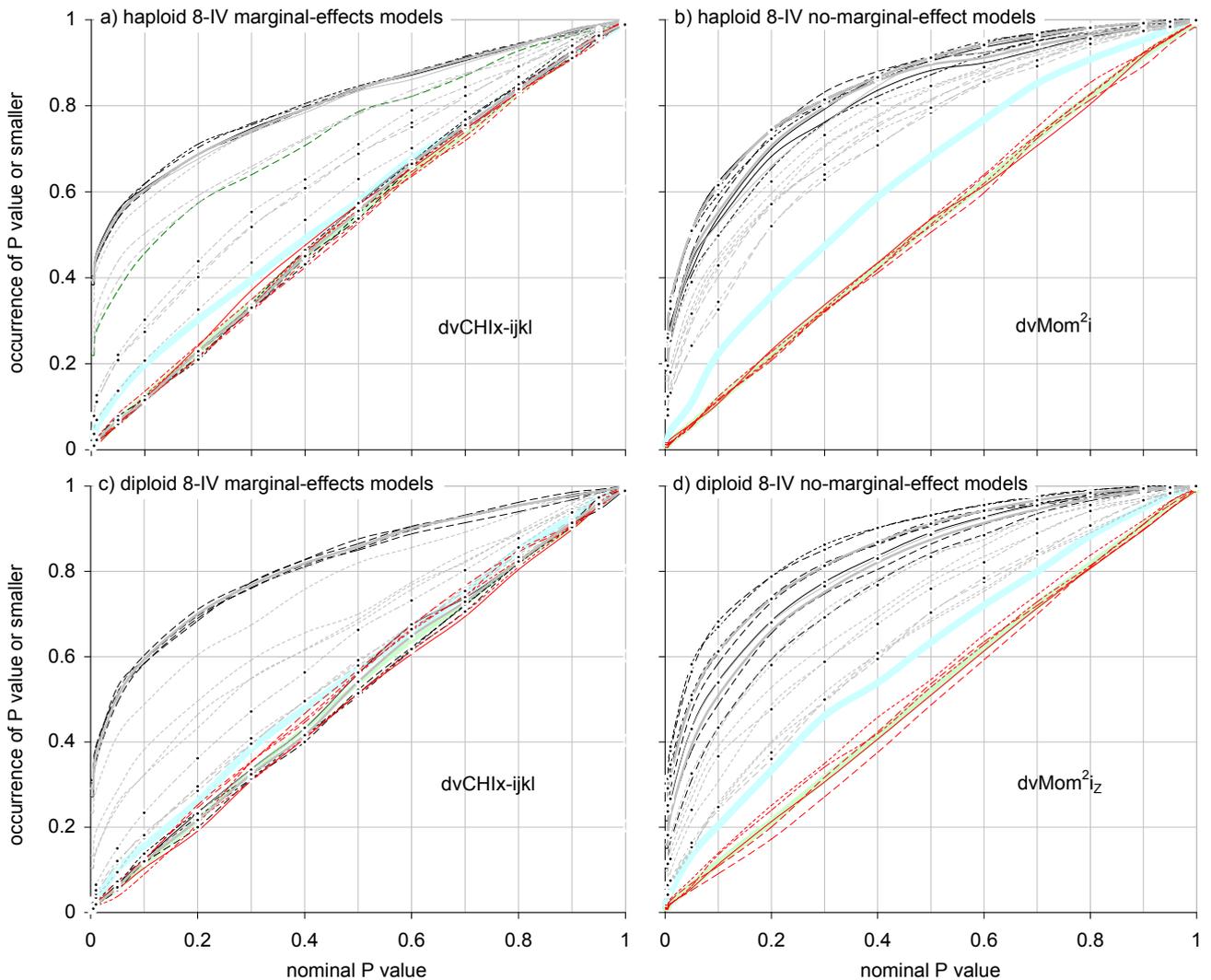

**Figure 35ii. Power and false positives of selected dvPASs when a marginal- and a no-marginal-effect model of 8-IV DV association co-occur independently in a DM with blocks: Four model IVs per block.** Repetition of Fig.35i but the 16 model IVs are placed four per block in four blocks. **Labels** like in Fig.35i. In (b,d) marginal effects with standard P value 0.0023 and 0.0015, resp.(1 and 2 d.f.; see Fig.28) are erased to bring false positives down to the diagonal. The DMs in (a,b,c,d) have 1'550, 3'600, 670, and 1'950 rows, resp., delivering 60% detection with plot-specified P value$\leq$0.1 of the two reference model IVs of the panel's focal model; these samples are like those in Fig.32 for the models in isolation. Like in Fig.35i, model IVs without marginal effects behave like random IVs in (a,c). In (b) and (d) the IVs of the models with erased marginal effects are detected 56 and 51%, resp., with P value$\leq$0.1, i.e., much like the IVs of the no-marginal-effect models. The marginal-effects model IVs with different marker frequency are detected about equally in (a,b,c) whereas in (d) the no-marginal-effect model IVs with H&W frequency 0.5 and 0.4 are better detected (like in Fig.32). Model-linked IVs at pos.2s are detected like model IVs in (a,b,c,d) and specially in (a,c), i.e., more than pos.89 and pos.64 ones, all like in Fig.32, suggesting again that fine- and coarse-mapping are more effective in (a,c) and (b,d), resp.

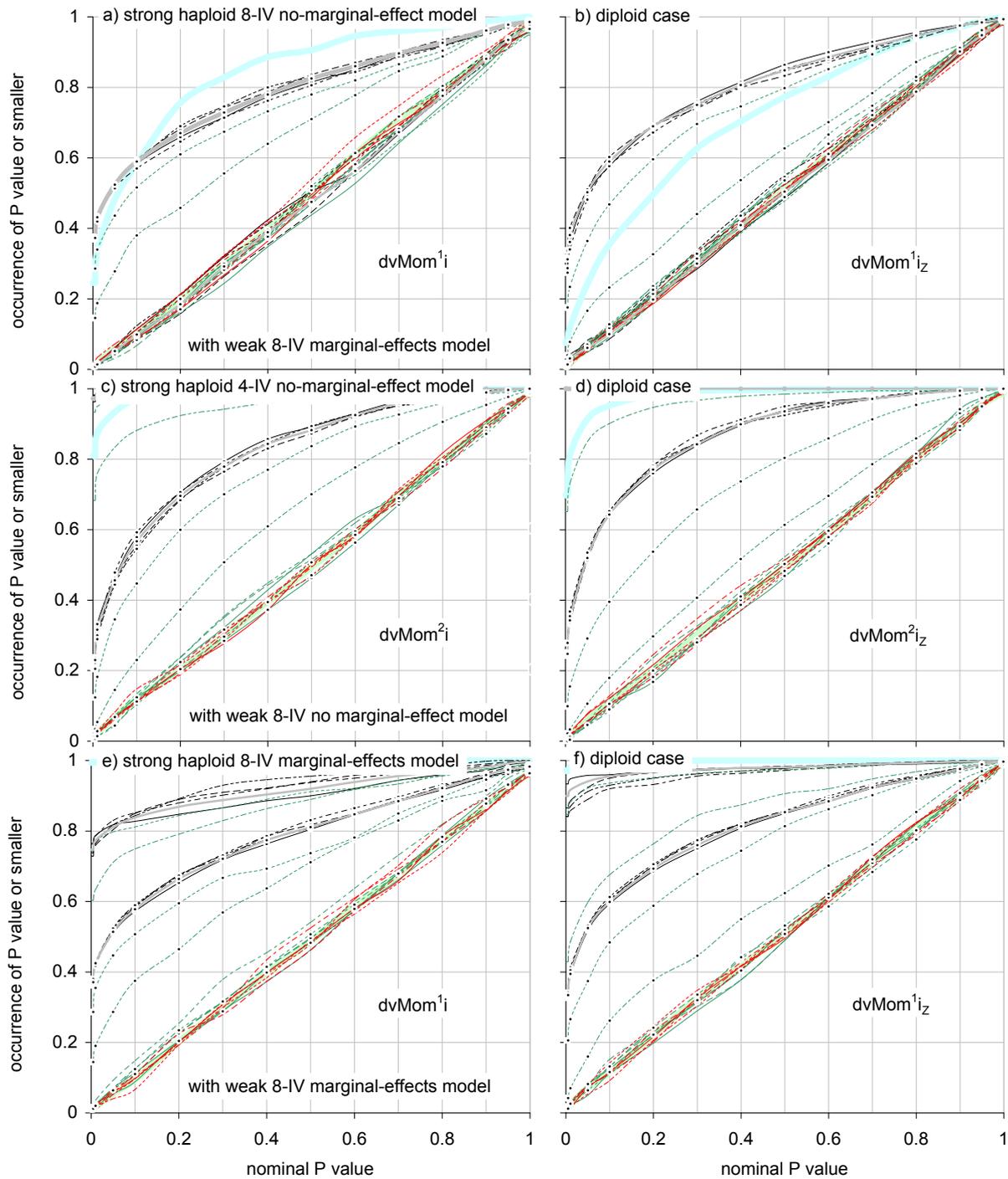

**Figure 36. Interference when a strongly and a weakly detected randomly encountered model co-occur independently in a DM with blocks: Remaining cases**. Power and false positives of dvMom$^n$i P values when further cases of co-occurrence of binary or trinary 8-IV models of DV association are simulated in DMs with blocks and model IVs are placed one per block (Figs.34,35). In all cases 100-row first-model DMs and 800-row second-model DMs are expanded and inserted in 100-block DMs with enough rows for 85%+ of the two reference columns of first-model DMs to have dvMom$^n$i P value$\leq$0.1 while ~60% of those of the plot-focal second-model DMs are so-detected. In (a,b), (c,d), and (e,f) are (binary,trinary) results when the paired 100-row first- and 800-row second-models are i) no-marginal-effect models with marginal-effects ones, ii) no-marginal-effect models with no-marginal-effect ones, and iii) marginal-effects models with marginal-effects ones, resp. The simulated DMs have (6'150 2'500), (9'200 5'400), and (6'500 2'800) rows, resp. In every case 500 model pairs are simulated, each pair once. **Labels** are like in Fig.35. Note that dvMom$^1$i and dvMom$^1$i$_Z$ are used in both (a,b) and (e,f) rather than dvCHIx-ijkl (see text). In all cases the detection samples of each plot's focal second-model IVs, the c.d.f.s of second-model IVs with various marker frequencies, and the c.d.f.s of false positives are like in isolation. P values are estimated with 100 permutations of DV markers.
Fig.36n
Fig.36n

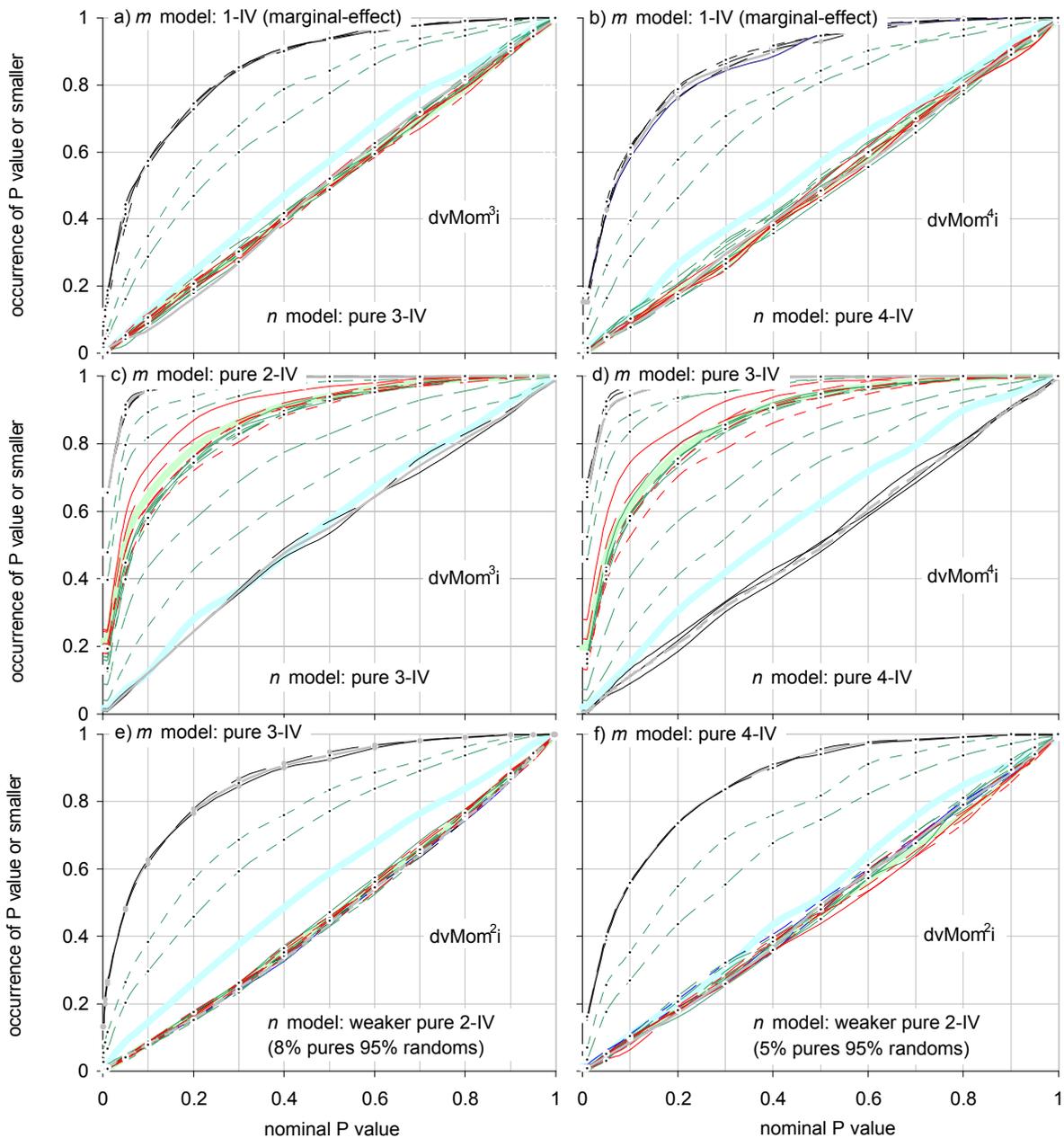

**Figure 37. Interference when two models of binary pure $m$- and $n$-IV DV association co-occur in a DM whose IVs form blocks: Selected cases**. An $m$- and an $n$-IV model of pure $m$- and $n$-way DV association co-occur independently in a DM whose IVs form blocks and whose rows suffice for ~60% detection with dvMom$^n$i P value$\leq$0.1 of the IVs of the $n$-IV second model. Model IVs are placed one per block and simulated DMs have 100 100-IV blocks (10'000 IVs total). Model-block pos.49s have marker frequency 0.5 and non-model-block ones are force-sampled to have o12345 markers (M&M). In (a,c) and (b,d) DMs have 1'000 and 13'000 rows, resp., while in (e,f) DM rows are 11'400 and 27'000. Given these rows, the $m$ IVs of first models are 95%+ detected with dvMom$^m$i P value$\leq$0.1 in every panel (not shown). Marginal effects with standard P value$\leq$0.0009 and 0.001 (1 and 2 d.f.) are erased in (a,b) to bring false positives down to the diagonal. Panels (a,b,c,d) are for $(m,n)$ pairs (1,3), (1,4), (2,3), and (3,4), resp. In (e,f) the $(m,n)$'s are (3,2) and (4,2) but the 2-IV second models are "diluted" so their IVs are only ~60% detected with dvMom$^2$i P value$\leq$0.1. Results for $(m,n)$ pairs that are essentially (1,2) and (2,1) are in Figs.35,36 (1-way: marginal effect).
**Labels** like in Fig.35. The 1-IV first models in (a,b) inflate the false positives of dvMom$^n$i when $n$ is 3 and 4, i.e., possibly for any $n>1$, but $m$-way models 2-way and higher do the same only for $n=m+1$ as shown for $(m,n)$ equal (2,3) and (3,4). The c.d.f.s of the pure 2- and 3-IV first models in (c,d) match the diagonal, showing that these IVs' DV associations make them unable to sink each other's dvMom$^3$i and dvMom$^4$i P values, albeit they sink those of the IVs from blocks without model IVs; also pos.89 and pos.64 first-model-linked IVs tend towards the diagonal in (c,d) and as much as their model-IV-like behavior in Figs.30-33 suggests they should. The excesses in false positives in (c,d) must disappear if all strong pure 2- and 3-way effects, resp., are erased since there are no such excesses when pure 3- and 4 IV models of DV association are detected in isolation (with and without IV blocks). P values are estimated using 100 DV permutations; 500 replications of each pair.

Fig.37 INTERF

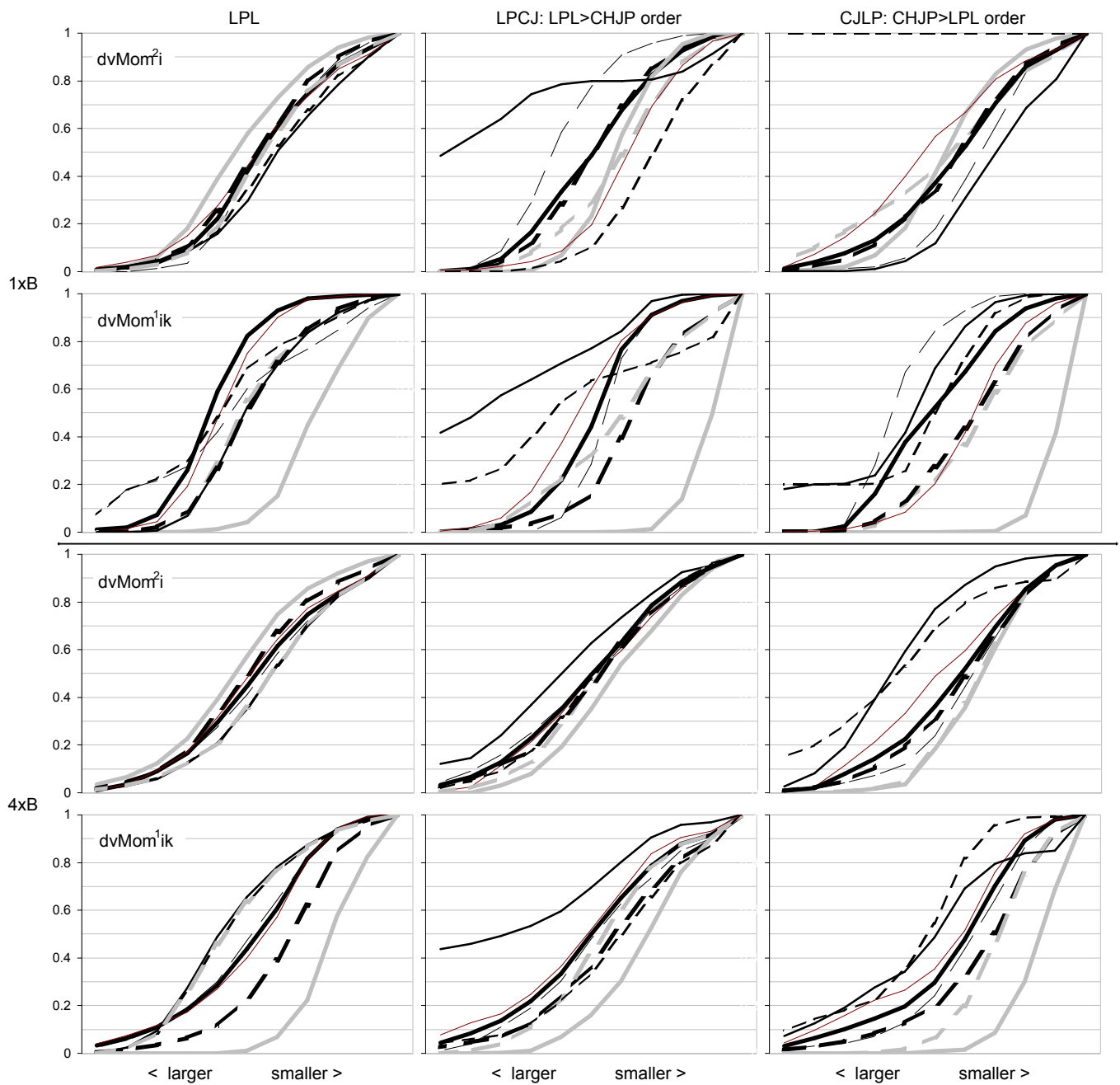

**Figure 38. C.d.f.s of dvMom$^2$i and dvMom$^1$ik in binary DMs with two types of IV blocks when two models co-occur.** Distributions of dvMom$^2$i and dvMom$^1$ik (not their P values) when the first 100 binary 100-row marginal- and no-marginal-effect 8-IV model DMs from Fig.33 co-occur independently in DMs with one or two types of blocks and to which 500 random IVs are independently "user-added" (see text). The 16 model IVs are placed one per block or four per block (top 1xB and bottom 4xB half). Marginal effects with standard P value$\leq$0.003, 0.002, 0.0008, and 0.002, from top to bottom, are erased on the left (1 d.f.) and those with P value$\leq$0.007 elsewhere. Leftmost "LPL DMs" have 100 LPL blocks but mid and right LPCJ and CJLP DMs have five and eight block tandems in the 1xB and 4xB half, resp., each with an LPL and a 1000-IV "CHJP" block in either order, such that LPL and CHJP blocks harbor first- and second-tetrad model IVs (4xB case) or odd and even model IVs (1xB case), and viceversa. DMs have 6'000 and 4'000 rows (top,bottom), resulting in 90%+ detection with dvMom$^1$ik P value$\leq$0.1 of the model IVs. **Labels**: The c.d.f.s of the eight marginal- and no-marginal-effect model IVs (pooled as two groups) are thick segmented and solid black, those of odd- and even-block model IVs are thin segmented and solid black, and those of five model-linked IVs in the first block with one or four model IVs of either kind are mid-thick segmented and solid black, resp. The c.d.f.s of 16 IVs from as many random blocks pooled are thick segmented grey and those of five user-added random IVs pooled are thick solid grey. The dvMom$^2$i's of marginal- and no-marginal-effect model IVs are similar but the dvMom$^2$ik's of no-marginal-effect model IVs are larger, specially in the 1xB case in which, e.g., 90%+ of the dvMom$^1$ik's of the latter IVs are larger than the 10%-largest dvMom$^1$ik from amongst the user-added independent IVs, nearly matching detection by P value and fully unlike vs.random-block IVs. The dvMom$^1$ik's of model-linked IVs can be larger than those of model IVs, specially in LPCJ DMs (exceptions: top-left and bottom-right plots). Block types affect the c.d.f.s of model IVs in the 1xB case, pointing to interactions with LPL and CHJP sequences. Results with dvMom$^2$i seem disorienting but see text.

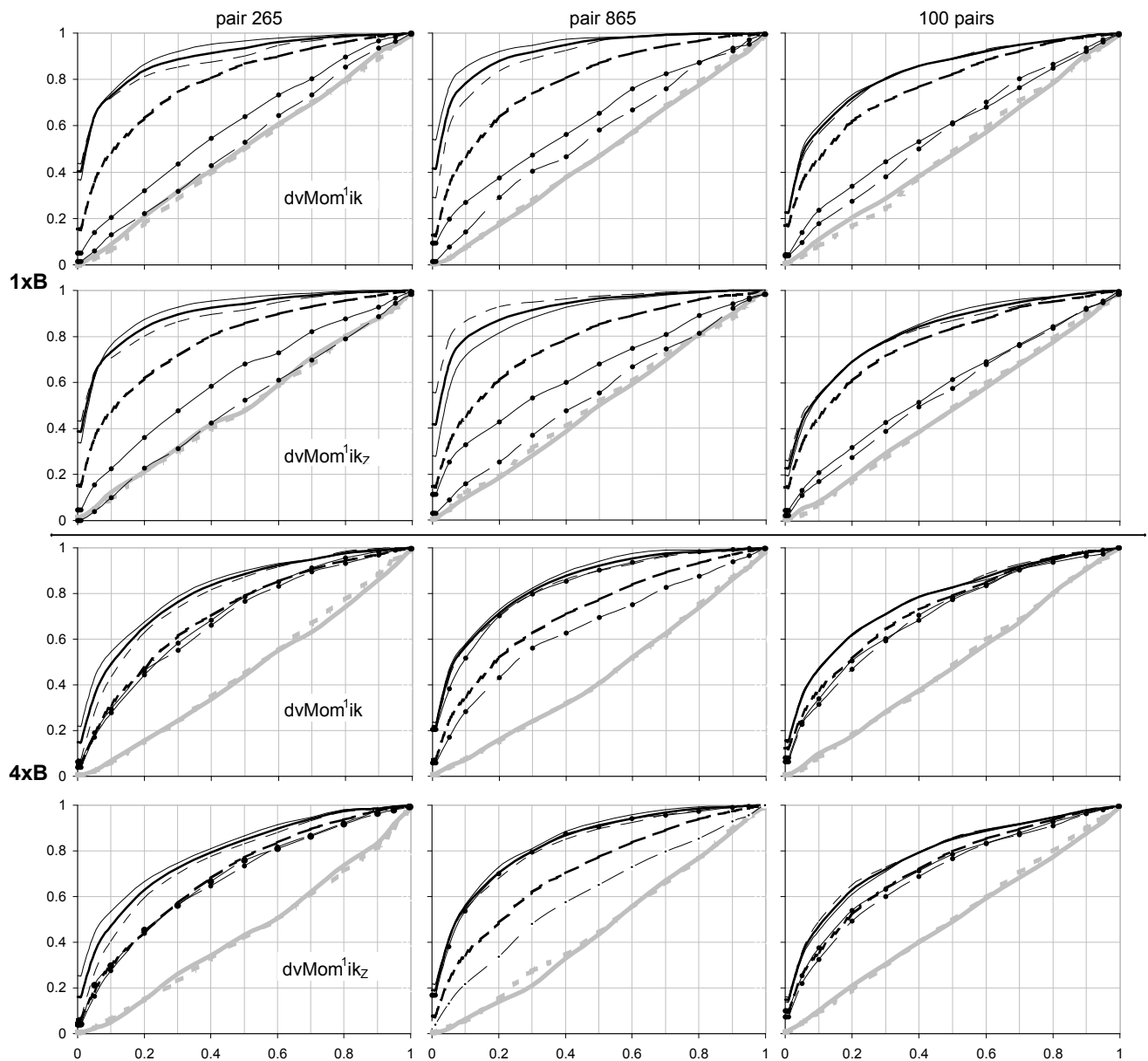

**Figure 39. C.d.f.s of the dvMom$^1$ik and dvMom$^1$ik$_Z$ P values of model and non-model IVs when two binary model pairs chosen for individual-DM analyses are simulated in binary DMs with LPL blocks.** The 265$^{th}$ and 865$^{th}$ binary 100-row model pairs from Fig.33 are each simulated in 100 DMs to each of which 500 random IVs are independently user-added (see text). Model IVs are placed one and four per block in the top and bottom half, resp.(1xB,4xB) where DMs have 100 LPL blocks and 1'300 and 1'700 rows, resp. Results for pair 265 and pair 865 are on the left and middle, resp., while on the right are results for the first 100 Fig.33 100-row model pairs pooled. **Labels**: c.d.f.s of the P values of dvMom$^1$ik and dvMom$^1$ik$_Z$ (1$^{st}$,3$^{rd}$ vs.2$^{nd}$,4$^{th}$ rows) are shown for, pooled, the eight model IVs with and without marginal effects, five model-IV-linked IVs of each type (from the first applicable block), five random-block IVs, and five user-added random IVs, in thick segmented and solid black, thinnest dot-segmented and dot-solid black, and thick segmented and solid grey, resp. C.d.f.s of the no-marginal-effect model IVs from odd and even blocks are shown in thinnest solid and segmented black, resp. Therefore the 1xB and 4xB halves constrast the detection of odd and even no-marginal-effect model IVs and first- and second-tetrad ones, resp.(see text). P values are estimated using 100 DV permutations. Marginal effects with standard P value$\leq$0.002, 0.004, 0.005, and 0.005 (1 d.f., dvMom$^1$ik results; from left) and 0.002, 0.005, 0.004, and 0.004 (1 d.f., dvMom$^1$ik$_Z$) are erased, letting the false positives of both user-added random IVs and random-block IVs become uniform-(0,1)-distributed, as intended. Detection with P value $\leq$0.1 of pair-865 odd- and even-block no-marginal-effect model IVs is ~14% different with one model IV per block (1$^{st}$ and 2$^{nd}$ left plots from the top), and that of pair-265 first- and second-tetrad model IVs is ~12% different (3$^{rd}$ and 4$^{th}$ mid plots).

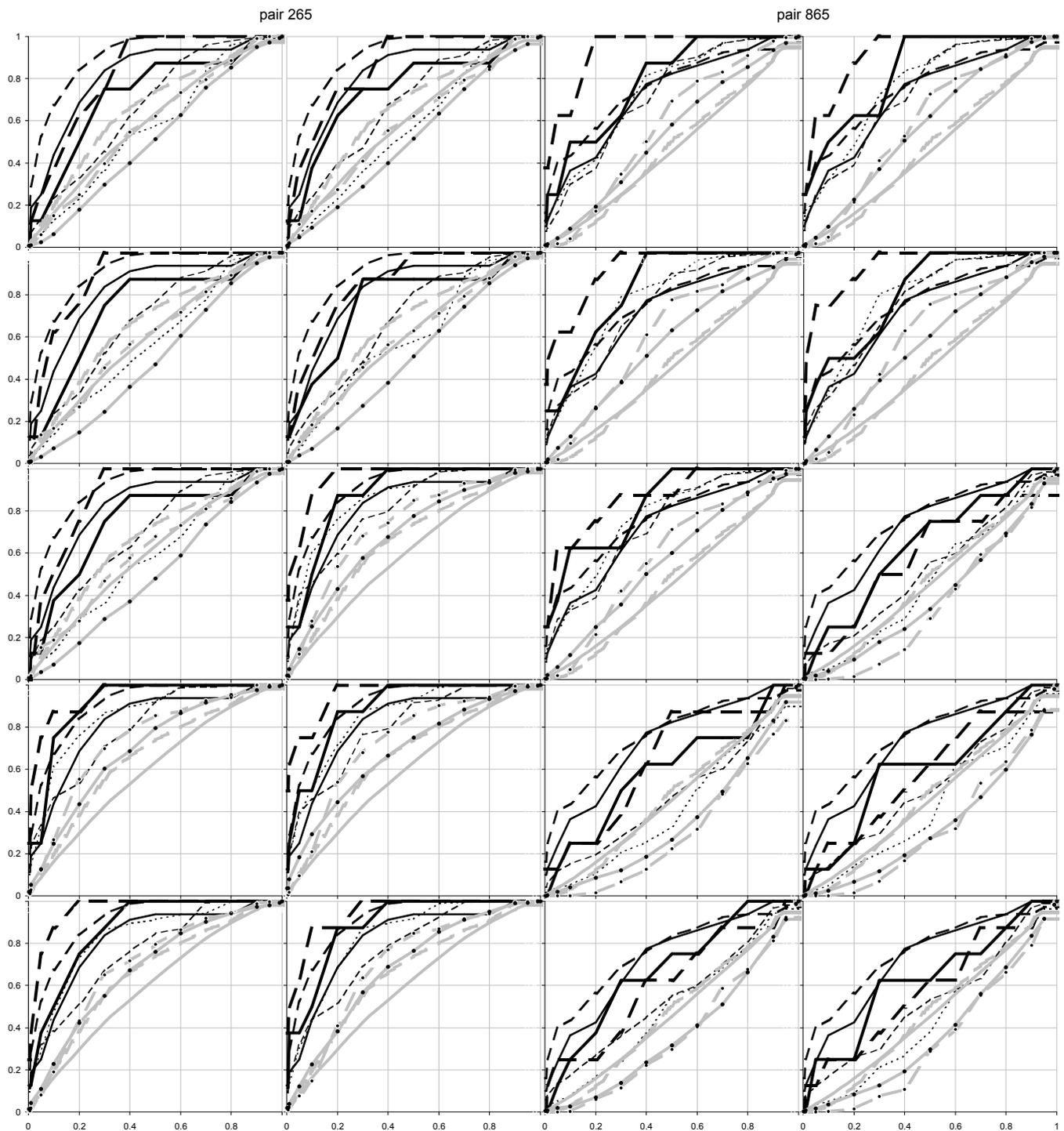

**Figure 40a. Same-DM c.d.f.s of dvMom $^1$ik P values when IVs form blocks and two binary 8-IV models of DV association co-occur independently, one with and one without marginal effects: One model IV per block.** The 265$^{th}$ and 865$^{th}$ 8-IV model pairs from Fig.35 are simulated to generate two independent 1'300-row 100-block "empirical" DMs for each pair (eDMs; LPL blocks). Five analysis runs are made of each of the four eDMs. Independently each run 500 random IVs are added to the eDM at stake and marginal effects with standard P value$\leq$0.001 are erased (which lets false positives at user-added random IVs become uniform-(0,1)-distributed when studying the first pair-265 eDM; 1 d.f.). For each of the 20 DMs, the dvMom $^1$ik P values of all the IVs in the 16 blocks with a model IV, all those in 14 blocks with no model IVs, and the 500 user-added random IVs are estimated using 100 DV permutations. Results for the runs with model pairs 265 and 865 are in the left and right half, resp., from the top in the order (1,2), (3,4), and (5,_) and then (_,1), (2,3), and (4,5). **Labels**: The thick black c.d.f.s of the P values of marginal- and no-marginal-effect model IVs pooled as two groups are segmented and solid, resp., with the c.d.f.s of each run being thicker. The thinner black c.d.f.s of the non-model IVs in blocks with marginal- and no-marginal-effect model IVs are dashed and segmented, resp. The across-5-runs "overall" thick gray c.d.f.s of the IVs from blocks without model IVs and the 500 random IVs, resp.(the five runs in either vertical half pooled) are segmented and solid, resp., with dots labelling the two c.d.f.s of each run. Other details like in Fig.35. Note i) the similarity of each run's null c.d.f.s, e.g., when they depart from the two across-5-runs c.d.f.s or the diagonal; and ii) that model-IV c.d.f.s rise and sink together with each run's two null c.d.f.s, keeping so their departure from these.

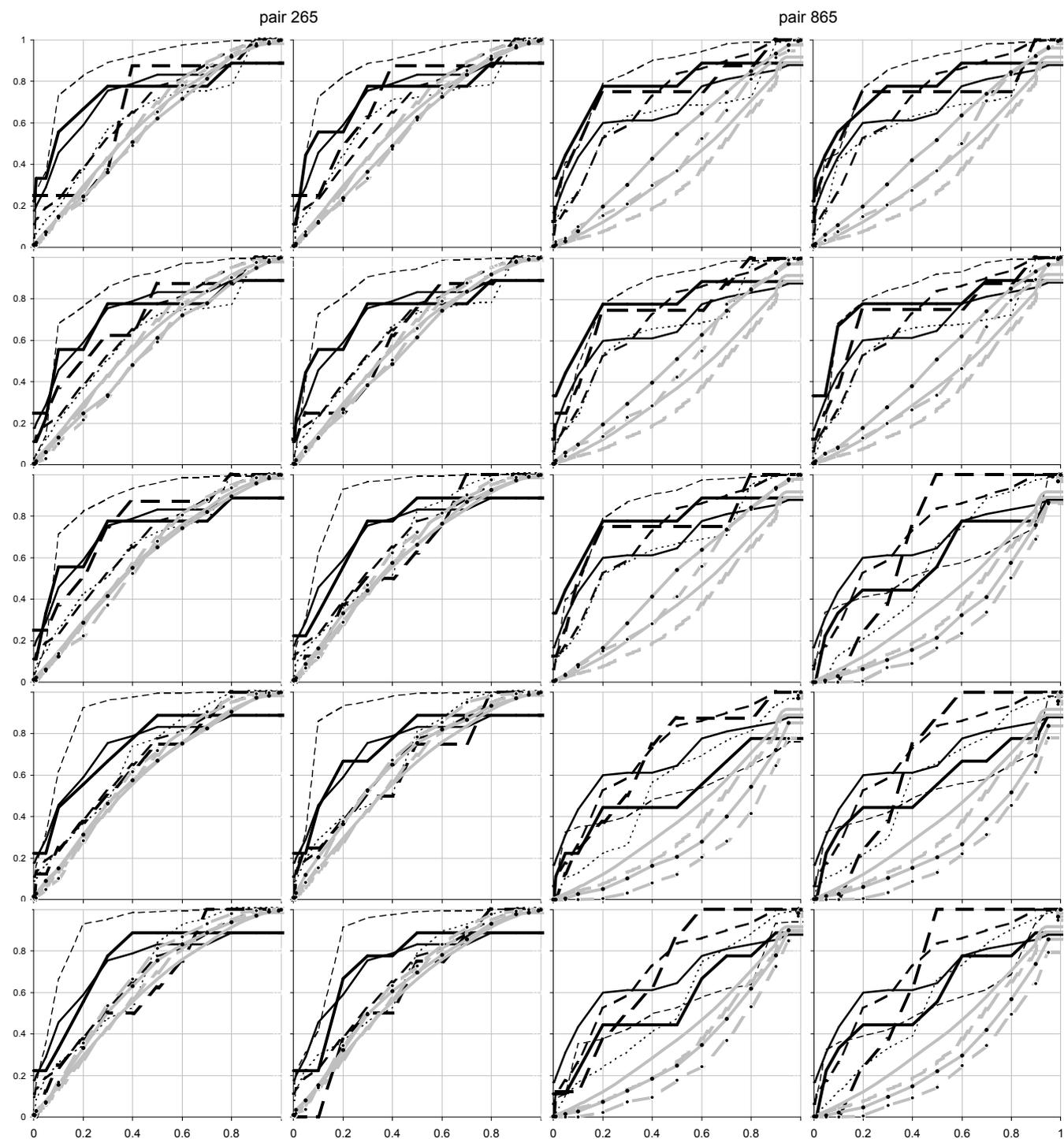

**Figure 40b. Repetition of Fig.40a but with four model IVs per block.** All other details as in Fig.40a. The 0.001 P value for erasure of marginal effects coincides with Fig.40a's; it too was estimated over 100 runs of the first pair-265 eDM each with an added batch of 500 independently generated random IVs. Here again i) the c.d.f.s of the random-block and user-added-random IVs of each run depart in concert from the applicable across-5-runs "overall" null c.d.f.s and the diagonal, and ii) model-IV c.d.f.s rise and sink in concert with the two null c.d.f.s of the run, keeping so more or less their departure from these null c.d.f.s.

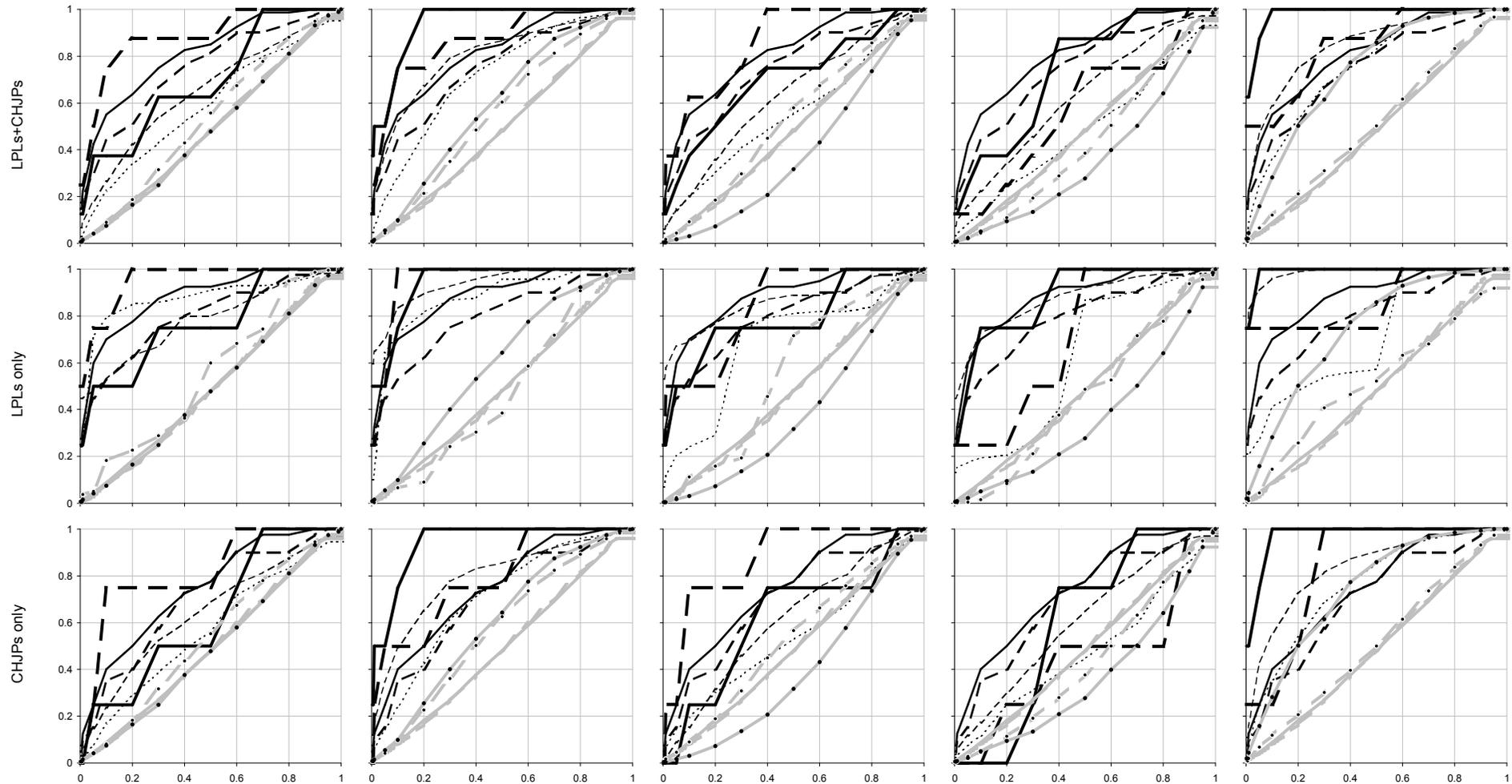

**Figure 41a. Same-DM c.d.f.s of dvMom $^1$ik P values of model and non-model IVs when IVs form two types of blocks and two binary 8-IV models of DV association co-occur independently, one with and one without marginal effects.** The $265^{th}$ binary 8-IV model pair from Fig.35 is simulated to generate five 450-row 20-block eDMs, to each of which 500 random IVs are added independently. The five runs are shown from the left. The 16 model IVs are placed four per block in four different blocks. Blocks form "LPCJ tandems" each with an LPL block and a 1000-IV CHJP block that repeat in that order from the left (see text and Fig.38). **Labels** and other details as in Fig.40. Each run includes randomly erasing marginal effects with standard P value $\leq$0.001 (1 d.f.), which lets the false positives at user-added random IVs and IVs in blocks without model IVs become uniform-(0,1)-distributed over the five runs of each eDM pooled. All of the IVs are evaluated. Results when pooling the IVs in LPL and CHJP blocks are in the top row, whereas for only the LPL IVs and only the CHJP IVs they are in the middle and bottom rows, resp. The c.d.f.s of each run's random-block IVs and added random IVs behave similarly, e.g., in departing from the across-5-runs c.d.f.s of the two types of false positives. The similarity is less marked for the IVs in random LPL blocks (see also Fig.41b,c,d). Also here model-IV c.d.f.s rise and sink together with those of each run's random-block and user-added random IVs, they too maintaining so their departure from the latter null c.d.f.s.

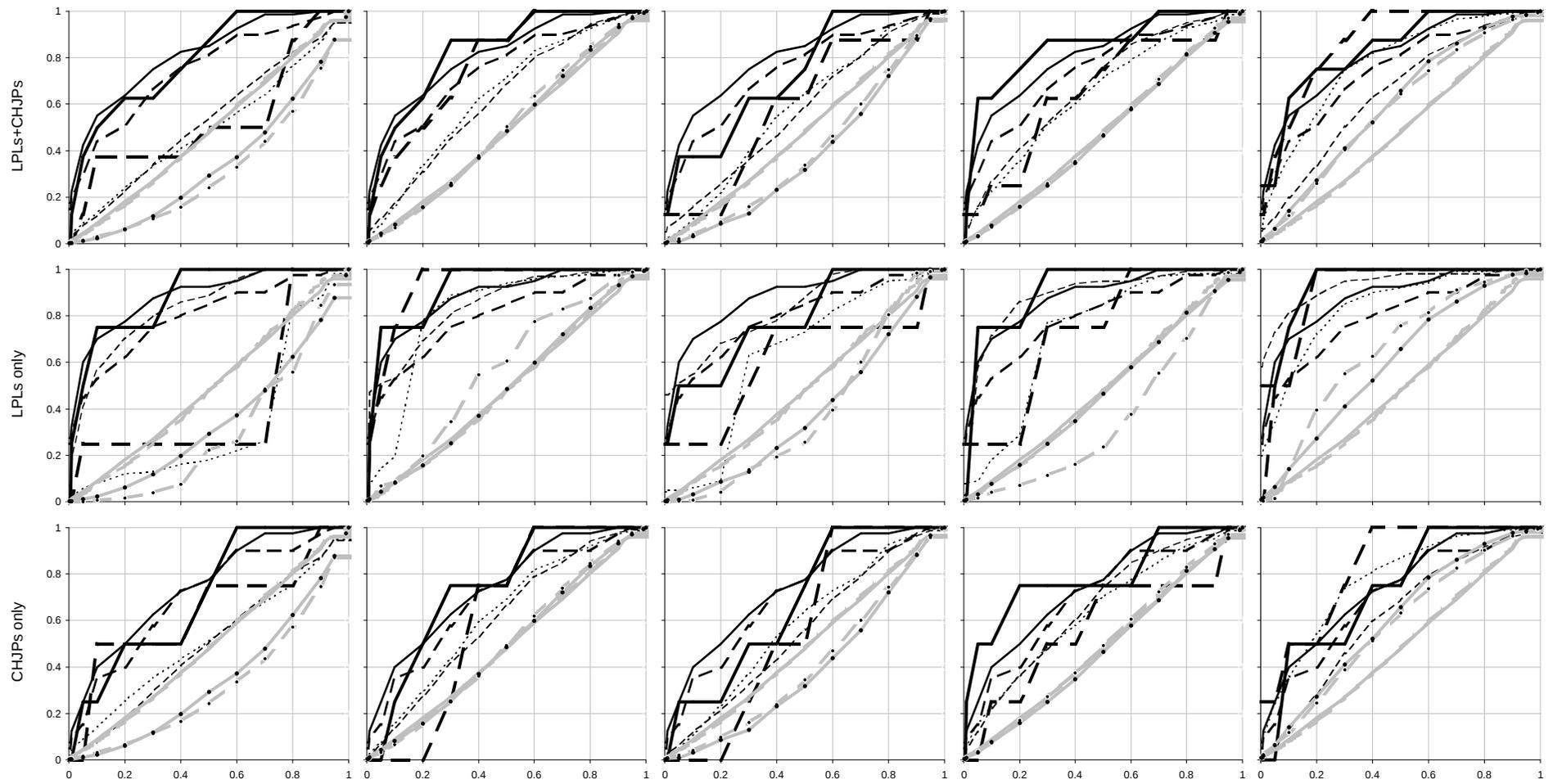

**Figure 41b. Repetition of Figure 41a but with model pair nr.865.** Details like in Figure 41a.

Fig41b 4xB LPCJ 865

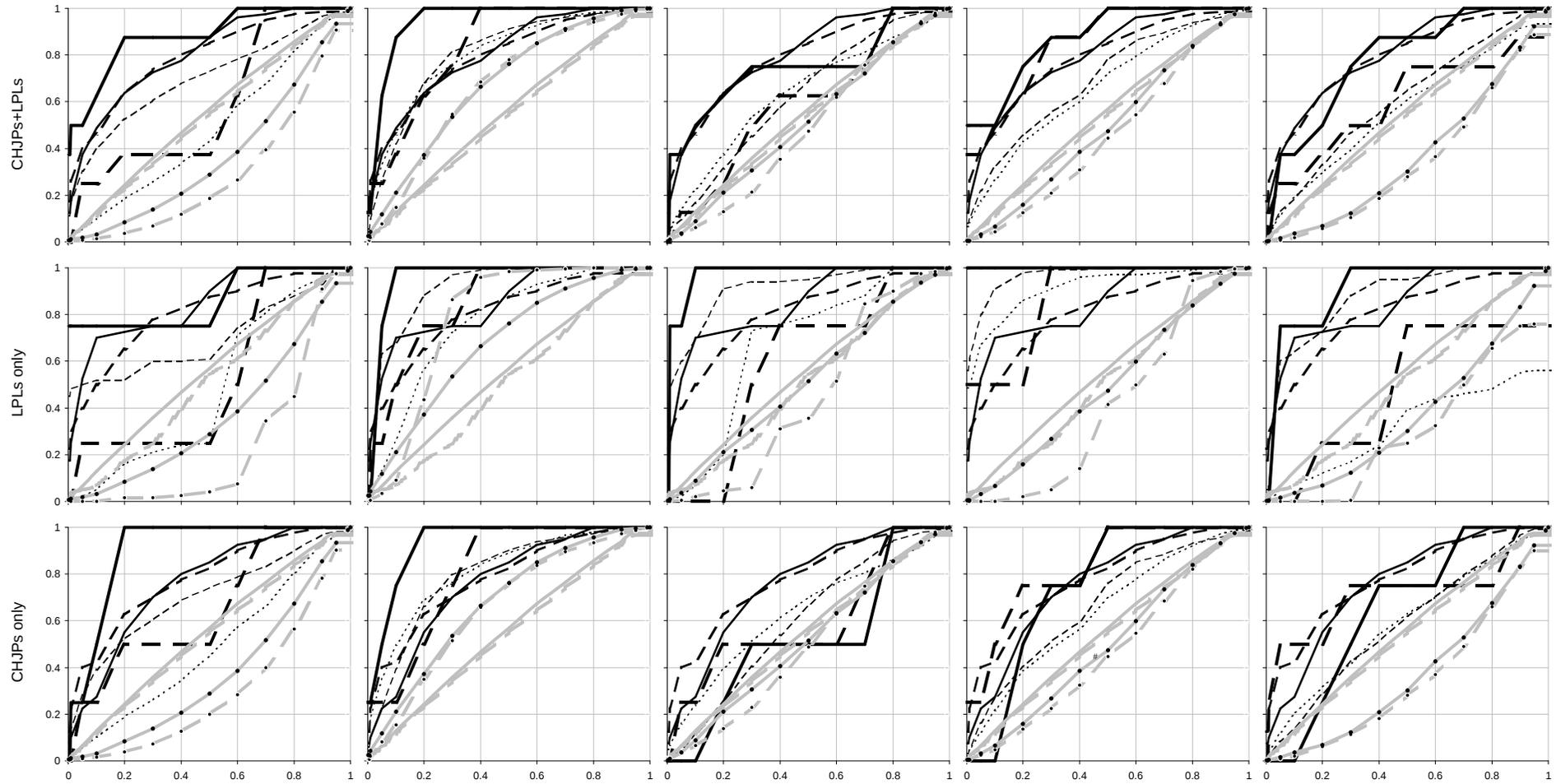

**Figure 41c. Repetition of Figure 41a with model pair nr.265 but with CJLP block tandems.**

Fig41c 4xB CJLP order 265

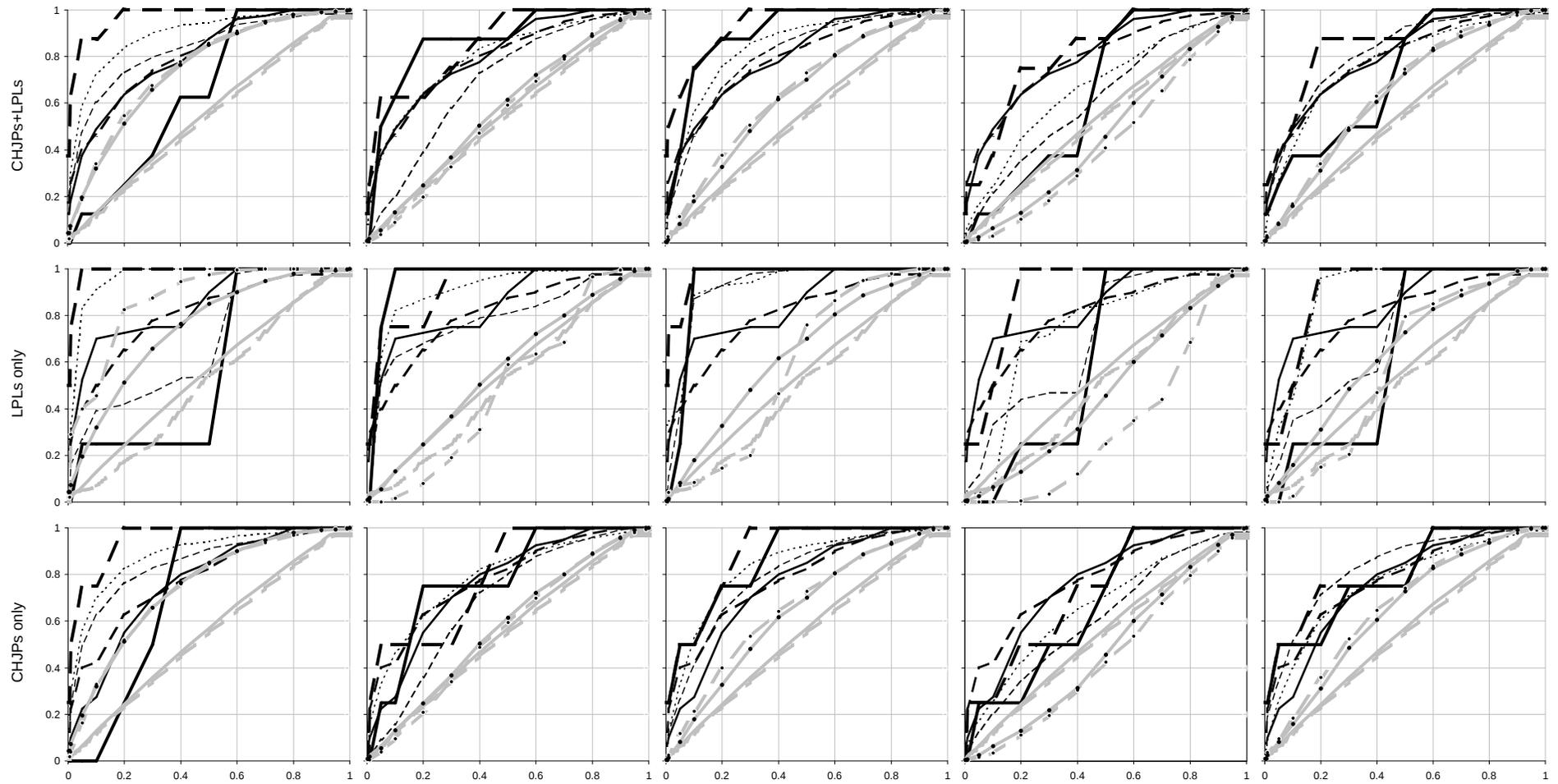

**Figure 41d. Repetition of Figure 41b with model pair nr.865 but with CJLP block tandems .**

Fig41d 4xB CJLP order 865

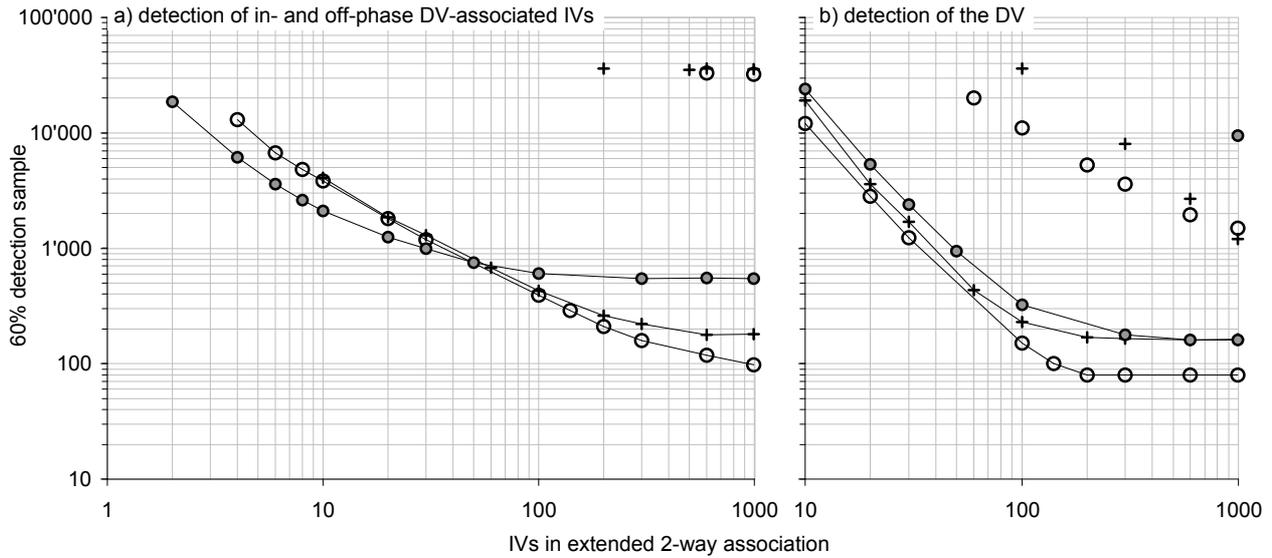

**Figure 42. The power of dvMom$^2$i P values when detecting models of cumulated partial 2-IV DV association.**
The number of DM rows required to detect 60% of the time with dvMom$^2$i P value$\leq$0.1, in **panel (a)**: increasing numbers of IVs that are in partial 2-way association with the DV either in-phase extendedly as runs (Fig.2) or off-phase as independent pairs; and in **panel (b)**: the two types of so-associated DVs. Partial 2-way associations are generated at each involved IV pair by adding 2-marker sequences in perfect 2-way association, namely 10 and 20% percent when binary-marker frequency is 0.5 and 0.1, resp., and 10% in the trinary 0.5 H&W case. DMs have a total of 1'000 IVs of which the non-model ones are independent random IVs with o12345 binary-marker frequencies or the trinary H&W thereof. **Labels:** Grey and empty symbols label freq.0.1 and freq.0.5 results, resp.(including the freq.0.5 only H&W case); lines label the in-phase results; and +'s label the H&W results (only freq.0.5). False positives are uniform-(0,1)-distributed and the power of dvMom$^1$ik P values is nearly identical (neither is shown). As the IVs in extended 2-way DV association increase to 100, their detection samples decrease quickly, plateauing at extraordinarily few DM rows. The samples for the DV are about one order of magnitude smaller when IVs are in extended vs.off-phase DV association. As expected (Fig.12), the detection samples of the IVs in the pairs in off-phase DV association do not react to the number of IV pairs. In the off-phase binary and trinary freq.0.5 cases the DV-detection sample sinks linearly but even given 990 such DV-associated IVs the two samples are ~19 and ~8 times larger, resp., than when as many IVs are in-phase DV-associated.